# Autonomous Restructuring of Asteroids into Rotating Space Stations

David W. Jensen, Ph.D.
*Technical Fellow, Retired*
*Rockwell Collins*
*Cedar Rapids, IA 52302*
*david.jensen@alumni.iastate.edu*

## Abstract

*Asteroid restructuring uses robotics, self replication, and mechanical automatons to autonomously restructure an asteroid into a large rotating space station. The restructuring process makes structures from asteroid oxide materials; uses productive self-replication to make replicators, helpers, and products; and creates a multiple floor station to support a large population.*

*In an example simulation, it takes 12 years to autonomously restructure a large asteroid into the space station. This is accomplished with a single rocket launch. The single payload contains a base station, 4 robots (spiders), and a modest set of supplies. Our simulation creates 3000 spiders and over 23,500 other pieces of equipment. Only the base station and spiders (replicators) have advanced microprocessors and algorithms. These represent 21st century technologies created and transported from Earth. The equipment and tools are built using in-situ materials and represent 18th or 19th century technologies. The equipment and tools (helpers) have simple mechanical programs to perform repetitive tasks. The resulting example station would be a rotating framework almost 5 kilometers in diameter. Once completed, it could support a population of over 700,000 people.*

*Many researchers identify the high launch costs, the harsh space environment, and the lack of gravity as the key obstacles hindering the development of space stations. The single probe addresses the high launch cost. The autonomous construction eliminates the harsh space environment for construction crews. The completed rotating station provides radiation protection and centripetal gravity for the first work crews and colonists.*

**Keywords:** Space station, asteroid, autonomous, automaton, productive replicator, anhydrous glass

## 1 Asteroid Restructuring - Introduction

Many researchers identify the high launch costs, the harsh space environment, and the lack of gravity as key obstacles hindering the development of space stations. Two of these obstacles produce detrimental effects on human workers (radiation and low gravity). In this paper we overview an approach to use robotics, self replication, and mechanical automatons to autonomously restructure an asteroid into a multiple-floor rotating space station. Researchers have designed space station for over fifty years. In the restructuring process, we use robotics to completely automate the building process. Other than a modest seed package of materials, we use only the bulk material of an asteroid to build our station. We show in Figure 1-1 a rendering of an envisioned station with its remaining asteroid and a moon.

Individuals familiar with previous space exploration and habitat construction studies [Johnson and Holbrow 1977] [Globus et al. 2007] [Metzger et al. 2012] often have preconceived notions of how this restructuring effort will work. Table 1-1 contains bullet summaries to help define the restructuring process. One column outlines what the restructuring process "produces" and the other column outlines what the restructuring process "does not produce."

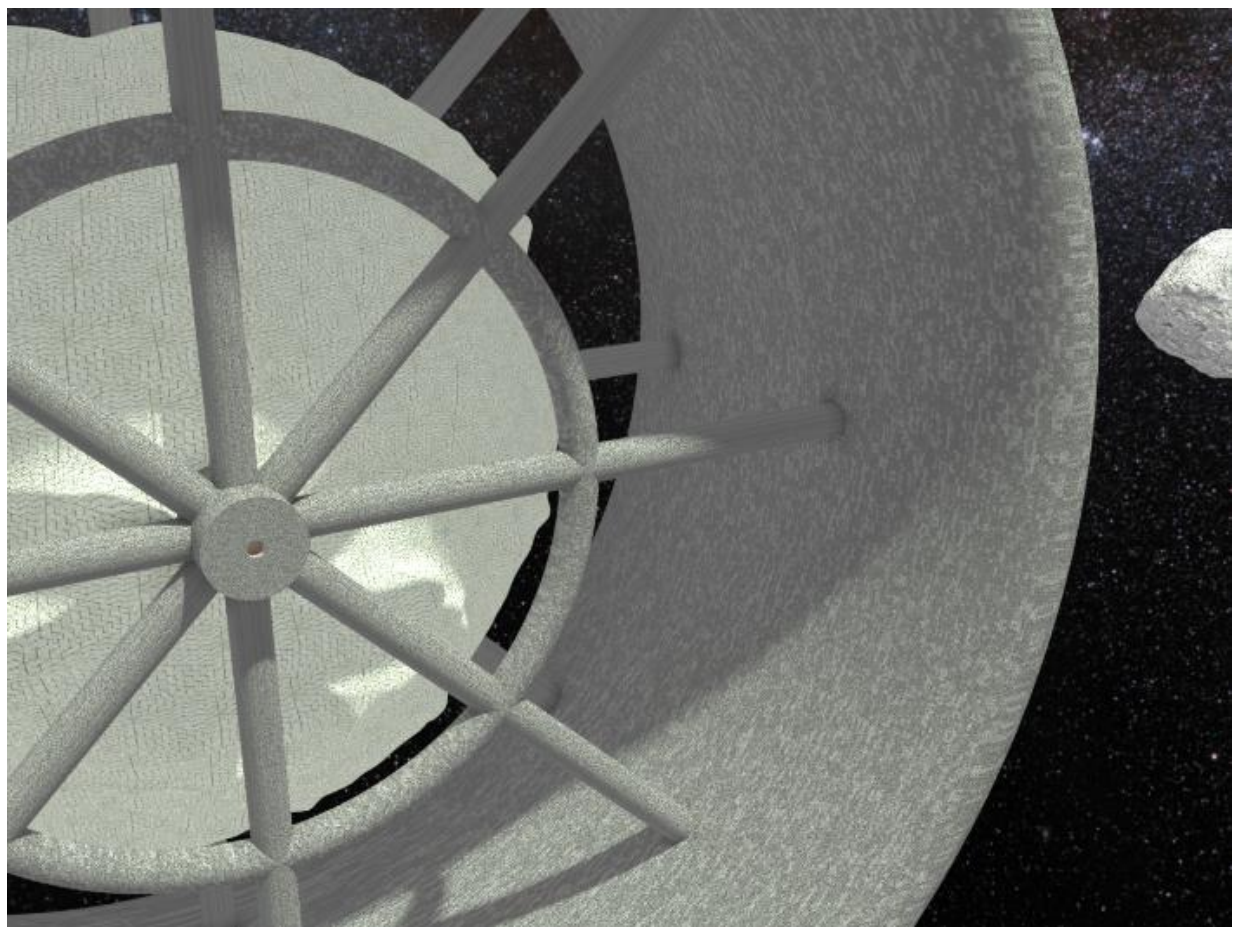

*Credit: Self produced with Blender using Background Milky Way: ESO/Serge Brunier [Brunier 2009] [CC BY-4.0]; Doug Ellison model [Ellison 2018] [CC BY-4.0] modified/rescaled to appropriate Atira and Moon dimensions*

**Figure 1-1 – Large Rotating Space Station**

Our goal is to land on an asteroid and restructure it to become the enclosed framework of a space station with a wealth of inventoried supplies. Using a modest seed package of materials and tools, robotic workers use the asteroid material to create copies of themselves, tools, vehicles, and automata.

| Table 1-1 – Restructuring Process Definition | |
|---|---|
| What Restructuring ***Produces***: | What Restructuring ***Does Not Produce***: |
| • Space station framework construction<br>• Multiple story/floor habitat design supporting large populations with abundant living space<br>• Earth like gravity and radiation protection<br>• Relatively low cost – low risk – automata intensive<br>• Inventoried valuable materials – metals & volatiles<br>• Single launch seed probe – completely autonomous – minimal human interaction<br>• No personnel support in space - minimal ground support required once launched<br>• Historic 18th and 19th century tools, technologies, and materials<br>• Processing with thousands of small systems<br>• Near 100% closure coverage using in-situ materials and initial seed<br>• Advances using 21st century robotics and artificial intelligence | • Finishing touches on space station<br>• Efficient small living quarters geared for short transportation distances<br>• Earth like air and temperature environment<br>• High cost – high risk – human labor intensive<br>• Mining business<br>• Multiple launches – semi-autonomous – human in the loop<br>• Launches of personnel, human base stations in space, extensive ground support<br>• Advanced 20th and 21st century tools, technologies, and materials<br>• Processing with a large monolithic system<br>• Requirement for continuous flow of support material and multiple launches<br>• Requirements for an Apollo or Artemis class program |

The initial probe and seed package are built with state-of-the-art 21st century technology. The materials, tools, vehicles, and automata produced on the asteroid will be more like 18th and 19th century technology. The seed package includes several thousand 21st century circuit board modules to replicate spiders; however, the spider framework, legs, connectors, and covers will ultimately be built with in-situ asteroid material and 19th century processes.

An initial set of four spiders use these modules and processes to construct thousands of robots and mechanical automata. With thousands of spiders and tens of thousands of mechanical automata, our simulations show that a large asteroid could be completely restructured into a usable space station framework in about twelve years. Within a decade, some parts of the space station would be ready for habitation. The rotating space station provides Earth-like gravity. A thick shell with regolith fill provides radiation protection. The mining and manufacturing process also produces excess metal and frozen volatiles. The spiders use sensors (e.g., testing jigs, optical, and base station) to measure and identify metals and volatiles. The identified products are categorized and stored in the station. This inventory information will be relayed to Earth and support the planning of future manned missions. The first manned mission will arrive at the rotating space station and be able to use these excess materials to enhance the framework with air, energy, heating, cooling, and light. The restructuring process overcomes the detrimental effects of space on human workers and the high cost of launching building material.

Asteroid restructuring mandates understanding technology details on space stations, asteroids, and robotics. We provide a section on each of these topics. Although somewhat lengthy, these sections are simply overviews and exclude much available detail. We first present a section on asteroids. This provides the background and information required to select an asteroid for the restructuring mission. We next present a section on space stations. This provides the background and information on the type of space station to create from the asteroid. The third section covers robotics. This provides the background and information describing the process to convert the asteroid into the space station. We include a fourth section to discuss system and construction concepts. These concepts include technology aspects from all three subject areas. The paper ends with a section covering future thoughts. This section includes final thoughts on station geometry, shuttle traffic, early colonist activities, asteroid moons, and future projects for restructuring. This section ends with our conclusions for asteroid restructuring.

# 2 Asteroid Restructuring – Asteroids

The purpose of this section is to review asteroid characteristics and select one for the restructuring process. We organize this section with subsections on background, analysis, results, and a summary.

## 2.1 Asteroids – Background

The goal of the restructuring process is to convert an asteroid to an enclosed space habitat framework. We offer in this background section an overview on asteroids. We introduce examples of asteroids, spectral classification, asteroid mining, and surface characteristics. We also overview asteroid resources and applications from those resources.

### *2.1.1 Asteroid – Background Credit*

Asteroids exist throughout the solar system and vary in composition, shape, size, and surface features. These various attributes help define the restructuring process and the asteroid selection. For our background on asteroids, we present key asteroid characteristics and research. In our later analysis and results sections, we present the selection process using those characteristics.

- Researchers use multiple approaches to determine asteroid types and materials. We use approaches and results developed by [Kesson 1975] [Ross 2001] [Mazanek et al. 2014] [Angelo 2014] and [Metzger 2015]. We use data from meteorites, asteroid spectral data, and lunar sample analysis to define our asteroid metrics.
- Terrestrial, lunar, and asteroid mining processes have been studied by many researchers. We use mining and beneficiation of lunar ores from [Williams et al. 1979], mining concepts from [Vandenbos 2006], and lunar resource processing from [Kayser 2011], and solar furnace concepts from [Sanders and Larson 2012].

- Missions to orbit and survey asteroids are occurring more frequently. We use concepts and parameters from various space missions and literature [Maynard and Sevier 1966] [Doody 2011] [Bradley et al. 2012] and [Fritz and Turkoglu 2017] to plan our mission to a selected asteroid.
- Most asteroid mining studies focus on retrieving and processing metals and/or volatiles. A good example study is [Mazanek et al. 2014]. We use the asteroid oxides as building materials and that changes the focus and material priorities. We plan to use these oxides as rods or tiles for our construction structures. It appears that these oxides could be formed into anhydrous glass.
- Anhydrous glass has been found to have remarkably high tensile strength [Bell and Hines 2012]. The asteroid vacuum environment will work well to create the anhydrous glass. We use concepts and results from [Blacic 1985], [Carsley, Blacic, and Pletka 1992], [Bell and Hines 2012] and [Soilleux 2019]. Rapidly cooling the glass will increase its strength [Yale 2013]. We envision that thin plies of anhydrous glass would be stacked, pressed, and sintered into high strength laminates.
- The asteroid restructuring process will produce excess metal and frozen volatiles. We do not discard these products; instead, they are categorized and stored in the station. This inventory will support future manned missions.

### *2.1.2 Asteroid Examples*

There are over 150 million asteroids larger than 100 meters diameter in our solar system. We want an asteroid with enough material to build a large space habitat. There are almost one thousand asteroids over 1 kilometer in diameter that are considered Near-Earth Objects (NEOs). Twenty-one known NEOs are over four kilometers in diameter. Each has enough material to construct a space station that could house a million people. We show in Figure 2-1 a set of asteroids ranging in radius from 256 meters to 2580 meters. We select this set of asteroids to evaluate a range of sizes. Šteins is the largest in that set and Bennu is the smallest.

Size is not the only important criteria. The location is important too. Ideally, we want the asteroid (and station) close to Earth to better support tourism and trade. We want it to be relatively close to the sun to provide solar power for manufacturing and energy. A station close to the orbit of Earth would reduce round-trip travel time. Apollo and Aten class asteroids cross the Earth's orbit. Amor asteroids always stay outside Earth's orbit and Atira asteroids stay inside Earth's orbit. Šteins is in the inner edge of the asteroid belt. The other asteroids of our set reside inside the orbit of Mars.

The asteroids shown in Figure 2-1 have had significant analysis and observation. The European Space Agency (ESA) OSIRIS mission photographed 2867 Šteins in 2008. Atira is a Near-Earth Object that has been radar imaged by the Arecibo Observatory [Rivera-Valentin et al. 2017]. The asteroid Moshup is a Near-Earth Object that has been radar imaged by the Goldstone Observatory. The JAXA Hayabusa2 mission explored the asteroid Ryugu. The NASA OSIRIS-REx mission explored the asteroid Bennu.

We include in Figure 2-1 known metrics from the JPL Small-Body Database. The metrics for Šteins came from [Deller 2017]. A few metrics came from various Wikipedia pages. Finally, we have computed our own estimates for missing metrics. These metrics are important for the analyses and simulations of this effort.

Besides the asteroids of Figure 2-1, we also considered 433 Eros, 25143 Itokawa, 65803 Didymos, 3753 Cruithne, and the comet 67P/Churyumov-Gerasimenko. All these bodies have had significant studies. They provide excellent examples of features that our restructuring process will need to take into consideration. Space missions have explored some of these asteroids. Others have been surveyed with Earth based radar or space-based telescopes. Most have been spectrum-analyzed to provide insights on their surface materials. We illustrate ten of these bodies in Figure 2-2. The bubble chart organizes the bodies along their distance from the Sun (semi-major axis) on the x-axis. It organizes them along their diameters on the y-axis. The bubble size also represents their diameters. The chart also includes coloring of the bubbles

| | 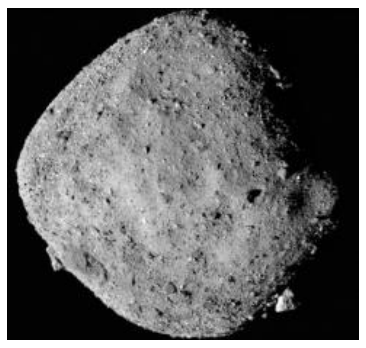 | 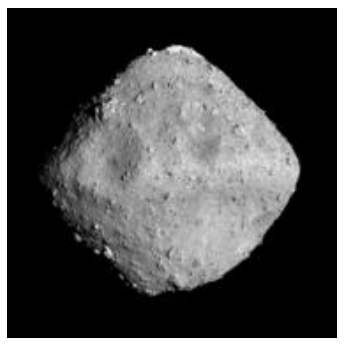 | 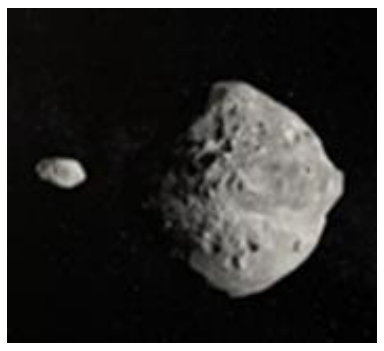 | 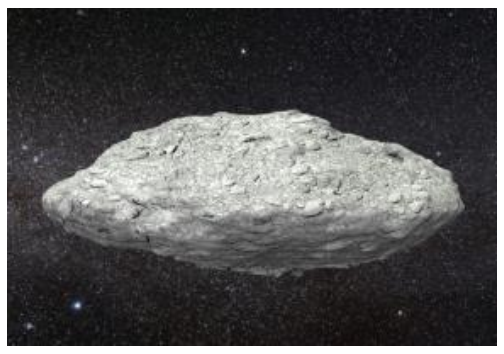 | 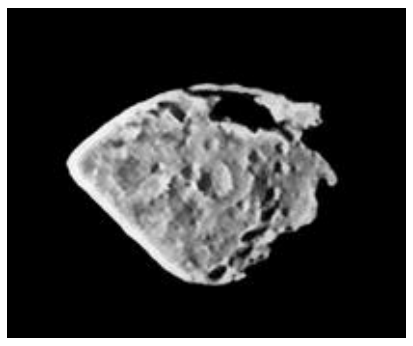 |
|---|---|---|---|---|---|
| **Name** | **101955 Bennu** | **162173 Ryugu** | **66391 Moshup** | **163693 Atira** | **2867 Šteins** |
| Diameter (km) | 0.482 | 0.896 | 1.53x1.50x1.35 | 5.2x4.8×2.3 | 6.8×5.7×4.4 |
| Volume ($m^3$) | 6.15e7 | 3.4e8 | 1.62e9 | 21.9e9 | 76.0 e9 |
| Mass (kg) | 7.33e10 | 4.5e11 | 2.49e12 | 4.11e13 | 1.98e14 |
| Material ($m^3$) | 1.54e7 | 8.50e7 | 4.05e8 | 5.48e9 | 19.8e9 |
| Density (g/$cm^3$) | 1.19 | 1.32 | 1.54 | 1.87 | 1.9 |
| Orbit (AU) | 1.13 | 1.19 | 0.64 | 0.74 | 2.36 |

*__Bennu:__ Credit: NASA/Goddard/University of Arizona; [NASA Image Public Domain]. __Ryugu:__ Credit: JAXA, University of Tokyo & collaborators; [CC BY-NC-ND 4.0]. __Moshup:__ Credit: ESA; [ESA Standard License] Modified: Cropped Image. __Atira:__ Credit: Self Produced with Blender using Doug Ellison model; [Ellison 2018] [CC BY-4.0] Modified: scaled to match Atira dimensions. __Šteins:__ Credit: Data from ESA 2008 OSIRIS Team, processing by T. Stryk [Stryk 2008]; [CC BY-SA 2.0].*

**Figure 2-1 – Asteroid Preview – Key Characteristics – Documented and Estimated Values**

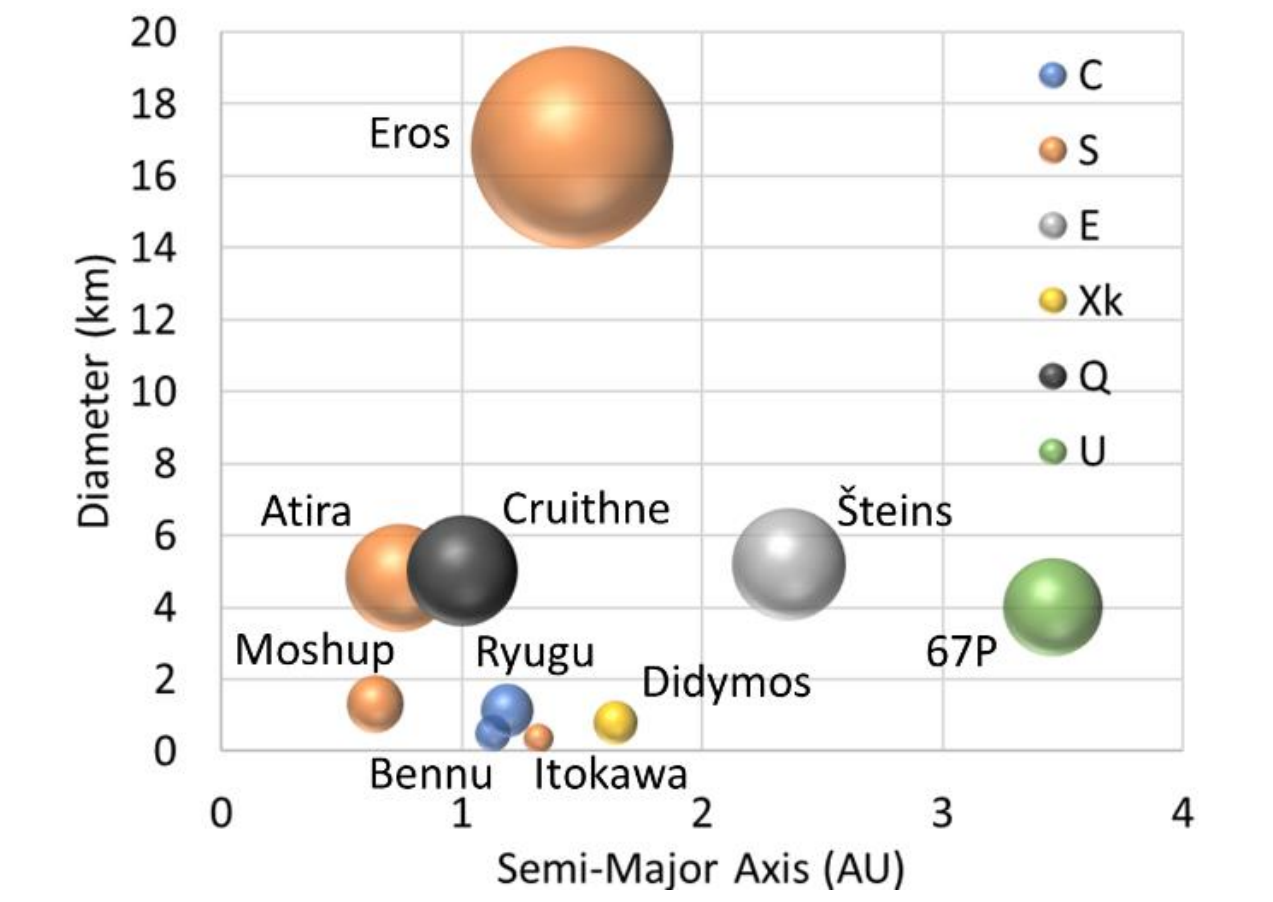

*Credit: Self produced using data from JPL Small Body Database Search Engine; [JPL SBD Search Engine] [Public Domain]*

**Figure 2-2 – Asteroids Reviewed**

based on the asteroid's spectral type. The asteroids covered five different spectral types and we include a U-Type as the undefined type of comet 67P. Comets are not typically given a type and have an icy composition. The shapes of six of the asteroids were ellipsoid, diamond, or somewhat spherical. Three of the bodies were irregular or peanut shaped with two major lobes. Three of the asteroids had moons.

This short review covered a broad set of asteroids. Their diameters ranged from 350 meters to 16.8 kilometers. Their orbits ranged from 0.65 AU to 3.46 AU. There are dozens of other asteroids with similar levels of detail. There are thousands of asteroids with minimal physical and orbit information. This wealth of information will guide the selection of an asteroid for our restructuring effort.

### *2.1.3 Asteroid Spectral Classification*

We need the composition of a target asteroid for our restructuring effort. The structure of the asteroid will help determine the mining and construction. The composition will help determine the building material and construction approaches.

Significant research has been done to identify asteroid types. Historically, observations of asteroids have been limited to ground-based telescopic sensing of the visible, near infrared, and radio spectrums. The surface of different asteroids reflect light uniquely. Scientists have used this reflected light to understand the asteroid surface chemical composition. Recently, ground-based spectral readings have been augmented with satellite and space probe sensor measurements. Space probes have visited several asteroids and obtained close measurements. In the last decade, samples from a comet and two asteroids have been returned to Earth. Spectral analysis provides the foundation for current asteroid classification schemes.

Because most meteorites come from asteroids, their taxonomies have relevance to understanding the composition of asteroids. Just like meteorites, the asteroids are classified into three main groups. They are often classified as C-Type (Carbonaceous), S-Type (Stony or Siliceous), and M-Type (Metal); see Figure 2-3a. More than 75% of asteroids are type C carbonaceous, which now includes similar types B, D, and F. C-Type asteroids are more common in the asteroid belt and D-Type asteroids become more common in the outer solar system. Less than 17% of the asteroids are type S siliceous and this now includes similar types R, V, and O. The S-Type are more common in the inner solar system and inner asteroid belt. The other 8% of asteroids are metallic type M, which also includes similar types P, X, and E.

We provide in Figure 2-3b three major asteroid types. We show the estimated composition for the asteroids computed from three sources [Ross 2001] [Angelo 2014] [Mazanek et al. 2014]. For most of the restructuring analysis, we use an S-Type asteroid. A C-Type has similar percentage of materials. In both cases, they are composed of over 80% oxides. The S-Type asteroids have more free metal. The C-Type asteroids have more volatiles and water. The M-Type is quite different in composition. The costs to extract and refine metals is higher than simply melting bulk oxide material. As such, we want either an S-Type or C-Type and this affects our final selection. Based on the sample from the JPL database, almost 90% of the asteroids will be acceptable for our restructuring effort.

A key take-away from Figure 2-3 is that most asteroids are C-Type or S-Type, and those asteroids are comprised

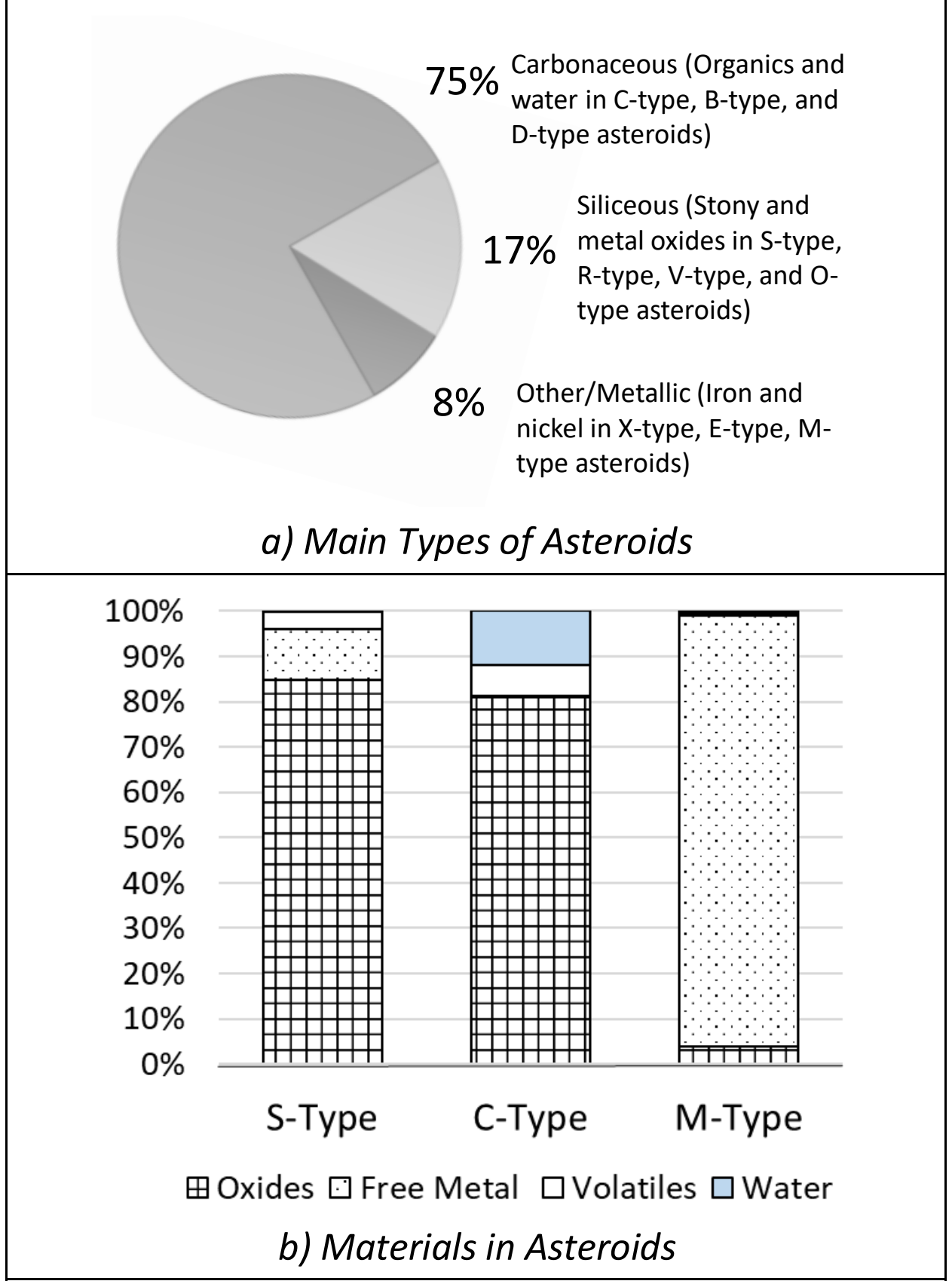

*Credit: Self produced using data computed from [Ross 2001] [Angelo 2014] [Mazanek et al. 2014] [Facts]*

**Figure 2-3 – Main Groups of Asteroids**

primarily of oxides. The S-Type asteroid material is moderately friable and more easily crushed than M-Type asteroid metal materials. The C-Type asteroid material is weakly friable and the easiest to crush for processing. For this study we use the oxides as our building material. The stony asteroids (S-Type) are dominated by silicates and are good candidates for processing into our construction elements.

We note two other types of NEAs that are valuable from a resource standpoint are designated as D-Type and P-Type. Few of these types exist as NEAs because they likely originate from the outer main belt or beyond. These are believed to be composed of organic-rich silicates, carbon, and anhydrous silicates, possibly with water ice in their interiors [Mazanek et al. 2014]. Mining these in the future will be important but not for our immediate restructuring goals.

### *2.1.4 Asteroid Mining*

Asteroid mining has been seriously considered for less than 50 years [O'Leary et al. 1979]. The major steps of terrestrial mining are relevant to asteroid mining. The terrestrial mining process can be organized with six major steps: prospecting, excavating, processing, extracting, fabricating, and storing. The restructuring process uses all these steps.

We provide one high-level view of the restructuring regolith processing in Figure 2-4. This view focuses on the excavation, processing, and extraction. We do not include many details in this summary. From the initial regolith, three groups of resources are produced: free metal, volatiles, and oxides. Many products can be created using these three resources.

A rubble pile asteroid will have loose surface material to initially process. Small robotic systems will first bring regolith dust, grains, and pebbles to the base station. Small pebbles could be further fragmented with an impact crusher. Crushers should reduce the fragments to grains and separate aggregates of ice, silicates, and metals. Jaw crushers will reduce cobbles and the product will be sized with screens. Larger fragments will be crushed again. Mechanical systems will use jack hammer chisels to reduce larger boulders. Monoliths and slabs may require tunneling and blasting.

Magnetic beneficiation should extract most of the free ferromagnetic metal grains (iron, nickel, and cobalt) from the silicate and carbonaceous grains. With additional complexity and higher strength magnetic fields, we could extract the paramagnetic metals too (platinum, titanium, zirconium, and magnesium). These metals will be separated, inventoried, and stored. Some of the iron-nickel grains will be used for 3D printing of metal parts.

Mirror systems and Fresnel lenses will focus sunlight to process the material. Low temperature solar heating will release much of the free volatiles from the regolith grains. A typical melt temperature of 1200°C is often referenced for oxides. There will be volatiles chemically bound to the minerals that will be released in this process. Cyclones can be used to separate the volatile gas from dust carried with the gas [O'Leary et al. 1979]. Staged cooling and condensing of the volatiles will separate the gasses for inventory and storage. Carefully lowering the temperature will condense gasses in an order like water (100°C), hydrogen sulfide (-60°C), carbon-dioxide (-78°C), methane (-161°C), oxygen (-183°C), carbon monoxide (-191°C), nitrogen (-196°C), hydrogen (-253°C), and helium (-269°C).

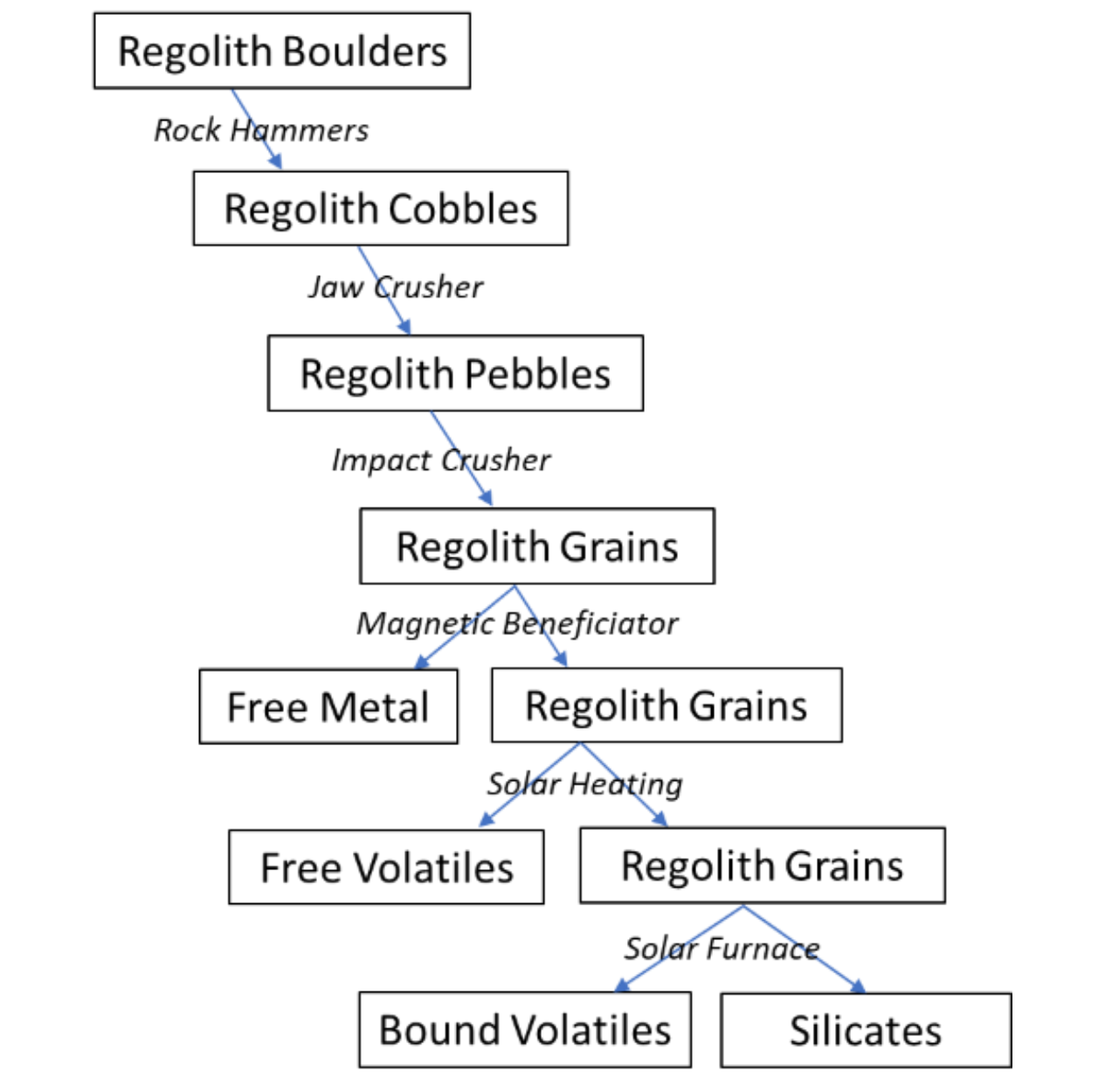

**Figure 2-4 – Restructuring processing of regolith**

A solar furnace will be used to melt the regolith grains to cast ceramic tiles and to produce anhydrous glass tiles. Initially the tile production from the molten regolith will be in the base station system. Casting is one option and could use a handful of molds brought on the restructuring mission. New molds can be produced using the asteroid resources. Instead of molds, we prefer the concept of continuous casting of the regolith. Anhydrous oxide strips will be rapidly cooled and used to produce strong laminate plies. Those laminate plies will be stacked as they cool and sintered under pressure. The laminate will be nearly as strong as the individual plies and provide the structural strength required in our construction.

### *2.1.5 Asteroids Surface Characteristics*

Classifying the various surface features will be important for programming the autonomous exploration, mining, and processing of the asteroids. Surface features to consider include craters, plains, terraces, cliffs, sink holes, boulders, and equatorial ridges. Each of these surface variations have a unique impact on mining and navigation.

### *2.1.6 Asteroids Resources and Applications*

Our effort is not the first to consider how to use asteroid material. We include Table 2-1 to review a set of resources and potential applications for our asteroid material. This was adapted from a seminar entitled *Construction with Regolith* [Mueller 2017]. The focus of this seminar was on processing regolith from the moon. The seminar did briefly consider Mars and asteroids. Most of these regolith resources are also available from asteroids. The applications include products

| Table 2-1 – Asteroid Resources and Applications |
| --- |
| **Some asteroid resources and their uses** |
| • Oxygen: propellant, life support<br>• Iron, aluminum, titanium: structural elements<br>• Magnesium: less strong structural elements<br>• Oxides: sintered blocks, concrete, glass<br>• Water: Ice blocks, molded ice |
| **Potential applications** |
| • Structural beams, rods, plates, cables<br>• Cast shapes (e.g., anchors, fasteners, bricks, and rods)<br>• Solar cells, wires for power generation and distribution<br>• Pipes and storage vessels for fuel, water, and other fluids<br>• Roads, foundations, shielding<br>• Spray coatings or linings for buildings<br>• Powdered metals for rocket fuels, insulation<br>• Fabrication in large quantities can be a difficult engineering problem in terms of materials handling and heat dissipation |
| *Credit: Self produced using data and concepts from [Mueller 2017] [NASA Report Public Domain]* |

for initial mining, early construction, and occupied space habitats. Their inclusion of oxide asteroid resources and the structural beam applications are consistent with our goals.

## 2.2 Asteroids – Analysis

In the previous subsection, we introduced asteroid background research and identified a set of 10 potential asteroids for our restructuring effort. In the following subsections we build on that background in order to select a candidate asteroid for restructuring. We cover the topics of asteroid materials, construction materials, material production, and mission cost. We also cover a technique to compute a return on investment (ROI) for a restructuring mission. These topics provide the criteria to select the asteroid.

### *2.2.1 Asteroid Materials*

Our restructuring effort uses the bulk oxide material from an asteroid to create the habitat framework. Most asteroid mining efforts focus only on valuable metals or water. These valuable products represent a small percentage of the asteroid. These mining efforts typically consider the bulk oxides to be waste. We advocate using the bulk oxides to create rods and tiles. Suddenly 80% to 90% of the asteroid is valuable. We use those rods and panels to build trusses, siding, and panels. Some leftover materials like volatiles and metals will be identified, inventoried, and stored for future use. Other bulk materials not suitable for construction can be used for fill.

We include Table 2-2 to illustrate how our asteroid restructuring process has the opposite priorities compared to typical asteroid mining approach. We use details from [Mazanek et al. 2014] as a typical asteroid mining approach. The table provides the high, medium, and low product priorities for the two approaches. We show the expected products and the processes used to obtain those products. The table shows the expected percentage of the products and the relative costs associated with obtaining those products. For our restructuring process, we consider an S-Type asteroid for the material composition percentages. A C-Type asteroid has similar percentages. The table results show the traditional and restructuring approaches have almost opposite product priorities.

### *2.2.2 Construction Material*

Glass can be used for structural applications (bricks, slabs, beams, windows) [Haskin 1992]. Glass oxides can easily be cast into structural elements for construction. Feasible elements include beams, columns, slabs, shells, blocks, and cylinders. End products such as floors, sinks, pipes, and electrical insulators can also be fabricated from these materials. Cables can also be made from high strength glass fiber [Ruess, Schaenzlin, and Benaroya 2006].

We plan that the primary product produced in our restructuring process will be rods and tiles for constructing trusses and panels. The vision is to melt the silicate resource and form the rods and tiles. The S-Type asteroid has about 85% oxide material. An example breakdown of these oxides on an S-Type asteroid would include SiO2 at almost 40%, MgO at 25%, and FeO at 10%. In the *Construction with Regolith* report [Mueller 2017], they considered building with basalt. Basalt is an igneous rock comprised of minerals such as pyroxene and olivine. Compositionally this is like the oxides on the S-Type asteroid. They used a Hawaiian basalt with $Si_2O$ at 50%, MgO at 7.2%, $Fe_2O_3$ at 12%, and $Al_2O_3$ 13.5% [Mueller 2017]. They compared the sintered basalt rocks to concrete. They found basalt rock density is similar to concrete and the compressive strength is better. Table 2-3 contains metrics for several basalt products. For reference, we also include values for concrete, conventional glass, and for anhydrous glass.

Glasses and ceramics generally work well under compressive loads but not well under tension stress. It may be possible to reinforce glass structures with asteroidal nickel-iron steel and enhance them to withstand a wide range of both tension and compression. This complexity may not be necessary with the additional tensile strength produced with the anhydrous, vacuum-produced glass [Blacic 1985] [Prado and Fraser 2018]. Glass produced in the absence of hydrogen or water has significantly better mechanical properties [Prado and Fraser 2018]. It may be possible to substitute this glass for structural metals [Blacic 1985] [Carsley, Blacic, and Pletka 1992] [Soilleux 2019]. This glass is called anhydrous glass. On Earth, water and hydrolysis weakens the strength of the silicate bonds by about an order of magnitude [Blacic 1985]. Production on the asteroid will be in a hard vacuum and water is extremely limited. Quickly cooling this glass could further increase the tensile strength [Yale 2013]. Researchers believe it may be possible to substitute this anhydrous glass for structural metals in a variety of space engineering applications [Blacic 1985]. They note that this glass would be competitive or superior to metals [Carsley, Blacic, and Pletka 1992]. A study in 2012 found that anhydrous glass could attain a tensile strength of 13,800 MPa [Bell and Hines 2012]. A 2019 report claims a bending strength of over 100

| Table 2-2 – Material Importance in Asteroid Mining and Restructuring | | | | |
|---|---|---|---|---|
| Priorities | Typical Asteroid Mining [Mazanek et al. 2014] | | Asteroid Restructuring | |
| | Products | Process | Products | Process |
| Top Priority | Regolith – Oxygen from oxides. Extract platinum group metals<br>Available Oxides: 75%<br>Valuable metals: 10% | Refine the regolith to extract the oxygen and metal products<br>Extracting oxygen cost: High<br>Extracting metals cost: Medium | Bulk Material – Oxides such as Olivines and Pyroxenes.<br>Available Oxides: 85% | Melt and form into rods, tiles, and sheets<br>Processing bulk material cost: Low |
| Secondary | Water and other volatiles<br>Available Volatiles: 15% | Refine the regolith to extract water and volatile products<br>Refining costs: Low | Water and other volatiles: Total: 4%; Water 0.13% | Identify materials with product and store for later processing<br>Refining costs: Low |
| Lowest Priority | Bulk material for shielding and construction<br>Available material: 50% | Excess from the refining process<br>Refining costs: Low | Free metal and platinum group metals<br>Free metals: 11% | Identify materials with product and store for later processing<br>Identification and storage cost: Low |
| *Credit: Self produced using data and concepts from [Mazanek et al. 2014] [NASA Report Public Domain]* | | | | |

| Table 2-3 – Basalt Product Comparison | | | | | |
|---|---|---|---|---|---|
| Metric | Terrestrial concrete | Basalt rock | Sintered basalt regolith | Conventional glass | Anhydrous glass |
| Density | 2500 - 2900 kg/m3 | 2630 +/- 140 kg/m3 | 2650 to 2900 kg/m3 | 2500 kg/m3 | 2700 kg/m3 |
| Compressive strength | ~20 - 40 MPa | ~144 - 292 MPa | 206 MPa | 1000 MPa | 1000 MPa |
| Tensile strength | 2 – 5 MPa | 11.2 – 17.8 MPa | 7.29 MPa | 45 MPa | 3,000 – 13,800 MPa |
| *Basalt rocks can be 4-7 X stronger in compression than normal terrestrial concrete [Mueller 2017]. Sintered basalt regolith can be 5X stronger in compression than normal terrestrial concrete [Mueller 2017]. Anhydrous glass can typically provide 3 GPa [Blacic 1985] or ideally up to 13.8 GPa [Bell and Hines 2012].* | | | | | |
| *Credit: Self produced using data and concepts from NASA [Mueller 2017] [NASA Report Public Domain] [Bell and Hines 2012] [Blacic 1985] [Facts]* | | | | | |

MPa for the anhydrous glass [Soilleux 2019]. We intend to use properly designed trusses to exploit the high tensile strength and compensate for the weaker bending strength.

For our space station structure, we assume anhydrous glass could be used to produce high strength rods and tiles. We plan to use the anhydrous glass as structural beams and skin panels, regolith as fill, and basalt fibers as cables.

### *2.2.3 Construction Material Production*

The researcher Schoroers states "*A glass can have completely different properties depending on the rate at which you cool it. If you cool it fast, it is very ductile, and if you cool it slow it¹s very brittle*" [Yale 2013]. It appears that temperature control will be key, and quickly cooling the glass will be necessary. Obviously, thin plies of anhydrous glass can be cooled more rapidly than thick slabs. Stacking and pressure sintering those thin plies of anhydrous glass could produce high strength laminates for our tiles and beams. The laminate should maintain the tensile strength of the individual anhydrous glass plies.

Continuous strip casting would be ideal to produce sheets of basalt to use as panels. Terrestrial continuous casting systems continuously melt feedstock, extrude that molten stock, and form a sheet of cooling material [Tosaka 2008]. The molten substance travels downward, solidifies, and increases in length. Molten material is fed into the tundish mold at the same rate as the solidifying casting exits the system. Gravity is required to flow molten regolith and form panels. The low gravity and vacuum of the asteroid will mandate modifications to the terrestrial continuous casting process. Rotating a continuous casting system could generate some gravity to cause the molten material to flow. The typical size and weight of this system may make a rotating version impractical. Operating on the outer rim structure would be possible once it begins to rotate. It may be possible to rotate smaller versions of the continuous casting system. We have begun an alternative smaller design to produce the laminated anhydrous glass tiles. The smaller design will produce smaller tiles at a slower rate; however, building the smaller design will take much less time and material. A multitude of small units could outperform the single monolithic unit.

In continuous casting, using augers or pressure plates to move the material is another option. The vacuum helps prevent contamination of the molten material in the tundish. Unfortunately, the vacuum reduces the cooling speed of the material with low-rate radiative cooling. We envision using large rollers to provide faster cooling with their larger surface area and direct contact with the anhydrous plies. Similar rollers would provide the pressure to sinter the plies into a laminate. Anhydrous oxide strips will be stacked and sintered with pressure assistance as they rapidly cool.

### *2.2.4 Mission Cost*

Delta-v is a measure used to quantify the cost to transfer from one orbit to another. It often represents the velocity change needed to achieve a new trajectory. Delta-v is typically measured in meters per second or kilometers per second. Mission designers use delta-v as the measure of the energy needed to carry out a space mission. We use delta-v as part of the selection criterion for picking an asteroid for restructuring. Our focus for this effort is to reach and restructure a near-Earth asteroid. The orbits of near-Earth asteroids can bring them within about 0.5 astronomical units of the Earth. It turns out that many of these asteroids require less fuel and lower delta-v than required to reach the Moon or Mars.

There will be multiple trajectory paths to reach our asteroid. Some use 2 impulses, others use 3 impulses or more. Low energy trajectories and slingshots offer additional means to reach an asteroid. For missions from Earth, asteroids in orbits with a semimajor axis less than 2.5 AU and inclination less than 10 degrees tend to minimize delta-v. Numerous studies exist showing how to optimize the trajectory to an asteroid. An earlier example is an asteroid retrieval mission [Bender et al. 1979]. We advocate using slingshots or low energy paths. Those methods seem important to reach asteroids on inner Earth orbits and with potentially higher inclination.

### *2.2.5 Return on Investment*

Planetary Resources owns the Asterank website and database containing ranked asteroids for mining [Webster 2020]. Asterank estimates the costs and values of mining asteroids. Value estimates are based on the mass of a given asteroid and its spectral type. Accessibility estimates are based primarily on delta-v. Profit and Return on Investment (ROI) calculations are a combination of accessibility and value.

We adapted the Asterank approach to evaluate asteroids for our restructuring. We assign a cost metric of the mission to reach an asteroid using delta-v (dV). The values come from the JPL Small-Body Mission-Design Tool [JPL SB Mission Design Tool] and we use the Low Thrust value and the minimum Total Delta-v. For comparison purposes, the values provide a first estimate of the travel cost to the asteroids.

We assign a value metric to the asteroid. The diameter of the asteroid provides a value measure of the available asteroid material. We also experimented with using the asteroid volume as the value metric. The asteroid type and material composition would have a significant effect on the value of the material. We are using only the oxide material from S-type (or C-type) asteroids. Instead of using the asteroid composition, using the diameter (or volume) is reasonable as the value metric because of the large percentage of oxide material in those types of asteroids.

We use the asteroid diameter and the mission delta-v to compute a return-on-investment (ROI). We used this ROI in the selection process of an asteroid for our restructuring process.

## 2.3 **Asteroids – Results**

We presented in previous subsections our asteroid background and analysis approach. In this asteroid subsection, we offer results from applying that analysis approach. The approach criteria helps to winnow the selection to a single asteroid. We present our selection process and the final selected asteroid. We also include details on that asteroid and the expected material to be harvested.

### *2.3.1 Asteroid Selection*

There are many millions of asteroids in our solar system that could be considered for restructuring. There are estimates of over 3 million asteroids in the inner solar system to the Jupiter orbit. We show in Figure 2-5 the selection process we used to select an asteroid. This is like the approach used to select the asteroid Bennu for the OSIRIS-REx mission [Enos 2020]. The JPL Small Body Database lists 22,122 asteroids as members of the Aten, Apollo, Amor, and Atira asteroid classes as of February 2020. This represents only a fraction of the estimated 200,000 sizeable asteroids in those classes.

We restricted the asteroids by limiting the inclination and orbits to be more Earth like. This would help reduce mission costs and improve future access for colonization and trade. We used an inclination less than 26 degrees, an eccentricity less than 0.4, a perihelion greater than 0.4 AU, an aphelion less than 2.2 AU, and a semimajor axis less than 1.5 AU. We used the following JPL SBD search constraints: asteroids and orbital class (IEO or ATE or APO or AMO) and i < 26 (deg) and e < 0.4 and Q < 2.2 (au) and q > 0.4 (au) and a < 1.5 (au). This reduced our choices to 5,570 asteroids.

We then restricted the asteroids using size and rotation speeds. We first eliminated small asteroids with an absolute magnitude parameter less than 20. We used the following JPL SBD search constraints: H < 20 and rot_per > 3 (h). This reduced our choices to 228 asteroids. Some of those asteroids did not have diameters in the database. We used the absolute

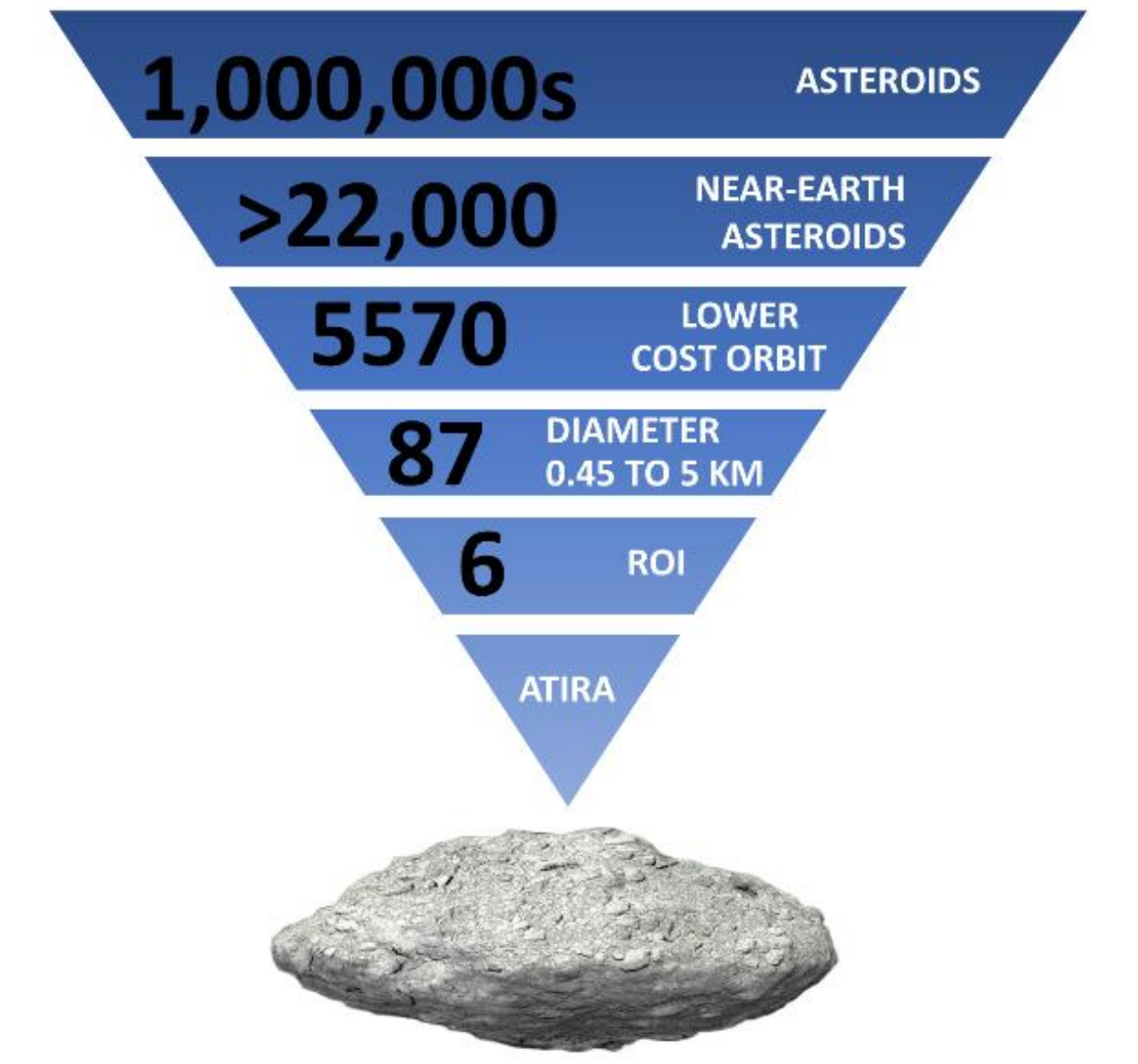

Selection Process for Atira Asteroid

- Asteroids: 1,000,000s
- NEO: 22,122 asteroids; orbital classes (IEO or ATE or APO or AMO)
- Orbit Cost: 5570 with i < 26 (deg) and e < 0.4 and Q < 2.2 (au) and q > 0.4 (au) and a < 1.5 (au)
- Diameter: 288 with H<20 (size)
- Diameter: 87 with size 0.45 to 5 km
- ROI: Best 6 for delta-v mission cost and asteroid value
- Final: Atira

*Credit: Self produced; Image Concept: Heather Enos, [Enos 2020], NASA/Goddard Space Flight Center, [NASA Report Public Domain]; Atira Image from Doug Ellison model [Ellison 2018], [CC BY-4.0] modified/rescaled to asteroid dimensions*

**Figure 2-5 – Overview of Asteroid Selection Criteria**

magnitude and albedo to computed diameters and selected only those in the range of 0.45 kilometers to 5.0 kilometers. This left 87 asteroids.

We use our simple estimate of the return on investment (ROI) for each of the candidate asteroids. That ROI is the ratio of the asteroid value over the travel cost. The diameter provides the asteroid value measure. The delta-v provides a travel cost measure. For the largest asteroids, the delta-v costs ranged from 4.6 to 16.7 meters per second. With the delta-v constraint, we found 55 asteroids being identified as potential candidates. The spectral type of many of the asteroids have not been determined. Sixteen of the 55 asteroids have spectral types of C or S (or similar). Both spectral types are expected in this region of the solar system. Figure 2-6 shows the ROI for each of the candidate asteroids. We show a line graph for all the asteroids and include the column bars for well-defined asteroids (i.e., spectral type available).

We separate the best six return-on-investment (ROI) asteroids from Figure 2-6. Those six asteroids are shown in Table 2-4 with their diameter value and their mission delta-v cost. From those 6 asteroids, the best ROI was for the asteroid 163693 Atira.

**Table 2-4 – Best Return on Investment Asteroids**

| Asteroid Name | dV (Cost) | Diam-eter | Spec Type | q | Q | i | ROI |
|---|---|---|---|---|---|---|---|
| 163693 Atira (2003 CP20) | 16.70 | 4.81 | S | 0.502 | 0.980 | 25.62 | 0.288 |
| 175706 (1996 FG3) | 4.71 | 1.20 | C | 0.685 | 1.422 | 1.99 | 0.254 |
| 65679 (1989 UQ) | 4.60 | 0.92 | B | 0.673 | 1.157 | 1.30 | 0.200 |
| 2063 Bacchus (1977 HB) | 7.67 | 1.08 | Sq | 0.701 | 1.455 | 9.43 | 0.141 |
| 162173 Ryugu (1999 JU3) | 6.26 | 0.85 | Cg | 0.963 | 1.416 | 5.88 | 0.135 |
| 2062 Aten (1976 AA) | 10.00 | 1.10 | Sr | 0.790 | 1.143 | 18.93 | 0.110 |

*Credit: Self produced using [JPL SBD Search Engine] [Public Domain]*

### 2.3.2 Harvested Material

The Atira asteroid is an S-Type asteroid. We assume it is comprised of free metal (11.1%), volatiles (4.0%), and oxides (84.9%). We have computed its volume using a mesh grid (2.19e10 cubic meters) and as an ellipsoid (3.01e10 cubic meters). We computed its mass of 41 trillion kilograms using the orbit period of its moon. We derived that the asteroid has a porosity of 48.3% and assume there is a 24.4% loss when processing the oxide. Table 2-5 contains these processing values and metrics. As a preview, the table also includes the amount of material required to build the example Atira space station with a major radius of 2116 meters and elliptical minor axes of 334 and 1003 meters. We found that we can extract enough building material for the station by harvesting 269 meters of regolith from the surface of the asteroid. The Atira asteroid has a mean radius of 1928 meters; harvesting only the top 14% of the asteroid is conservative. This volume of harvested material represents 30% of the total volume of the Atira asteroid.

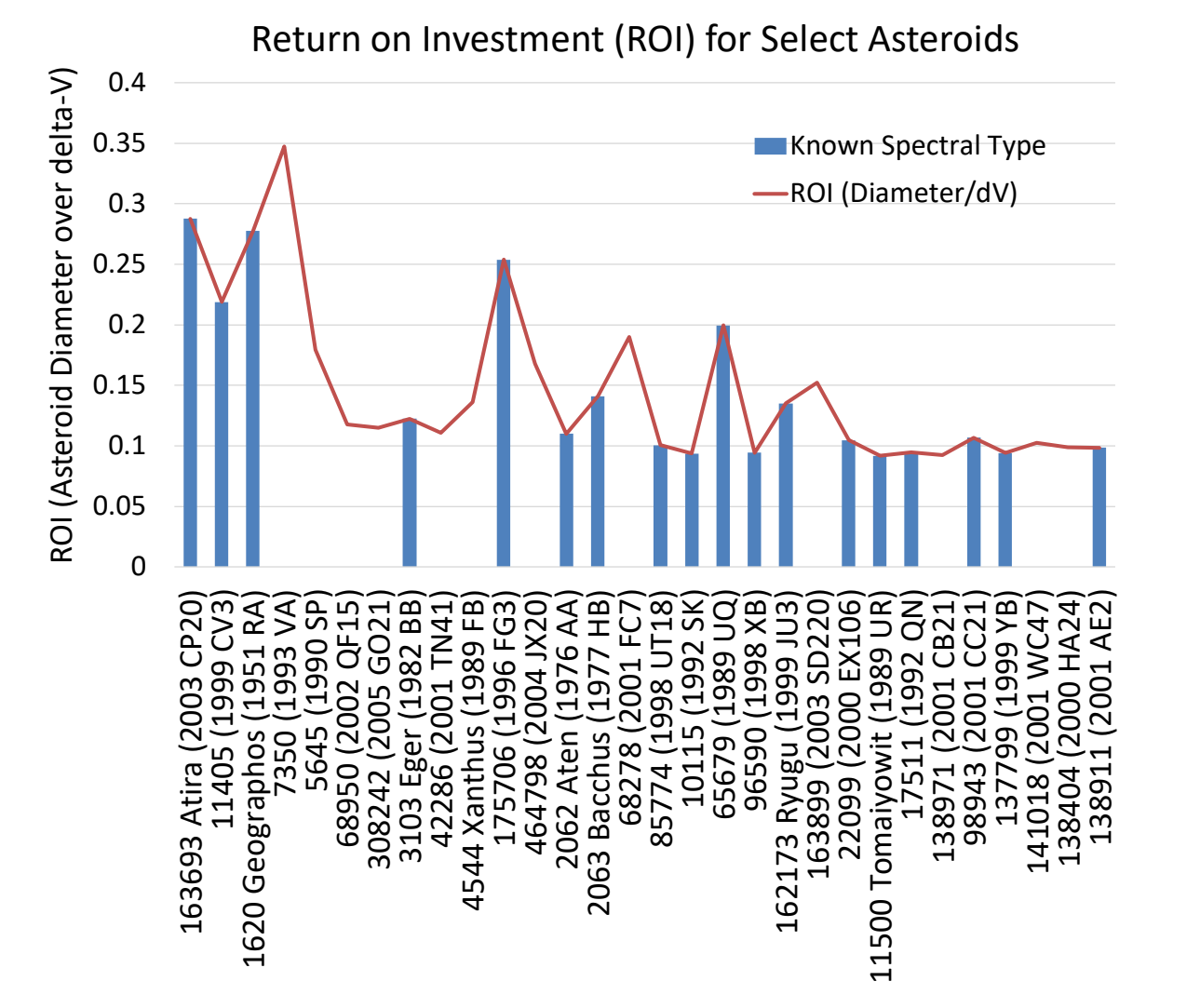

*Credit: Self produced using JPL Small Body Database Search Engine Results: 87 matching objects (Top 30 Largest Diameters); Constraints: asteroids and orbital class (IEO or ATE or APO or AMO) and H < 20 and rot_per > 3 (h) and i < 26 (deg) and e < 0.4 and Q < 2.2 (au) and q > 0.4 (au) and a < 1.5 (au); Data: [JPL SBD Search Engine] [Public Domain] and [JPL SBD Small Body Lookup] [Facts]*

**Figure 2-6 – Return on Investment for Candidates**

## 2.4 Asteroids – Summary

The purpose of this section was to review asteroid characteristics and select one for the restructuring process. We have looked for one that has enough material to construct a large station. We want an asteroid between 1 kilometer and 5 kilometers for this effort. We also want the asteroids to be in the Goldilocks zone near the Sun – not too hot and not too cold. A location near the Earth orbit will also reduce the mission costs. Our selection process has considered a return-on-investment metric using the mission cost and the available asteroid material. From our decision criteria, we selected the asteroid Atira.

# 3 Asteroid Restructuring – Space Stations

Asteroid restructuring mandates understanding space stations, asteroids, and robotics. The previous section covered the subject of asteroids. In this section we delve into the historic and foundational concepts and supporting technologies for space stations. The purpose of this section is to understand the type of space station to be built from a selected asteroid. We again organize this section with subsections to cover background, analysis, results, and a summary.

**Table 2-5 – Asteroid Material Summary**

| Material | Metric | Volume ($m^3$ millions) |
|---|---|---|
| Atira | Mesh Model | 21,900 |
| Harvested | 269 meters deep | 6,572 |
| Packed | 51.7% | 3,397 |
| Loss | 24.4% | 829 |
| Regolith | Processed | 2,569 |
| Building | 84.9% | 2,181 |
| Metals | 11.1% | 285 |
| Volatiles | 4.0% | 103 |
| Station | Calculated | 2,181 |

## 3.1 Space Stations – Background

We offer in this background section a brief overview on space station designs. We introduce key concepts on the space stations geometries, the rotation produced artificial gravity, and the effects of gravity on humans.

### *3.1.1 Space Stations – Background Credit*

Space stations have been researched and designed for over 150 years. Thousands of books and journal articles have been written about the design and construction of space stations and provide a foundation for our restructuring study.

We present a series of key technology areas and some of the associated researchers. In our later analysis and results subsections, we present some extensions to these space station technology areas. We also provide more details in later subsections on some extensions that may be innovative.

- **Station Geometry:** NASA Studies [Johnson and Holbrow 1977] [O'Neill et al. 1979] and Globus [Globus 1991] [Globus et al. 2007] present a solid foundation for station geometries. We review and extend this foundation to include large, rotating, multiple floors, and balanced space stations.
- **Rotational Imbalance:** Rotational stability is a constraint on space station designs [Brown 2002] and imposes new limits on their geometry sizes [Globus et al. 2007]. We apply this constraint to four rotating space station geometries.
- **Gravity Ranges:** Rotating the stations to provide centripetal gravity has long been a part of space station designs [Oberth 1923]. We review and consider the gravity ranges with multiple floors on the human physiology [Hall 1999] [Globus and Hall 2017].
- **Runways:** Most rotating station designs use the central hub to support shuttle arrivals, servicing, and departures. This single central hub can become a bottleneck for passengers and trade. We evaluate landing on runways built on the exterior of the rotating station.
- **Lighting:** The O'Neill cylinder [O'Neill 1976] [Johnson and Holbrow 1977] windowed half of its exterior to provide light. More recent station studies eliminate those windows and use internal lighting [Globus 1991]. Solar panel materials, light efficiencies, and prices have steadily improved since the 1970s. Power requirements for light generation have dropped by 4x using LEDs. Using narrow spectrum light generation further drops the power requirement by 7.5x [Wheeler 2017] [Soilleux and Gunn 2018]. We have also seen improved light transmission concepts through light pipes, fiber, chevrons, and light shelves [Johnson and Holbrow 1977] [Savard 2012] [Janhunen 2018].
- **Space Allocations:** The 1977 NASA Report [Johnson and Holbrow 1977] detailed the allocation of space station floor space to purposes such as open space, support infrastructure, agriculture, industry, and residence. We have enhanced these values to more modern allocations.
- **Agriculture:** NASA Studies [Johnson and Holbrow 1977] [Bock, Lambrou, and Simon 1979] and more recently [Fu et al. 2016] [Soilleux and Gunn 2018] [Stanley 2018] have evaluated the use of agriculture and plants to create a closed system environment. We review their findings and include recent advancements. Floor space requirements for agriculture in space stations have been dropping steadily since the 1970s. We have found details on recent advancements in hydroponics, aeroponics, and new biological approaches [Kersch 2015] [Cornall 2021]. Those recent advancements offer even more reduction to the agriculture space requirements.
- **Structural Material:** O'Neill [O'Neill 1974] and McKendree [McKendree 1995] present concepts on the maximum structural radius of rotating stations. We summarize and extend these structural values with new materials. A material of interest for our station is anhydrous glass [Blacic 1985] [Carsley, Blacic, and Pletka 1992] [Bell and Hines 2012] [Soilleux 2019].
- **Equipment:** Most terrestrial mining and construction equipment will not work in the low gravity and vacuum asteroid environment. Design modifications have been researched [Eisele 2001] [Schrunk et al. 2008]. Our restructuring equipment will need many of these modifications.

### *3.1.2 Space Station Geometries*

There are four common space station geometries: Sphere, Dumbbell, Torus, and Cylinder [Johnson and Holbrow 1977]. We include four artworks to help visualize these stations in Figure 3-1. These shapes have symmetry to support spinning and the production of centripetal gravity. They also have hollow regions to hold atmosphere. These space station geometries have been documented for over 150 years. We review those four geometries in the following paragraphs.

**Dumbbell:** The dumbbell is typically the smallest geometries considered for space habitats. The dumbbell consists of two small modules connected with a tether or trusses; see Figure 3-2a. The modules rotate around a common axis and produce an artificial gravity inside the module. The length of the tether and the rotation speed determines the artificial gravity. The centripetal gravity is equal to the rotation radius times the angular rotation speed squared. To prevent motion sickness, most dumbbell designs use a tether that is several 100 meters in length [Mordanicus 2014]. Long tethers can provide Earth-like gravity with fairly slow rotation rates. The tether structure connects each pair of modules and would need to withstand the centripetal force of rotation [García et al. 2016]. With dumbbells, the rotation radius and speed are independent from the size of the two dumbbell modules. This provides the advantage of a small module shape requiring less material and atmosphere than many of the other shapes [Mordanicus 2014]. The drawings in Figure 3-2a illustrate single dumbbells and a composite shape with multiple dumbbells [Johnson and Holbrow 1977]. The basic dumbbell shape provides a building block for composite shape stations. The composite structure is advantageous in that early pioneers could live in the simplest configuration as additional dumbbells are constructed and attached [Mordanicus 2014].

| NASA Dumbbell Station | Interior Spherical Space Station | Cylinder Space Station | Torus Space Station |
|---|---|---|---|
| 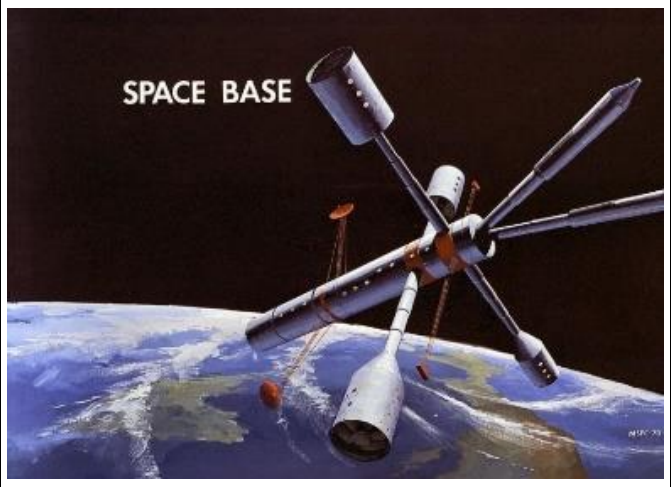 | 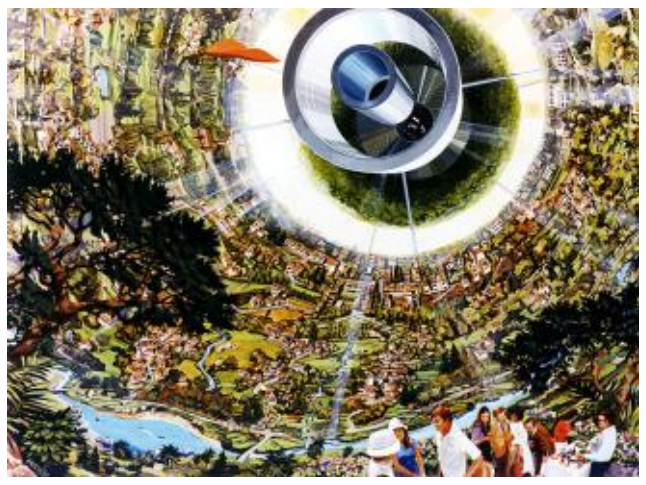 | 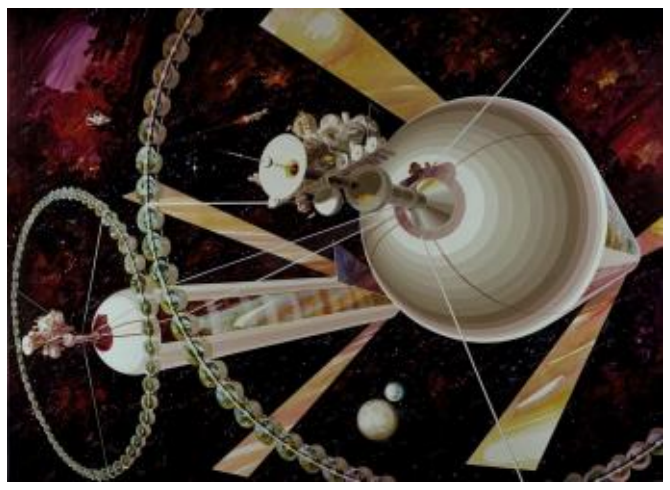 | 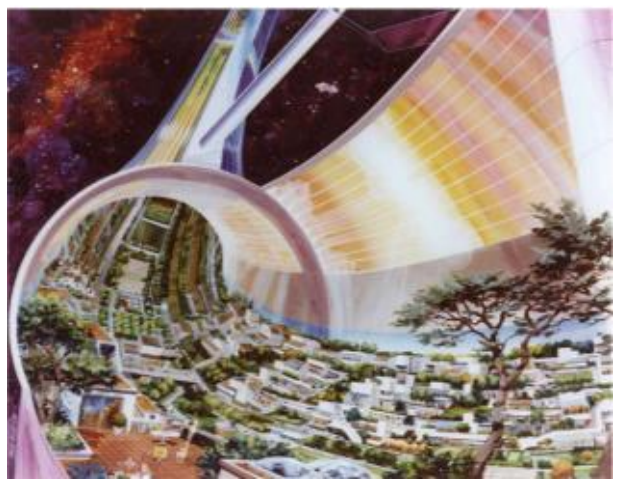 |
| *Image Credit: NASA, Marshall Space Flight Center, 1 January 1970; [NASA Image Public Domain]* | *Image Credit: NASA, ARC, Artist: Rick Guidice: 1976-04-01; [NASA Image Public Domain]* | *Image Credit: NASA, Artist: Rick Guidice: 27 May 1975; [NASA Image Public Domain]* | Credit: NASA Ames Research Center. Artist: Rick Guidice*; [NASA Image Public Domain]* |

**Figure 3-1 – Space Station Geometries – Artwork from 1970s**

**Sphere:** A space colony could reside inside a sphere or ball shaped structure. Spherical space stations have been proposed since the 1880s. Examples include Konstantin Tsiolkovsky's spherical spaceship in 1883 [Tsiolkovsky 1883], a Dyson Sphere in 1960 [Dyson 1960], and Gerard O'Neill's Island Two in 1976 [O'Neill 1976]. O'Neill wrote that "*spherical geometries for space colonies ranked highest in simplicity, ruggedness, economy, and safety among Earth-like colony designs*" [O'Neill 1976]. Spheres have a strong structure because stresses are evenly distributed over the entire surface. For a given wall thickness, a spherical vessel has twice the strength of a cylindrical vessel. Thinner walls are considered valuable because thickness directly affects material weight and launch costs.

We show the basic spherical shape in Figure 3-2b. The sphere would rotate to produce Earth-like gravity at the outer shell equator. Gravity would decrease as one moves toward the rotation poles. A rotating sphere design has the risk of being imbalanced. We also include an oblate ellipsoid shape in Figure 3-2b. Unlike the sphere, the oblate ellipsoid geometry would be rotationally balanced. The ellipsoid minor axis length should be less than 0.8165 times the major axis length to passively balance the rotating ellipsoid station.

**Cylinder:** Another geometry for a space colony is a cylindrical structure. Cylinder space stations have been considered since the 1920s and have ranged from 30 meters to several kilometers in radius. The radius and rotation speed are typically chosen to provide one Earth-like gravity (1g) at the outer shell. A rotation speed of less than 2 revolutions per minute (rpm) is usually acceptable to avoid motion sickness and disorientation. A radius of 224 meters revolving at 2 rpm produces 1g at the outer shell.

We show basic cylindrical shapes in Figure 3-2c. The cylinder includes end caps that could be flat, hemispherical, or other shapes. We show the hemispherical and flat end caps in Figure 3-2c. Longer cylinder lengths produce more habitable square footage and support higher population. Multiple reports from the 1970s used lengths that were 10 times the radius. O'Neill and his students connected two counter-rotating cylinders to eliminate gyroscopic effects and precession [O'Neill 1976]; see Figure 3-1. This design arrangement would properly point the station at the sun during the station's circumsolar orbit. Recently, authors have imposed a limit on the cylinder length to passively control the imbalance of the rotating structure. Globus and his group found the cylinder length should be less than 1.3 time the radius [Globus et al. 2007]. This design produces a station looking more like a hatbox; see Figure 3-2c. Longer length cylinders would require some active balance technology to prevent the risk of possibly tumbling.

**Torus:** Another geometry for a space colony is a torus or wheel structure. Torus structures have been described by researchers and science fiction writers since the early days of space science [Noordung 1929]. The torus sizes in our historic review range from 50 meters to 30,000 meters in radius. The station is typically rotated at a speed to provide one Earth-like gravity at the outer rim. We show torus examples in Figure 3-2d. The torus can have a cross-section of a circle, ellipse, or extended cylinder. Alternatively, the torus shape can be composite with many smaller spheres or nodes. Early researchers envisioned connecting 22 emptied rocket stages end-to-end to produce a 160-foot-radius, rotating, wheel-shaped space station [Koelle and Williams 1959].

The torus has an advantage that the tube structure size is independent of the center rotation radius. The habitable region of the torus can be located far from the center of rotation. A larger rotation radius provides lower rotation rates to generate Earth-like gravity. Lower rotation rates reduce stresses and the required material tensile strength. Independent of the rotation radius, the radius of the habitable region can be made larger or smaller based on available construction material.

### *3.1.3 Artificial Gravity*

As a part of our space station background, we review artificial gravity. Even in the 1920s, researchers were rotating space stations to provide centripetal gravity [Tsiolkovsky 1920] [Oberth 1923] [Noordung 1929]. We review in this subsection the forces in a rotating station that produce the artficial gravity. We cover the issue of rotation causing disturbances in the inner ear, which leads to motion sickness and disorientation. We cover microgravity health problems to end this review of artificial gravity.

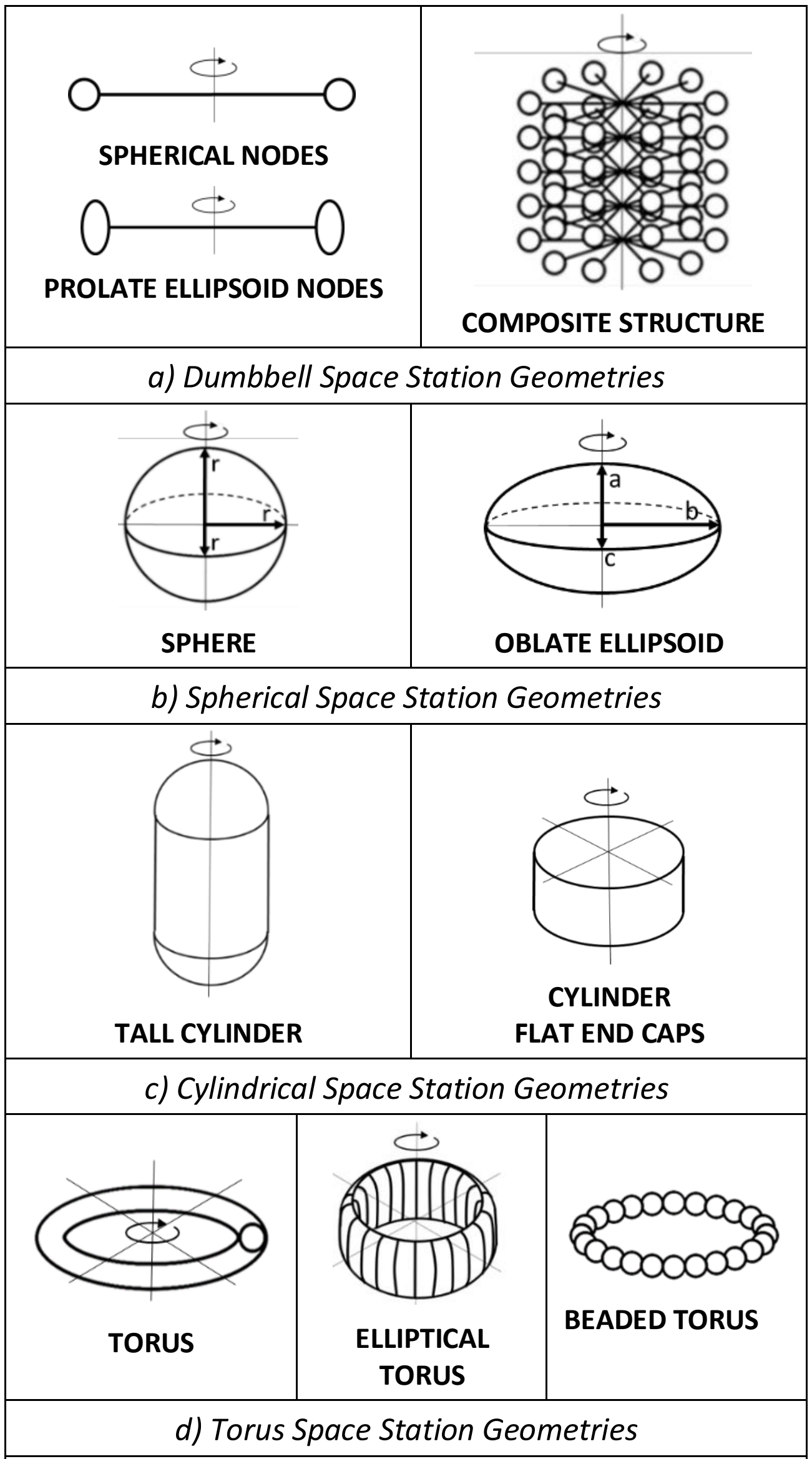

*Images Credit: Self Produced using Images and Concepts from [Johnson and Holbrow 1977][NASA Report Public Domain]*

**Figure 3-2 – Space Station Geometries**

**Forces in Rotating Station**: We review the three forces felt in the rotating station: centripetal (inward), centrifugal (outward), and Coriolis (movement) forces. The word centripetal means center-seeking. It is from the Latin word *centrum* meaning center and *petere* meaning to seek. Similarly, the word centrifugal means center-fleeing. It is from the Latin word *centrum* meaning center and *fugere* meaning to flee. The Coriolis force is named after Gaspard-Gustave de Coriolis. Almost 200 years ago, he defined these forces in a rotating frame of reference [Coriolis 1835].

Only the centripetal force is a “real” force. The centripetal force is always directed radially towards the center of rotation. The Coriolis and centrifugal forces are typically called fictious, inertial, or supplementary forces. When Newton's laws are transformed to a rotating frame of reference, the inertial forces appear [Hand and Finch 1998]. The equation of motion for a rotating body takes the form:

$$\underbrace{\boldsymbol{F}}_{\substack{Apparent\\Forces}} = \underbrace{m\boldsymbol{a}}_{\substack{Motion\\Force}} + \underbrace{m\frac{d\omega}{dt}\times\boldsymbol{r}}_{\substack{Euler\\Force}} + \underbrace{2m\boldsymbol{\omega}\times\boldsymbol{v}}_{\substack{Coriolis\\Force}} + \underbrace{m\boldsymbol{\omega}\times(\boldsymbol{\omega}\times\boldsymbol{r})}_{\substack{Centrifugal\\Force}}.$$

In that equation, F is the sum of the forces acting on the object in the rotating frame, m is the mass of the object, and ω is the angular velocity. Multiple parameters describe the object relative to the rotating frame: ***a*** is the acceleration, ***r*** is the position, and ***v*** is the velocity. In the rotating frame, the inertial or fictitious forces act as additional forces and contribute to the object acceleration. The Euler force only applies with changing angular velocity, ω. In a rotating space station, the rotation speed is constant and the term dω/dt is zero. The Coriolis force is function of the object velocity. The centrifugal force is a function of the object position, ***r*** (distance from the center of rotation). Multiple sources are available to provide details on these forces [Hall 1991] [Hand and Finch 1998] [Lucas 2019].

**Large Station Example:** We offer one example to illustrate the effect of large stations on the artificial gravity. We show in Figure 3-3a a ladder oriented with the rotation of the station. This diagram shows the Coriolis and centrifugal accelerations. The individual is ascending the ladder at a constant velocity, $v_r$. Centripetal force maintains the person’s position in the rotating system and pulls towards the rotation center. The apparent centrifugal force is equal and opposite to the centripetal force and pulls the individual downward to the base of the ladder. The Coriolis force pulls the ascending individual in the spinward direction. When ascending, the climber leans antispinward to counteract the Coriolis force. When descending, the climber leans spinward to resist the opposite direction Coriolis force [Queijo et al. 1988]. The angle of the lean would be the arctangent of the ratio of $A_{cor}$ over $A_{cent}$ and simplifies to atan($2v_r/v_t$).

We also include a chart showing the lean angle as a function of the station radius; see Figure 3-3b. The x-axis of the chart is logarithmic and shows the radius of the station. The stations are rotating to produce a 1g centripetal gravity at the outer rim. The y-axis is also logarithmic and shows the lean angle. The chart includes lean angles for three velocity ladder climbers. The data of the chart shows that the lean angle increases with climber velocity and with smaller station radii (faster rotation). Fast climbers would ascend a ladder at 0.5 meters per second. We show a building code limit of 0.5 degree lean in the chart. This is below the terrestrial building code limits for cross slopes, parking stalls, and stairway treads [ICC 2009] [ADA 2010]. Slower climbers on most rotating stations would experience less than 0.5 degree lean on the ladder from the Coriolis force. Another study defined lean angles less than 1.8 degrees would be neglible in a rotating environment [O’Neill and Driggers 1976]. The results in this chart shows the lean angle will be barely noticeable on our large rotating stations.

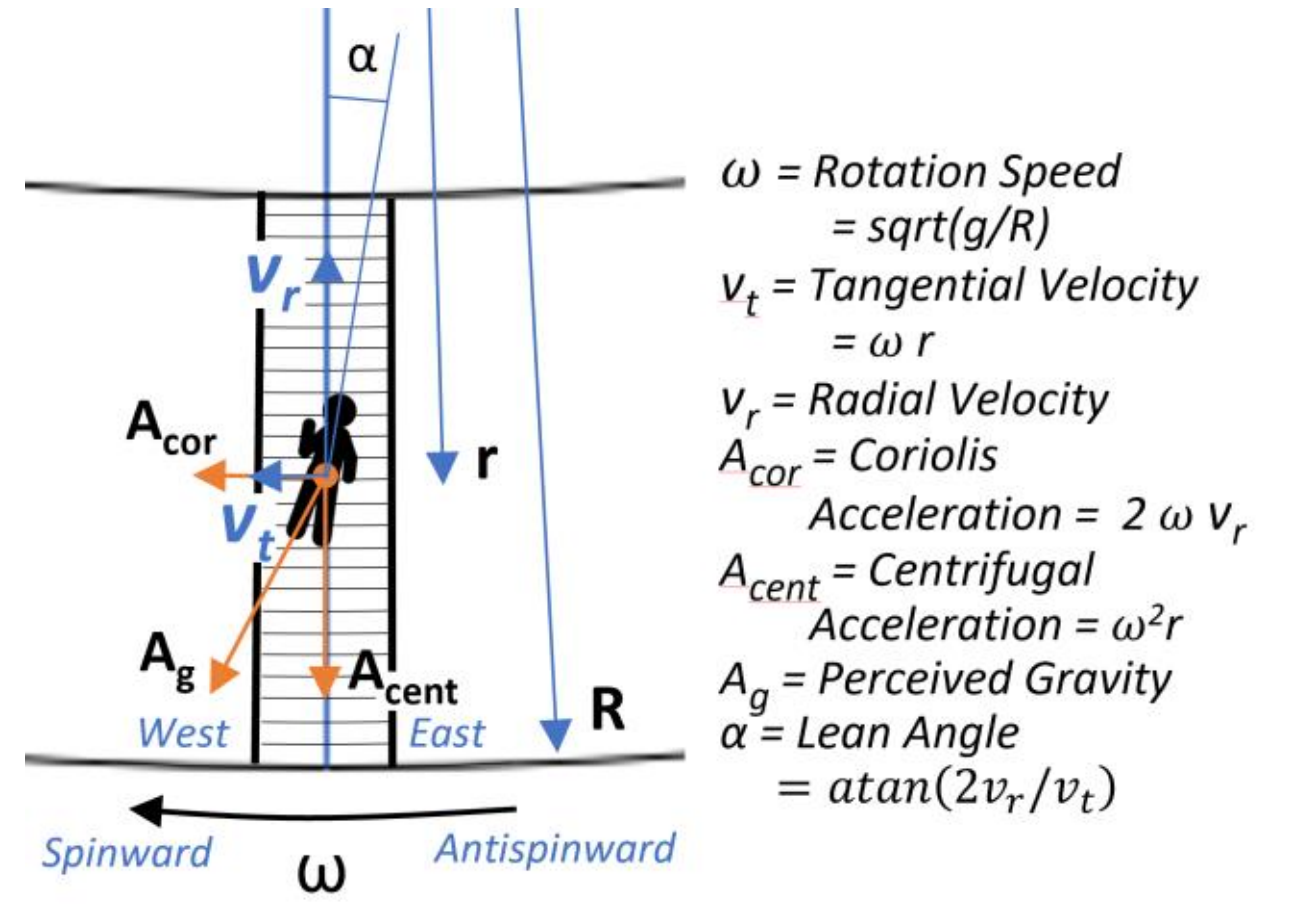

*a) Lean Angle Climbing Ladder*

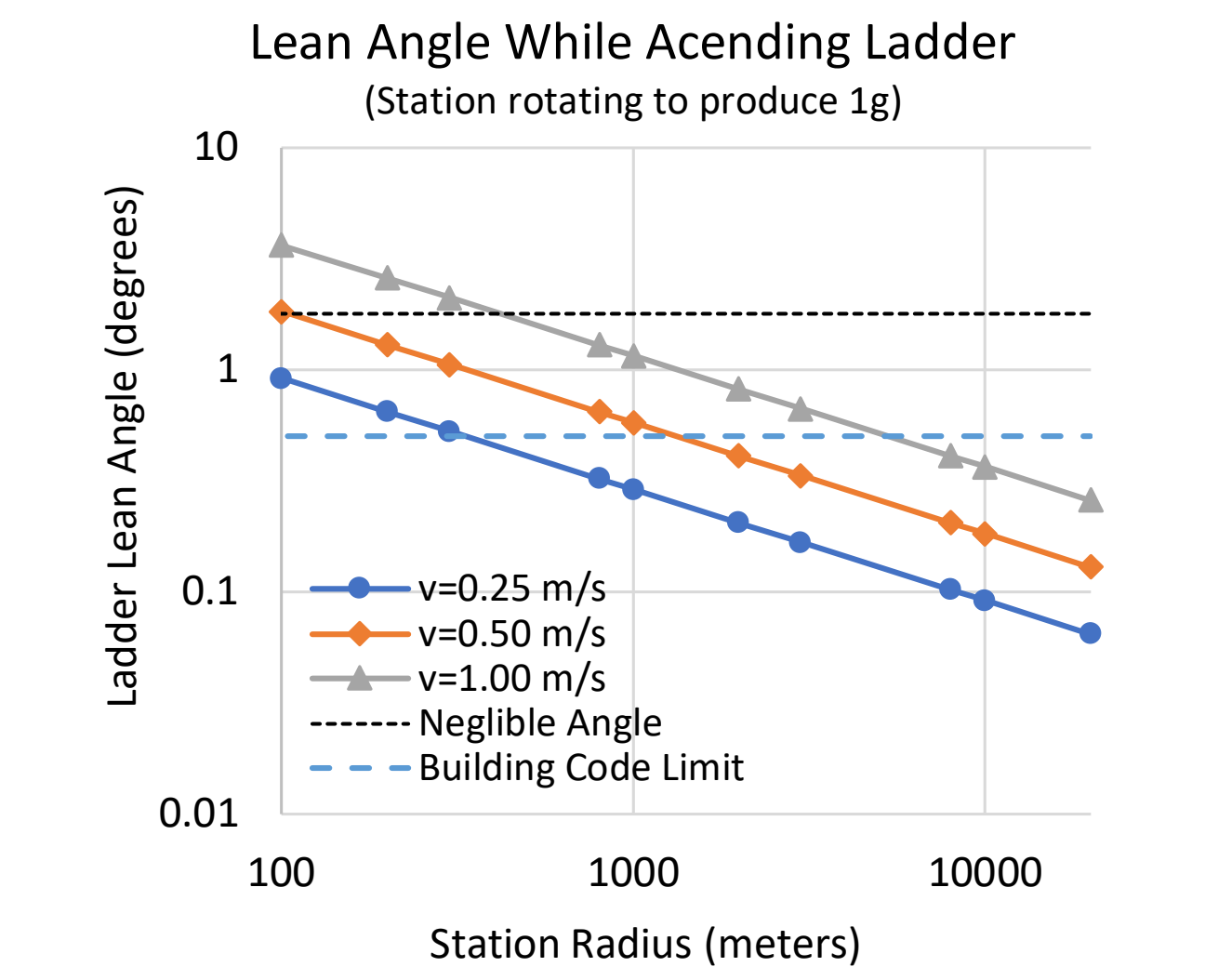

*b) Lean Angle Function of Radius*

**Figure 3-3 – Climbing Ladder Oriented with Rotation**

In addition to this ladder climbing example, we have further explored these forces and their effects. We considered scenarios with objects in motion and people in motion in larger stations. Objects in motion scenarios included dropped objects, thrown objects, hopping objects, free fall, high speed vehicles, and fire arms. The people in motion scenarios include walking, climbing ladders, stairs, ramps and paths, and elevators. Evaluation parameters include drop displacement, extra weight, lean, and elevator velocity. Our exploration has reviewed and consolodated multiple studies [Hall 1991] [Hand and Finch 1998] and [Lucas 2019]. We extend their results to larger stations and to new scenarios. The combination of forces creates anomalous movements of the objects and people. Striving to attain Earth-like gravity with moving objects or people requires reducing the Coriolis force. Our exploration leads us to agree with Hall who ultimately finds "*that it is impossible to design away the gravitational distortions inherent in rotating environments. They can be kept arbitrarily small only by keeping the radius sufficiently large*." [Hall 1993].

**Human Rotation Tolerance:** In a rotating space station, centripetal gravity replaces the normal Earth gravity. If the rotation rate is too fast (radius too small), a Coriolis effect will cause disturbances in the inner ear and lead to motion sickness and disorientation. Rotation rates below 2 revolutions per minute (rpm) are acceptable to most people [Globus and Hall 2017]. This is a threshold revolution speed used by many researchers.

Theodore Hall provides a thorough review and analysis of five historic guidelines [Hall 1997]. He composites the five studies into one "comfort chart." He identifies a comfort zone in a station radius and rotation rate chart. He notes that the boundaries of these charts can be influenced by tasks and other environment criteria [Hall 1997].

Hall uses five boundaries [Hall 1997] [Hall 2006] to limit the consolidated comfort area and produces a pentagon shaped comfort zone. The pentagon is bounded by gravity values, head-to-foot acceleration gradient, rotation rate, and tangential velocity. Recently Al Globus and Theodore Hall used this comfort chart in their paper on rotation tolerance [Globus and Hall 2017]. It may be comfortable (and fun) to be in zero or low-gravity zones for a short time; however, long term there are serious health problems from microgravity. It is more appropriate to call these tolerance charts rather than comfort charts. We are looking for long-term living tolerance limits, not comfort zones.

We use five boundary points in our tolerance chart; see Figure 3-4. The chart shows the station radius on the y-axis from 100 to 25,600 meters on a logarithmic scale. The x-axis shows the rotation in revolutions per minute. We label the same five boundary points to form a pentagon like Hall. The maximum gravity tolerance we show in Figure 3-4 is 1.2g and the minimum gravity is 0.8g. In most of our designs and analysis, we conservatively limit the habitable region of the rotating space station between 0.95g and 1.05g. Results from future experiments aboard the International Space Station such as [JAXA MHU 2019] may refine these boundary ranges. We limit the spin tolerance to 2 rpm to prevent motion sickness. Like Globus and Hall, we include a limit using a large ratio of Coriolis to centripetal force (Acor/Acent =25% at velocity=5 meters per second). For our large stations, the Coriolis distortion and the motion sickness limits have minimal effect on our tolerance zone. Unlike Hall, we limit the top boundary points (① and ⑤) to a maximum hoop stress radius. In a later section, we define a maximum rotation radius of 10 kilometer considering material strengths. The stresses of the top boundary points are determined from this maximum. This stress radius is a structural limit and not a human tolerance measure; however, exceeding a structural limit could have a significant impact on the health of the inhabitants.

We update the definitions of the pentagon boundary points:

- Top Boundary (points ⑤ and ①): Maximum hoop strength. Near-Earth-like normal gravity.
- Angle Right Boundary (points ① and ②): High gravity limit with increasing Coriolis effects.

- Bottom Right Boundary (points ② and ③): Motion sickness limit from maximum high angular velocity.
- Bottom Left Angle Boundary (points ③ and ④): Distortion from high Coriolis to centripetal forces.
- Top Left Angle Boundary (points ④ and ⑤): Low gravity limit with decreasing Colriolis effects.

We highlight 1 rpm in Figure 3-4 as a maximum rotation velocity when using an attached shield [Johnson and Holbrow 1977]. This implies a minimum rotation radius of almost 900 meters. Newer materials appear strong enough to permit attached shields at higher speeds and we allow a maximum rotation speed of 2 rpm in our tolerance chart. We strive to keep the rotation speed lower than 1 rpm in our designs.

**Microgravity Health Problems**: Today it is known that microgravity can negatively affect human health. One of the more common effects is called the Space Adaptation Syndrome and includes nausea, vomiting, anorexia, headache, malaise, drowsiness, lethargy, pallor, and sweating [Hall 1997]. Longer term issues have been found such as cardiovascular changes, muscle damage, bone damage, and possible genetic changes.

We know that the human body has no problem with Earth gravity; we know that it has a plethora of problems with zero gravity. What is not known is if the human body can tolerate long-term gravity between 0g and 1g [NASA 2004]. Studies are planned to investigate these partial gravities. Lower gravities are of interest because of planned long missions and colonies on the Moon and on Mars. We include in Figure 3-4 the Mars gravity of 0.38g and the moon gravity of 0.17g for reference. Those gravities are outside the tolerance zone. There is debate on whether those gravities could prevent the

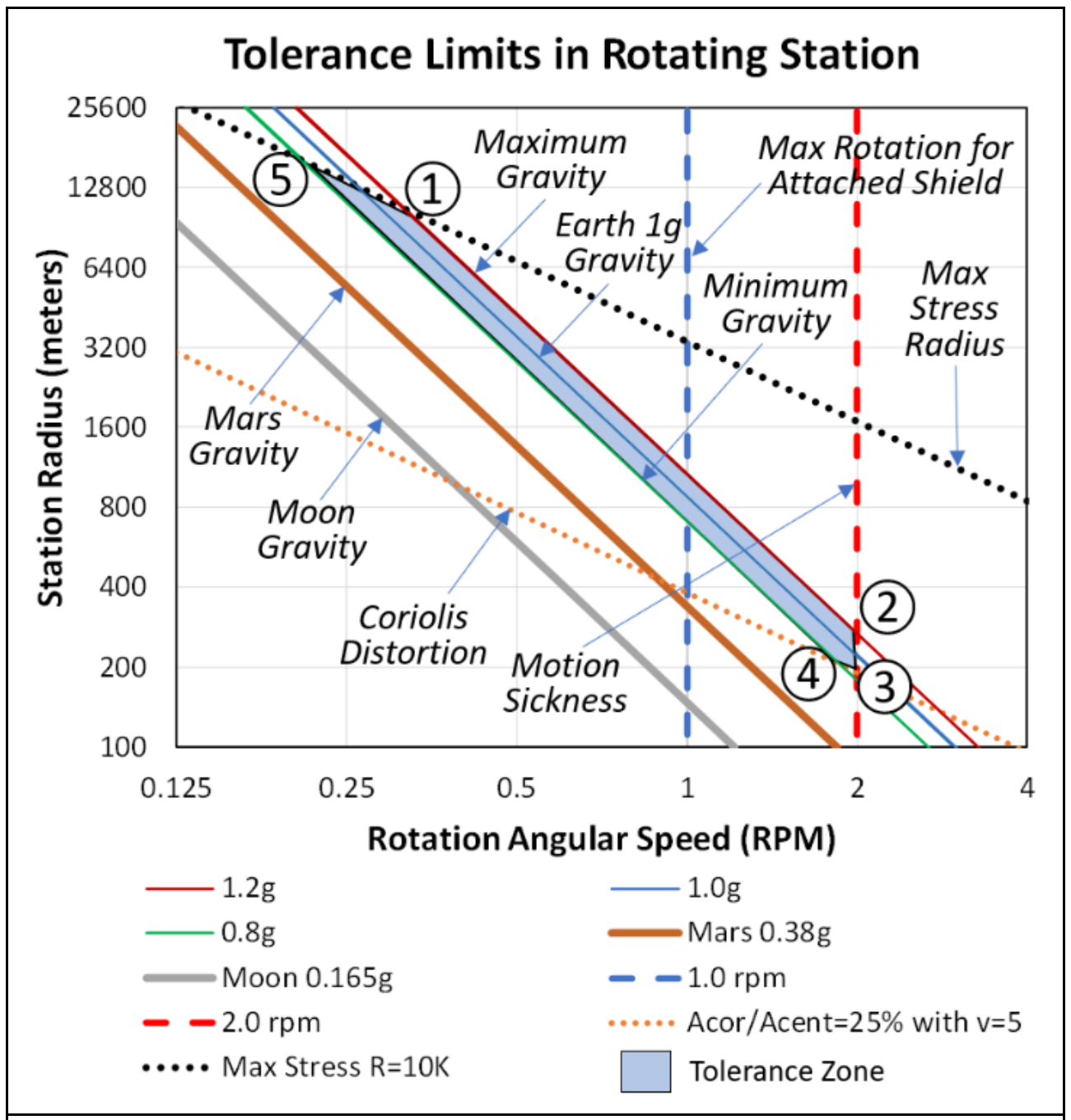

*Credit: Self Produced using [Hall 1993] concepts [Facts]*

**Figure 3-4 – Tolerance Zones in Rotating Space Station**

low-gravity physical problems. Boyle [Boyle 2020] writes that, despite more gravity than microgravity, the *"long-duration visitors will still experience some of low gravity's deleterious effects."*

We hope to avoid those deleterious effects by maintaining a near-Earth gravity range in the habitable areas of the large rotating space stations. For most of our restructuring analysis we limit the artificial gravity range in the station between 0.8g to 1.2g. We strive to limit the primary long-term residential areas to a gravity range between 0.95g to 1.05g.

**Artificial Gravity Summary:** Centripetal gravity addresses microgravity health issues but introduces anomalous effects on moving objects and people. Multiple papers provide our foundation for artificial gravity [Johnson and Holbrow 1977] [Hall 1997] [Hall 2006] and [Globus and Hall 2017]. In our research, we have reviewed and extended their results to larger stations. The large stations envisioned with the restructuring process minimize these anomalous effects.

## 3.2 Space Stations – Analysis

We extend the space station findings introduced in the previous background section. We cover the broad topics of station characteristics, rotational stability, and station gravity ranges. We review rotational stability and gravity ranges for the four geometries. We cover the geometry adaptations to support these topics. We investigate the required station mass for the design.

### *3.2.1 Station Characteristics*

As a part of our space station analysis, the following paragraphs contain more details on the station geometries, station size, structure stress, population, multiple floors, station mass, and floor allocation usage of our envisioned space station. Large station sizes and using almost half of the station volume as multiple floors are important details and considered in all the following subsections.

**Station Geometries:** The literature contains four types of space station geometries: spheres, torus, dumbbells, and cylinders. From the 1900s to today, the preferred space station geometry has varied from the torus, to spheres, and to cylinders. In our studies, we have found there are many ways to evaluate and select a station geometry. For maximum volume, the sphere geometry was superior [O'Neill 1976]. To minimize the mass for a given population, the cylinder geometry was superior [O'Neill 1976] [Globus et al. 2007]. To minimize mass for a given rotation radius, the torus geometry was superior [Johnson and Holbrow 1977] [Misra 2010].

These assessments and selections were typically based on thin space station shells. In our asteroid restructuring, we use thick shells to provide structural integrity, radiation protection, and safety from debris collisions. These historic assessments were also typically based on a single projected floor. We design our station with many floors to support greater populations. We also explored slightly different geometries than earlier studies. We use ellipsoids instead of spheres; short cylinders instead of long cylinders; elliptical cross-section toruses instead of circular cross-section toruses; and avoid composite dumbbell structures.

**Station Sizes:** Historically, space station designs have ranged from tiny cans to huge habitats. With the near unlimited resources of a large asteroid, one wonders how large to make the space station. There are multiple ways to structurally analyze this problem [O'Neill 1974] [McKendree 1995]. Using their approaches, we replicate their set of maximum radius values in Table 3-1. We extend their estimates with additional materials. The exterior station shell could be over 20 kilometers in radius by using melted asteroid material (basalt rods). We include anhydrous glass data [Blacic 1985] [Carsley, Blacic, and Pletka 1992] [Bell and Hines 2012]. The tensile strength of anhydrous glass suggests a station of over 40 km could be feasible.

O'Neill's envisioned rotating cylinders with a radius of 16 kilometers. We see that this does not require esoteric materials such as graphene or nanotechnology. A recent paper [Soilleux 2019] uses anhydrous glass as a shell material. The paper does not review maximum radii for tensile strengths; however, it does introduce bending strength. We realistically do not expect perfect results from our restructuring (somewhat primitive) manufacturing techniques. We include a filled shell structure using anhydrous trusses and processed regolith fill in our estimates. The structure does have lower tensile strength; however, the low density increases the station radius to over 20 kilometer in Table 3-1. We conservatively intend to aim for a much smaller radius closer to 3 or 4 kilometers.

**Station Stress:** We provide a brief review of the stresses produced in the rotating space station. Stresses are produced from air pressure and from centripetal forces. We compare these stresses in Figure 3-5 for a rotating cylinder. The logarithmic x-axis shows the rotation radius from 100 meters to 10,000 meters. The logarithmic y-axis shows the stress ranging from 1 kilopascal to 1 gigapascals. The cylinder shell is 20 meters thick. The shell rotates at a speed to produce one Earth gravity on the outer rotation radius. The shell centripetal hoop stress, σc, is the largest and is equal to gρR where g is the centripetal gravity, ρ is the material density, and R is the rotation radius. In this design, the circumferential air pressure, σa, is the next largest and is equal to P R/t where P is the internal air pressure, t is the shell thickness, and R is the rotation radius. We use the furnishing stress values from the 1977 NASA study [Johnson and Holbrow 1977]. The radial centripetal force and the radial air pressure force are both minimal in the overall stresses in the cylinder; see Figure 3-5. We assume a truss-like structure made of anhydrous glass filled with crushed regolith. The tensile strength of this structure is defined as 1500 MPa and the density is set to 1720 kilograms per cubic meter.

We present in Figure 3-6 the working stress for various materials. The working stress is the sum of the stresses introduced in Figure 3-5 and includes the stresses from the air pressure, the centripetal forces on the shell, and the centripetal forces on the internal structures and furnishings. The chart shows the working stresses in megapascals ranging from 1 to 10,000 on the logarithmic y-axis. The x-axis shows the outer rim radius of the rotating station ranging from 100 to 20,000 meters. The station is rotating at a speed to produce 1g at the outer rim. The chart includes the material stresses for four materials. The anhydrous glass has the largest tensile strength and aluminum has the smallest. Steel, with the highest density, creates the largest working stress. The filled structure, with the lowest density, creates the smallest working stress. Except for steel, all the materials support their working stress below the rotation radius of 10,000 meters. We use this radius as a maximum value because of material strength.

**Table 3-1 –Materials and Space Habitat Radius**

| Material | Tensile Strength (MPa) | Density (g/cm3) | Radius (km) |
|---|---|---|---|
| Molecular Nanotechnology | 50,000 | 3.51 | 343.9 |
| Anhydrous Glass (max) | 13,800 | 2.70 | 123.4 |
| Anhydrous Glass | 3,000 | 2.70 | 42.8 |
| Basalt fiber | 3,000 | 2.67 | 27.1 |
| Basalt rod (7mm) | 2,471 | 2.79 | 21.4 |
| Filled Structure | 1,500 | 1.72 | 21.1 |
| O'Neill Future | 2,068 | 3.12 | 16.0 |
| Titanium | 1,450 | 4.50 | 7.8 |
| Steel | 1,240 | 7.80 | 3.8 |
| Aluminum | 352 | 2.65 | 3.2 |
| Iron | 275 | 7.20 | 0.9 |
| Glass | 7.0 | 2.50 | 0.1 |

*Credit: Self produced using data from [O'Neill 1974] [McKendree 1995] [Bell and Hines 2012]; [Facts].*

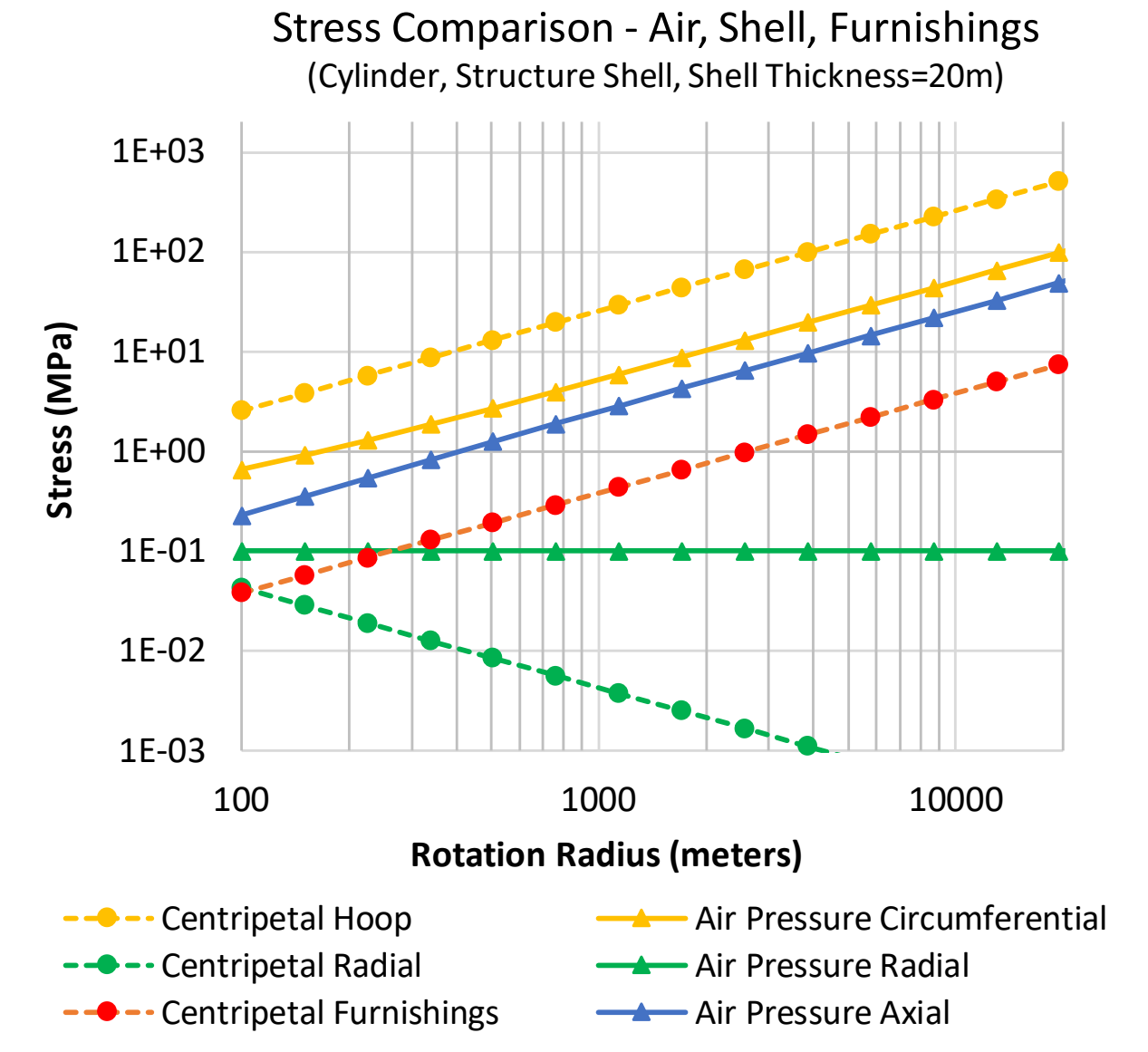

**Figure 3-5 – Space Station Stress Comparisons**

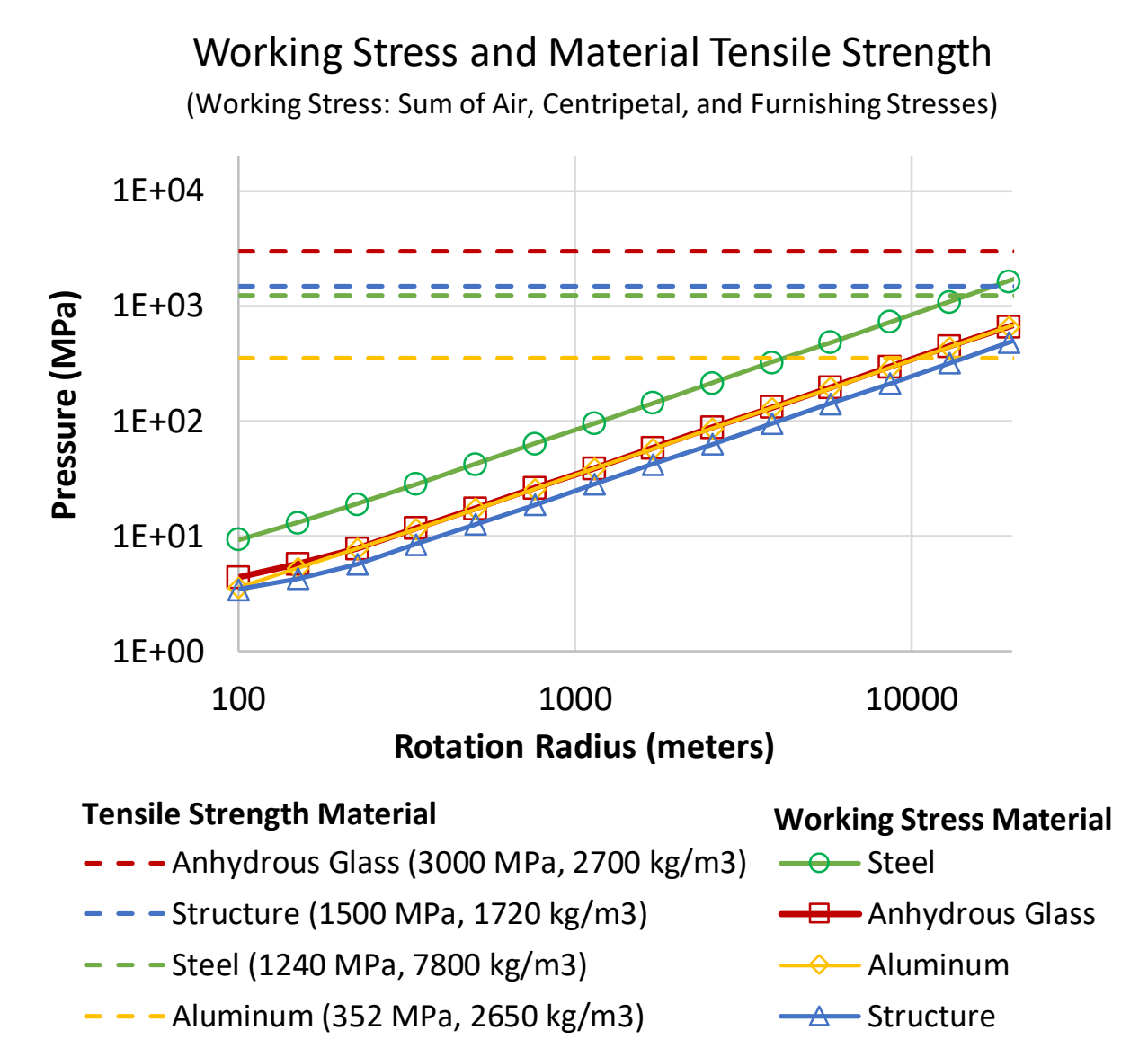

**Figure 3-6 – Space Station Stress Comparisons**

As shown in Figure 3-6, the material densities and tensile strengths affect the working stress. We include a chart in Figure 3-7 to show the effect of the shell thickness on the station stresses. We show the two largest stresses from Figure 3-5 – the shell centripetal stress and the air pressure stress. The chart shows the stresses on the y-axis and the shell thickness on the x-axis. The chart includes data from a torus station and a cylinder station. The torus is similar to the Stanford Torus with a major radius of 830 meters and a minor radius of 65 meters. The cylinder is similar to the O'Neill C-3 design with a radius of 1000 meters. With a higher ceiling and thicker atmosphere, the cylinder air pressure stress is greater than the torus air pressure stress. We use standard sea-level air pressure at the outer rim. In both designs, the air pressure stress decreases with the increasing shell thickness (σa is equal to P R/t). The Stanford torus was designed to use ½ standard sea level air pressure; as such, its air pressure marker is lower than torus air pressure line. The Stanford Torus has a thin shell of 1.68 centimeters [Johnson and Holbrow 1977]. We design the C-3 cylinder with a thick shell of 20 meters. The centripetal shell stress is slightly greater for the torus than the cylinder. This is because the cylinder is made using the filled structure shell and the torus is made with an aluminum shell. The centripetal stress is proportional to the density and aluminum is denser than the anhydrous structure. Additional analysis is recommended; however, this brief review of structure stress suggests that large stations with thick structure shells are viable.

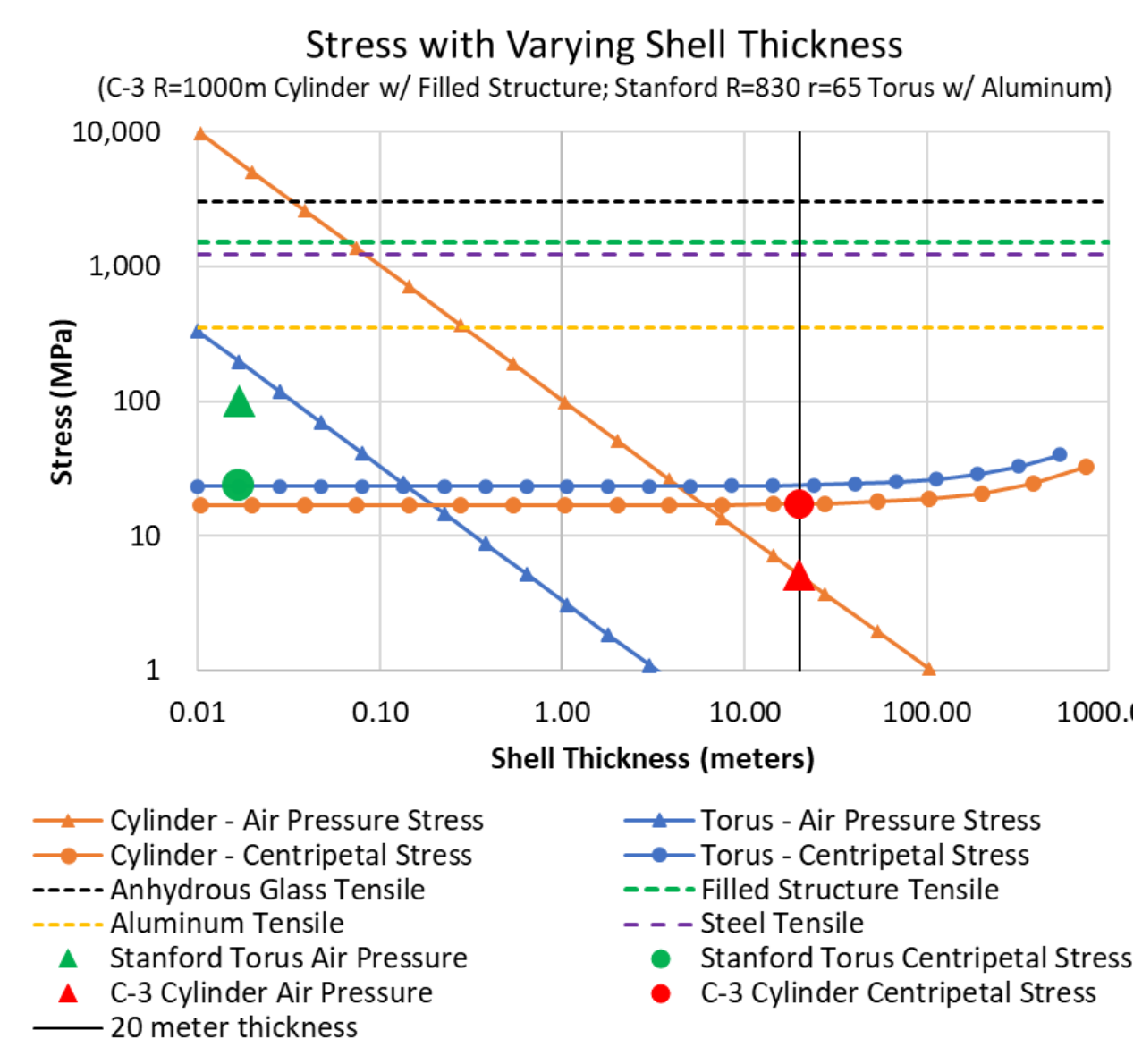

**Figure 3-7 – Space Station Stress Comparisons**

**Population:** We include in Figure 3-8 population estimates from various reports [Johnson and Holbrow 1977] [O'Neill et al. 1979] [Globus et al. 2007] [Brody 2013]. Everyone in a space station needs space to support their living, working, industry, and agriculture needs. Technical estimates of that space ranges from 35 to 200 square meters of projected area per individual. Historically, most designs use only projected floor values using only a single floor on the outer perimeter of the station for living space. The populations in this chart have been normalized to a projected surface area allocation of 67 square meters per person, which comes from a NASA study [Johnson and Holbrow 1977].

Figure 3-8 includes four O'Neill cylinder models (C-1 through C-4) [Johnson and Holbrow 1977]. The lengths of these cylinders are 10 times the radius. Figure 3-8 includes the Kalpana One cylinder, which is a rotating cylinder with a radius of 250 meters and a length of 325 meters [Globus et al. 2007]. This cylinder is not on the cylinder line in the chart because the Radius to Length (R/L) ratio for Kalpana One is 250/325 or about 3/4 instead of 1/10. The O'Neill cylinders use two counter rotating cylinders to reduce precession. The designers of Kalpana One used only one cylinder and reduced the length for rotational stability [Globus et al. 2007].

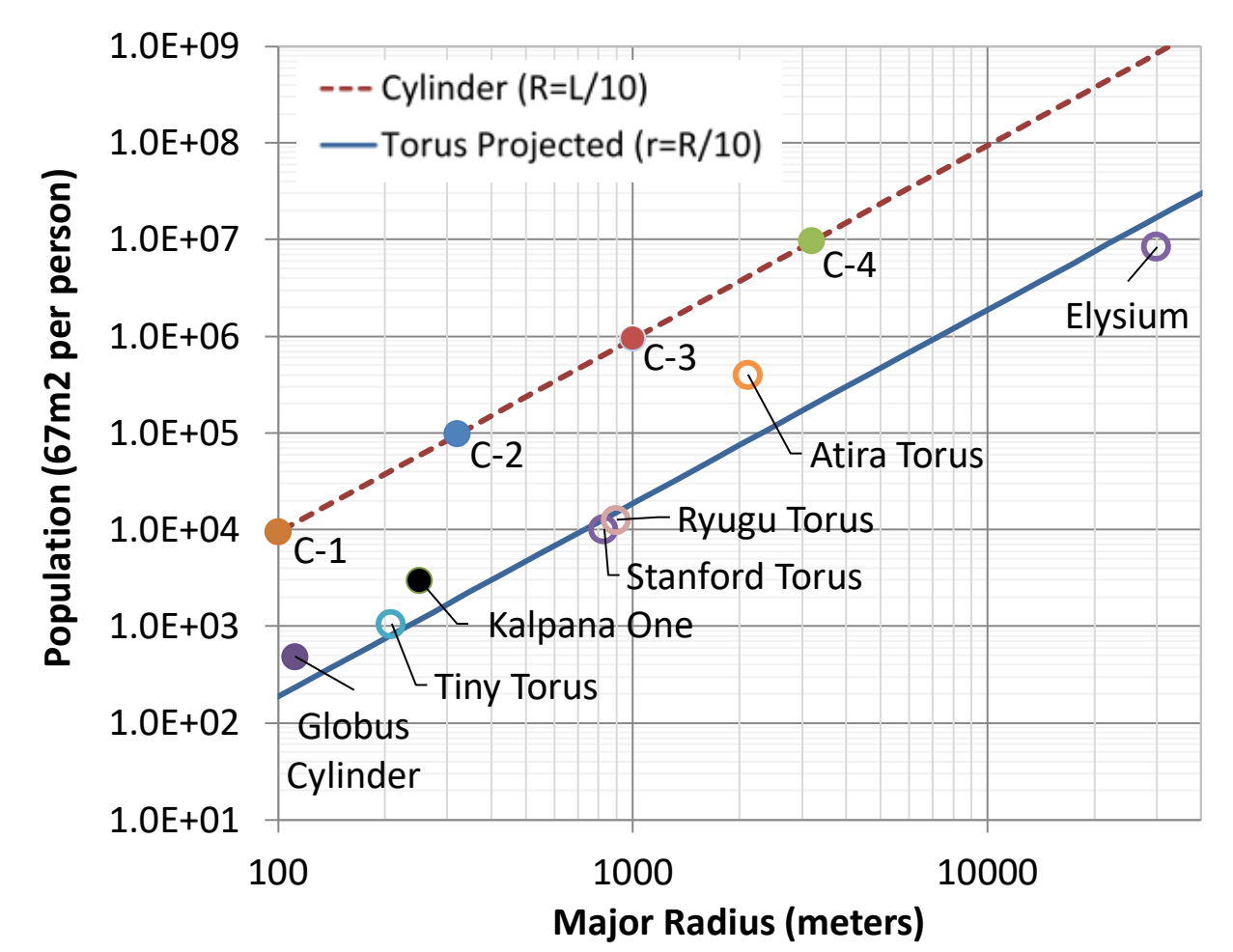

*Radius and population of space habitats*

**Figure 3-8 – Space Stations Radius and Populations**

The chart in Figure 3-8 also includes various torus designs for comparisons. We include the Stanford Torus and a Tiny Torus from a NASA study [Johnson and Holbrow 1977]. A 1974 Stanford study detailed a rotating torus design with minor radius of 65 meters and a major radius of 830 meters [Johnson and Holbrow 1977]. The Stanford Torus and the Tiny Torus both fall along the torus line in Figure 3-8. We include several of our study's restructured asteroid torus stations in the graph for reference. These results are for the single main floor of the torus. The Atira torus is above the torus r/R line because of its elliptic cross section. We also include the station from the movie Elysium [Brody 2013]. For consistency, the Elysium population in Figure 3-8 uses the same population density metric (67 meters squared per person). We plan to use multiple floors and significantly increase the available living space of our stations.

**Multiple Floors:** Adding floors to a structure greatly increases the available floor space. Examples of multiple floor structures include underground cities, submarines, cruise ships, and skyscrapers. Historic and modern underground cities exist [Garrett 2019]. Entrepreneurs have begun to convert abandoned military missile silos into multiple floor homes and underground cities [Garrett 2019]. Limited space and costs in urban environments promote high rise living.

We contemplated whether there are other issues with creating and using many floors. Researchers and developers believe the biggest problems for underground cities are not technical but social [Garrett 2019]. Studies have found that people living in highrises suffer from greater mental health problems, higher fear of crime, fewer positive social interactions, and more difficulty with child rearing [Barr 2018]. Luckily, literature and the lessons learned from highrise public housing failures offer knowledge and experience to address such issues. NASA studies have considered the oppressive closed-quarters ambience of a space station to be a risk to the colonists' psychological well-being [Johnson and Holbrow 1977] [Keeter 2020]. Researchers offer approaches to address these risks. With proper planning and space allocation, it seems that using many floors is acceptable for space station habitation.

Figure 3-9 illustrates the cross-sections and the floor structures for an inner and outer torus. The inner torus has a circular cross-section in Figure 3-9a. The outer torus has an elliptical cross-section in Figure 3-9b. We use this level of detail in our analysis to compute the mass of the station and the floor surface area. We hope to use this level of detail in a future stress analysis.

For large space stations, we feel the number of floors becomes excessive. The material in an Atira sized asteroid can support the construction of a circular torus with a major radius of 5.5 kilometers and a minor radius of over 800 meters. With a 5-meter spacing between floors, this torus could have 80 stories underground. We feel many of the lower floors would not be desirable for residential use. Over the course of our study, the cross-section of our torus shaped station evolved from a circle to an ellipse. This reduced the number of floors. It increased the surface area and improved the vista on the main floor. We believe the elliptical cross section will provide much of the structural strength of a circular cross section. We also lowered the main floor – again reducing the number of floors and improving the vista. The outer torus in Figure 3-9b illustrates an example station with the ellipse cross-section and a lowered main floor. For scale, the spoke of the elliptical torus is the same diameter as the inner torus.

**Multiple Floor Population:** We use the space station geometry to compute the available surface area. In a torus, the main floor surface area is the floor width (twice the minor radius) times the station circumference (two pi times the major radius). As we go deeper into the torus, the floor width decreases as the square root of the depth. The floor circumference increases linearly with the radius (depth). We sum these floor areas to compute the total square footage for the station. We produce a population estimate by dividing that total by a floor space allocation metric. We use the same analysis approach on the other station geometries.

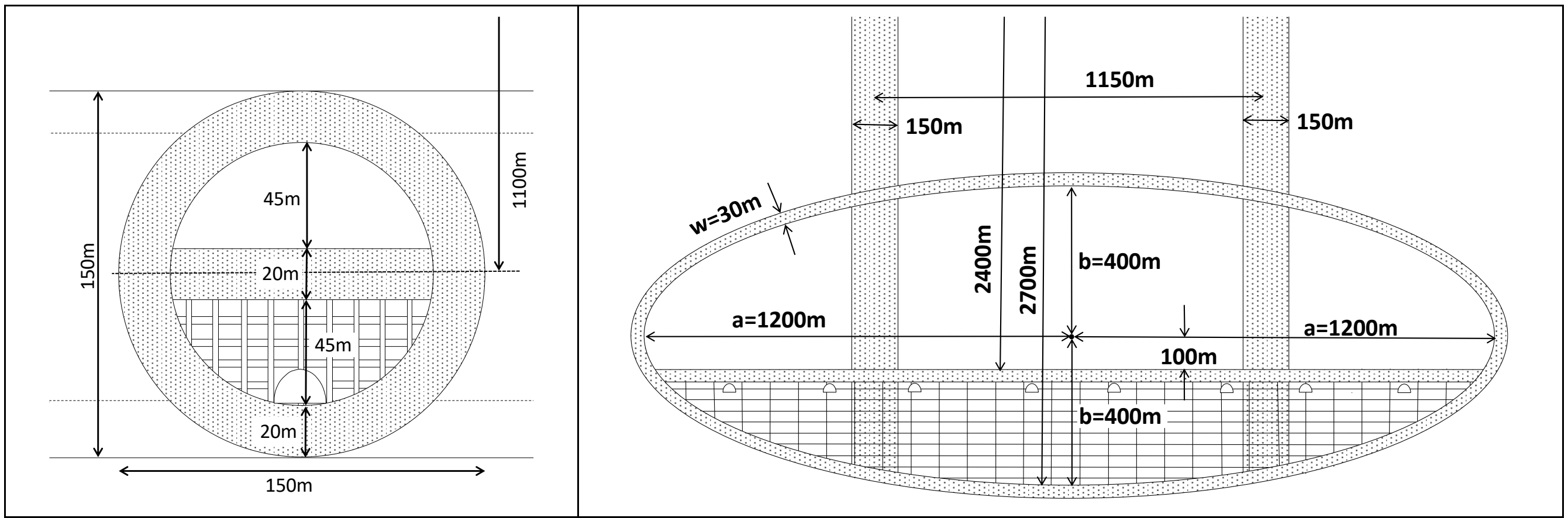

*a) Cross-section of inner torus* | *b) Cross-section of outer elliptic torus*

**Figure 3-9 – Torus Cross Sections – Multiple Floors and Elliptical Design**

We compare the population of stations with a single floor and with multiple floors in Figure 3-10. This chart includes the potential populations for some of literature's cylinder and torus geometry space stations [Johnson and Holbrow 1977] [Globus et al. 2007]. We show the major radius of the station along the horizontal axis in Figure 3-10. This chart extends the data from Figure 3-8 and for consistency we continue to use the population density of 67 square meters per person. The torus stations assume the outer half of the torus is filled with floors. We limit the number of cylinder floors to better compare to the torus geometries and to retain the open view in the cylinder. For both the cylinder and torus geometries, adding floors dramatically increases the square footage to support larger populations. The elliptic Atira torus design has a major radius (R) of 2116 meters and minor radii of 1003 meters and 334 meters (r). Because it has an elliptical cross section and r=R/6.33, the station supports more population than the r=R/10 line. The top floor supports almost 400,000 people using the 67 square meters per person. It supports a maximum population of 21.5 million and a realistic population of 3.5 million.

We again include the station from the movie Elysium. Elysium is a torus with a major radius of 30,000 meters and a minor radius of 1500 meters [Brody 2013]. In the movie, Elysium has a population of 500,000 and that provides 1131 square meters of space for each individual. Using only 20 of the multiple floors possible, Elysium would increase its floor area from 565 square kilometers to 11,852 square kilometers. With 67 square meters per person, the 20 floors of Elysium would support 177 million people. Using the generous 1131 square meters per person used in the movie, it would still support over 10 million people.

**Multiple Floor Visualization:** To help visualize these floors, we include a cut through diagram in Figure 3-11. This torus station has major radius of 2300 meters. The torus has an elliptic cross section with minor radii of 400 meters and 1150 meters. The top floor is 100 meters below the major radius distance. There are 17 floors under the top floor with 15 meters between each floor. There are 82 floors in the skyscraper towers with 6 meters between each floor. The shell is 20 meters thick. The ceiling in the torus is 500 meters above the top floor. This provides an open vista of 6 kilometers over the curved top floor. The top floor in this station would have 33.6 million square meters of floor space. The 18 floors in such a station would have 487 million square meters of floor space. Using 155.2 square meters per person [Johnson and Holbrow 1977], the floors of this station could support over 3 million people. The projected floor area density of 67 square meters per person would support over 7 million people. Other constraints (such as gravity ranges and psychological well-being) will impact the floor space and reduce the population to a lower and more realistic value.

**Station Mass:** For our analysis, we use the masses of both the station and the asteroid. Our analysis leads to charts showing population as a function of mass and station geometry. The station mass uses the volumes of the towers, spokes, outer shells, floors, fill, and the shuttle bay. We derived the mass of those components from material densities and volume of individual panels, tiles, rods, and fill. In some analysis, we use different densities for the components and materials. The basalt tile density is assumed to be 2,790 kilograms per meter cubed. The asteroid densities vary depending on composition and porosity and ranges from 1,192 to 2,566 kilograms per meter cubed. The fill material uses that asteroid material and is packed and includes melted regions for stability. We set the fill material density to 1,721 kilograms per meter cubed. This is in the density range of gravel. We use that as the density of our fill structure too. In some analyses, we compute and use the density of complete structures in the station. We include example densities of the

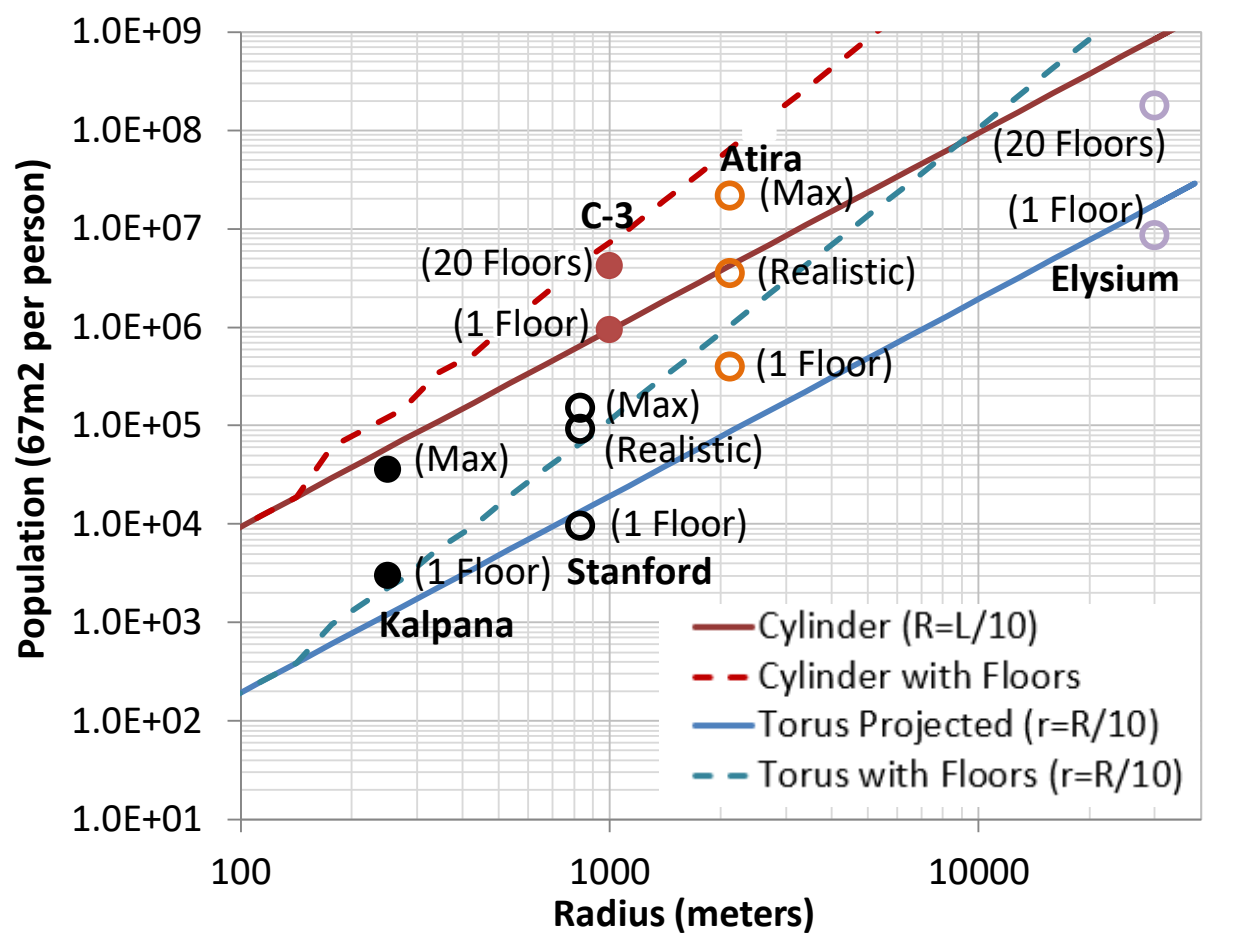

*Population shows significant gains with multiple floors*

**Figure 3-10 – Torus and Cylinder Populations**

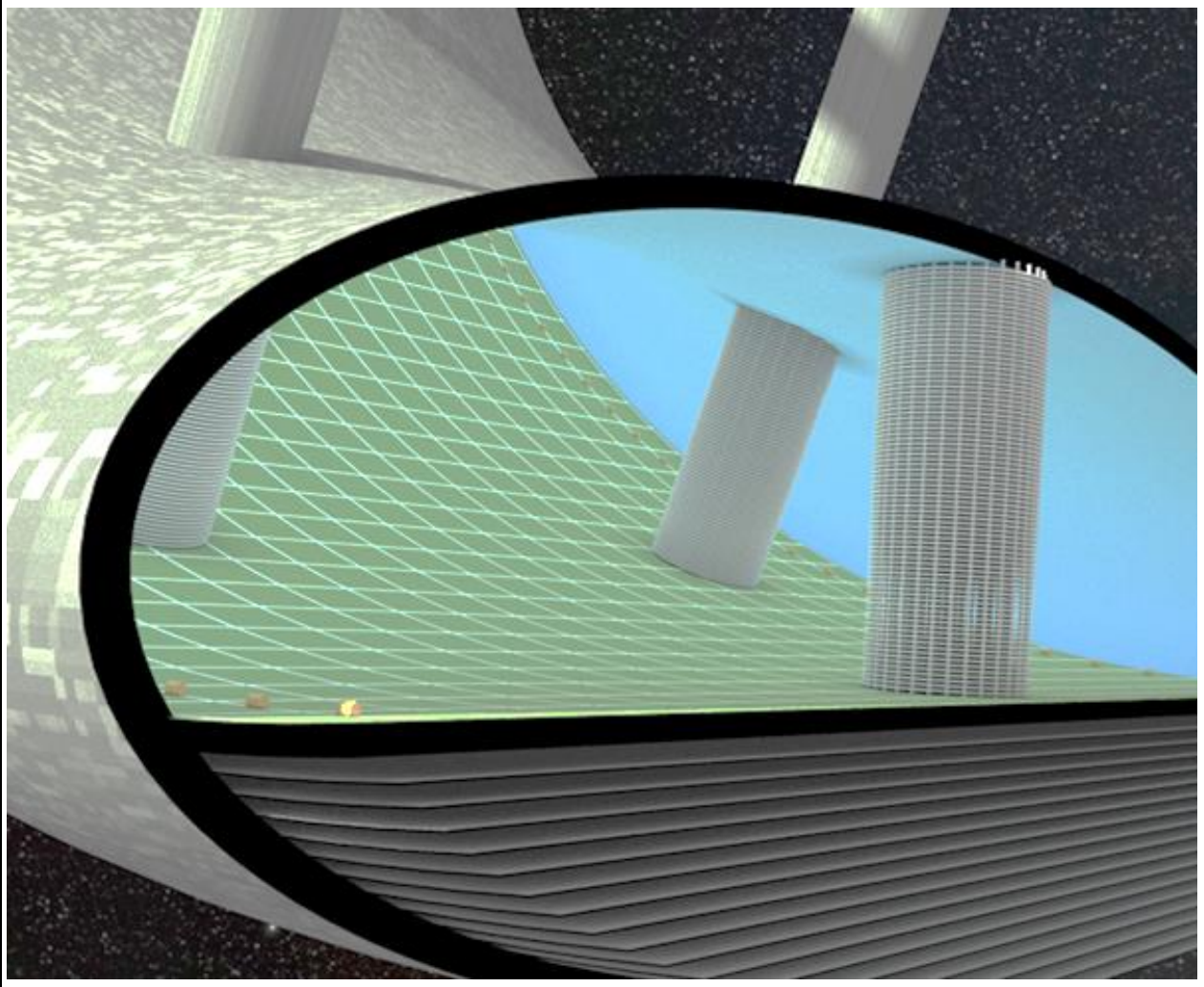

*Credit: Self produced with Blender using Background Milky Way: ESO/Serge Brunier, [Brunier 2009] [CC BY-4.0]*

*Large Torus Space Station: Cut through diagram showing 17 subfloors and 82 floors in spoke towers*

**Figure 3-11 – Exploiting Floors in Space Stations**

shuttle bay, spokes, and different floor spacings in Table 3-2. We sum the volume and mass of the individual pieces in the components of the station. Figure 3-9 illustrates the level of detail for this analysis. The mass includes the material needed to construct the trusses, to fill the exterior walls with regolith, and to cover the exterior and floors with panels.

We show in Figure 3-12 the summed masses from this analysis. This includes mass estimates for the four station geometries. All four of the geometries use a thick filled outer shell. This chart shows the radius along the x-axis and the station mass along the y-axis. It is close to a linear relationship between the radius and the mass. It is not linear because our analysis varies the dimensions of structures (such as shells and tethers) with the changing radius. These dimension changes provide additional support, strength, or radiation protection. Additional structures (such as spokes and towers) are also added to provide strength in large stations.

**Floor Allocation Usage:** A NASA design study included surface usage metrics such as 49 square meters per person for residential, 10 for open spaces, 12 for transportation, and 61 square meters per person for agriculture [Johnson and Holbrow 1977]. Their total surface area was 155.2 square meters per person of surface area. Some structures, including houses, used multiple stories to support that requirement. Using multiple stories results in 67.0 square meters of projected surface area per person. Another NASA study allocated 157.1 square meters per person and 70.5 square meters of projected area [O'Neill et al. 1979].

We include this space usage distribution in Figure 3-13. We organize their categories into open areas, support, agriculture, industry, and residential. This chart also includes our estimates for different floors spacings. With a fixed floor spacing of 15 meters, our updated space allocation provides each individual a similar 65.5 square meters per person. Using this metric on previous example station, the 33.6 million square meters of the first floor by itself would support a population over 500,000 individuals.

Like the NASA studies, we also use multiple story structures to support our usage requirements. We also account for the multiple floors in our station designs. We consider floor spacings of 5, 10, and 15 meters. Different usage metrics are shown for the categories and floor heights; see Figure 3-13. NASA used various heights for the different space categories (an average height of 11.2 meters).

Even with the different floor spacings, many of our detailed categories use the same volume; as such, their space requirement simply scales with the changing floor distance. As an example, the agriculture allocation increases from 13 to 39 square meters per person as the floor distance decreases from 15 to 5 meters. The total agriculture volume remains constant at 195 cubic meters per person.

We increase the open space in the NASA study from 10 square meters per person to 18.7 square meters per person. This includes some of the public areas in the support category and adds an extra 5 square meters per person. The top floor is more suited for open space, tourism, and recreation. The top floor in our example station could meet the open space requirement for over 1.8 million people. Lower floors may also be used for residence and leave more of the top floor open to help meet the open space requirement of 18.7 square meters per individual. In large stations, the top floor can often meet the open space requirement. There is also opportunity to be creative and use low and high gravity regions to create additional open areas.

**Table 3-2 – Densities of Station Components**

| Density | Value (kg/m3) | Density | Value (kg/m3) |
|---|---|---|---|
| $\sigma_{basalt}$ | 2790 | $\sigma_{spoke}$ | 337.4 |
| $\sigma_{rod}$ | 2790 | $\sigma_{floor5}$ | 32.2 |
| $\sigma_{panels}$ | 2790 | $\sigma_{floor10}$ | 17.0 |
| $\sigma_{bay}$ | 2291 | $\sigma_{floor15}$ | 11.2 |
| $\sigma_{fill}$ | 1721 | $\sigma_{air}$ | 1.23 |

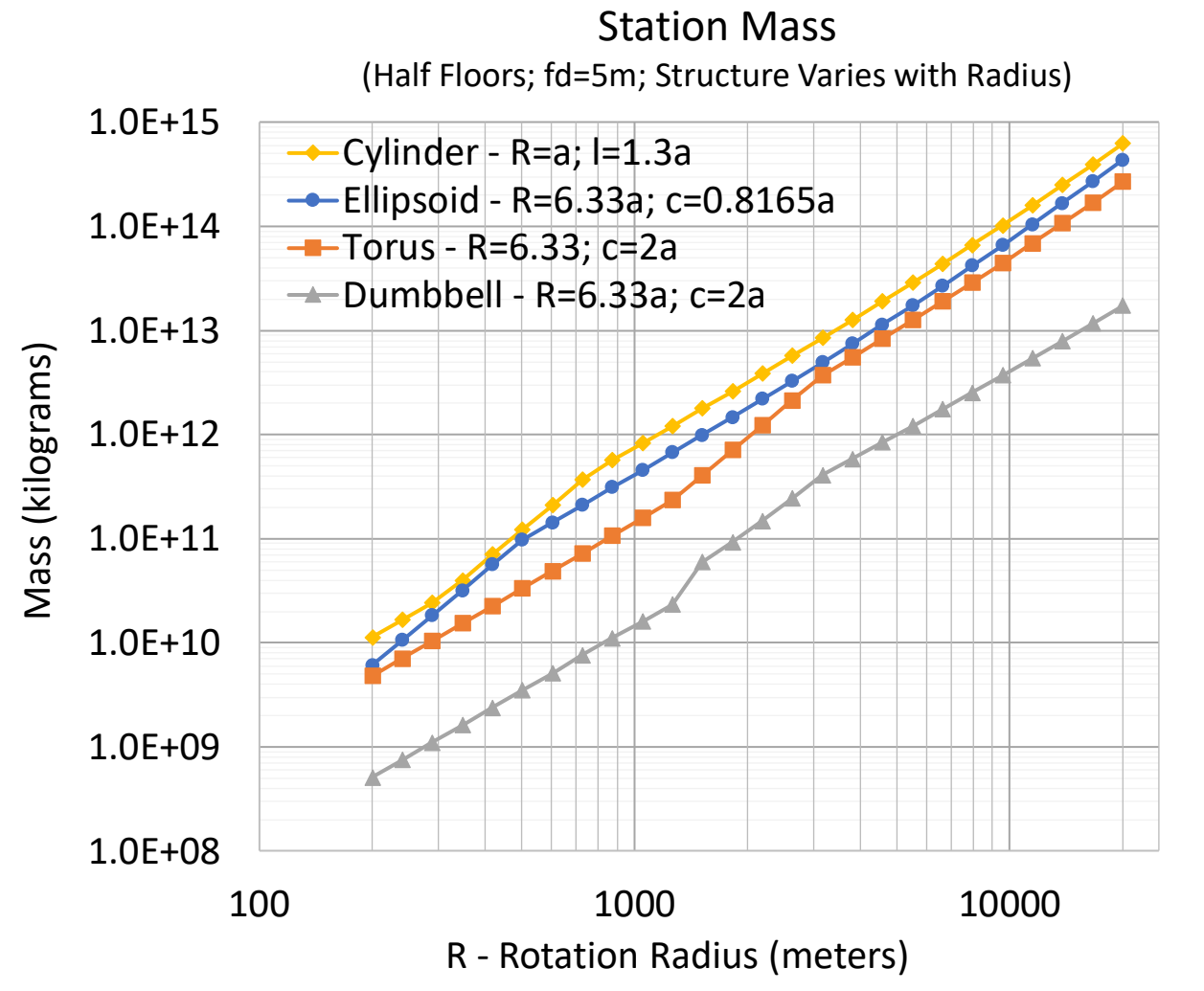

**Figure 3-12 – Station Mass Summed from Individual Station Components**

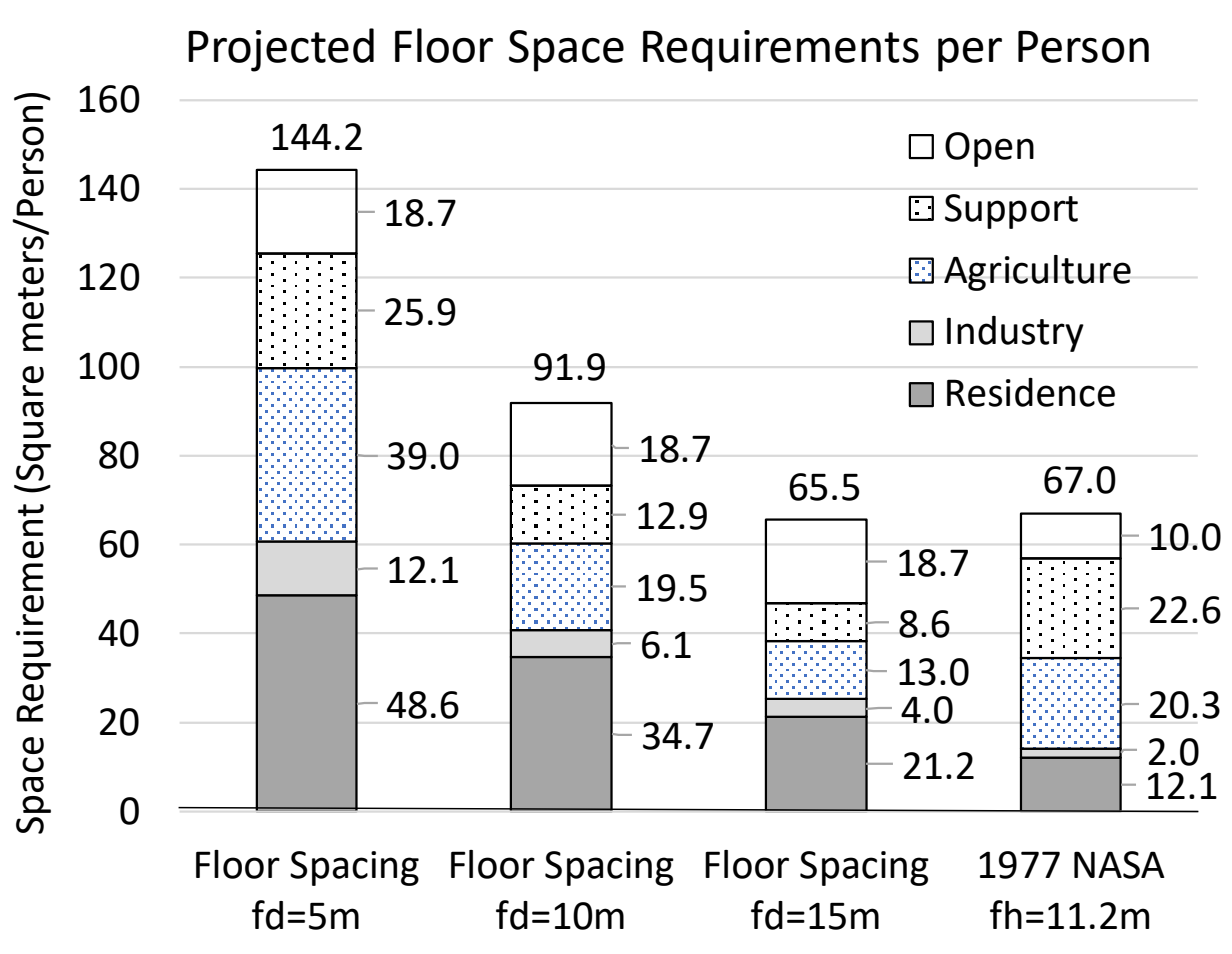

**Figure 3-13 – Floor Space Requirements per Person**

The average home size has increased by 50% since the 1970s; as such, we increased the requirements for the residence. We increased the industry area requirements. This will be important to make the station self-sufficient with export products.

NASA studies [Johnson and Holbrow 1977] [Bock, Lambrou, and Simon 1979] and more recent studies [Soilleux and Gunn 2018] [Fu et al. 2016] have evaluated the use of agriculture and plants in a closed system environment. We decreased the requirement for agriculture using those recent metrics. The biggest change is in the height of the agriculture areas. The NASA studies used conventional farming and assumed heights of 15 meters. We use short stacked shelves of growing areas. We can even stack multiple shelves of agriculture in the 5 meter floor spacing. These changes reduced the required agriculture volume from 915 to 195 cubic meters. There are also even more recent advancements in hydroponics, aeroponics, and new biological approaches [Kersch 2015] [Cornall 2021]. Those advancements offer more reduction to the crop growing requirements. We conservatively do not include those most recent reductions in our estimates.

**Floor Space Allocation Examples:** We evaluate an example torus station with a major radius of 2300 meters and an elliptic cross-section with minor radii of 400 meters by 1150 meters. It also includes two inner tori, each with a radius of 1100 meters and a cross section radius of 75 meters.

This design has over 34 square kilometers of floor space on the elliptic main floor alone. With a floor spacing of 15 meters, there are 18 floors with 487 square kilometers in the outer torus. The 15-meter spacing provides room for some multistory buildings. This design has two inner torii and creates 8 dividers, 8 towers, 16 spokes, and two shuttle bays; see Figure 1-1. Including all surface area on all the floors and structures there is a total of 1070 million square meters of projected floor space.

Some regions of the space station are not conducive to long term habitation. Other regions are ideal for parks and recreation. We show in Figure 3-14 the space available in the major components of the Atira station. With the total of 1070 million square meters and a floor space allocation of 65.5 square meters per person, we find that the station could hold a maximum of 16 million people.

We show a bar chart in Figure 3-15 as a final preview of the station floor space allocation. The chart continues with the same example torus station. The chart example uses the entire space station to support a large population. With the floor spacing of 15 meters, there are 18 floors. We assign the top 6 floors to be the habitable region and the bottom 12 floors to be the lower ellipse region. By adjusting the portion used in each of the station components, the chart in Figure 3-15 shows this station could comfortably support 10 million people. Most of the population lives on the first six floors of the station. This region has the most Earth-like gravity. We expect most of the top floor would provide the open recreation regions for psychological well-being. Below the top floor, facilities such as aquariums and botanical gardens could create more open regions. Agriculture is likely to be in the lower portions of the station. Industrial manufacturing is highly automated and requires few people to support and could reside in the least desirable regions such as the higher-gravity lowest floors and the low gravity spokes. Even with 10 million people, there is still significant space available for storage. This storage would initially be used for the inventoried metals and volatiles from the asteroid restructuring process. It could be later used for more growth, tourism, industry, or agriculture. Conservatively, we aim to support a population of 700,000. It is comforting to see that this station could support many more people.

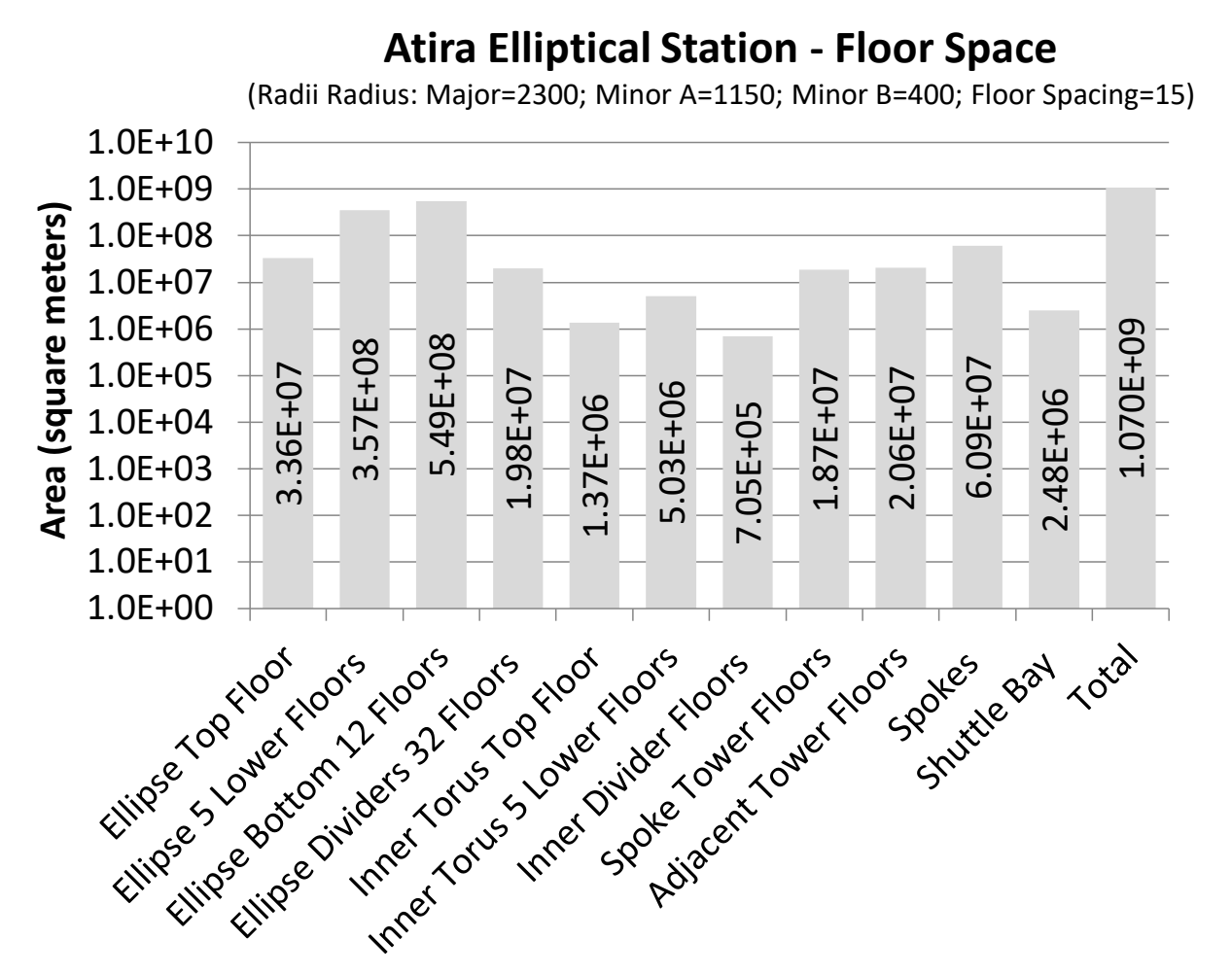

**Figure 3-14 – Space Available on Station Components**

### 3.2.2 Geometry Rotational Stability

In the *Elements of Spacecraft Design*, Charles Brown recommends that the desired axis of rotation should have an angular moment of inertia (MOI) at least 1.2 times greater than any other axis to provide rotational stability [Brown 2002]. This constraint is important in developing the geometry sizes. A recent paper investigated the rotational instability of long rotating cylinder space stations [Globus et al. 2007].

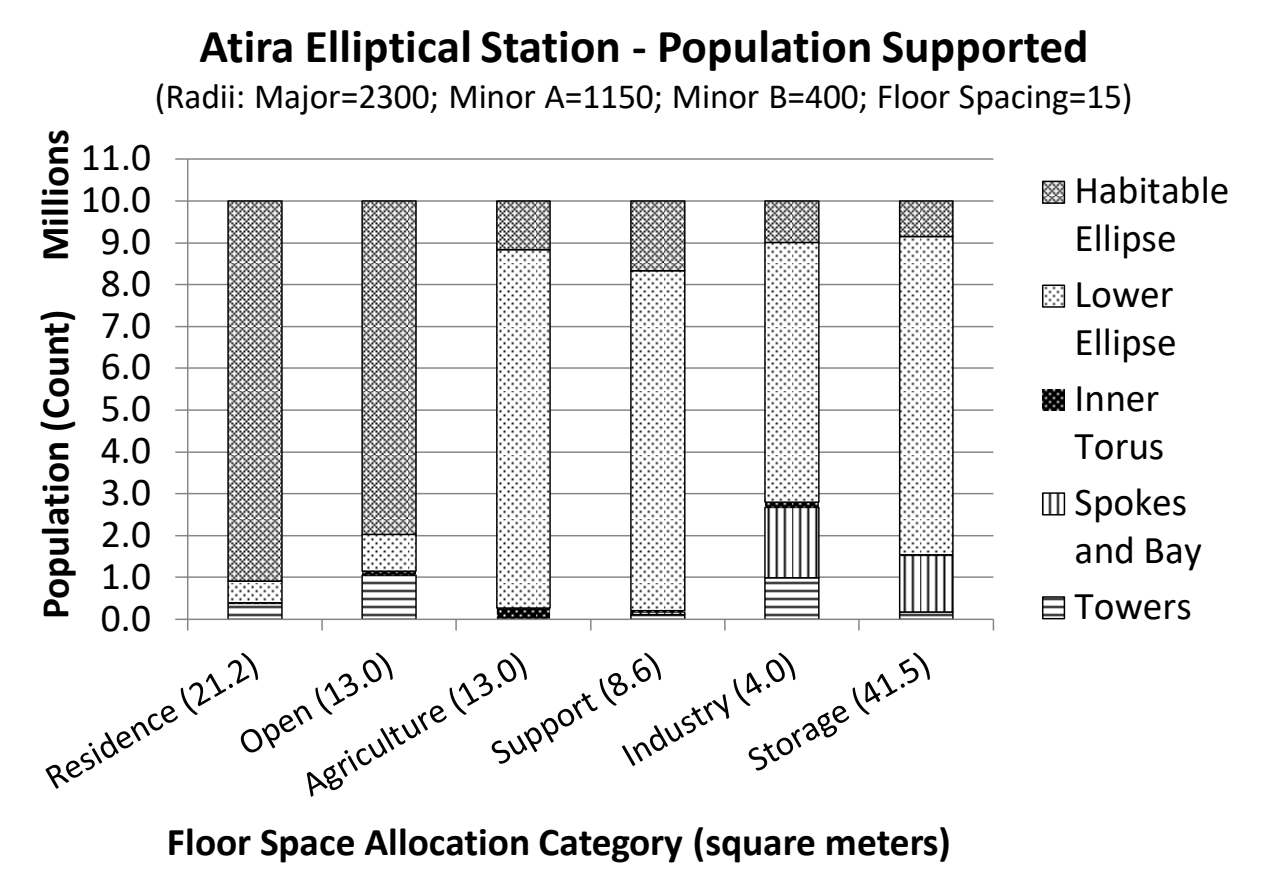

**Figure 3-15 – Atira Station – Space Allocation**

That study imposed a limit on the cylinder length to passively control the rotational stability. The study also mentioned that a perfect sphere has a similar issue with rotational stability.

We have extended the Global cylinder analysis to three other geometries. We show the cross sections of the four geometries in Figure 3-16. The diagrams show the floors half filling the station with labels on the radii and other dimensions. We have analyzed using thick and thin shell stress equations. We present our results using thin shell hollow geometries. Our 20 meter thick shells are "thin" compared to our large station radii. We provide terse overviews of the Globus Cylinder Station analysis [Globus et al. 2007] and our Oblate Ellipsoid station analysis. We include only final summary statements for the Torus and Dumbbell geometries.

**Flat-capped cylinder:** The cylinder is short and squatty (hatbox) and has a rotation radius (R) larger than the height (h); see Figure 3-16a. The moment of inertia (MOI) for the cylinder uses the mass (M), radius (R), and height (h). Along the longitudinal axis (the x-axis in Figure 3-16a) the inertia is $Ix=MR^2$ for a cylinder without endcaps. Along the other two axes, the inertias are $Iy=Iz=(1/2)\ MR^2 + (1/12)\ M\ h^2$. To be stable, Ix >= 1.2 Iy and we find h<=2R. With endcaps, we must include the MOIs for those end disks. Globus uses a thin shell and assumes the disks and shell have the same thickness and density. For the total system $Ix = M\ R^2\ (1 + R/2h)$ and $Iy = M\ (R^2/2 + h^2/12 + R^3\ /(4h) + Rh/4)$. The end caps have a detrimental effect on stability. Globus finds the system would be stable when h=1.3R with the endcaps instead of 2R without the endcaps [Globus et al. 2007].

**Oblate Ellipsoid:** An oblate ellipsoid has two major axes that are the rotation radius length (a=b) and a third minor axis with a length (c) that is shorter than the other two; see Figure 3-16b. The moment of inertias for the oblate ellipsoid uses the mass (M) and the rotation radii (a, b, and c). The ellipsoid rotates about the minor axis. The MOI of the ellipsoid around the x-axis is $Ix = (2/3)\ M\ a^2$ Along the other two axes, the inertia is $Iy = Iz = (1/3)\ M\ (a^2 + c^2)$. To be stable, Ix >= 1.2 Iz and we find the length of the minor axis c must be less than 0.8165 times the major axis (a or b) length. This mathematically validates that a sphere would not be stable.

**Dumbbell:** We analyzed a dumbbell with spherical nodes; see Figure 3-16c. The rotation x-axis and the y-axis have identical MOIs (Ix and Iy). The z-axis MOI is much smaller than the rotation x-axis. Having Ix much greater than Iz would mistakenly imply the dumbbell is stable. There is no reason why the station would prefer to rotate about either of the perpendicular axes to the z-axis. Fortunately, with two equal MOIs the instability is algebraic and not exponential [Fitzpatrick 2011] [Fitzpatrick 2023]. As such, although the station is formally unstable, its rotation would settle into a limit cycle where the rotational axis wobbles slightly. For typical designs, the dumbbell will have a rotational wobble. The impact from this characteristic requires more analysis.

**Torus:** We also analyzed tori with circular and with elliptic cross sections. The drawing in Figure 3-16d shows the elliptic cross section. We have found torus designs would always be rotationally stable for typical dimensions of elliptic and circular cross section.

Table 3-3 contains a summary of our current stability equations for the four station geometry types. These equations represent moments of inertia for thin-shell hollow geometries. We evaluated hollow and solid geometries and found minimal (or no) difference in the MOI results. This suggests that the half-filled station will have the same stability MOI characteristics. We have begun a similar analysis using the shell, air, and floor densities (see Table 3-2) with these multiple floor geometries (see Figure 3-16).

### 3.2.3 Station Gravity Ranges

We cover three aspects of centripetal gravity in the following paragraphs. We review the health impact of gravity, viable ranges of gravity in the multiple floor rotating stations, and a specific station example for its centripetal gravity.

**Gravity Health Impact:** Earth's surface gravity ranges from 0.996g to 1.003g. Our human bodies have adapted to this narrow range. We need to consider what range of gravity in the space station will be acceptable for the colonists' health.

Astronauts in the International Space Station (ISS) experience microgravity. This microgravity negatively affects human health by weakening bones and muscles. This increases the risk of osteoporosis and cardiovascular problems. Planned habitats on the moon will subject the residents to 0.16g. Habitats on Mars will subject residents to 0.38g. These gravities may be insufficient to prevent the related health problems [Boyle 2020].

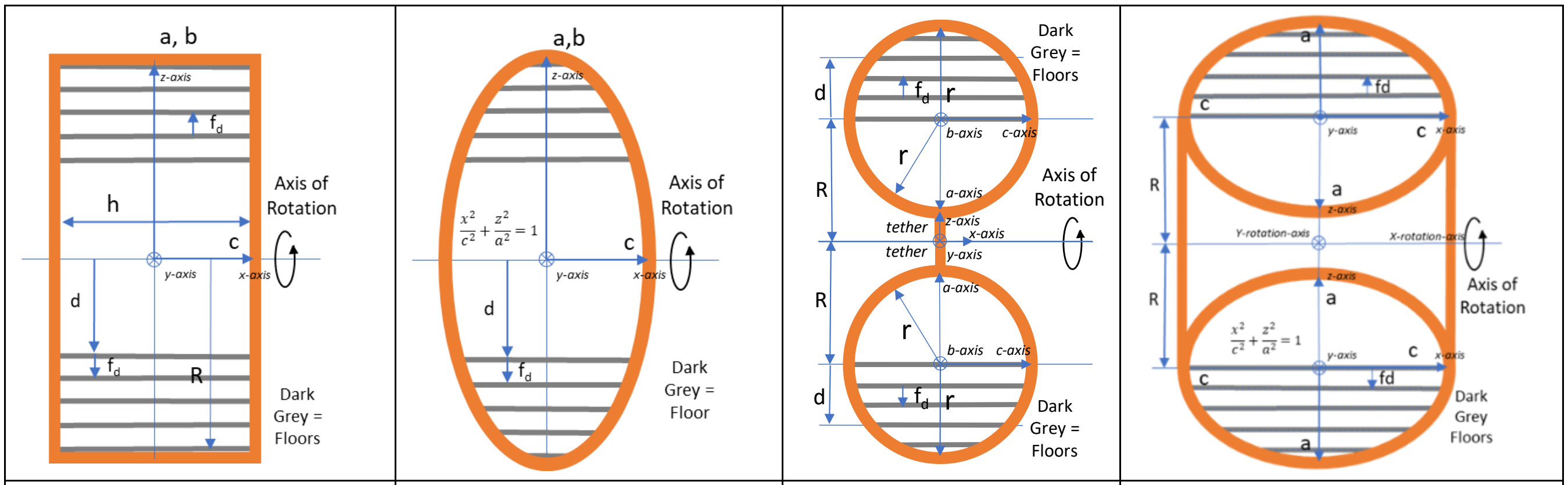

a) Cylinder Station | b) Ellipsoid Station | c) Dumbbell Station | d) Elliptic Torus Station

**Figure 3-16 – Station Cross Sections – Floors Half Filling Station**

| Table 3-3 – Geometries and Rotational Stability | | | |
|---|---|---|---|
| **Geometry** | **Key Stability Factor** | **Rotational Stability** | **Notes** |
| Cylinder | c < 0.65 a | Hatbox cylinders can be stable | Flat endcaps |
| Ellipsoids | c < 0.8165 a | Oblate ellipsoids can be stable | Sphere stations are not stable |
| Dumbbell | Ix/Iz=1.2 when $R^2 \geq \frac{1}{3}(1.2c^2 - 0.8a^2)$<br>Ix/Iy is less than 1.2 for all R | Dumbbells are formally unstable but only experience a rotational wobble. | Tether mass and radius are much less than the nodes. |
| Elliptical Torus | $R^2 > 1.5 (c^2\text{-}a^2)$ | Elliptical tori are stable for our designs where R>=6.33a and c=2a or c=3a | Torus only – inner docking station and spokes not included |
| *Credit: Self produced by extending cylinder concepts from [Globus et al. 2007] [Facts]* | | | |

There could be issues with high gravity too. A recent paper determined a muscle strength upper limit of 1.1g [Poljak, Klindzic and Kruljac 2018]. They also found gravity maximum tolerance could be increased by a factor of four to 3g or 4g (with the strength of a few elite athletes).

**Geometry Gravity Ranges:** We can design the station with different radii and rotation rates. This will vary the gravity values over the station. The 1977 NASA study used a gravity range to 0.9g to 1.0g and offered a relaxed constraint range of 0.7 to 1.0g [Johnson and Holbrow 1977]. In the paper Advanced-Technology Space Station for the Year 2025 [Queijo et al. 1988], the authors designed a torus station with a gravity range from 97% to 103% Earth gravity. They felt this range would not have any significant influence on human physiology and performance. Our main concern is the gravity range over the habitable portions of the station. We focus on the gravity values from the top floor to the outer shell of rotating stations.

To provide multiple floors, we also advocate relaxing the gravity range constraint. We typically design a gravity range of 0.95g to 1.1g for most of our half-filled multiple-floor design in the large restructured asteroid stations. The top floors are where most of the population will spend most of their time. We want these floors to have the most Earth-like gravity. We aim for a gravity range of 0.95g to 1.05g over those floors to minimize the health risk to the residents. We note that some "higher" regions on the curved floor of spheres, spoke towers, cylinder endcaps, and torus dividers will provide lower gravity environments. Only the "lowest" floors near the outer rim of the designs will have the higher gravities. We recognize the larger gravity values on the outer rim will require stronger materials and structures in the station design.

We consider two classes of rotating stations. One class rotates about an axis outside the habitable region and includes the torus and dumbbells. The other class rotates about an axis inside the habitable region and includes the cylinder and ellipsoid. We also vary the number of floors in the station volume. In one case we fill the habitable volume halfway and the other case we fill the volume completely. We include Figure 3-17 to illustrate these classes and cases. Figure 3-17a shows a half-filled volume rotating about an external axis. Figure 3-17b illustrates a half-filled volume rotating about an axis internal to the habitable volume. Figure 3-17c shows a fully filled volume rotating about the external axis. The circular outer shells in the figures should be squished to represent the elliptical cross section for the torus or the prolate ellipsoid for the dumbbell. The circular outer shell in Figure 3-17b should be squared to represent the cylinder.

We first consider the half-filled stations in Figure 3-17. We have taken liberty in the term "half-filled." The geometry shape and providing a habitable gravity range determines the actual fraction filled. In Figure 3-17a and b, we see that the major radius is R to the floor at the center of the space station. The centripetal gravity at the outer shell will be $g_{max}$ and will be equal to $(R+h)\,\omega^2$, where ω is the station rotation rate in radians per second. The minimum gravity, $g_{min}$, will be at the top floor and will be equal to $R\omega^2$. We create a relationship between the distance h and the distance R. The radius R is a multiple, m, of the distance h; i.e. R=mh. On a torus, this will be the ratio of the major and minor axes. We first find the minimum and maximum gravities for half-filled volumes as shown in Figure 3-17a and b as $g_{max} = (R + \mathrm{h})\omega^2$ and $g_{min} = R\omega^2$. Reorgnizing we find:

$$\frac{mh\omega^2}{g_{min}} = \frac{(m+1)h\omega^2}{g_{max}}$$

Divide each side by h and $\omega^2$ and reorganize, we find the scaling factor would be:

$$R/h = m = \mathrm{g}_{min}/(\mathrm{g}_{max} - \mathrm{g}_{min})$$

For the fully filled station with the external rotation axis in Figure 3-17c, the maximum centripetal gravity would be at the radius R+h and the minimum at the radius R-h. With a similar analysis, we find for the Figure 3-17c the scaling factor would be:

$$R/h = m = (\mathrm{g}_{max} - 2\mathrm{g}_{min})/(\mathrm{g}_{max} - \mathrm{g}_{min})$$

We do not include a drawing of the full-filled internal rotation axis in Figure 3-17. The minimum gravity, $g_{min}$, would be at the center on the rotation axis and be 0g. The floor on the outer rim would be $g_{max}$ and be equal to $R\omega^2$. The scaling factor, m, would not be defined because h would be equal to zero. The low gravity near the center would make this region

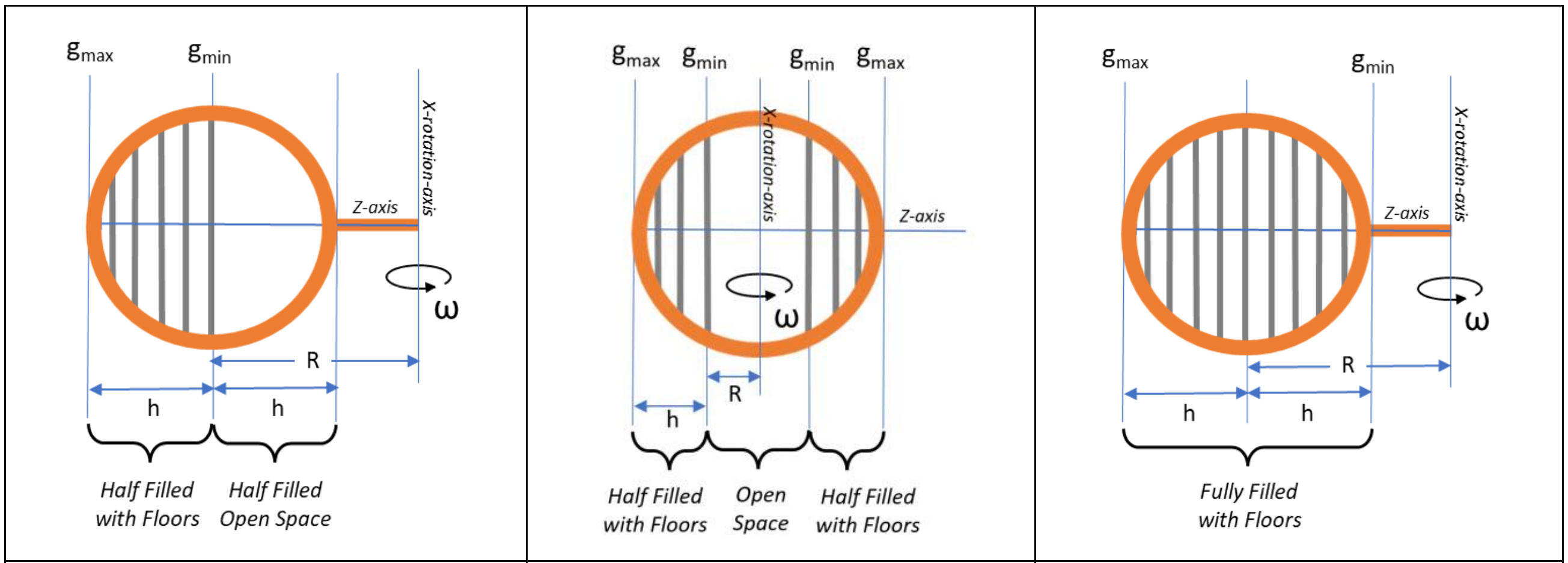

*a) Half Filled External Rotation* | *b) Half Filled Internal Rotation* | *c) Full Filled External Rotation*

**Figure 3-17 – Gravity Ranges in Half and Full Filled Volume**

not habitable for long term residence. Acceptable gravity regions can be determined using the equations and the scaling factor of the half-filled station analysis.

We include Table 3-4 to illustrate the variation of the scaling factor, m, for different minimum and maximum gravities. This chart is for the half-filled stations shown in Figure 3-17. We show the minimum gravity values across the top of the chart and the maximum gravity values along the left side. The gravity in the floors areas of our designs is between 0.95g and 1.05g. For a half-filled station, we find that the scaling factor, m, would be equal to 0.95/(1.05-0.95)=9.5 and the rotation radius of the station would be 9.5 times the distance h. In Table 3-4, the scaling values range from 2 to large values when the $g_{min}$ and $g_{max}$ are almost the same value. We highlight two values in the table: 9.5 and 6.33. These represent a minimum of 0.95g on the top floor and a maximum of 1.05g or 1.1g on the outer rim. We use those values and gravity ranges extensively throughout our research.

**Example Station Centripetal Gravity:** We consider a station design where the main floor of the outer elliptic torus is 2400 meters from the center of the station. The main floor of the inner torus is 1100 meters from the center of the station. The station will rotate once about every 1.6 minutes. This will produce the sensation of Earth like gravity on the main floor of the large outer elliptic torus. Centrifugal force from the rotating frame of reference produces this artificial or pseudo gravity. We show in Figure 3-18 the centripetal gravity in this rotating space station. On the right side of the chart, we show the gravity for the Earth, Moon, Mars, and Atira.

The bottom floor of the outer torus has the highest gravity of 1.06g. In the outer torus, the divider top floor and the tower top floor has a gravity of 0.89g. The main floor of this design has a gravity of 1.0g and only increases to 1.03g on the 20th floor below the main floor. Most of the 700,000 residents will live and work in this region between 1.0g and 1.03g. Figure 3-18 also shows the inner torus will provide a gravity greater than Mars but less than Earth. The small shuttle bay at the center will provide a gravity more like the Moon.

**Station Gravity Summary**: When possible, we aim for a gravity range of 0.95g to 1.05g in the most occupied areas of the station. We believe this gravity range should minimize the health risk to the residents. We reserve the upper-half, low-gravity regions of the space station to serve as open space. This open space provides good vistas and beneficial aesthetics for the residents and visitors. Regions with gravity near Earth-like (1g) are used for living, recreational, and working quarters. The centripetal gravity increases with depth into the station. Higher-gravity regions could provide high strength training, filtration systems, and higher gravity research. A recent space station design used an entire deck as a large ventilation duct [Soilleux and Gunn 2018]. We would place this ventilation deck (or two) in the higher-gravity region. This saves space on the 1g regions to support a larger population. Other functions could be placed in the lower, high-gravity, decks including machinery, storage, and agriculture. With today's understanding, working in the higher and lower gravities should be kept to a minimum and avoided or perhaps done in shifts. Our hope is the chosen gravity range and limited exposure to lower and higher gravities will have no impact on the inhabitants' health.

**Table 3-4 – Half Filled Floor Scaling Factor for Various Centripetal Gravities**

| m | gmin | Scaling factor (m) where Major Radius (R)=m*Minor Radius (h) | | | | | | | |
|---|---|---|---|---|---|---|---|---|---|
| gmax | 0.8 | 0.825 | 0.85 | 0.875 | 0.9 | 0.925 | 0.95 | 0.975 | 1.00 |
| 1.00 | 4.00 | 4.71 | 5.67 | 7.00 | 9.00 | 12.33 | 19.00 | 39.00 | N/A |
| 1.025 | 3.56 | 4.13 | 4.86 | 5.83 | 7.20 | 9.25 | 12.67 | 19.50 | 40.00 |
| 1.05 | 3.20 | 3.67 | 4.25 | 5.00 | 6.00 | 7.40 | 9.50 | 13.00 | 20.00 |
| 1.075 | 2.91 | 3.30 | 3.78 | 4.38 | 5.14 | 6.17 | 7.60 | 9.75 | 13.33 |
| 1.1 | 2.67 | 3.00 | 3.40 | 3.89 | 4.50 | 5.29 | 6.33 | 7.80 | 10.00 |
| 1.125 | 2.46 | 2.75 | 3.09 | 3.50 | 4.00 | 4.63 | 5.43 | 6.50 | 8.00 |
| 1.15 | 2.29 | 2.54 | 2.83 | 3.18 | 3.60 | 4.11 | 4.75 | 5.57 | 6.67 |
| 1.175 | 2.13 | 2.36 | 2.62 | 2.92 | 3.27 | 3.70 | 4.22 | 4.88 | 5.71 |
| 1.2 | 2.00 | 2.20 | 2.43 | 2.69 | 3.00 | 3.36 | 3.80 | 4.33 | 5.00 |

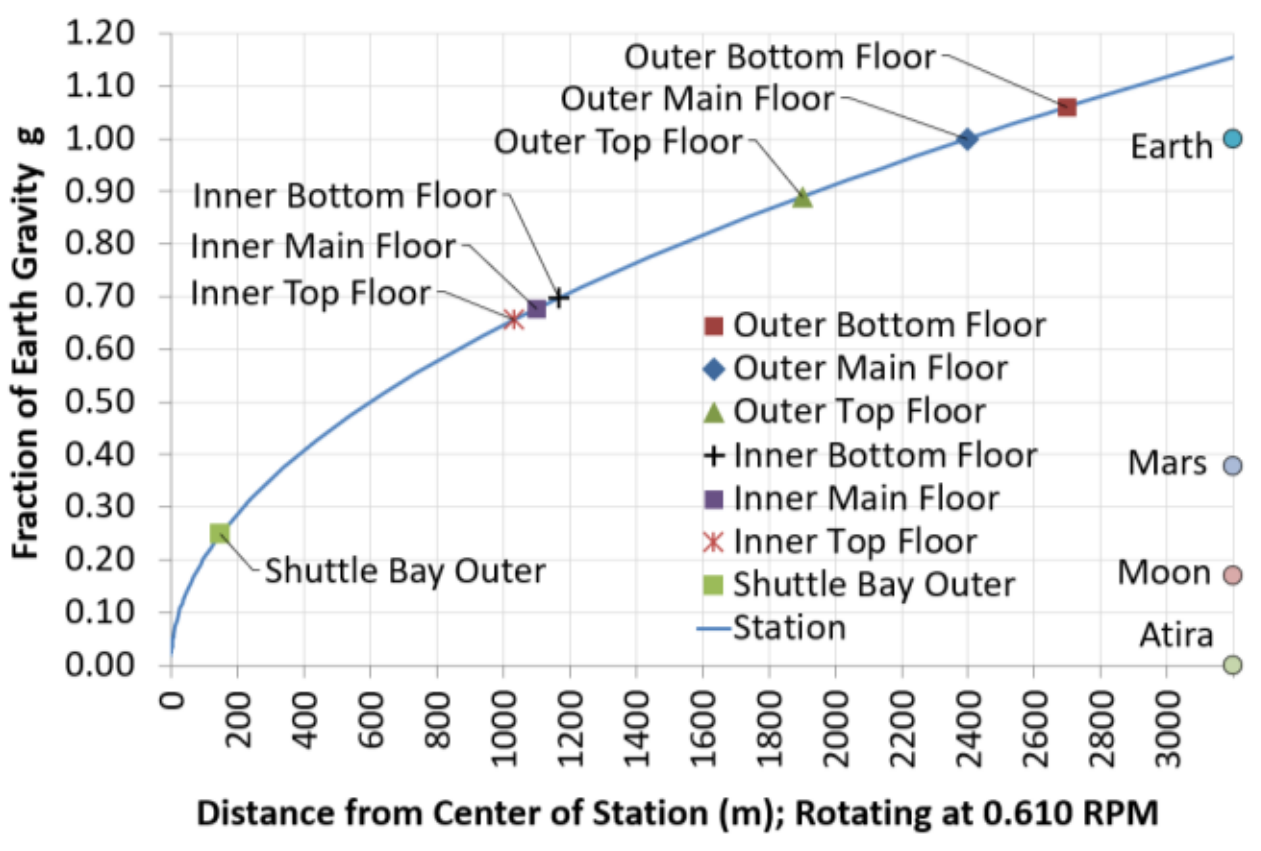

**Figure 3-18 – Centripetal gravity in Atira Station**

### *3.2.4 Geometry Adaptations*

Most historic studies use only the outer shell of their geometry for living space. We review in this subsection our station geometry adaptations when using multiple floors and providing a reasonable gravity range over those floors. The chart in Figure 3-10 illustrated the significant gains from using multiple floors on cylinders and tori. We include in Table 3-5 eight example stations from the literature [O'Neill 1976] [Johnson and Holbrow 1977] [Bock, Lambrou, and Simon 1979] [Globus et al. 2007]. We selected two example stations for each of the four geometries. The top half of the table presents the original floor surface area, the population density, and the population for the eight stations. The bottom half of the table presents the multiple floor version of those stations. The following paragraphs focus on one example from each of the four geometries.

**Dumbbell Adaptations:** We consider a station covered in the 1977 NASA study [Johnson and Holbrow 1977]. In that study, they use a projected surface area equal to the center cross-section of the dumbbell sphere. Their baseline dumbbell station design had two sphere nodes. Each had a 65-meter radius and a 13,273 square meter cross-section. They allocated 67 square meters for each individual and this station could support 396 people.

The dumbbell sphere nodes could have multiple floors. If we leave the top half of the sphere open for aesthetic reasons, the bottom half could have 13 floors separated by 5 meters. The dumbbells would have 243,002 square meters of floor space. Allocating 67 square meters for each person, we find that the two spheres with 13 floors could hold a population of 3627 people – almost 10 times the single floor version. A more comfortable density is 144.2 square meters per individual; see Figure 3-13. At this density, the dumbbell station would support a population of 1685 people. If we were to rotate the two 65-meter-radius spheres at 1.2 RPMs about a radius of 618 meters, the cross-section floor would experience 0.95g and the outer floor would experience 1.05g.

We see in Table 3-5 that the multiple floors increase the available surface floor area of the spheres in the dumbbell designs. Each of the small spheres increase from 13,273 square meter to 121,501 square meters. The 1977 NASA study [Johnson and Holbrow 1977] also describes a single-floor dumbbell with a radius of 316 meters to support a population of 10,000 people. We include in Table 3-5 a multiple floor dumbbell design to support 10,000 people. Compared to the 316-meter radius design, the multiple floor design has a smaller radius of 118.6 meters. These spheres would rotate at a radius of 1127 meters and produce gravities at 0.95g and 1.05g at the top (center) floor and bottom (outer rim) floor.

**Sphere Adaptations:** Researchers and writers have considered spherical space stations for over 100 years. Gerard O'Neill proposed two spherical stations in his book *The High Frontier: Human Colonies in Space* [O'Neill 1976]. The first was called Island One, had a diameter of 500 meters, and could support 10,000 people. The second was called Island Two, had a diameter of 1800 meters, and could support 75,000 people. Even though spherical space stations are rotationally unstable [Globus et al. 2007], our analysis shows that ellipsoid space stations can be rotationally stable. In the following paragraphs, we provide a brief comparison of the spherical station to a similar ellipsoid station.

We use an oblate ellipsoid, and it rotates about the minor axis. The minor axis length is less than 0.8165 times the length of the major perpendicular axes to provide the rotational stability. For Island Two, we keep the diameter of the major axes at 1800 meters and set the minor axis diameter to 1469.7 meter. This ellipsoid would be rotationally stable. The outer surface would be at a radius of 900 meters, and we set the top floor at 814.3 meters. The gravity would range from 0.95g to 1.05g with the station rotating at 1.02 RPMs. There would be 18 floors with a total surface area of 40.2 million square meters. O'Neill wrote that Island Two could support a population of 75,000. This assumed 35 square meters per person and all agriculture was in an adjacent banded torus "crystal palace" structure. With 18 floors, the Ellipsoid Island Two station could support 600,000 people at 67 square meters per person. Using the 144.2 square meters per person density would eliminate the separate agriculture structure and still support 278,693 inhabitants.

**Table 3-5 – Restructuring Improvement Examples for Space Station with Four Geometries**

| | | Torus | | Cylinder | | Sphere | | Dumbell | |
|---|---|---|---|---|---|---|---|---|---|
| Single Floor Design | Units | Tiny | Stanford | Kalpana One | Model 3 | Island One | Island Two | NASA 65 | NASA 10K |
| Major Axis (R) | meters | 190 | 830 | 250 | 1,000 | 250 | 900 | 650 | 3,002 |
| Minor Axis or Length | meters | 33.5 | 65.0 | 325 | 10,000 | 250 | 900 | 65 | 316.0 |
| Population | count | 10,000 | 10,000 | 3,000 | 2,000,000 | 10,000 | 75,000 | 396 | 10,000 |
| Floor Area | meters$^2$ | 276,485 | 678,000 | 510,508 | 125,663,600 | 350,000 | 2,625,000 | 26,546 | 670,000 |
| Population Density | m2 / person | 27.6 | 67.8 | 170.2 | 62.8 | 35.0 | 35.0 | 67.0 | 67.0 |
| Max Ceiling Height | meters | 67.0 | 130.0 | 500.0 | 2,000.0 | 500.0 | 1,800.0 | 130.0 | 632.0 |
| Multiple Floors | Units | Tiny | Stanford | Kalpana One | Model 3 | Ellipsoid One | Ellipsoid Two | NASA 65 | Pop=10K |
| Floors | count | 7 | 12 | 5 | 20 | 4 | 18 | 13 | 24 |
| Minor Axis or Length | meters | 33.5 | 65.0 | 325.0 | 1,300.0 | 204.1 | 734.9 | 65.0 | 118.6 |
| Top Floor Radius | meters | 190.0 | 830.0 | 226.2 | 904.8 | 226.2 | 814.3 | 617.5 | 1,126.7 |
| Population | count | 13,715 | 70,051 | 16,723 | 2,047,560 | 6,546 | 278,693 | 1,685 | 10,000 |
| Floor Area | meters$^2$ | 1,977,703 | 10,101,300 | 2,411,457 | 295,258,152 | 943,981 | 40,187,494 | 243,002 | 1,442,000 |
| Population Density | m2 / person | 144.2 | 144.2 | 144.2 | 144.2 | 144.2 | 144.2 | 144.2 | 144.2 |
| Max Ceiling Height | meters | 33.5 | 65.0 | 452.4 | 1,809.6 | 452.4 | 1,628.6 | 65.0 | 118.6 |

We see in Table 3-5 that the multiple floors increase the Island One floor area from 350,000 to 943,981 square meters. The Island Two size station population increases from 75,000 to 278,693 people. Using the same major radius and a smaller minor radius, the exteriors of these oblate ellipsoid stations are a little smaller than the O'Neill spherical station designs. The population density is improved considerably from 35 to 144.2 square meters per individual. There is still reasonable openness and vistas on the top floor.

**Cylinder Adaptations:** Multiple reports from the 1970s and 1980s used cylinder lengths that were 10 times the radius. A recent paper investigated the rotational instability of long rotating cylinder space stations [Globus et al. 2007]. That study imposed a limit on the length to passively control the imbalance of the rotating structure [Globus et al. 2007]. To meet this imbalance metric, they found the cylinder length should be less than 1.3 times the radius.

We show artwork of the Model 3 O'Neill cylinder space station [O'Neill 1974] in Figure 3-1. This cylinder was 10,000 meters long and had a radius of 1000 meters. O'Neill expected it hold 200,000 to 2 million people. Two counter-rotating cylinders were attached to eliminate gyroscopic effects and precession. Using the Globus imbalance metric, a rotationally stable version of a single cylinder with a radius of 1000 meters would have a length of 1300 meters. We would design the bottom floor (outer cylinder) at 1000 meters radius and the top floor (inner cylinder) at 904.8 meters radius. This would support 20 floors spaced 5 meters apart. The station would rotate at 0.97 RPM and provide 1.05g at the bottom floor and 0.95g on the top floor. This version of cylinder would not have the windows or complex mirrors like the Model 3 station. The top floor would house 51,250 people at 144.2 square meters per person. The 20 floors would support 2,047,560 people. With its multiple floors and with a more comfortable population density, this single cylinder would hold as many people as the much bigger and double O'Neill Model 3 cylinders.

We compare in Table 3-5 the single floor and a multiple floor version of the Kalpana Station. Globus and his team designed the Kalpana One space station and found it could support 3000 people [Globus et al. 2007]. The Globus team mentions multiple floors doubling the surface area. We find adding 5 floors, each separated by 5 meters, would increase the floor space from 510,508 square meters to 2,410,000 square meters. The station would rotate at 1.94 RPM and the outer hull would have 1.05g and the top floor would have 0.95g. Given 144.2 square meters per person, the multiple floor Kalpana station would house 16,723 people.

**Torus Adaptations:** Table 3-5 includes the rotating station from a 1974 Stanford study. The Stanford Torus has a major radius of 830 meters and a minor radius of 65 meters [Johnson and Holbrow 1977]. This torus rotates at 1 rpm to produce the effect of one Earth gravity on the outer surface. Their study suggested 47 square meters of projected area for each inhabitant. They also determined that agriculture requires an additional 20 square meters for each person and brings the total population density is 67.8 square meters per resident. The Stanford Torus surface area would be 678,000 square meters and support a population of 10,000.

Our multiple floor Stanford Torus still uses a circular cross section with an 830-meter major radius with 65-meter minor radius. There could be 12 floors spaced 5 meters apart in the outer half of the torus. The station would rotate at 1.01 RPM and the top floor would have 0.96g while the bottom floor would have 1.03g. The top floor would support 4,412 people at 144.2 square meters per person. Using the 67 square meters per person metric, the top floor would support 9,496 people. We find the maximum population would be 70,051 people including the area of all 12 floors, spokes, and shuttle bay. A more realistic population would be 42,465 and would primarily have inhabitants living on the top 10 floors.

Table 3-5 also includes a Tiny Torus from the 1979 O'Neill Study [O'Neill et al. 1979]. This torus had a major radius of 190 meters. It had 276,485 square meters of floor space and could support 10,000 people [Bock, Lambrou, and Simon 1979]. Using 7 floors, the floor space increases to almost 2 million square meters.

To better compare the single floor and multiple floor torus, we use a consistent population density of 144.2 square meters per person. We see the supported population increase from less than 2,000 to more than 13,000 people on the tiny torus station with only 7 floors. The population increases from 4,701 people to over 70,000 people with a Stanford Torus size station with 12 floors. We note the perceived height to the ceiling would be cut in half for the inhabitants. The vista in the rotation direction would remain about the same. The ceiling height reduction would be noticeable for small stations; however, the ceiling height reduction would be hard to perceive in larger stations. In our opinion, the gains in population and the improved population density outweigh the loss in ceiling height.

**Restructuring Population Improvements:** The examples in the previous paragraphs and in Table 3-5 illustrate the benefits from using multiple floors in the space stations. It is possible to maintain acceptable gravity levels from the top floor to the outer rim (0.95g to 1.05g). A station designed to be rotationally stable typically has a smaller dimension on one of the axes. Multiple floors increase the population and compensate for that smaller dimension. With the increased floor space, the station design can be simplified by bringing the agriculture in the station (instead of an external "crystal palace"). Even with the simpler design and the internal agriculture, the stations can support greater populations than a single floor design.

### *3.2.5 Station Mass*

The previous subsection detailed the impact of using multiple floors with the four geometries using historic stations. In all cases there were significant improvements in the population supported. We now overview the relationships between population, station radius, and mass.

**Example Asteroids**: We consider the effect of construction material mass and geometries on the supported population. This assessment uses the same four geometries, thick shells, and the station volume half filled with floors. In the previous asteroid section, we chose 5 asteroids with a good range of sizes for our study; see Figure 2-1. We use these same 5 asteroids to illustrate the available oxide building material. We show in Table 3-6 their characteristics and the resulting processed material. In our building material estimate we account for asteroid porosity, processing losses, and asteroid composition. The asteroid density varies with the asteroid size and type. We derived porosity and processing metrics of 48.3% and 24.4%. The asteroids are assumed to contain 84.9% oxides, 4.0% volatiles, and 11.1% metals. We designed the stations to use 30% of the volume of asteroid oxides as building material. This results in using only 10% to 13% of the total asteroid mass to create the station construction material.

**Mass and Radius:** We show Figure 3-19 the radius of the same four station geometries and their required construction material. We show along x-axis the amount of building material in kilograms. This is the processed oxide material from the asteroids. We show the 5 example asteroids along the axis at the mass of their station oxide building material. We show along the y-axis in Figure 3-19 the radius of the station that can be constructed with each of the geometries.

We include a line for the minimum radius at 224 meters, which when rotating at 2 revolutions per minute produces 1g Earthlike gravity. We include a line at the maximum radius (10 kilometers) in Figure 3-19. We included in Table 3-1 different materials and the maximum rotating station radius given material density and strength [McKendree 1995]. O'Neill envisioned stations up to 16 kilometers using future materials [O'Neill 1974]. Basalt rods and fibers could be that type of future material and could support a station radius over 20 kilometers. We envision using anhydrous glass tiles and beams to support even larger station radius. We reviewed the properties of these materials in Table 3-1. We conservatively set 10 kilometers as a maximum radius on this graph. We earlier found that most of our considered materials can support our station working stresses with this radius.

We see in the chart that the dumbbell station has a much larger radius given the same amount of material. This makes sense because the other geometries encircle the center of rotation. The dumbbell has two nodes at the radius distance from the center. Given the lower and upper limits, we see in Figure 3-19 that dumbbell geometries are viable for small to medium sized asteroids. Dumbbell geometries are not viable for large asteroids because their rotation radius would exceed our maximum 10-kilometer metric. The building material from the asteroid Šteins would create a huge dumbbell station with a large radius exceeding that limit. We also see in Figure 3-19 that cylinders and ellipsoids are not viable for small asteroids because their radius would be smaller than our minimum 224-meter radius. The building material from the asteroid Bennu would create a cylinder station with too small of radius.

**Mass and Population:** The charts in Figure 3-20 compare the population supported by the four geometries. For reference, we show the mass of five asteroids along the horizontal axis of the chart. Many parameters such as the radius, floor count, and support structures vary with the available mass. The charts in Figure 3-20 combine all those parameters. It combines many assumptions, requirements, and design decisions. Its value is more for comparative than absolute results.

| Table 3-6 – Available Building Material | | | | | |
|---|---|---|---|---|---|
| **Asteroid** | **Bennu** | **Ryugu** | **Moshup** | **Atira** | **Šteins** |
| Mass (kg) | 7.3E+10 | 4.5E+11 | 2.5E+12 | 4.1E+13 | 1.5E+14 |
| Volume (m3) | 6.2E+07 | 3.4E+08 | 1.6E+09 | 2.2E+10 | 7.6E+10 |
| Mean Radius (m) | 241 | 448 | 659 | 1928 | 2580 |
| Asteroid Type | C-Type | C-Type | S-Type | S-Type | E-type |
| Density (kg/m3) | 1192 | 1324 | 1537 | 1875 | 1908 |
| Packed (m3) | 3.2E+07 | 1.8E+08 | 8.4E+08 | 1.1E+10 | 3.9E+10 |
| Processed (m3) | 2.4E+07 | 1.3E+08 | 6.3E+08 | 8.6E+09 | 3.0E+10 |
| Oxides (m3) | 2.0E+07 | 1.1E+08 | 5.4E+08 | 7.3E+09 | 2.5E+10 |
| Volatiles (m3) | 9.6E+05 | 5.3E+06 | 2.5E+07 | 3.4E+08 | 1.2E+09 |
| Metals (m3) | 2.7E+06 | 1.5E+07 | 7.0E+07 | 9.5E+08 | 3.3E+09 |
| Station (m3) | 6.1E+06 | 3.4E+07 | 1.6E+08 | 2.2E+09 | 7.6E+09 |
| Mass Used (%) | 12.3% | 11.7% | 11.3% | 10.8% | 10.1% |

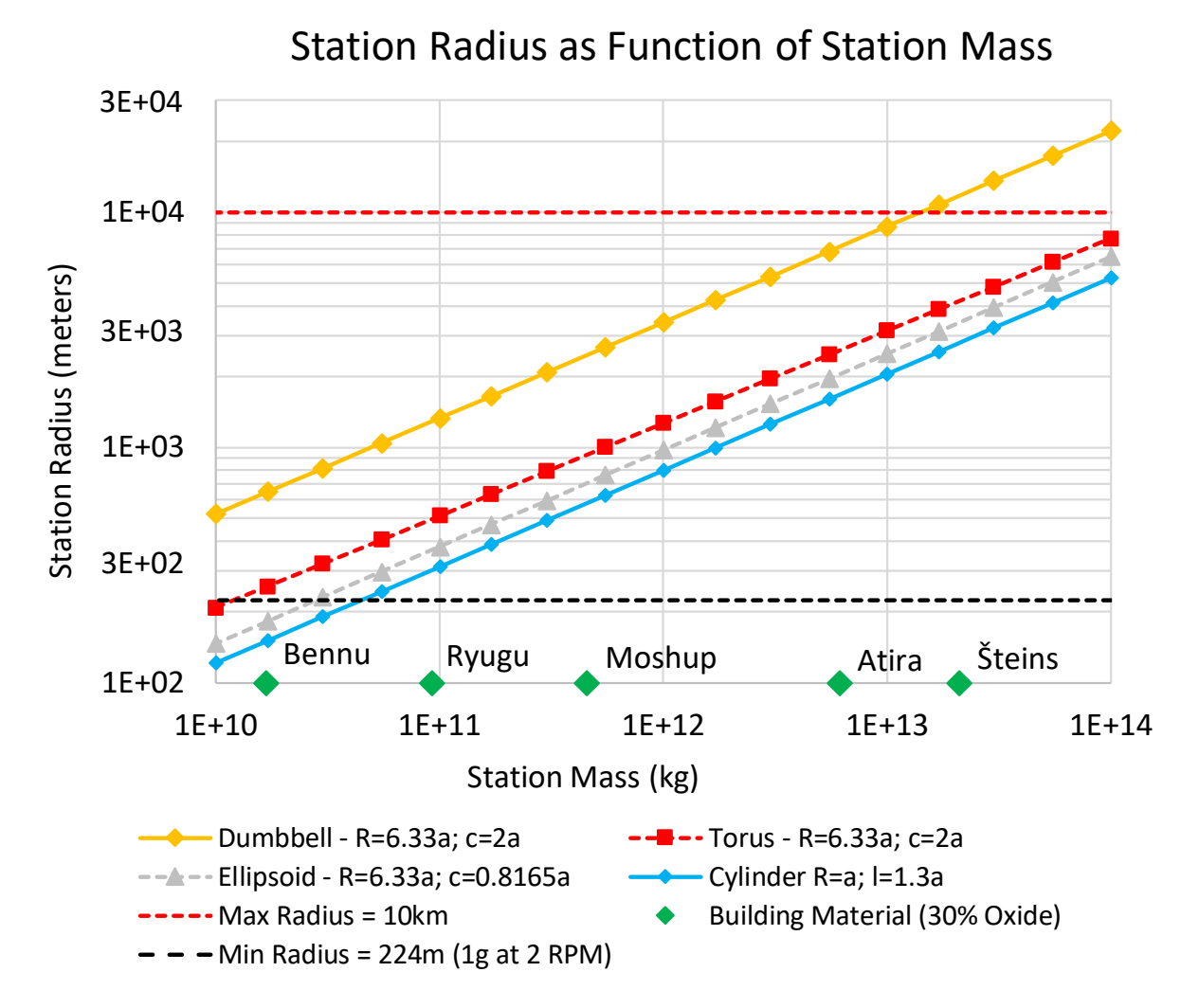

**Figure 3-19 – Preview - Radius and Mass Comparison**

The chart in Figure 3-20a shows a logarithmic horizontal x-axis measuring mass in kilograms. We show two masses for the 5 example asteroids along the x-axis in Figure 3-20a. One is the mass of the asteroid and the other is the mass of the constructed station. The difference would include losses, margin, and surplus. The logarithmic vertical y-axis shows the supported population of the four station geometries. For the five asteroid masses, the maximum populations range from 8,000 to 200 million. At logarithm scales, two groups of two geometries appear to support nearly the same population for a mass of material. For a given mass, the cylinder and ellipsoid geometries support less population than the torus and dumbbell. To better compare the different geometries, we normalize the populations to the torus geometry population; see Figure 3-20b. The chart shows that, except for the smallest masses, the dumbbell geometry supports the largest population compared to the other geometries for a given mass of building material.

**Station and Population Details:** We include in Table 3-7 population details for stations constructed from the same five asteroids. We include an asteroid station Stanford in the list for comparison to the O'Neill Stanford Station. The table contains the computed radii and populations for torus stations. The radius column shows the major radius and the minor radii. Smaller stations only have one torus, and they have a circular cross section. Larger stations have an inner torus and an outer torus. For the larger stations, we only show the three outer elliptical torus radii in the table. The floor count assumes the torus is half filled with floors. We use a space of 5 meters between each of the floors. We also design the top floor with additional thickness to support soil and vegetation.

Our maximum population is generated using all floor space in the station and a density of 144.2 square meters per person. The "realistic" population uses only a small percentage (5%) of the spoke, inner torus, and shuttle bay for habitation. We also include the population of the top floor of the station in Table 3-7. This top floor population provides a comparison to historic single floor designs.

**Population Details:** Figure 3-21 shows the populations for those six space habitats and uses the data from Table 3-8. We include results showing the maximum population, a more realistic population, and the population supported on the

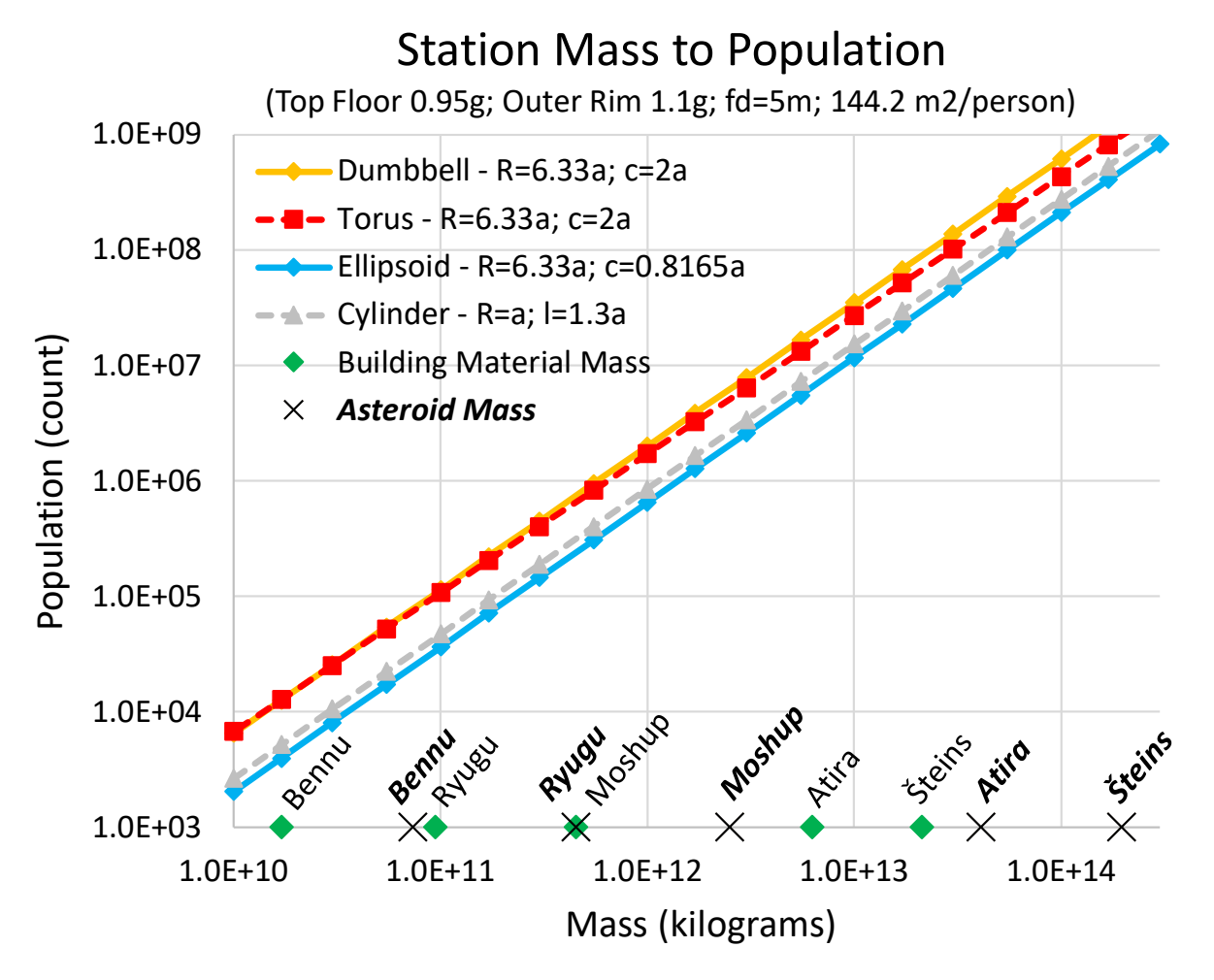

*a) Station Population versus Mass*

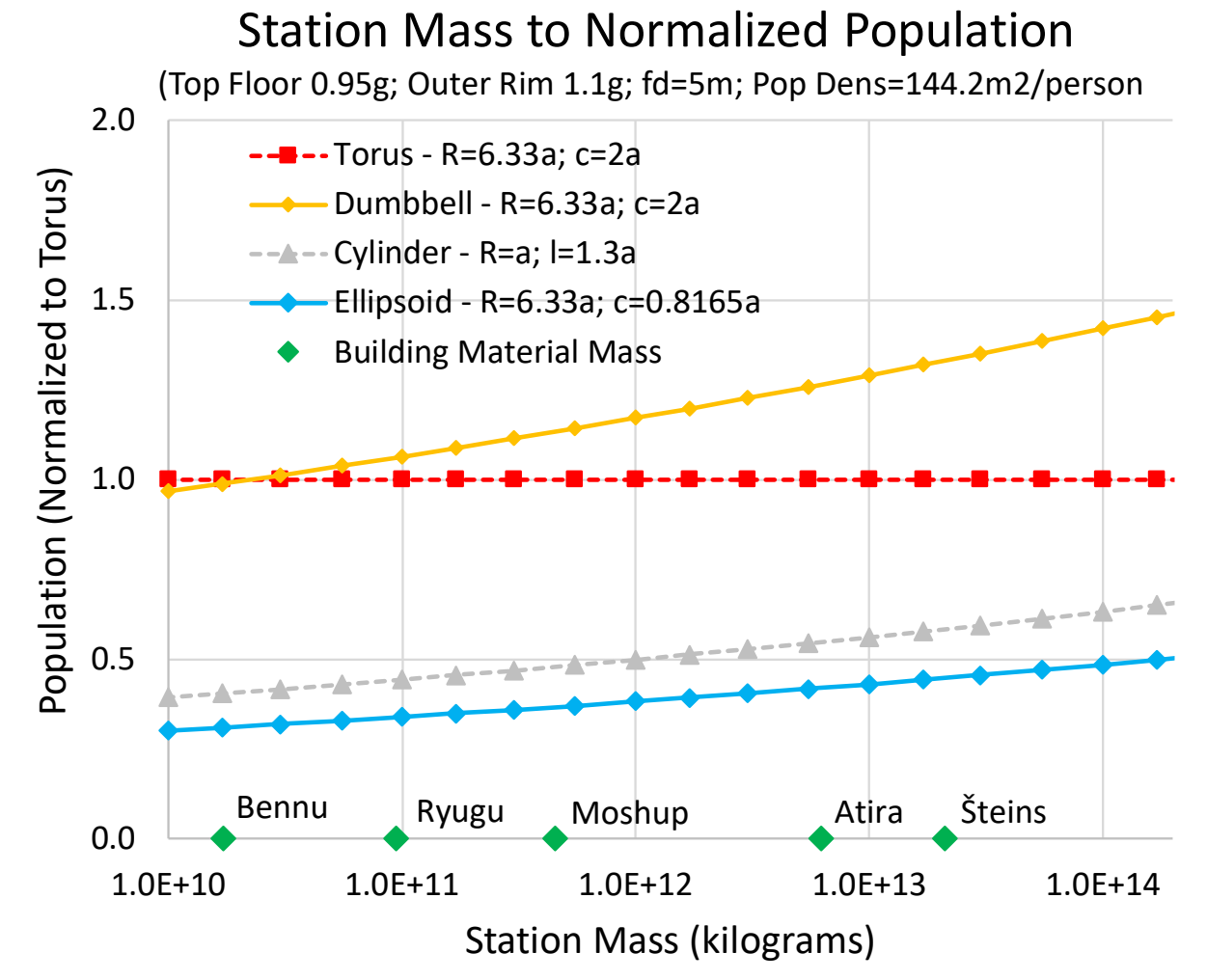

*b) Station Population (Normalized) versus Mass*

**Figure 3-20 – Preview - Population Comparison**

middle (top) floor of the torus. The maximum population uses all the available space in the stations. This includes space on lower floors, in spokes, and in the shuttle bay. These population estimates use a density of 144.2 square meters per person. Using this density and Bennu as an example, we find the station would support a maximum of 8,273 people in its torus. The Ryugu station would support 33,150 people. Constructing and using floors in the Stanford Torus, we find it could support a maximum of 70 thousand people. The Atira station could support 10 million people. The Šteins station would support a maximum of 70 million people.

**Table 3-7 – Example Radius and Populations for Torus Stations from Asteroids**

| Asteroid Station | Radius (meters) (Major & Minor Radii) | Floor Count (fd=5m) | Maximum Population | Realistic Population | Top Floor Population |
|---|---|---|---|---|---|
| Bennu | (205,32) | 5 | 8,273 | 2,771 | 450 |
| Ryugu | (354,56) | 10 | 33,150 | 16,475 | 1,531 |
| Stanford | (830,65) | 12 | 70,051 | 42,760 | 4,412 |
| Moshup Inner | (937,148) | 28 | 371,267 | 123,418 | 11,096 |
| Atira Double | (2116,1003,334) | 60 | 10,009,838 | 1,625,065 | 184,912 |
| Šteins | (4042,1915,638) | 120 | 69,769,913 | 5,790,393 | 674,557 |

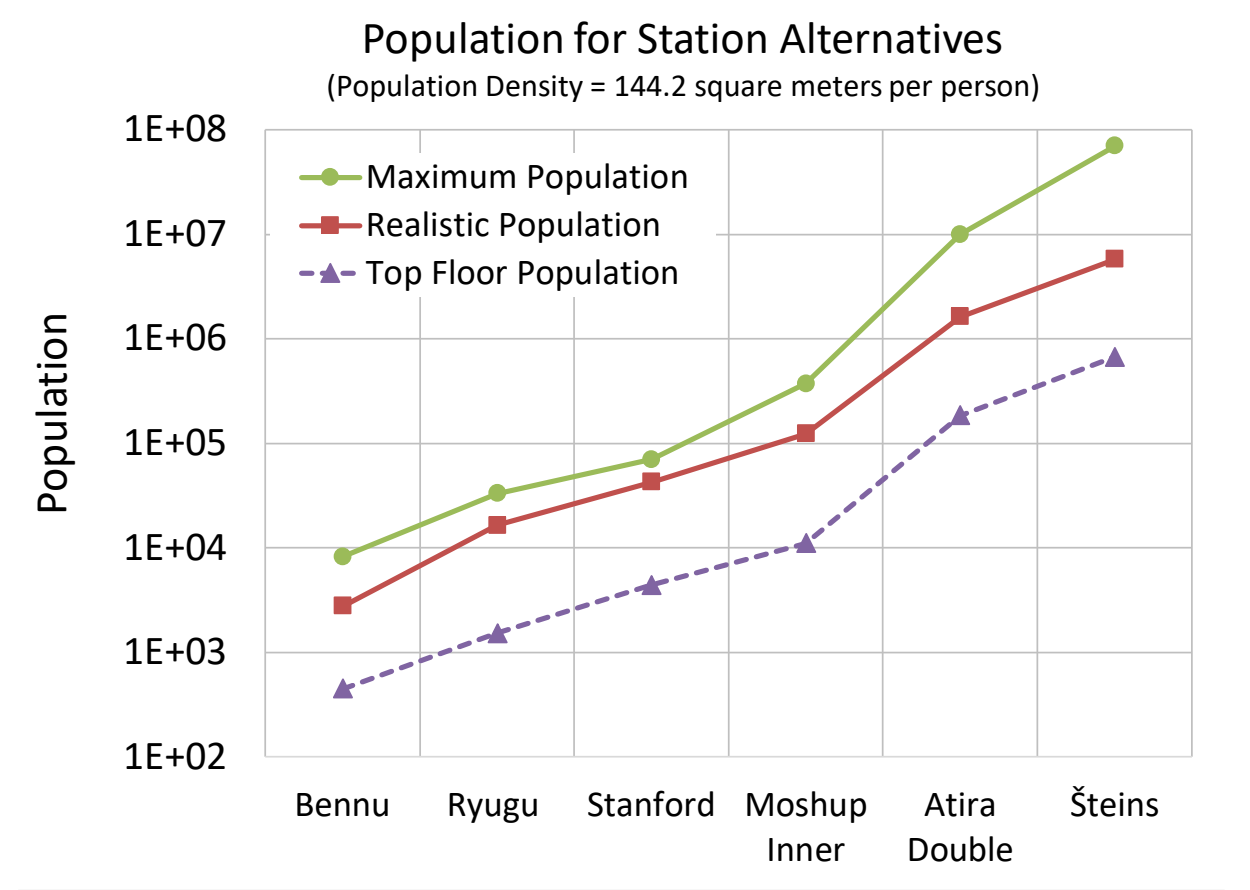

**Figure 3-21 – Populations with Station Alternatives**

We include a more "realistic" population metric in Figure 3-21. Low gravity and high gravity regions are excluded from the population estimate. The Bennu and Ryugu stations can realistically support about 2,771 and 16,475 people. Larger stations have more than 50 floors and it is unlikely that people will want to live in the lowest levels. The realistic population metric limits the living region to roughly the top 10 floors. It also uses a small percentage (5%) of the spoke, inner torus, and shuttle bay for habitation. The Atira station would support a realistic population of 1.6 million people. The large Šteins station with a double set of spokes and inner tori can support a realistic 5.8 million people.

We also include the population estimates using the Stanford Torus dimensions. We include the population supported using only the top floor of the torus. With the Stanford design, this top floor nearly matches the projected surface area estimates of earlier NASA studies [Johnson and Holbrow 1977]. The Stanford Torus was expected to support a population of 10,000 people. We find our top floor of the Stanford Torus supports 4,412 people using the 144.2 square meters per person. This would be 9,496 people with a density of 67 people per square meter. There can be a dozen floors in outer half of the Stanford Torus. This would support a maximum population of 70,051 people and a realistic population of 42,760 people. These examples continue to highlight the value of using floors to increase the available population on stations.

**Population and Radius Details:** We include additional data in Table 3-8 from our population analysis. This includes populations for torus and dumbbell geometries. This table includes the five asteroids with their mean radius and mass. Again, we include a fictitious asteroid Stanford in the list for comparison with the Stanford Station. We show the amount of building material used in each of the stations. This represents 30% of the volume of the available construction material (oxide) of the asteroids; see Table 3-6. We computed the population using a density of 144.2 square meters per person from Figure 3-13. The station mass uses the volumes of the towers, spokes, outer shells, floors, fill, and the shuttle bay. We include the maximum and realistic populations for the torus and the dumbbell geometries. There are many estimates, curve fittings, and extrapolations in these population estimates. These estimates should be considered a first approximation and mainly for comparative purposes.

**Population and Mass Equations:** We have developed equations that estimate the radius and the population of stations based on the available asteroid mass. The Station Mass subsection of *§3.2.1 Station Characteristics* detailed our approach to compute the station mass. We derived and summed the volume and mass of the individual pieces in the components of the station. We derived the population in this same analysis by summing the floor spaces.

We took the results of these detailed estimates to create simple equations to estimate the population, radius, and mass. We fit a power curve to the data in Figure 3-12 to derive radius to mass equations. The data for Figure 3-12 was generated using earlier, slightly different station metrics. We have also fit power curves to the data in Table 3-8. We extrapolated results from both set of power curves to create our equations. We show those equations for the four geometries in Table 3-9. These equations subsume many parameters and reduce the estimates to a single variable – the station mass.

We define that the station is filled with multiple floors. In these cases, the equations were generated with floors separated by 5 meters and the population density is 144.2 square meters per person. The geometry design metrics in Table 3-9 produce rotationally stable space stations and Earth-like gravity ranges over the floors. The stability and range concepts were detailed in previous subsections. The major rotation radius is typically scaled from one of the minor rotation radii. The dumbbell is designed with two prolate ellipsoid nodes and the ellipsoid station is designed as an oblate ellipsoid. The equations show that the mass from the radius equation is not a quite a volume relationship. We see that the station mass is a 2.4 to 2.6 power function with the thick shells, the half-filled station, and the open regions between the floors. The station mass is between a surface area (quadratic) and a volume (cubic) relationship with the station radius. As one would expect, the population is a volume or cubic relationship with the station radius. The population of the station increases faster than the mass of the station with increasing radius. We see that the population is a 1.20 to 1.25 power function of the mass.

**Table 3-8 – Estimate Radius and Maximum Population for Station Geometries Asteroids**

| Asteroid | Mean Radius (m) | Mass (kg) | Station Material (kg) | Torus Radius (m) | Torus Maximum Population | Torus Realistic Population | Dumbbell Radius (m) | Dumbbell Maximum Population | Dumbbell Realistic Population |
|---|---|---|---|---|---|---|---|---|---|
| Bennu | 263 | 7.33E+10 | 9.03E+09 | 205 | 8,273 | 2,771 | 245 | 7,766 | 2,521 |
| Ryugu | 448 | 4.50E+11 | 5.28E+10 | 354 | 33,150 | 16,475 | 433 | 32,614 | 15,871 |
| Stanford | 502 | 7.02E+11 | 8.31E+10 | 830 | 70,051 | 42,760 | 502 | 85,400 | 54,324 |
| Moshup | 659 | 2.49E+12 | 2.81E+11 | 937 | 371,267 | 123,418 | 659 | 396,325 | 126,903 |
| Atira | 1928 | 4.11E+13 | 4.45E+12 | 2,116 | 10,009,838 | 1,625,065 | 1,736 | 11,944,304 | 1,822,202 |
| Šteins | 2580 | 1.98E+14 | 1.98E+13 | 4,042 | 69,769,913 | 5,790,393 | 2,628 | 88,939,570 | 6,778,960 |

| Table 3-9 – Equations to Estimate Population and Mass | | | | |
|---|---|---|---|---|
| **Geometry** | **Cylinder** | **Ellipsoid** | **Torus** | **Dumbbell** |
| Design Metrics | *R=a; l=1.3a* | *R=6.33a c=0.8165a* | *R=6.33a; c=2a* | *R=6.33a; c=2a* |
| Mass (m) from Radius (r) | $m = 80681\ r^{2.4430}$ | $m = 52811\ r^{2.4340}$ | $m = 12391\ r^{2.5471}$ | $m = 2121\ r^{2.4551}$ |
| Population (p) from Mass (m) | $p = m^{1.2533} / 1.284E+9$ | $p = m^{1.2533} / 1.676E+9$ | $p = m^{1.2017} / 1.538E+8$ | $p = m^{1.2435} / 4.165E+8$ |
| Population (p) from Radius (r) | $P=r^{3.0618}/1187.7$ | $P=r^{3.0506}/1546.8$ | $P=r^{3.0609}/1854.7$ | $P=r^{3.0530}/30414.3$ |
| Station Metrics: Mass (m) in kilograms; Population (p) in count; and Radius (r) in meters | | | | |

Using these equations, we can now estimate the radius and population of a station given the mass of the asteroid. Again, we note that many parameters such as the radius, floor count, floor spacing, population densities, and support structures can be varied. Material densities, asteroid porosity, and construction processes will all affect the results. This section has introduced the supporting details on those parameters. Given the multiple extrapolations and sources of data, these equations should be used for comparative purposes only.

**Station Mass Summary:** This brief assessment does not consider many issues. A qualitative issue example is that dumbbells do not have the open views that the other geometries offer. A construction and scheduling issue example is that small asteroids have limited surface area for building construction equipment. We plan additional evaluations of these types of issues.

The data of this subsection serve to highlight two important observations. First, using multiple floors produces stations that can support large populations even with small asteroids. Second the different geometries support similar populations for the same mass of building material. For a given mass, dumbbells support the largest population. The torus geometry supports the next largest population. Cylinders and ellipsoids support similar but smaller populations than the torus and dumbbell geometries. The dumbbells may provide the largest population; however, they have other limitations that need to be considered.

## 3.3 Space Stations – Results

We refined slightly the set of typical space station geometries to address rotational stability and gravity issues. We use ellipsoids instead of spheres; short cylinders instead of long cylinders; elliptical cross-section torus instead of circular cross-section torus; and a single dumbbell instead of a composite dumbbell structure. We design the space stations with multiple floors to significantly increase the supported population. The design provides near Earth-like gravity over the habitable floors.

In this section we focus on the details and geometry of a specific space station for a restructuring mission. We provide details on the selected space station and cover its characteristics, construction materials, and population.

### *3.3.1 Specific Space Station*

As previously reviewed, there are many ways to evaluate and select a station geometry. These earlier assessments and selections were typically based on thin space station shells. In our asteroid restructuring, we use thick shells to provide structural integrity, radiation protection, and safety from debris collisions. We use the abundant asteroid material to build the thick shell.

These historic reviews were also typically based on a single projected floor. We design our station with many floors and support greater populations. Our assessment uses these geometries with thicker shells and many floors. A preview of this assessment was in Section 3.2.5 and results were included in Figure 3-20. The charts provided the population as function of the available asteroid mass and the different station geometries. That section provided some of the underlying assumptions and we offer more assumptions and details in this section.

As shown in Figure 3-20, in general, the dumbbell geometry supports the largest population compared to the other geometries for a given mass of building material. The next best geometry is the torus. Like other studies, we have been excluding the dumbbell geometry stations because they are typically a small design with limited vistas. The limited views of small dumbbells could create an oppressive ambience and be a risk to the psychological well-being of the colonists [Keeter 2020]. There could also be structural issues with large dumbbell nodes at the end of long tethers.

Ultimately, we need to consider qualitative and structural metrics in selecting a station geometry. Based on our current analysis, we suggest that an elliptical cross-section torus is the preferred geometry for our restructuring effort.

### *3.3.2 Station Characteristics*

We have reviewed historical studies on space stations. We have explored many shaped and sized space station with our simulations and our analysis. In this section, we focus on a rotating torus as our habitat. We use an elliptical cross section instead of circular. The elliptical shape reduces the excessive number of floors in large stations, increases the station floor space, and maintains a good station vista. This example torus has a major radius of 2400 meters and an elliptical cross-section with radii of 1200 meters and 400 meters. The cross section is half-filled with floors. The spacing between the floors ranges from 6 meters to 15 meters in different station components. We presented an internal drawing of a similar station in Figure 3-9. The external view of this station is like the one shown in Figure 1-1. This version of the station has two inner tori. Each of the inner tori has a radius of 1100 meters. This station rotates about once every 100 seconds and produces an Earth-like gravity on the main outer-torus floor.

We show in Table 3-10 an earlier set of population estimates for this example torus. The table includes the number of floors, the surface area, and population for the station components. On some of the floors, we use multistory construction to increase the usable surface area. Even using the full surface area allocation of 157.1 square meters per person from [O'Neill et al. 1979], we find the station could support 6 million people. Even larger populations could be supported using the allocation of 67 square meters per person.

We do not recommend living in low gravity or high gravity regions for medical reasons. We also do not recommend living in the deep bowels of the station for physiological reasons. In the rightmost column of Table 3-10, we show a population estimate with colonists living only on the first five floors of the outer elliptic torus, and in part of the dividers and the skyscrapers. The main top floor is used exclusively for openness, public areas, and recreation. As stated in the NASA SP-413 study [Johnson and Holbrow 1977], habitats with large living space *"would be settled with much lower population densities, so as to permit additional "wild" areas and parkland."* A population density metric of 547 square meters per individual can support a population of 700,000. There is still significant usable space despite the generous density metric. This extra space could support recreation, tourism, and other *"wild"* uses.

### *3.3.3 Station Construction*

The first asteroid restructuring task is to construct additional spiders and a web of trusses to access raw materials. We illustrate in Figure 3-22a many trusses over the surface of the asteroid. Each truss is 1 meter wide and 20 meters long. The web provides a stable and less varying surface for the vehicles. Navigation is much simpler on this web compared to the natural surface of the asteroid.

There are multiple station components to build. The external view of this station is like the one shown in Figure 1-1. We show in Figure 3-22b an interior view of a similar space station. We label the major components of the station in this rendering. This is a view of almost 2 kilometers distance and 3 kilometers across. The spokes and dividers in the distance have 32 stories. The rendering tool was programmed to draw a 100-meter grid on the main floor of the torus for reference. In our rendering code, we include almost 100 parameters to define the station and the asteroid. Changing these parameters produce different renderings. The rendering tool also outputs a summary report with statistics about the surface area, volumes, heights, and counts for the major components. We use multiple tools in our investigation to create details and cross check results. The rendering tool and our simulations allow us to evaluate different size stations. We can vary many parameters on the station elements including their radii, depths, widths, and thicknesses.

### *3.3.4 Station Materials*

To simulate the restructuring of the asteroid, we need to understand the amount of construction material available from the asteroid. We also need to understand the amount of construction material required to build the station. Previous subsections previewed our analysis of the available asteroid material. We detail in this subsection the amount of construction material in the station. We described in the previous subsection how we analyzed the structure as shown in Figure 3-9. We have developed estimates for the material needed to construct the trusses; to fill the exterior walls with regolith; and to cover the exterior, walls, and floors with panels.

**Table 3-10– Population Estimates for Atira Ellipse Station**

| Station Component | Part | Floors | Floor Surface Area (m2) | Height (m) | Useable Surface Area (m2) | Count | Total Surface Area (m2) | Max Population (157 m2) | Target Population (547 m2) |
|---|---|---|---|---|---|---|---|---|---|
| Inner Torus | Main Floor | 1 | 684,239 | 55 | 684,239 | 2 | 1,368,478 | 8,711 | - |
| | Lower Floors | 5 | 2,515,401 | 8 | 2,515,401 | 2 | 5,030,802 | 32,023 | - |
| | Dividers | 8 | 352,657 | 6 | 352,657 | 2 | 705,314 | 4,490 | - |
| Shuttle Bay | Center | 8 | 1,238,541 | 10 | 1,238,541 | 2 | 2,477,082 | 15,768 | - |
| Spoke | Floors (Beyond Inner Torus) | 158 | 1,503,299 | 6.67 | 1,503,299 | 16 | 24,052,778 | 153,105 | - |
| | Floors (Inside Inner Torus) | 242 | 2,302,780 | 6.67 | 2,302,780 | 16 | 36,844,482 | 234,529 | - |
| Outer Torus | Main Floor | 1 | 33,973,140 | 470 | 33,973,140 | 1 | 33,973,140 | 216,252 | - |
| | Lower Floors | 5 | 161,732,128 | 15 | 323,464,256 | 1 | 323,464,256 | 2,058,970 | 591,802 |
| | Lowest Floors | 11 | 248,683,950 | 15 | 497,367,899 | 1 | 497,367,899 | 3,165,932 | - |
| | Dividers | 32 | 1,238,470 | 15 | 2,476,940 | 8 | 19,815,520 | 126,133 | 36,254 |
| Spoke Skyscraper | Floors | 77 | 2,339,240 | 6 | 2,339,240 | 8 | 18,713,920 | 119,121 | 34,239 |
| Adjacent Skyscraper | Floors | 82 | 2,576,110 | 6 | 2,576,110 | 8 | 20,608,880 | 131,183 | 37,705 |
| Population Density | Square meters per individual | | | | | | | 157.1 | 547 |
| Total | | | 459,139,954 | | 870,794,502 | | 984,422,551 | 6,266,216 | 700,000 |

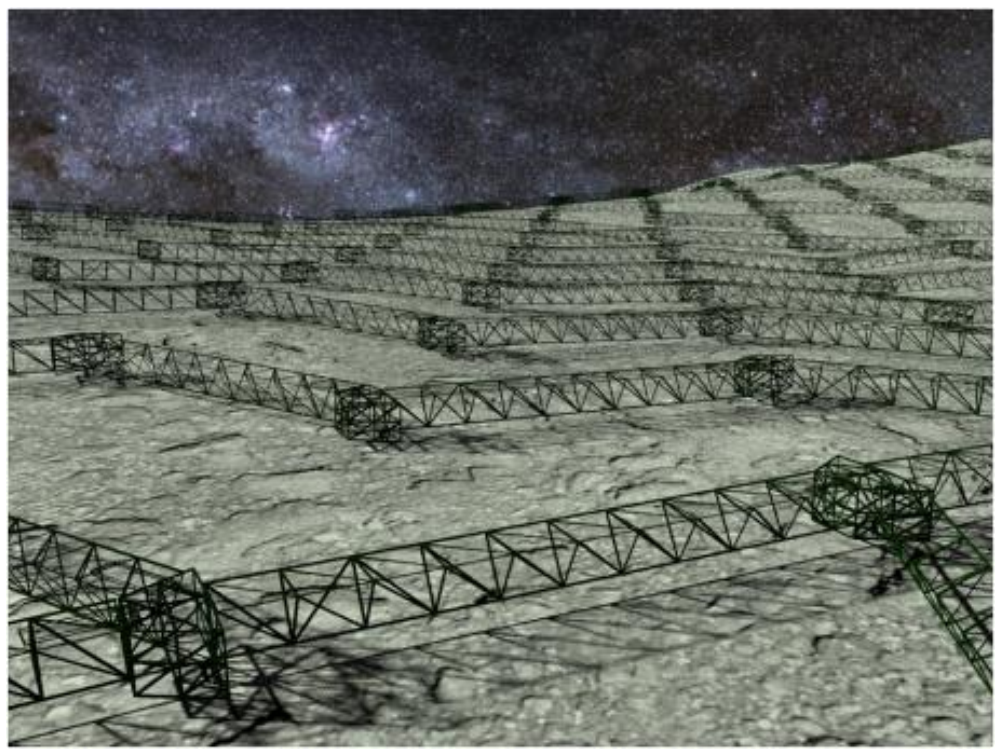

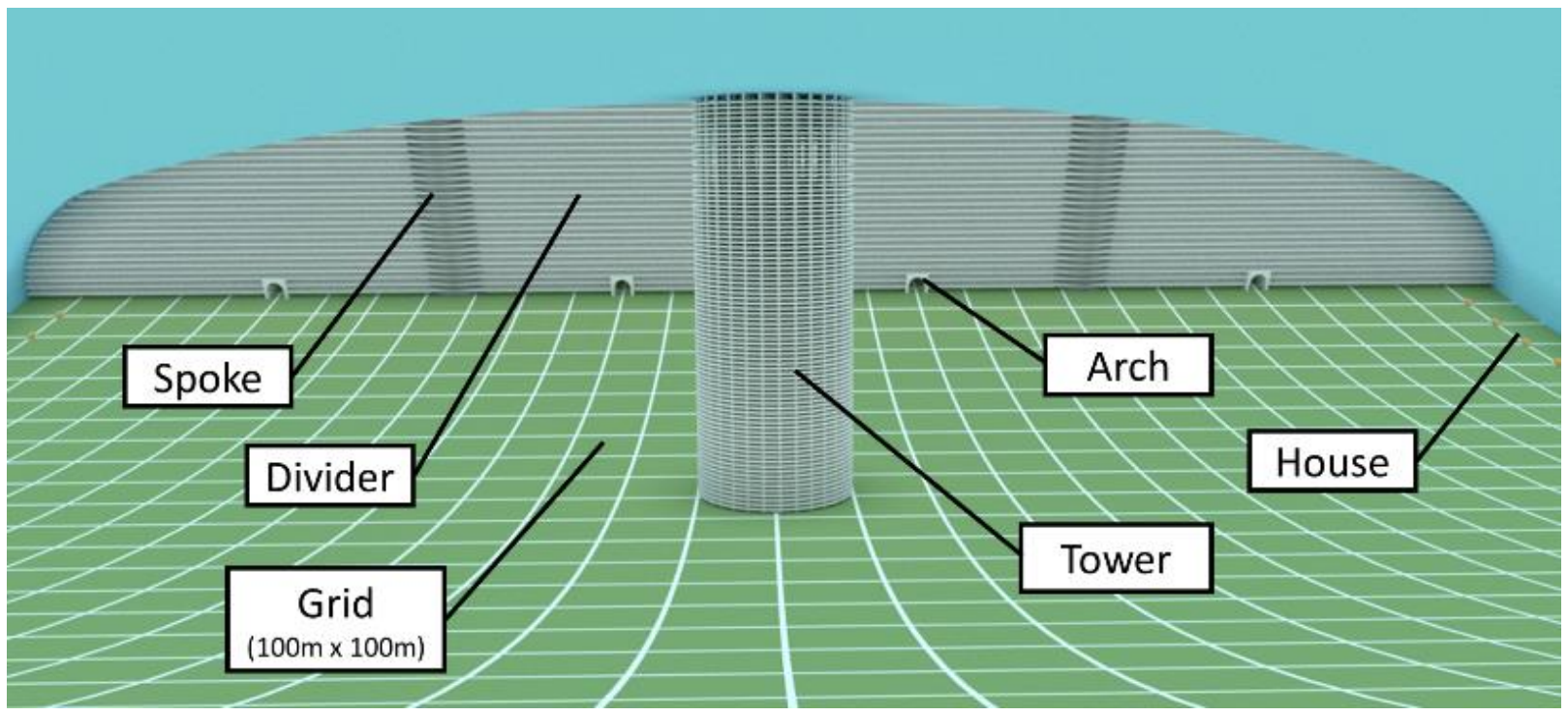

| *a) Web trusses on Atira* | *b) Inside station – tower, divider, and spokes* |
|---|---|

*a) Credit: Self produced using Blender, Image ESO/Serge Brunier [Brunier 2009] [CC BY-4.0], and Doug Ellison model [Ellison 2018] [CC BY-4.0]*
*b) Credit: Self produced using Blender*

**Figure 3-22 – Atira Space Station – Web Development and Interior View**

We compare the material provided by the asteroid and used by the station. The asteroid is about 4.8 kilometer in diameter. The total volume of material in the Atira asteroid is almost 22 billion cubic meters. By our estimates, it would produce about 8 billion cubic meters in usable raw material. For comparison, we use the same previous example torus station. Details on the material supply and requirement is shown in Figure 3-23. That station requires 2.6 billion cubic meters of the oxide building material. The fill and structure of the large outer torus uses nearly all that processed oxide material. We include estimates of the metal and volatiles that would be harvested from the asteroid. Not all of the asteroid would be mined to produce the required building material. Another additional 5 billion cubic meters of building material could be produced from the unused surplus of the asteroid. Our plan is to use the oxides as the building materials and not the metal. Only 9000 cubic meters of metal is needed to construct the station. There should be almost a billion cubic meters of metal in the Atira asteroid and 341 million cubic meters of metal will be harvested. The station has over 2.5 billion cubic meters of storage space. The unused metals will be inventoried and stored in the station for future colonists.

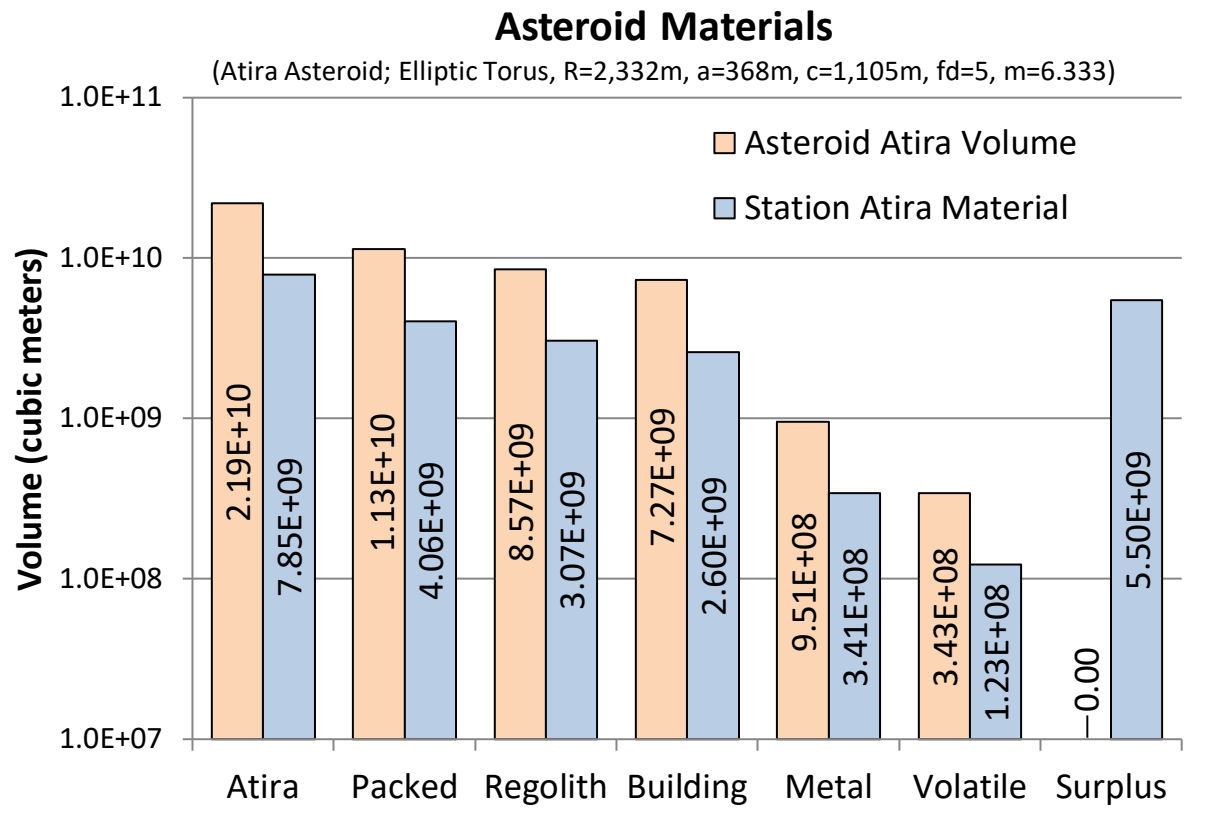

*a) Material from Atira Asteroid*

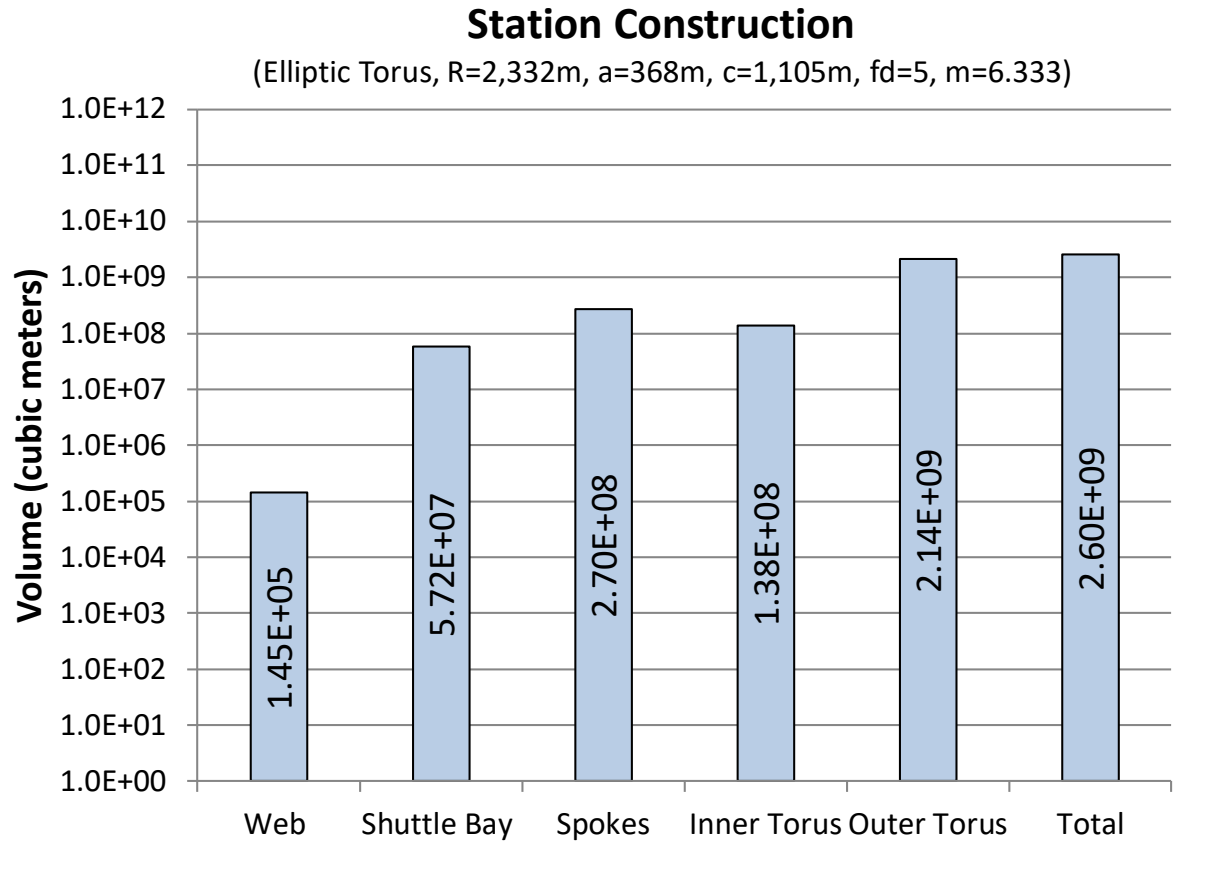

*b) Material Used by Station*

**Figure 3-23 – Building Material for Space Station**

### *3.3.5 Station Building Materials*

We show in Figure 3-24 the estimated volume of material used in an Atira torus station design as we vary the major radius. The horizontal axis shows the major radius from 1200 meters to 4000 meters. The minor radii of the elliptic outer torus are proportional to the major radius. The smaller minor axis length is about 1/6 the size of major axis. The other minor axis length is 3 times the smaller minor axis. This design includes an inner torus that is positioned halfway between the shuttle bay and the outer torus. The inner torus has a minor radius of 75 meters. This simulation assumes 20-meter-thick exterior walls. The Atira asteroid could produce 7.3 billion cubic meters of material for construction. This is after compressing out the porosity and assuming over 20% loss in processing. This only includes the bulk stony material and does not include the metal or volatiles. The vertical axis of

Figure 3-24 shows the required construction material ranging from 0.8 billion cubic meters to 7.3 billion cubic meters. This represents from 11% to almost 100% of the available construction material. The chart shows the outer torus uses the most material of all the station elements. The volume includes the material needed to construct the trusses, to fill the exterior walls with regolith, and to cover the exterior and floors with panels. Our space station structure will use anhydrous laminates in structural beams and as skin, regolith as fill, and basalt fibers as cables for additional strength and rigidity. We find that the structure is 95% basalt fill (like gravel), 4.7% anhydrous glass beams and panels from the basalt materials, and 0.3% basalt fiber cables. Most of the material is used to fill the exterior shell of the outer torus.

### *3.3.6 Station Population*

Figure 3-25 shows the population supported as the major radius of the station is changed. The maximum population assumes all interior surface areas are used, the floor spacing is 5 meters, and each person is allocated 144.2 square meters. For a 1200-meter radius station we see the maximum population is over 2 million. For a 4000-meter radius station, the maximum population is about 64 million.

Many areas of the station are not desirable for long term habitation. Figure 3-25 includes a more realistic population estimate where smaller percentages of the surface areas are used. Most individuals will reside in areas with near Earth-like gravity. As an example, few people will live in the higher gravity depths of the outer torus nor in the lower gravity between the top of the outer torus to the shuttle bay. In Figure 3-25 we used the top 10 floors of the outer torus surface area, 5% of the inner torus surface area, 5% of the spokes surface area, and 2% of the shuttle bay surface area. We see the population varies from 570,000 to 5.660,000 individuals. With the increasing radius, the area used by the realistic population estimate ranges from 30% to 10% of total available surface area. We envision that the unused surface area could be used for tourism, manufacturing, agriculture, and storage.

## 3.4 Space Stations – Summary

In this section, we reviewed space station background information and introduced features for our restructuring. These features included the multiple floors, rotationally balanced geometries, and Earthlike gravity ranges. To support these restructuring features, we refined slightly the set of typical space station geometries to address rotational stability and gravity issues. Spherical stations become ellipsoidal stations; long cylindrical stations become short hatbox stations; and circular cross section torus stations become elliptical cross section torus stations. We found dumbbells have a rotational wobble that requires additional analysis. We also reviewed features of the interior environment such as the surface area allocation to living quarters, public areas, industry, and agriculture. The analysis included computing the floor surface areas, the volume of construction material, and the mass of the station. Sizes, densities, and quantities of the individual pieces of the major station components were summed to produce the details for this analysis. We recognize that there could be structural issues with some of these stations. Ultimately, we need to consider qualitative and structural metrics in selecting a station geometry. Based on our current analysis, we believe that an elliptical cross-section torus is the preferred geometry for our restructuring effort. We finished this section with more details on an example of that station.

# 4 Asteroid Restructuring – Robotics

Asteroid restructuring mandates understanding space stations, asteroids, and robotics. We have covered the asteroid and materials that will be used to construct the station. We have covered the type of space station that will be created by the asteroid restructuring process. In this section we cover robotics and the technologies that will be used to construct the station. We include multiple subsections to cover the robotics background, analysis, and results. We offer details on robots, autonomous systems, and self replication.

## 4.1 Robotics – Introduction

Our initial robotic workers will use 21st century technologies including solar cells, miniature efficient motors, advanced computing, robotic software, and state-of-the-art artificial intelligent software. In a lunar factory study, Metzger advocated to first produce tools and support equipment using technologies from the 1700's [Metzger et al. 2012]. We agree with that recommendation. Our initial robots will be able to make these tools and equipment with a modest set of supplies and without industrial infrastructure or operator direction. Our manufacturing techniques will reach a technology level of the 1800's. Most of the space station framework will be produced with lower quality material and with low precision. We have no goal to build advanced technologies such as solar cells or computer chips. The end goal is to build a rotating space station framework with a rich store of cataloged volatiles and metals. The rotation will produce centripetal gravity. The thick shell enclosing this framework will provide radiation protection. We envision this framework and catalog will attract and support follow-up missions with crews bringing more advanced manufacturing technologies.

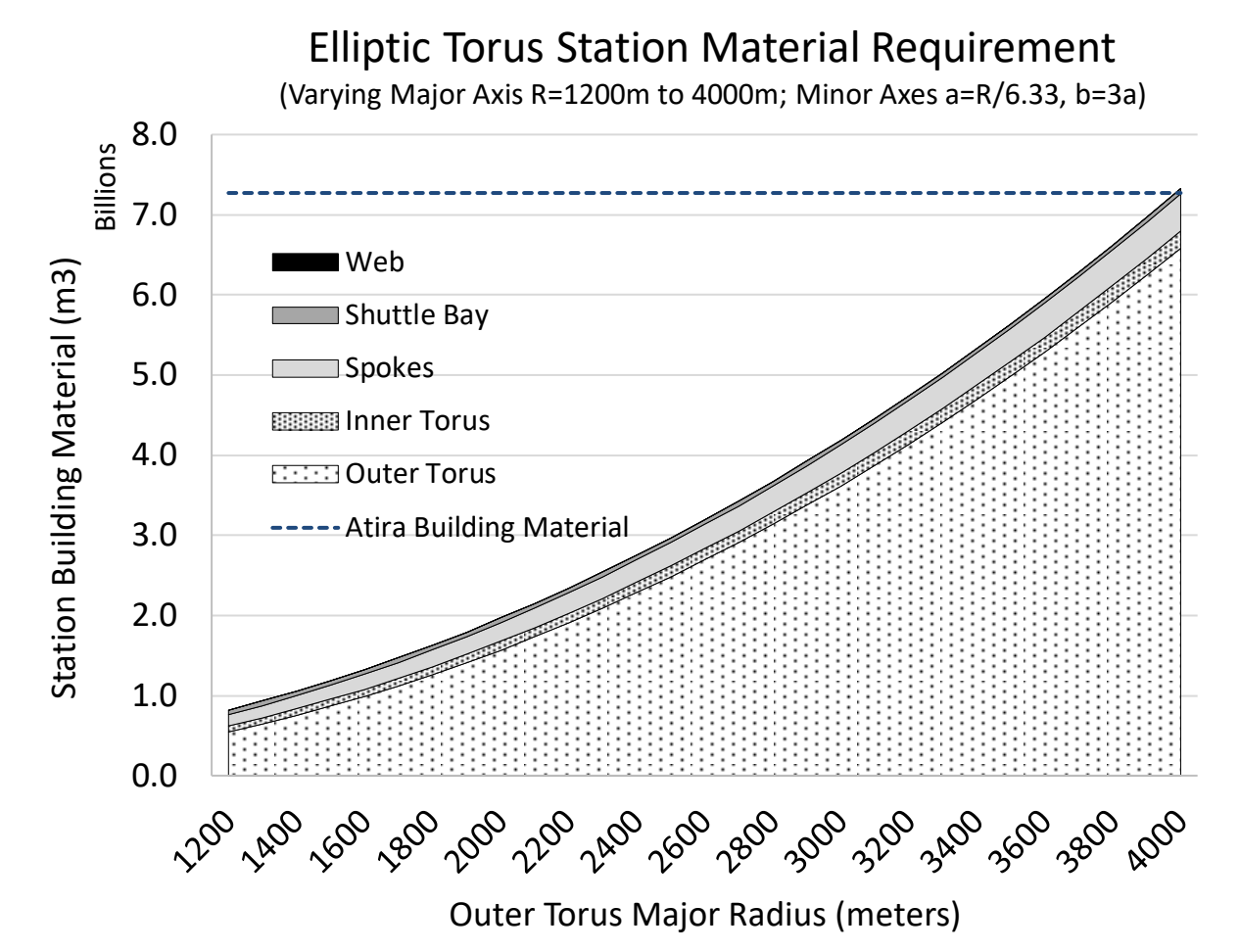

**Figure 3-24 – Construction Material Estimates**

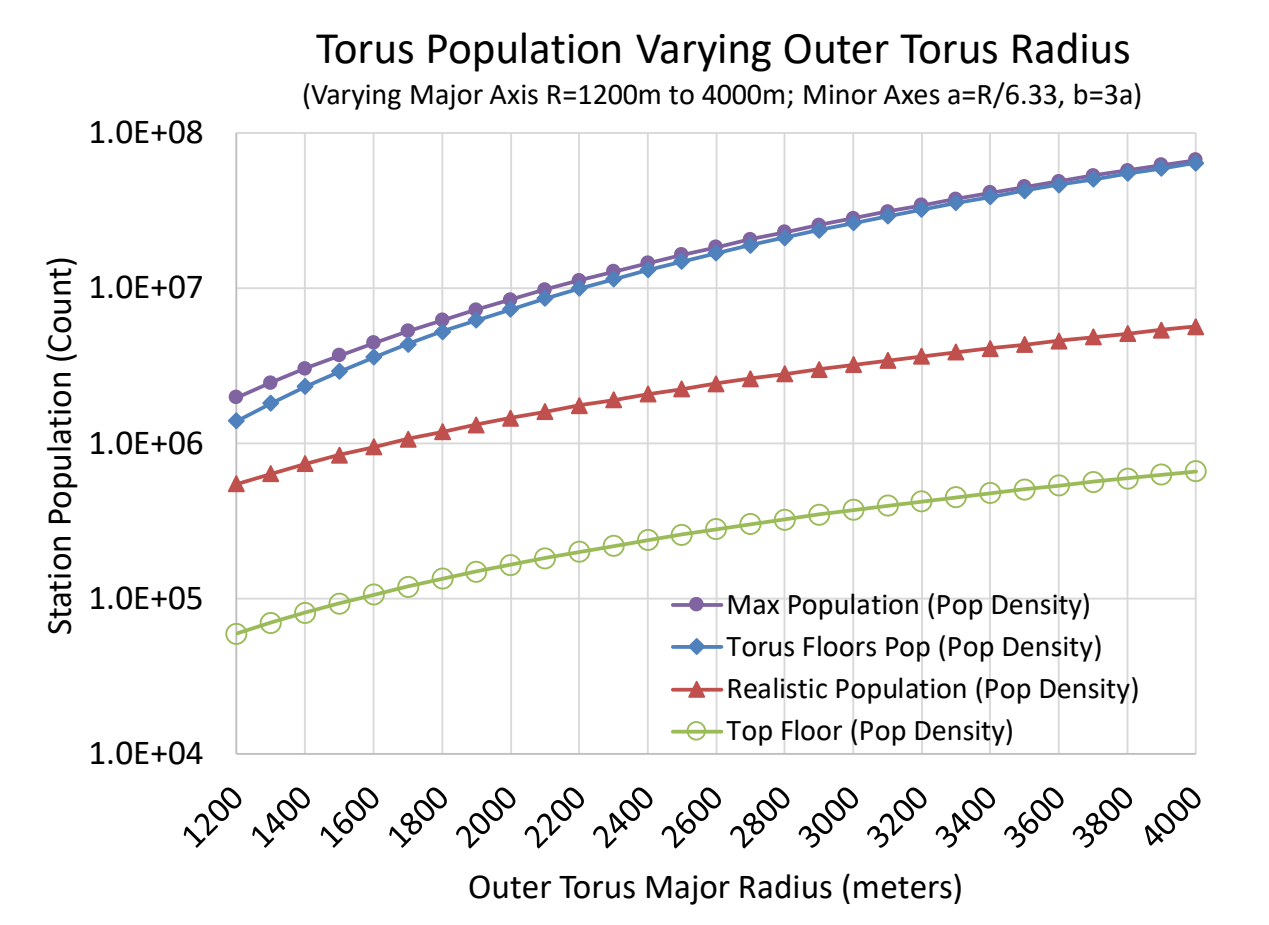

**Figure 3-25 – Station Population Estimates**

We provide a brief overview of the background and technologies of robots. This includes example systems for space exploration and mining. We provide a set of background credits supporting our robotic technologies. To operate in the asteroid environment and perform the complicated restructuring process, the robots will use autonomous system software. The restructuring process of an asteroid represents a huge construction task. To support this huge restructuring project, we will need the equivalent of millions of robots. It is too costly to ship the millions of robots; as such, we use a technology called self replication. We provided detail on a version called productive self-replication where robots (replicators) can make copies of themselves and produce products. We introduce the concept of having the robots make specialized tools (helpers) to improve their productivity. Many of the tools are mechanical robots (or automatons) programmed with simple state machines. We include our mathematical analysis of the performance of these replicators and helpers. We describe our simulations of these robots and systems. The results of these simulations show that asteroid restructuring is a viable approach to create a space station from an asteroid.

## 4.2 Robotics – Background

The restructuring process uses autonomous robotics instead of manual or teleoperated labor. Robots are comprised of six major components: Program, Sensors, Actuators, Effectors, Locomotion, and Power. Our asteroid restructuring robots requires all these components. To introduce robotics, we provide background examples of robots used or planned for space exploration and mining. Besides background and examples on robots, we also include background on important supporting technologies: autonomous system software and self replication.

### *4.2.1 Robotics – Background Credit*

For our background on robotics, we present important topics with their researchers and their work.

- **Robotics:** NASA projects provide a wealth of information for robotics. Lunar vehicles have evolved from the 1950s [Koelle and Williams 1959], 1970s [Nishioka et al. 1973], to recent ideas in the 2020s for Mars and Lunar missions [Artemis 2022]. We use this information as a foundation for our asteroid restructuring mission. We also credit concepts to RASSOR (Regolith Advanced Surface Systems Operations Robot) [Mueller et al. 2013] and to ATHLETE (All-Terrain Hex-Limbed Extra-Terrestrial Explorer) testbed [Volpe 2018b].
- **Space Technologies:** NASA has funded industry partners to advance robotic technology for space. The space industry company Made in Space plans to use their Archinaut spacecraft in Earth orbit to manufacture structures. Their technology includes 3D printing in the zero gravity of space and could be commercially operational by the 2020s [Wall 2017]. Honeybee Robotics was funded to develop the Spider Water Extraction System as part of the NASA World Is Not Enough (WINE) program [Honeybee 2019]. This system has demonstrated the ability to gather water from asteroid-like materials under thermal vacuum conditions.
- **Self Replication Projects:** NASA projects have detailed self-replication projects on the moon [Freitas and Gilbreath 1982] [Metzger et al. 2012]. Lewis-Weber describes a lunar self-replicating system [Lewis-Weber 2016]. We overview these projects' concepts, tools, schedules, and costs. The projects include concepts such as teleoperation, seeds, and closure.
- **Tracks for Navigation:** Several studies have considered using tracks for self-replication systems to simplify navigation and locomotion [Lackner and Wendt 1995] [Moses et al. 2014]. Our restructuring process builds a web of trusses over the asteroid to provide this simpler work environment instead of the natural asteroid terrain. This simplicity better supports our mechanical computing automatons.
- **Artificial Intelligence:** The replicators (and base station) must have the "intelligence" to meet assigned goals in a complex and dynamic environment. In the author's career, he studied and published multiple papers and proposals to apply cognitive processing to defense products [Jensen 1996]. We offer a short credit to the long history of artificial intelligence. In our studies, we have captured history and details of cognitive architectures from [Brooks 1985] [Boyd 1987] [Albus 1997] [Sifakis 2018]. The cognitive system for our asteroid restructuring might easily be extended from current self-driving autonomous systems [Liu et al. 2020]. We offer only a brief introduction of our intelligent system in this paper.
- **Mechanical Automatons:** Our restructuring process creates mechanical automaton using 18th and 19th century technologies and materials. One example of mechanical computers and automatons is a mechanical monk built in the 1560s by Juanelo Turriano using springs, cams, and levers [King 2002]. Another example is the Babbage Difference Engine built in 1822 [Copeland 2000]. A recent NASA funded study considered using mechanical computing for the harsh environment on Venus [Good 2017] [Sauder 2017]. NASA also funded another study project (RAMA) and found it is possible to build mechanical computing on an asteroid [Ackerman 2016] [Dunn and Fagin 2017]. The primitive mechanical control is simpler to manufacture than semiconductors and electronics.

We also present new concepts discovered during this restructuring research. In later sections, we provide more details on these robotic technology discoveries.

- **Parallelism and Specialization:** Our asteroid restructuring uses exponential growth of replicators and uses specialized tools to improve the process performance. Parallelism provides a speed-up in the execution time using multiple Replicators (Spiders) and Helpers (Tools and Equipment). Specialization provides a speed-up in the task execution time using the Helpers (Tools and Equipment). We recognize this growth as a form of self replication called productive replicators [Freitas and Merkle 2004]. The extensive use of parallelism and specialization may represent a somewhat novel approach to self-replication.
- **System Analysis Extension:** We review and extend analysis concepts from [Hall 1999] to our replicators (Spiders), helpers (Tools), and products (Construction Material and Structures). Hall did not explicitly cover multiple types of tools. Our productive self-replication creates different tools and different products at different rates. The Hall analysis evaluated a system where the self-replication and the product production were at the same rate [Hall 1999]. We apply and extend Hall's analysis to include different products and rates.

### *4.2.2 Robotic Examples*

There is good background on robotics for our restructuring effort. Multiple robotic systems have been proposed and designed to support lunar or asteroid mining. Simulations have been performed, papers published, and prototypes have been built. We offer in this section an overview of those systems. Our approach in the restructuring effort is to use existing and historic technologies. It is not our goal to create today's advanced manufacturing capabilities on the asteroid; instead, we will use a limited number of 21st century robots to construct 18th century mechanical automata. The following systems are examples of early 21st century robotic technology.

The NASA Jet Propulsion Laboratory (JPL) has multiple space robotic projects [Volpe 2018]. One of those robots is the LEMUR (Limbed Excursion Mechanical Utility Robot). The second version of this system, LEMUR IIa, consists of six limbs arranged around a hexagonal body platform [Volpe 2018a]. These limbs incorporate a feature that allows the rapid change of its end-effector tools [Volpe 2018a]. We show in Figure 4-1 two pictures of this robot. Our spider (robot) will be similar to this design with its multiple arms, replaceable end-effectors, sensors, and frame.

The NASA Innovative Advanced Concepts program funded a mechanical rover in 2015. Jonathan Sauder proposed the Automaton Rover for Extreme Environments (AREE) to operate in the harsh environment of Venus [Sauder 2017]. This clockwork mechanical robot would explore the Venus surface and only use primitive components such as levers and gears. The mechanical technologies could survive in the Venus conditions where most electronics would simply melt. We include an image of the AREE rover in Figure 4-2 [Good 2017] [Sauder 2017]. Our restructuring process takes advantage of mechanical computing because it can be built using impure materials with millimeter tolerances. Semiconductor electronics require more pure materials with micrometer tolerances.

NASA has been studying and documenting space missions since the 1950s. These reports offer great detail and concepts, often studied for years or decades before implementation. One historic 1973 document is entitled *Feasibility Of Mining Lunar Resources for Earth Use: Circa 2000 A. D.* [Nishioka et al. 1973]. This long memorandum details the technologies and systems required to establish the mining base, mine, refine, and return the lunar resources to earth for use. They also include gross equipment requirements such as weights and costs. We include a sketch of an automated miner vehicle from that study in Figure 4-3. Some of our robotic helpers have characteristics similar to this miner. We use many concepts from this wealth of NASA information throughout our restructuring process.

### *4.2.3 Robotics and Autonomous Systems*

The robots will have autonomous system software to operate in the asteroid environment and to perform the tasks in the dynamic and complicated restructuring process. The

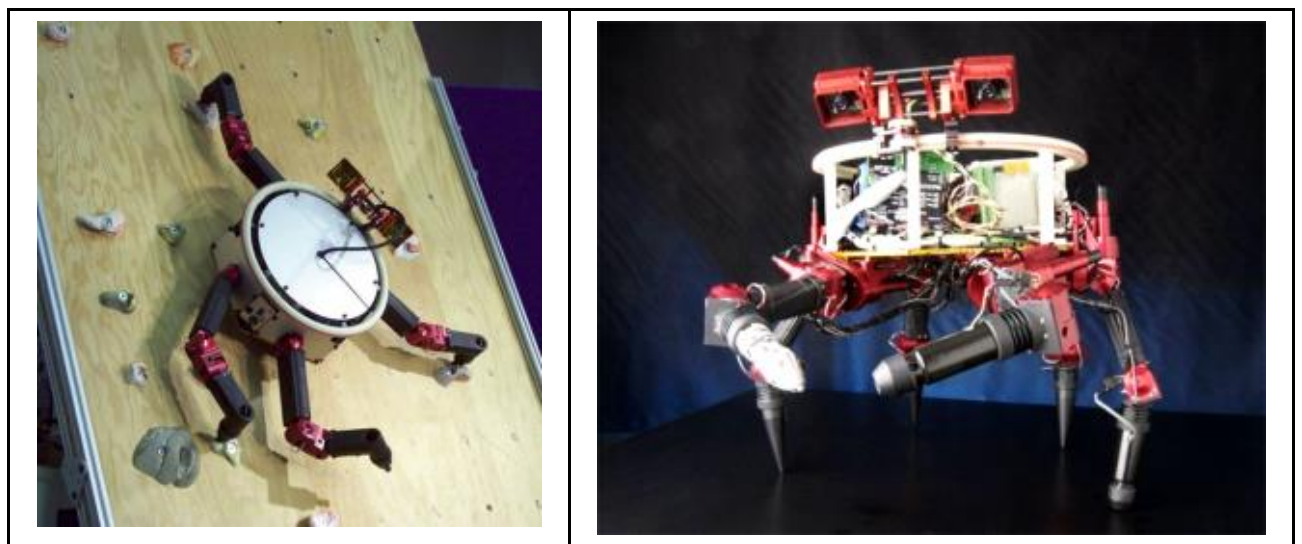

*a) LEMUR IIb climbing* | *b) LEMUR IIa*

*Image Credit: NASA, JPL, [Volpe 2018a] [JPL Image Public Domain]*

**Figure 4-1 – NASA JPL LEMUR II Robot**

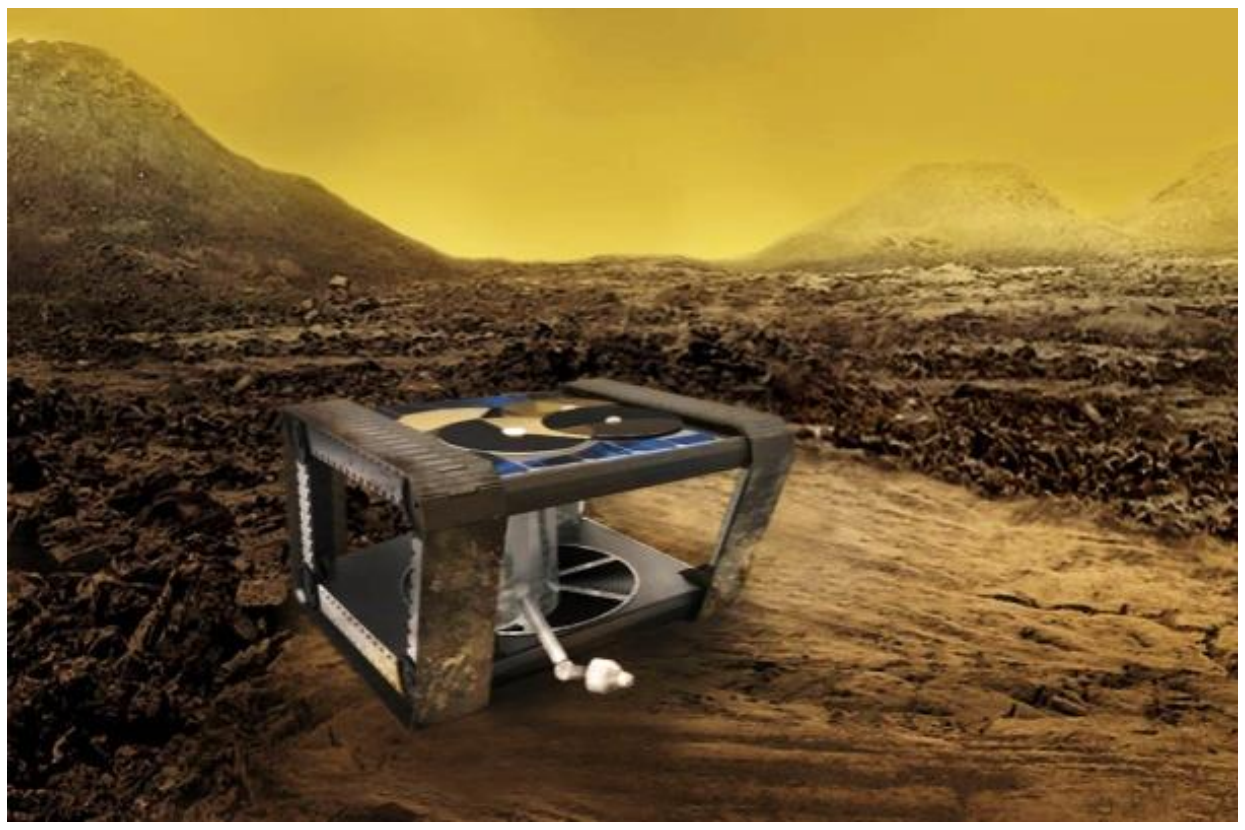

*Image Credits: NASA/JPL-Caltech [Good 2017] [JPL Image Public Domain]*

**Figure 4-2 – NASA JPL-Caltech Mechanical Rover**

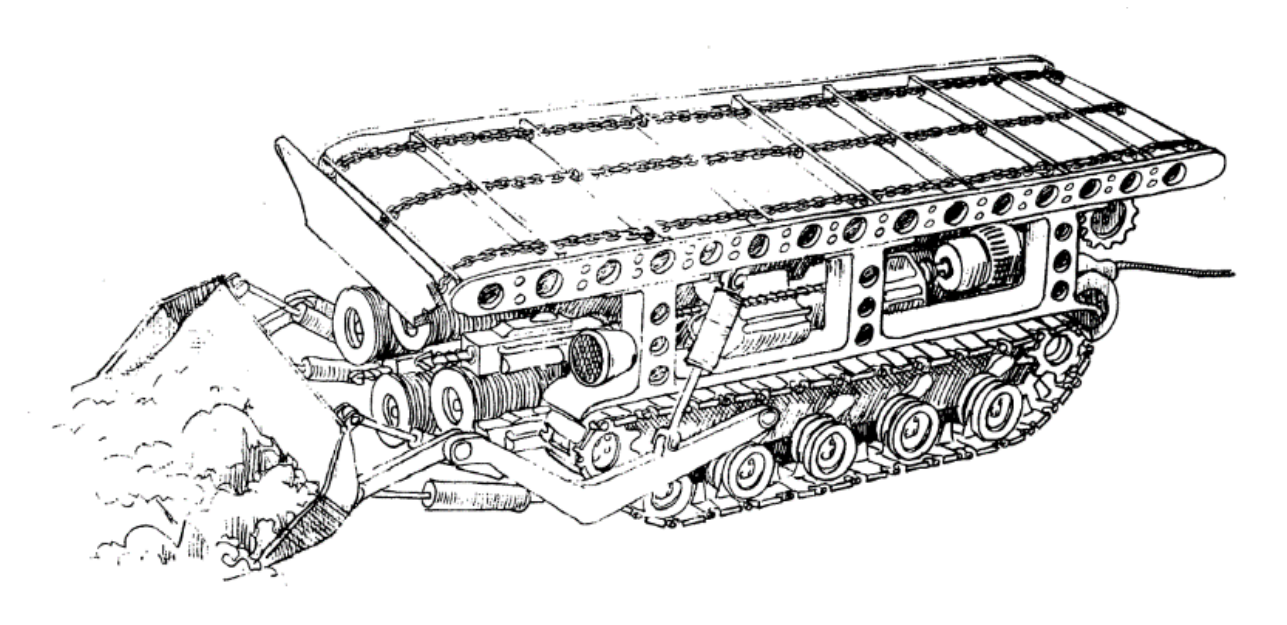

*Credit: NASA, [Nishioka et al. 1973] Figure 5-1 [NASA Report Public Domain]*

**Figure 4-3 – Automatic Lunar Miner**

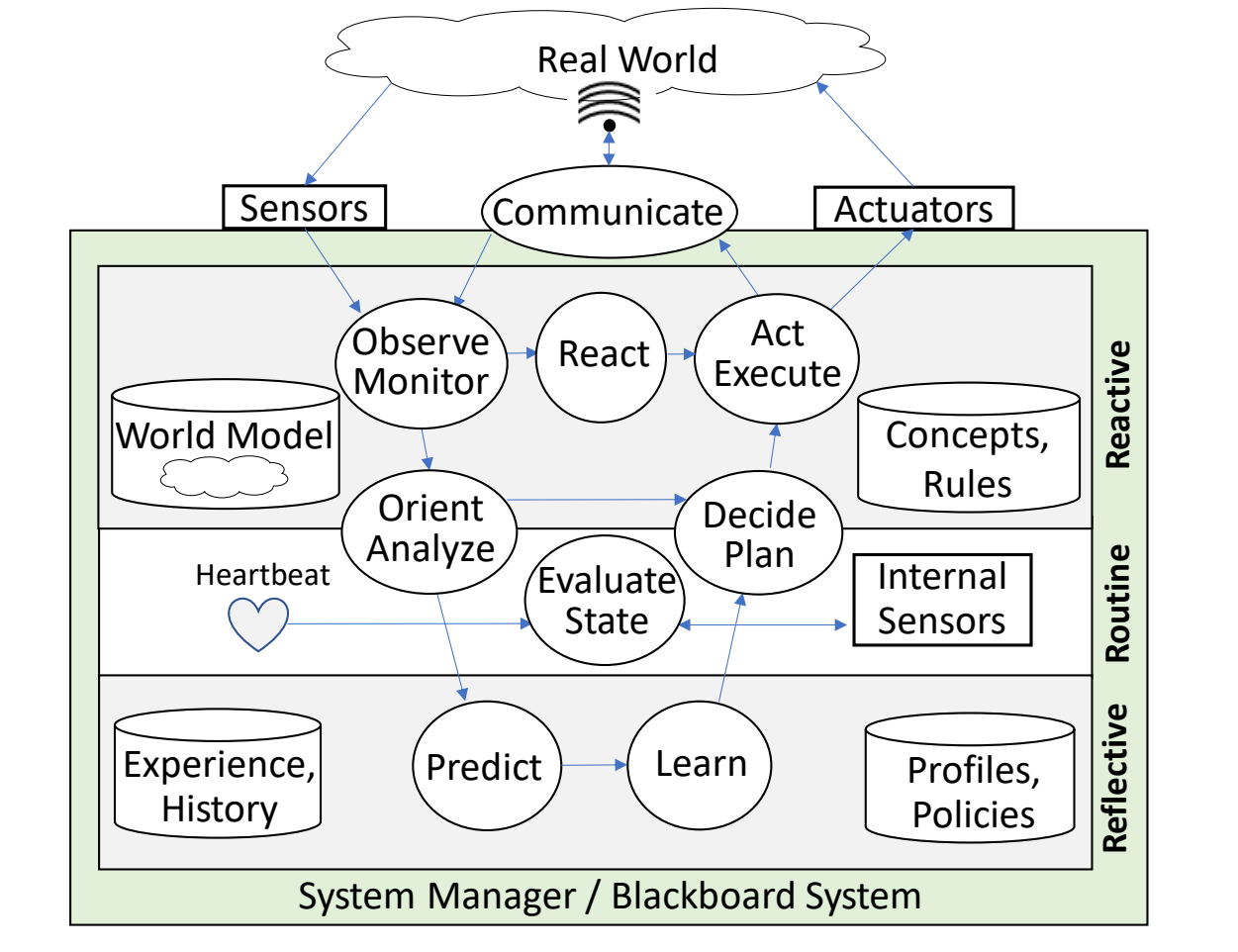

*Cognitive Architecture for Restructuring Mission*

**Figure 4-4 – Cognitive System Architecture**

software framework of the robots will integrate real time processing, artificial intelligence, image processing, knowledge bases, and more.

Our restructuring effort will need an embodiment of an autonomous system. Our plan is to integrate individual artificial intelligent algorithms and instantiate the building blocks of the autonomous system architectures. Our replicator (and base station) must have a complete plan and the "intelligence" to meet assigned goals. The autonomous system combines five complementary capabilities: Perception, Reflection, Goal management, Planning, and Self-adaptation [Sifakis 2018]. Such capabilities started with research in the 1980s [Brooks 1985] [Boyd 1987].

Many years ago, we architected an autonomous system as a military hierarchy of nodes [Jensen 1996]. There is value in such a hierarchy. The restructuring process software will communicate as a hierarchy of nodes executing through crews of spiders, lead spiders, the base station(s), an orbiting system, and terrestrial support computers. Eight years later we enhanced our 1996 design to become a cognitive system with building blocks of sensory processing, world modeling, behavior generation, and situation assessment. Design and proposal work was performed with planning experts [Nau et al. 2003] and with multi-agent and blackboard experts [Lesser and Corkill 2014]. The system in each node was to be built using multiple intelligent agents working through a common blackboard system.

Today, new software environments and standards exist to support our restructuring software system. The cognitive system for asteroid restructuring might easily be extended from current self-driving autonomous systems [Liu et al. 2020]. For reference, we have added building blocks to support communication and a heartbeat for self-awareness to our earlier autonomous system designs. We follow many of the architecture concepts from [Sifakis 2018]. We illustrate the architecture with those new building blocks in Figure 4-4. This architecture highlights the building blocks necessary to support a restructuring process. Agents represent those building blocks. These intelligent agents provide an interface to *physical devices* (such as sensors, actuators, and communication), *software algorithms* (such as simulations, planning, and optimization), *architecture databases* (such as world model, concepts and rules, experience, and policies), *architecture system modules* (such as observer, orient, decide, and act [Boyd 1987]), and *real time control capabilities* [Albus 1997]. The cognitive system architecture in Figure 4-4 includes hierarchy layers of building blocks. The layers represent a cognitive hierarchy with reactive, routine, and reflective levels [Sterritt and Hinchey 2005]. They also represent an autonomous system hierarchy with automation, autonomous, autonomic, and cognitive layers. We do not delve into the implementation details in this paper.

### *4.2.4 Robotics and Self-Replication*

Self replication will produce many robotic units. Buzz Aldrin stated: *"It's amazing what one person can do, along with 10,000 friends."* We find it's amazing what a single probe and 26,500 restructuring units can accomplish. It may not be launching a man to the moon, but we found those units can restructure a large asteroid into a rotating space-station framework in 12 years,

The original seed probe with its 4 spiders and a few supplies creates quite a legacy of equipment. The restructuring process builds thousands of robotic spiders, tools, and equipment using the seed package and the raw materials of the asteroid. Our baseline example simulation created almost 3000 spiders and over 23,500 other pieces of equipment. In about 12 years, an asteroid the size of Atira would be restructured into the enclosed framework of a space station.

It would be very costly to launch thousands of robots, mining equipment, and construction equipment to accomplish this goal. Self replication provides the capability for a system to create a copy of itself. The restructuring of an asteroid represents a huge construction task. Launching a probe with the right tools and building that equipment at the asteroid is essential from a cost standpoint. These thousands of units are produced with *productive replicators*. This is a form of self-replication where the replicators can make copies of themselves and produce products. In our restructuring system we include **Replicators** (*Spiders)*. They produce **Helpers** (*Tools and Equipment*). Combined they produce the final **Products** (*Construction Material and Station Structures*). Self replication allows us to not launch the thousands of pieces of equipment, and instead, build them at the asteroid.

**Self Replication – Introduction:** The goal of the restructuring process is to use a single rocket launch to send a seed probe to an asteroid. The restructuring process exploits self-replication to reduce launch costs and construction times. We use robotics, image processing, and artificial intelligence to completely automate the restructuring process. Other than a modest seed package of materials, we use only the bulk material of an asteroid to build our station. With that modest seed package of materials and tools, robotic workers use the asteroid material to create copies of themselves, create more tools, and other forms of vehicles and automata.

Our restructuring concept is a form of self-replication. Our seed probe includes a small set of 21st century robots along with fabrication and testing equipment. It includes sufficient supplies to support the construction of a few thousand robots and some materials to support the construction of other equipment. The robot frames are built from local asteroid material. The supply of robotic electronics and actuators are installed in those frames. By themselves, thousands of these robots would take many centuries to build a large space habitat. An innovation of the restructuring process is to provide the robots with many tools and equipment to complete the space habitat. These robots construct tools and equipment using processes available during the first industrial revolution. Many pieces of equipment will be powered with Stirling engines and/or clock springs. Heat and furnaces will be solar powered. Instead of large industrial complexes, overwhelming numbers of small mining and construction equipment restructure the asteroid into a space habitat.

Self replication and the restructuring process reduce the seed probe costs. We do not need to develop and bring processes to manufacture 20th or 21st century technologies such as semiconductors, solar cells, electrical wiring, and HVAC components. Instead, we need to develop processes to manufacture 18th or 19th century technologies like shovels, plows, wagons, trusses, and bridges. Using in-situ material to build the tools, equipment, and habitat reduces the weight of the seed probe and launch costs.

**Self Replication – History:** Original historic details on self replicating systems first appeared in the 1950s and 1960s by John von Neumann [von Neumann 1966] and Edward Moore [Moore 1962]. In the 1970s, Freeman Dyson developed concepts for four terraforming self-replicating systems [Dyson 1970]. NASA commissioned a study detailing a lunar self-replicating system in 1980 [Freitas and Gilbreath 1982]. This study describes a 100-ton probe with over a dozen systems including paving robots, mining robots, warehouses, computing, and fabrication. Robert A. Freitas Jr. and Ralph C. Merkle wrote an extensive book on Self-Replicating Machines [Freitas and Merkle 2004]. A study in 2012 describes a more modern version of a self-replicating lunar industry [Metzger et al. 2012]. Their study reduced the initial seed probe weight to 7.7 tons and includes a similar set of systems such as excavators, chemical plants, refineries, and solar cell manufacturing. This newer study includes 3D printers. The authors of that paper believe that this lunar industry will revolutionize the human condition. Those lunar self-replicating examples require teleoperation initially and periodic supplies from the Earth.

The restructuring process alone cannot accomplish everything envisioned by the 1980 NASA study [Freitas and Gilbreath 1982] or the 2012 bootstrapping study [Metzger et al. 2012]. Their designs resulted in manned stations with all the supporting environment and technologies. This is with many high-cost launches and with considerable astronaut labor in low gravity and unprotected radiation environments. The restructuring process can construct a space station framework at a lower cost and with less technical support. The restructuring process finishes the enclosed framework that rotates to provide centripetal gravity, has a thick shell to provide radiation protection, and includes an inventoried store of metals and volatiles. Follow-up missions are required to make the station habitable. The workers and then colonists of those missions will be able to work in this framework environment without negative effects from microgravity and radiation.

Our restructuring approach and study agrees with the concepts, direction, and benefits captured by Metzger, Freitas, Gilbreath, and their colleagues; however, our restructuring approach increases the amount of automation, eliminates the human teleoperation, and reduces the costs.

**Self Replication - Equipment and Tools**: We considered using and building a single type of robot to perform all tasks. This approach quickly ran into complexity, performance, and production bottlenecks. We found many of the restructuring tasks are simple, repetitive, and constrained to well defined areas. This led to the concept of building specialized tools to offload that repetitive work and reduce the required number of spiders (replicators). The launch cost limits the number of spiders while building simple tools and equipment is essentially free and unlimited using in-situ asteroid material. Our approach evolved over the course of this research to include multiple types of tools to manage larger volumes of material and specialization to perform specific tasks more efficiently. In the replicating process, these helpers (tools and equipment) are mechanical automata built from in-situ material and use 18th and 19th century processes. The mechanical automata are sophisticated state machines using hardware such as gears, levers, springs, and cams. The primitive mechanical control is much simpler to manufacture than semiconductors and electronics.

Eventually, the replicators (spiders) only need to load, monitor, unload, wind springs, and adjust settings on those helper pieces of equipment. These helpers are mechanically designed to perform repetitive operations. A spider would adjust levers and knobs on equipment to control and tune the operation. A recent NASA funded study considered using mechanical computing in the harsh environment of Venus [Good 2017] [Sauder 2017]. The RAMA asteroid project also plans to use a purely mechanical analog computer and use things like gears, rods, and levers linked by chain belts. Jason Dunn from Made In Space [Ackerman 2016] comments that "*This is all very old technology, but it's also well understood, reliable, and easy to build from simple parts.*" Project RAMA felt most of the fundamental technologies already exist. The NASA funded study found it is possible to build an asteroid spacecraft using mechanical computing [Dunn and Fagin 2017].

Electronics can survive in the asteroid environment and could be used. Electronics could perform many of the same operations as the mechanical computing. The advantage of mechanical computing for restructuring is the required manufacturing materials and tolerances. Building semiconductor electronics requires pure materials and micrometer tolerances. Mechanical computers could be build using impure metals or ceramics with millimeter tolerances. Our restructuring process builds in-situ mechanical devices to perform the repetitive restructuring operations. The large number of mechanical automatons provide an immense productivity gain. Building equipment on the asteroid with in-situ material dramatically reduces the launch costs.

Our spiders and equipment build and operate on a web of 1-meter-wide trusses. This web provides a simpler navigation environment for the mechanical computing than the natural asteroid surface. A similar set of tracks were considered for self-replication to simplify navigation and locomotion in a 1995 study [Lackner and Wendt 1995].

**Self Replication – Closure:** Closure is a metric that measures the ability of a replicator to gain access to the resources required for replication [Freitas and Gilbreath 1982]. A self-replicating system with 100% closure has access to all the materials require to self-replicate. We send sufficient supplies "vitamins" to produce 3000 replicators (Spiders) using only in-situ asteroid material. Those replicators build tools and equipment using only the asteroid material and very limited supplies from the initial probe. The replicators and helpers are primarily built from in-situ material (95% to 99%) with a small percentage of vitamins (1% to 5%). We have 100% closure using the few supplies from the single space probe until the "vitamins" are exhausted. Once completely exhausted, no additional tools can be built.

**Self Replication – Tool Production:** Our asteroid restructuring approach can be compared to the effort of early pioneers. A wagon heading west in America in the 1800s could not carry enough supplies to support a family for the journey or at the destination. These pioneers brought tools with them to be self-sufficient [Williams 2016]. Our restructuring relies on self-replication, but more important, on the production of tools to make the restructuring effort self-sufficient and sufficiently productive.

Our restructuring process relies on productive replicators. These replicators can make copies of themselves and can also make other products [Freitas and Merkle 2004]. The replicators are the "spider" robots. Four of these spiders are sent with the initial bootstrapping probe. The seed probe lands on one of the asteroid poles and becomes the base station for early operations. The base station can process regolith and produce structural rods, tiles, and panels. The probe contains enough supplies to support the construction of 3000 spiders. The supplies are small and lightweight; they include small electro-mechanical modules and solar cells. The housings and frameworks for these spiders and solar panels are built with local asteroid material. The spiders assemble the solar cells into frames to create solar panels. The modules and solar cells are designed to clamp easily to the frames. The seed probe includes sets of connectors. Spiders use these connectors to interconnect the electro-mechanical modules, jigs, and legs. Legs are made from structural rods. The four initial workers assemble the modules, regolith frames, rods, solar panels, and connectors to produce more spider robots.

The base station includes a few hundred jigs for the spiders to start the restructuring process. The spiders use these jigs to build these housings and frameworks. Testing jigs are used to measure material samples for inventory and storage. The base station also includes testing capabilities.

The spiders will also create other products (helpers/tools and equipment) to support the restructuring effort. Early simulations quickly determined that thousands of robotic workers would be insufficient to build a large station in a reasonable length of time. We found spiders were performing numerous, repetitive, time-consuming tasks in our simulations. We began to build helper systems to augment the initial robotic workers. Trucks can haul materials to and from processing stations. Diggers can work on breaking rocks and producing loose regolith. A Sun Tracker can aim mirrors for solar furnaces as the asteroid rotates. Under the spider supervision, simple clockworks and mechanical settings provide the control for these helper systems. The spiders build larger tools such as vehicles - trucks and movers. The spiders will also build mining equipment such as jaw crushers, filters, and solar furnaces. They build more complex manufacturing devices such as truss builders, harvest units, and tile placement units. Power for these helper systems comes from springs (for clockworks) and from Stirling engines (for motion).

Navigation on an asteroid's surface is difficult because of the low gravity and rugged diverse terrain. The robotic spiders build a web of trusses to provide a much simpler navigation surface. The spiders make the trusses using regolith rods or tiles manufactured initially by the base station. Spiders weld the tiles together using a Fresnel lens jig to form trusses and panels. Initially simple legs provide the locomotion on those trusses in the near zero gravity environment. The simple legs are augmented with wheels and rollers to adapt to the increasing centripetal gravity on the rotating station.

The base station manufacturing is eventually relegated to other pieces of equipment built by the robotic workers. Many new pieces of equipment are built to crush rocks, filter material, extract metal, and produce tiles and panels.

The spiders will use 21st century technologies including solar cells, miniature efficient motors, advanced computing,

robotic software, and state-of-the-art artificial intelligent software. We agree with Metzger that the first produced tools and support equipment should be more like technology from the 1700's or 1800's [Metzger et al. 2012]. The advanced spider workers will be able to make these tools and equipment without additional supplies, infrastructure, or operator direction. Most of the framework and tools will be mass produced with lower quality material and with low precision. With asteroid restructuring, we have no goal to build advanced technologies such as solar cells or computer chips.

Our restructuring goal is not to build a self-replicating system. Our goal is to build the framework of a space habitat. We do not attempt unlimited self-replication and we strive to minimize human direction. The end goal is to build a rotating space station framework with a rich store of cataloged volatiles and metals. We envision this enclosed framework will attract and support follow-up missions with colonists and/or robots. These groups will then develop more advanced manufacturing technologies.

**Self Replication – Summary**: It would be wondrous to send a self-replicating system to an asteroid. Multiple studies and science fiction stories promote self-replicating systems. A single self-replicating system creates create copies of itself. The number of units increases exponentially and would quickly provide the "labor" to restructure the asteroid. Multiple studies describe self-replication with units ranging in size from nanobot assembles (nanotechnology) [Drexler 1986] to von Neumann machines [von Neumann 1966] to large factories [Metzger et al. 2012]. We introduced self-replication with a review of a 1980 lunar self-replicating system [Freitas and Gilbreath 1982] and a more recent self-replicating lunar industry [Metzger et al. 2012]. Both these self-replicating systems require teleoperation initially and periodic supplies from the Earth. Our restructuring concept assumes complete autonomous operation and does not require additional supplies from the Earth. Our restructuring process uses units called Productive Replicators [Freitas and Merkle 2004]. We increases productivity using parallelism and specialization. We advocate that launching a probe with the right tools and building equipment on the asteroid is viable and very cost effective.

## 4.3 Robotics – Analysis

We provide in following subsections the results of our self-replication analysis. We first extend a historic mathematical analysis to model replicators, helpers, and products. We review a specific example of our replicator, helper, and product model. We then provide the restructuring results using production rates of the replicators and helpers. We include our approach to analyze and simulate the entire restructuring of the asteroid. We describe our simulators, the restructuring equipment, and the initial seed package. We conclude this section with a summary.

### *4.3.1 Mathematical Analysis*

Our initial probe lands on the asteroid Atira. The probe serves as a base station and has some communication, sensor, computing, and manufacturing capabilities. It only has 4 robotic systems (spiders). The robots have 21st century technologies like sensors, motors, solar panels, processors, advanced software, and communication capabilities. We do not intend on building these 21st century technologies on the asteroid. We include with our probe several thousand small modules with robotic electronics and actuators. We build frames for the robots using in-situ materials to support this module. We analyze the restructuring self-replication growth. In particular, we consider the growth using different production and self-replication rates

**Background:** A historic analysis evaluated such a system where the self-replication and the product production were at the same rate [Hall 1999]. This analysis produced an equation to define the time in generations to complete the self-replication and production task:

$$2^k S_u = S_p ln2.$$

In this equation, k is the number of generations; Su is the size of the universal constructor (Spider); Sp is the size of the product. Robots and tools will produce different products at different rates. The productive replicator can produce copies of itself or produce different products. In our restructuring process we have different production rates for building tools and for tools building products. We initially only consider two production rates. This produced an equation to define the time in generations to complete the self-replication and production task:

$$2^k S_u = \frac{\lambda_u}{\lambda_p} S_p ln2.$$

In this equation, we include λu and λp as the production rates of the constructor and the product. The minimum time occurs when the total construction volume ($2^k S_u$) is equal to total production volume times 69.3% and the ratio of the production rates λu/λp. Universal constructors self-replicate until generation, $k_{min}$, and then all replicators, $2^{kmin}$, work on producing the product $S_p$. This analysis is obviously a simple extension to the previous Hall equation.

**Productivity Gains:** Our restructuring process takes advantage of productivity gains from both parallelism and specialization. Parallelism, the number of replicators, is the traditional improvement of self-replication and provides only part of our productivity gains. The replicators (spiders) also offload tasks to helpers (specialized tools and equipment). The helpers are designed to perform simple and repetitive tasks much faster than the general-purpose replicators. This specialization speeds up the execution of the tasks, improves the productivity, and frees the replicators to perform more complex tasks and management functions. There are similarities in the analysis of this system's parallelism and specialization to our analysis of system-on-a-chips [Jensen 2008].

We have developed many tools to support the spiders and improve the space station production rate. We analyze the product production in our system. We can use spiders to build the product, or we can use a helper tool to build the product. We first analyze the production mathematically and then with simulations. The results of this effort illustrate the trade-off between using sophisticated robot spiders and limited capability mechanical tools.

**Mathematical Analysis:** We mathematically evaluated this problem assuming the restructuring effort first builds all the spiders, then builds all the helpers, and finally builds all the product. There are many options to overlap the building efforts. This sequential approach simplifies the analysis and provides a conservative estimate for the productivity improvements. We provide a summary of the analysis equations produced from this analysis in Table 4-1. The equations on the left side of the table define the length of time required to build spiders, helpers, and products. Our analysis has found the minimum times for each of those tasks and those equations are in the center column. We also include in the right column of the table the equations that define the quantity of produced spiders, helpers, and products.

We summarize the parameters with example values in Table 4-2. Our descriptions are for an early example of our restructuring system. This system describes spiders (replicators) self-replicating to produce other spiders. The spiders build truss building units (TBUs), which are specialized pieces of equipment (helpers). The spiders and TBUs build trusses (products). We consider the construction of all the trusses required for our Atira space habitat.

The first four parameters in Table 4-2 describe the time required to build the individual pieces of equipment. The next three parameters represent the number of spiders, helpers, and product. The next set of parameters are the production rates for the system. The next set of parameters define the size (or complexity) of the equipment. The last parameters in Table 4-2 define maximum counts for the system. There is good correlation to Hall's analysis and syntax of self-replicating systems [Hall 1999].

**Summary:** We have extended a historic system analysis [Hall 1999] to better represent the self-replicating spiders and tool construction environment of the restructuring process. Hall determined that replicators benefit more from specialization and pipelining than they do from parallelism [Hall 1999]. We agree with that insight. We keep the universal constructor (spider) conceptually simple in our restructuring process. The spiders cannot do everything efficiently, and they do not construct the entire station. They can build additional spiders using vitamins from the seed unit. They can build multiple tools and pieces of specialized equipment. Parallelism and specialization increase the restructuring productivity dramatically.

The performance can be calculated given the parameters of a system and a selection of the number of spiders to dedicate to the production. In the next subsection, we apply the developed equations using counts, maximums, rates, and sizes from an example restructuring system.

### *4.3.2 Replicator, Helper, and Product Example*

The previous subsection analyzed a system that self-replicated spiders, built helper tools, and produced product. We apply that analysis to a restructuring system example in the following paragraphs.

We use previous analytic results and the specific values from Table 4-2. We illustrate those production timing results for the spiders and the truss building units. The graph in Figure 4-5 shows the time spent producing trusses for the space habitat. The purpose of this chart is to illustrate the minimum production time as a function of the number of helpers. The space habitat uses 27,382,485 meters of trusses, which is 7,097,540 cubic meters of material. The y-axis shows the production time in hours. The x-axis shows the number of truss building units (TBUs) in the system. The chart includes the time taken to build the spiders (replicators), build the helpers, build the product, and the total time. These form the relationship of the main equation of the previous subsection, $t = t_s + t_h + t_p$. We use a maximum of 1000 spiders and the spider build time of 120 hours. We see in the chart that the time to build the spiders is a constant of 120 log2(1000) or 1196 hours. This example assumes all the spiders are built, then all the helpers, and then the product. A single spider would take 10,550 hours to build the Truss Building Unit (TBU) and we would have 1000 spiders building it. The build time for the Truss Building Unit (helper) is an average of 10.55 hours. The total helper build time, $t_h$, grows linearly from zero at hour zero at the rate of 10.55 hours per helper unit. The time to build the product (trusses), $t_p$, is the total amount of product divided by the production rates of the spiders and helper units. The number of spiders is constant, and the number of truss building units increases with the x-axis values (helpers). The production time, $t_p$, decreases monotonically with the increasing number of helpers. The chart also shows a black square at the minimum point on the total production time. For this example, the minimum occurs with 604 helpers and the entire production of trusses is built in 14,207 hours. This is only 1.6 years while previous chapters have indicated the space habitat would take 12 years to complete. This analysis only includes the building of the trusses. It does not include the mining, transportation, construction, and many other tasks associated with the truss building units. We later find that over 3500 TBUs are used in the full station restructuring simulation. This chart and approach illustrate the optimization potential for restructuring in general and validates our analysis with another simulation.

**Table 4-1 – Minimize Production Time with Helper Tool**

| Activity | Time Equation | Minimum Time Equations | Quantity Equation |
|---|---|---|---|
| Total to Build | $t_t = t_s + t_h + t_p$ | $t_{tmin} = t_s + t_{hmin} + t_{pmin}$ | $N_{units} = N_{smax} + N_h$<br>$N_p = S_{pmax}/S_p$ |
| Build Spiders | $t_s = T_{ss} log_2(N_s)$ | $t_s = T_{ss} log_2(N_s)$ | $N_s = N_{smax}$ |
| Build Helpers | $t_h = \frac{N_h T_{sh}}{N_S}$ | $t_{hmin} = \sqrt{\frac{S_{pmax}}{N_s}\frac{T_{sh}}{\lambda_{hp}}} - T_{SH}\frac{\lambda_{sp}}{\lambda_{hp}}$ | $N_h = \frac{N_s}{\lambda_{hp}}\left(\sqrt{\frac{S_{pmax}}{T_{sh}}}\sqrt{\frac{\lambda_{hp}}{N_s}} - \lambda_{sp}\right)$ |
| Build Product | $t_p = \frac{S_{pmax}}{N_h \lambda_{hp} + N_s \lambda_{sp}}$ | $t_{pmin} = \sqrt{\frac{S_{pmax}}{N_s}\frac{T_{sh}}{\lambda_{hp}}}$ | $N_p = \frac{S_{pmax}}{S_p}$ |

| Table 4-2 – Parameters for Spider with Truss Building Unit (TBU) Helper Tool | | | |
|---|---|---|---|
| **Description** | **Parameter** | **Unit** | **Example Value** |
| Time for spider to build spider | Tss | hours | 120 |
| Time for spider to build helper (TBU) | Tsh | hours | 10550 |
| Time for spider to build product (1m of truss) | Tsp | hours | 6.1 |
| Time for helper (TBU) to build product (1m of truss) | Thp | hours | 0.153 |
| Number of spiders | Ns | count | 4 to 1000 |
| Number of helpers (TBUs) | Nh | count | Optimization Variable |
| Number of products (meters of truss) | Np | count | 0 to 27,382,485 |
| Production rate of spiders making spiders | λss | $m^3$ / hour | 0.000153 |
| Production rate of spiders making helper equipment | λsh | $m^3$ / hour | 0.002737 |
| Production rate of spiders making product | λsp | $m^3$ / hour | 0.0425 |
| Production rate of helper making product | λhp | $m^3$ / hour | 1.700 |
| Size (complexity) of spider | Ss | $m^3$ | 0.0184 |
| Size (complexity) of truss | St | $m^3$ | 0.2592 |
| Size (complexity) of helper | Sh | $m^3$ | 28.88 |
| Size (complexity) of product | Sp | $m^3$ | 0 to 7,097,540 |
| Production quantity maximum | Spmax | $m^3$ | 7,097,540 |
| Spider quantity maximum | Ssmax | $m^3$ | 184 (Ss x Nsmax) |
| Maximum count of spiders | Nsmax | count | 1000 |

### *4.3.3 Production Rate Analysis Results*

The previous subsection only included the building of the trusses. It did not include the mining, refining, transportation, assembly, and many other tasks. To optimize these activities, it is important to match the input and output rates for each of the stages of processing. A previous study by the author of this paper found that:

> "*Finally, simple analysis of a proposed architecture traffic patterns can provide immediate insights into potential performance and bottlenecks. Using average device interface rates and uniform request distributions can identify many system constraints.*" [Jensen 1993]

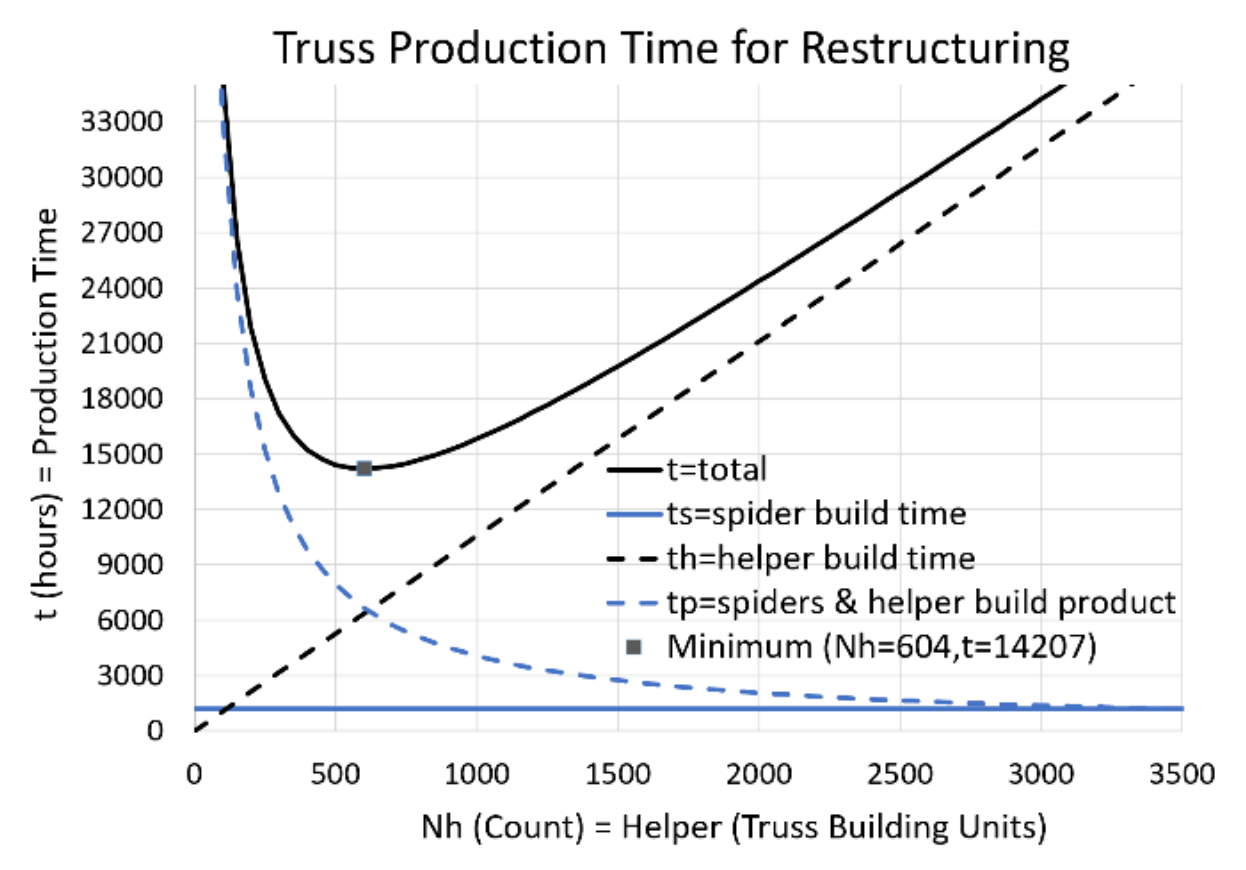

**Figure 4-5 – Minimize Truss Production Time using Spiders and Truss Building Unit Helpers**

Performing complex, detailed, and compute intensive simulations clearly provides valuable results; however, early analytic analysis and rate analysis can provide insights to guide those simulation efforts. We provide in this subsection an analysis of the self-replicating growth for the restructuring system. In this system, the spiders are self-replicating. They are constructed using a limited quantities of solar cells and electro-mechanical subunits.

**Production Rate Example:** As an example, we consider the initial Base Unit production. To maximize production, we need to maximize the Base Unit input and output rates. We capture in Table 4-3 the analysis of spiders digging and loading regolith into the base station, the base station producing tiles, and the spiders using those tiles to construct new spiders. We use some input and output rates to analyze the system. The simple analysis of the rates in Table 4-3 shows that we would need Nd=33.3 spiders digging and delivering regolith as input to the base station. It also shows we need Na=64.6 spiders using the output to assemble spiders.

This simple assessment illustrates the value of such analysis. We find that many spiders are required to keep the single base seed unit filled and processing. We also find that even more spiders are required to assemble new spiders and match the seed unit tile output rate. Initially, there are not enough spiders to keep the base unit loaded with regolith and operating continuously. Even if the base unit were operating at 100%, there would not be enough spiders to use all the output tiles to build new spiders. The rates give us a good estimate of the system at full operation. We see the base unit is producing 7 tiles per hour at full production. We now consider in a simulation how to reach full operation from the initial seed with 4 spiders.

**Initial Seed Simulation:** Our simulation begins with the base unit and four spiders. We assign two spiders to dig and deliver regolith to the base unit. The base unit processes the raw regolith and outputs tiles. The other two spiders assemble new spiders using those tiles and vitamins. We use the rates and volumes previously described.

To introduce the simulation, we include the count of spiders and their activity in Figure 4-6a. This chart shows along the x-axis the passage of time in hours and covers almost 3 months. The y-axis of the chart shows the count of spiders. The graph includes the total number of spiders in the system, the spiders assembling other spiders, the number of spiders digging and delivering regolith, and the number of idle spiders. Additional assembled spiders not being used are considered idle; however, in a more detailed simulation they would be working on other tasks.

In this example, as spiders are built, they are assigned to work on filling the base unit or building new spiders. The assignment simply considers the times being used currently on the filling and building tasks and assigns the new spider to the task using more time. The graphs in Figure 4-6 show at first the number of build-spiders (building new spiders) is increasing more rapidly than the number of fill-spiders (digging regolith and filling the base unit). All newly built spiders are assigned to work on digging and delivering regolith. We show the same simulation data in the more detailed chart in Figure 4-6b. After 300 hours we see that the number of fill-spiders begins to increase.

We see in Figure 4-6a that the number of fill-spiders plateaus at a count of 34 shortly after the time 1000 hours. The number of build-spiders plateaus at a count of 66 spiders at about time 950 hours. With these spider counts, the base unit is being filled and the tile product is being used at maximum rates. Shortly after time 1000 hours, we see more spiders are being built and those become idle spiders. At this point the base unit is operating at 100% capacity with maximum regolith input and tile output.

The simulation results closely match the counts analyzed and presented in Table 4-3. These simulations allowed us to delve into details of the spider and base unit activities. This kind of detail has helped to identify bottlenecks and determine what types of tools and equipment to create. With these simulations we could monitor many parameters. We monitor the surplus of regolith arriving at the base unit hopper, the regolith rate coming out of the base unit, and the regolith rate being used to build spiders. We could track the tiles being used to build the spiders. We measured the count of surplus tiles in a pile near the output of the base unit. Many additional simulations have been performed and results captured.

**Tile Production Simulation:** As an example, we were curious how much construction material could be produced with the Base Unit and Spiders. We adjusted the simulation to build 100 extra (idle) spiders and then start producing tiles; see Figure 4-7. The extra spiders could use the tiles to begin building other tools. As before, the 66 build-spiders and the 34 fill-spiders are complete shortly after the time 1000 hours. We see in the chart that the extra spiders are completed in less than 200 hours. The system is producing the spiders at the maximum rate of 0.5 spiders per hour. This rate is expected given the 66 build-spiders and a 120-hour spider construction time. The tiles are produced at a rate of 7 tiles per hour. Looking at Figure 4-7, this production rate seems great at first glance. After an initial setup period, the base unit and the fill-spiders and build-spiders are all working at 100% capacity. We see the number of tiles growing rapidly and we have 100 extra spiders to begin working on other equipment.

We consider how long it would take to harvest and convert the Atira asteroid to tiles for the station using the 100 extra spiders and producing 7 tiles per hour. Our example torus space station requires 2.62 billion cubic meters of material; see Figure 3-23. Each tile represents 0.01 cubic meters of material. It would take 37.4 billion hours (or over 4 million years) to produce the material at 7 tiles per hour. Given this

**Table 4-3 – Example Rate Analysis for Restructuring Process**

| Phase | Dig & Deliver | Fill Bin | Manufacture Tiles | Assemble Spiders |
|---|---|---|---|---|
| Activity | Obtain Raw Regolith | Input Raw Regolith | Output Tiles | Tiles and Vitamins |
| Details | 1 Shovel = 0.001 m3<br>1 Shovel per 20 minutes | Input Bin = 0.1 m3<br>One bin = 100 shovels = 33 spider hours.<br>1 Bin processed per hour | Output = 0.07 m3<br>Tiles = 2m x 0.1m x 0.05m = 0.01 m3<br>7 Tiles per hour<br>14 meters per hours | 1 Spider built with Na Spider in 120/Na hours<br>Na <= 4 Spiders overcrowding constraint<br>4 Spiders build 1 spider in 30 hours<br>1 Spider requires 26 m tiles = 13 tiles; 6 connectors; 1 electro-mechanical subunit; 8 solar cells |
| Rate | 0.003 Nd m3 per hour regolith dug and delivered | Inputting 0.1 m3 per hour maximum | Outputting 14 meters per hour or 7 tiles per hour maximum | 1 Spider / (120/Na) hours<br>(13 tiles * Na / 120) tiles per hour<br>Using 0.1083 Na tiles per hour<br>Output 7 tiles/hours = Build 0.1083 Na tiles/hour |
| Spider Count | Nd Spiders Digging | Nd = **33.3 Spiders** | Base Unit Production | Na = 7/0.1083 = **64.6 Spiders**; About 6 groups of 4 working on the tile |

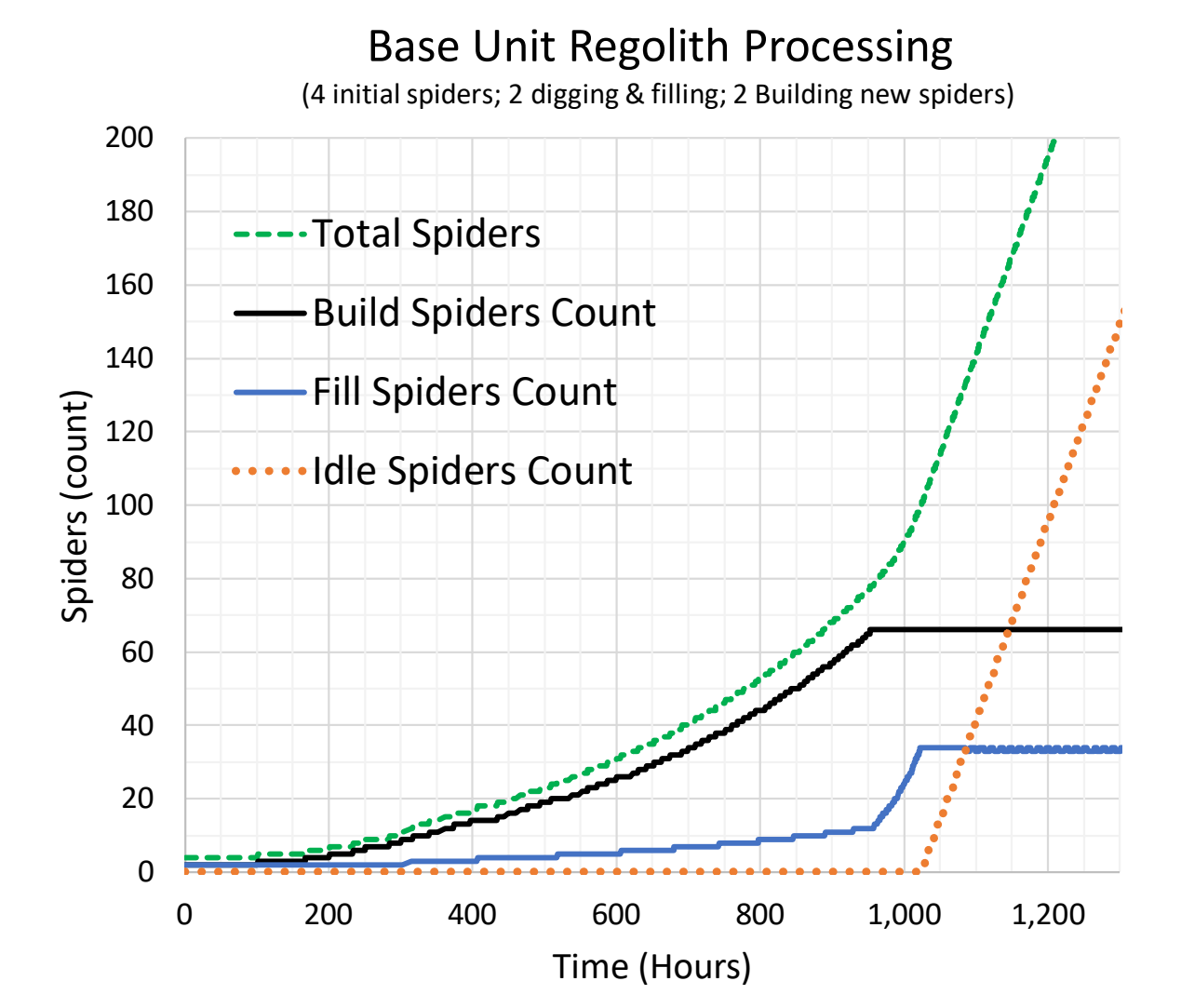

*a) Initial Spider Activity Matching Production Rates*

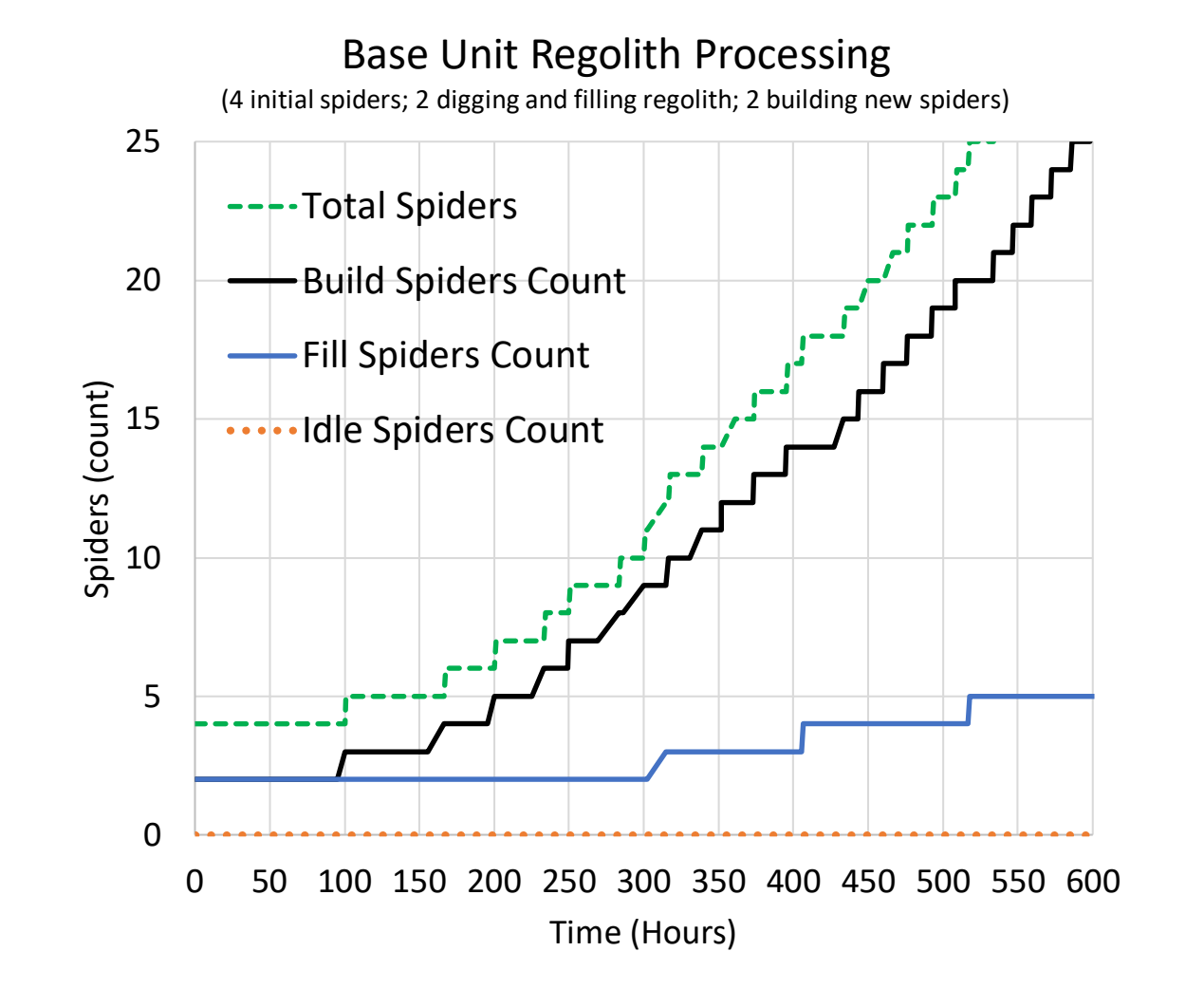

*b) Details on Spider Activity*

**Figure 4-6 – Spider Construction Analysis**

simple assessment, we see the base unit tile production is a severe bottleneck for the asteroid restructuring process.

Our original plan was to build rods (or tiles) of basalt; later we chose to build tiles of anhydrous glass laminates. We reference Rod Production Units (RPUs) in most of our analysis. Re-evaluating this analysis with Laminate Production Units (LPUs) instead of RPUs would be appropriate in the future. For the current time, we freely interchange tiles and rods and RPUs and LPUs. We do not expect significant changes in the restructuring approach or general results given this change.

Because the base unit tile production is a severe bottleneck, we need to build Rod Production Units (RPUs) to increase the tile production rate. With additional RPUs, we quickly find that the transport of regolith becomes time-consuming for the spiders. We address this issue by making trucks to offload that task from the spiders. Digging the regolith is also time consuming for spiders and we build harvest units to dig and load the trucks more efficiently. Finally, these tools will quickly deplete the availability of loose regolith pebbles and grains on the asteroid surface. We build Regolith Crushing Units (RCUs) to produce more pebbles and grains.

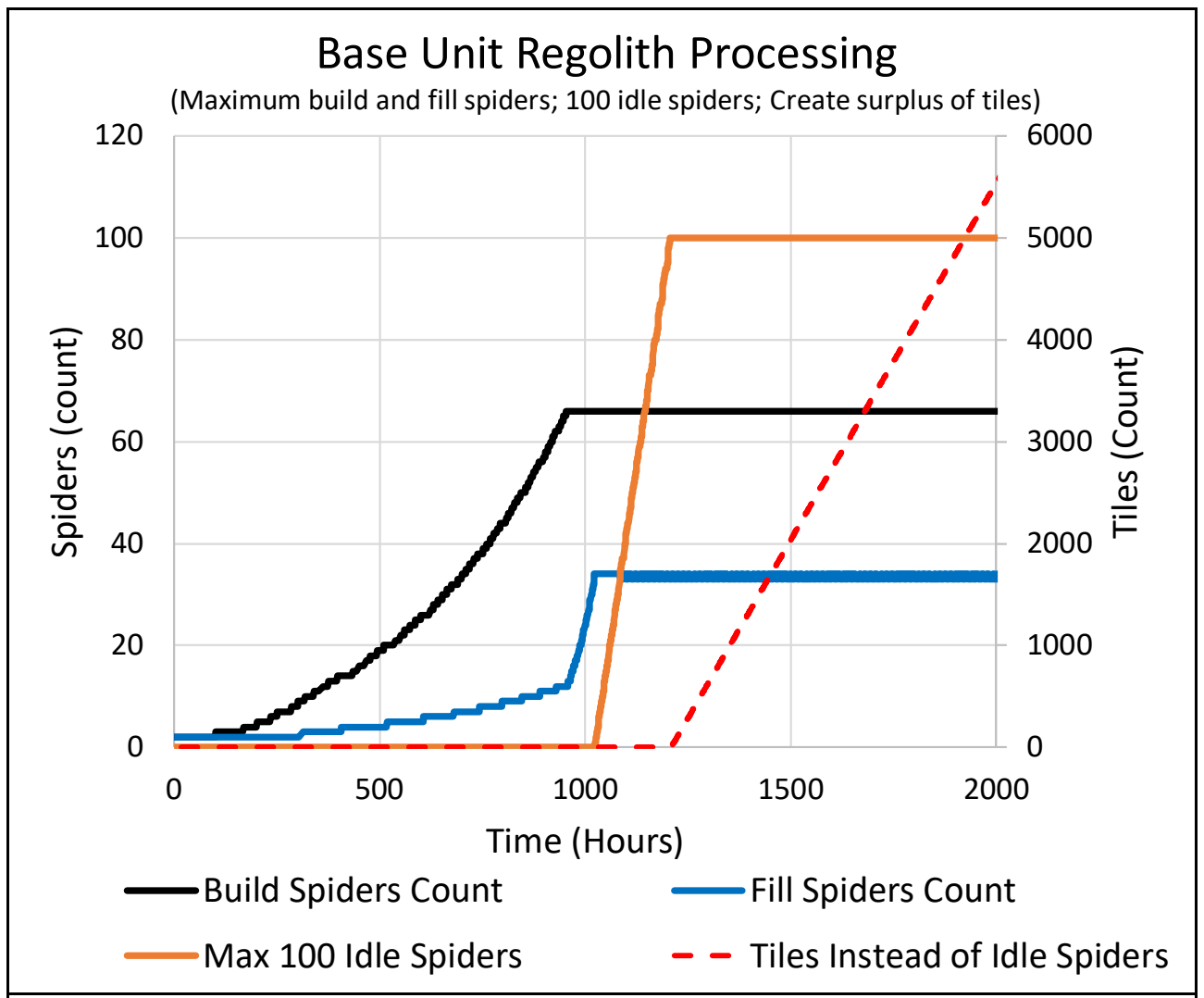

**Figure 4-7 – Surplus Tile Production**

**Restructuring Tools:** We show these tools with some of their construction metrics in Table 4-5. Tools are shown as acronyms in the table's left column. Later subsections and paragraphs provide acronym definitions and descriptions; see Figure 4-10. We estimate that some of these tools will take thousands of hours to build. Fortunately, they are complex and multiple parts can be built at the same time with many spiders. As an example, it takes 3,516 hours for one spider to build the subunits and construct and assemble the Regolith Crushing Unit (RCU). The material would take 140 hours to construct. With 32 spiders working on that tool, it could be complete in only 250 hours. The table also shows the quantity (length) of tiles required to build the tool. The Rod Production Unit requires 4,768 meters of rods (or tiles). At 7 tiles per hour or 14 meters per hour this would require 340 hours if they were produced with only the base unit. Taking over two weeks for the tiles seems excessive; as such, we need to improve their production. Having other RPUs producing tiles increases the tile production rate and reduces the material time correspondingly.

We include Table 4-5 and Table 4-4 to provide a summary of the construction time and material requirements for all the tools and subunits. These are our current estimates and only rough designs for most of these tools and subunits exist. We created engineering estimates for the construction material and time. We considered the general size estimates to evaluate the number of tiles and trusses required. We plan on using concentrated solar energy to weld the parts together. We could consider a basalt "rivet" to connect the tiles together after drilling (melting) a hole in the tiles with focused solar energy. To estimate the build time, we counted the number of welds required and conservatively assumed 30 minutes per weld. We also doubled that build time for most tools to provide extra margin on the estimate. Some of the more complicated tools have multiple Stirling engines and/or clockworks. We add the construction time and materials from those subunits to the total time. We also assume that it would require 120 hours to assemble, attach, and integrate a subunit to one of the tool housings. The total time also includes the time to build the material assuming a production rate of 7 tiles per hour. We divide the truss length estimates by 14 meters per hour to determine the material build time. We include the characteristics of the subunits in Table 4-4. These subunits have the same metrics as the tools. In these tables, we sometimes explicitly show the subunit build time and other times implicitly include the subunit build time in the equipment build time.

**Table 4-5 – Equipment used in simulations**

| Tools | Material | Build w/ Margin | Build Material | Detail Complexity | Stirling Engines | Clock-works | Mirror Units | Integrate Subunit | Build Time (Material & Tool) | Maximum Spiders to Build | Minimum Build Time |
|---|---|---|---|---|---|---|---|---|---|---|---|
| Metric | (m) | (hours) | (hours) | 60 hrs | 659 hrs | 217 hrs | 318 hrs | 120 hrs | (hours) | (count) | (hours) |
| Spider | 26.0 | 120 | 1.9 | 0 | 0 | 0 | 0 | 0 | 122 | 4 | 31.86 |
| Truck | 114.4 | 148.4 | 8.2 | 0 | 659 | 0 | 0 | 120 | 936 | 8 | 124.10 |
| Mover | 95.1 | 111.3 | 6.8 | 0 | 659 | 0 | 0 | 120 | 897 | 8 | 118.08 |
| Digger | 93.6 | 330 | 6.7 | 0 | 659 | 0 | 0 | 120 | 1,116 | 8 | 145.31 |
| RCU | 1965.2 | 2400 | 140.4 | 0 | 659 | 217 | 0 | 240 | 3,656 | 32 | 250.25 |
| HU | 622.8 | 2496 | 44.5 | 0 | 2636 | 0 | 0 | 480 | 5,656 | 16 | 395.24 |
| TPU | 2872.2 | 916 | 205.2 | 60 | 659 | 1519 | 1910 | 1680 | 6,949 | 32 | 415.91 |
| RPL | 2086.4 | 4142 | 149.0 | 120 | 1318 | 868 | 0 | 720 | 7,317 | 32 | 373.03 |
| TBU | 5071.2 | 1630 | 362.2 | 120 | 659 | 2387 | 3184 | 2640 | 10,982 | 64 | 528.16 |
| RPU | 4768.8 | 148 | 340.6 | 120 | 2636 | 2387 | 3184 | 3000 | 11,815 | 64 | 519.92 |
| SAU | 6967.2 | 3300 | 497.7 | 0 | 659 | 2387 | 3184 | 2640 | 12,667 | 64 | 687.81 |
| F&MU | 6967.2 | 5494 | 497.7 | 120 | 659 | 2387 | 3184 | 2640 | 14,981 | 64 | 723.96 |
| CPP | 7450.8 | 6520 | 532.2 | 120 | 2636 | 2387 | 3184 | 3000 | 18,379 | 64 | 811.05 |

**Table 4-4 – Tool Subunits Characteristics**

| Subunits | Material | Build w/ Margin | Build Material | Detail Complexity | Stirling Engines | Clock-works | Mirror Units | Integrate Subunit | Build Time (Material & Tool) | Maximum Spiders to Build | Minimum Build Time |
|---|---|---|---|---|---|---|---|---|---|---|---|
| Metric | (m) | (hours) | (hours) | 60 hrs | 659 hrs | 217 hrs | 318 hrs | 120 hrs | (hours) | (count) | (hours) |
| Fresnel Len Frame | 11.2 | 139.8 | 0.8 | 0 | 0 | 0 | 0 | 0 | 141 | 2 | 70.7 |
| Solar Panel Frame | 12.8 | 145.8 | 0.9 | 0 | 0 | 0 | 0 | 0 | 147 | 2 | 73.8 |
| Clockwork | 28.0 | 95 | 2.0 | 120 | 0 | 0 | 0 | 0 | 217 | 4 | 55.8 |
| Mirror | 397.0 | 290 | 28.4 | 0 | 0 | 0 | 0 | 0 | 318 | 16 | 46.5 |
| Stirling Engine | 56.0 | 138 | 4.0 | 300 | 0 | 217 | 0 | 0 | 659 | 8 | 85.9 |
| Connectors | 1.2 | 2400 | 0.2 | 60 | 0 | 0 | 0 | 0 | 877 | 6 | 146.2 |
| Sun Tracker | 484.0 | 293.8 | 34.6 | 120 | 0 | 217 | 0 | 0 | 772 | 8 | 96.5 |

### *4.3.4 Analysis Approach Using Simulations*

We present in this section our approach to analyze and simulate the entire restructuring of the asteroid. We describe our simulators, the restructuring equipment, the initial seed package, and expected technology advancements.

**Restructuring Simulations:** We have run hundreds of simulations to evaluate the restructuring concepts. These included orbit and transfer simulations using Verlet integration, simulations using production rate analysis, and discrete event simulations of the restructuring process. We gathered and encoded accurate mathematic formulae for the station physical characteristics. When possible we used comparable values or engineering estimates for consumption and production rates. We developed a tool to estimate the number of welds for the various equipment built during the restructuring process. In all cases, we used conservative estimates.

We used C++ and Python for various programming efforts. We use Excel to simulate and produce many of our results. We used Wolfram-Alpha to assist with mathematical analysis. We used Blender to render the images of our asteroid and torus station. The Blender station rendering was written in Python code and has size parameters for key components of the station. Not only does the Blender Python code render a station but it also produces a report with details including parameter values, component sizes, number of floors, and surface area. Using multiple tools provides cross checking of results to identify issues and correct discrepancies. Many of the simulation tool concepts were developed over the author's 40-year engineering and research career. The discrete event simulation software has its foundation from the 1990s [Jensen 1993]. The orbit analysis approaches were part of a more recent research effort. We used multiple approaches for the scheduling and system engineering of this effort. The analysis, decomposition, modeling, and building of complex systems has been a successful part of the author's career for multiple products and research projects.

The restructuring process includes mainly mining, processing, and construction. Our simulation effort evaluates the building and operation of equipment and tools. We do not claim to have every issue resolved; however, we feel sufficient progress has been made to present the restructuring process for review. Our evaluation and analysis have run into bottlenecks, but every time there has been an alternative solution. We have evolved and adapted our approach a dozen times during the development and execution of this effort.

**Equipment Overview**: Our simulations include the building and operation of equipment and tools. The asteroid environment will pose problems for existing terrestrial equipment. Design changes will be required for terrestrial equipment to work in the asteroid environment. The design and modeling of the restructuring equipment must consider a low gravity and vacuum environment.

Earth gravity is important for the operation of terrestrial mining equipment. Equipment tied to the earth or using cables to move buckets have the least dependency on strong Earth-like gravity. Anchoring equipment to the asteroid using harpoons or spikes helps overcome some of the issues with low gravity. Portable wheeled equipment used to move ore can operate well on a track with low gravity except for a risk of reaching escape velocity on small asteroids. Portable wheeled equipment on Earth that assert digging or scraping force on the surface relies on their mass and gravity to counteract that force. That equipment is highly dependent on gravity and will require new designs.

We recognize there are other issues in the asteroid environment. The lack of an atmosphere means that waste heat cannot be dissipated by atmospheric convection. No atmosphere means lubricants such as oil and grease will evaporate in the vacuum. No atmosphere means combustion engines would no longer work without a separate supply of oxygen. The extreme temperature changes will subject equipment to significant thermal stress. Another problem for the asteroid operations could be an abundance of dust from the regolith. Dust contamination can obstruct and abrade moving joints in equipment. We strive to address these issues but we recognize they have not been fully addressed in our study. These environment issues are problematic; however, solutions could enable new approaches for mining and processing ore. Additional research, testing, and development will address these issues and refine our designs.

Self-replicating systems can create millions of units to perform a large task. Because of launch costs we could not send millions of units or even the vitamins to build the millions of units. Because we did not have millions of devices, we found that our restructuring approach with replicators regularly hit bottlenecks in the processing or transporting of material. Our list of equipment for the simulations evolved from a single device to over twenty unique pieces of equipment and subunits. We have not completed the final formal design and drawings for these pieces of equipment. We have good engineering estimates on the size and required material. We include the primary 13 tools in Table 4-5 and the 7 support subunits in Table 4-4. The tables include for each piece of equipment its size in required length of regolith tiles and the time to build with spiders. The simulation of these equipment models includes other relationships such as the amount of metal required to build, the production rate, travel rates, and the equipment required to build. We hope to refine this simulation with additional detail and complete the final formal estimates, designs, and drawings of the restructuring equipment in the future.

**Initial Seed Package:** Our restructuring approach sends a probe and seed package to an asteroid. This seed package contains enough supplies to produce a limited number of spiders and many tools. A 1980 study defined a 100-ton seed package for a self-replicating system on the Moon [Freitas and Gilbreath 1982]. A 2012 study defined a more modern approach and estimated a total of 41 metric tons using six launches of seed packages to the Moon [Metzger et al. 2012]. Both of these self-replicating systems used teleoperation initially and periodic supplies from the Earth. We estimate that our total seed package weighs 8.6 metric tons and requires only a single rocket launch. We envision that our seed package is self-sufficient and does not need additional supplies or direction from Earth.

Our restructuring process will construct hardware made from in-situ materials. The produced tools will be more primitive than the original seed equipment. Our 21st century robots will be creating 18th and 19th century frameworks, trusses, tools, and engines. We send enough solar cells and computer modules to build the required number of spider robots. These robots only perform a small fraction of the total work – primitive trucks, diggers, crushers, and specialized units perform most of the heavy lifting for the restructuring effort. Our goal is to have full closure on the restructuring process – no additional materials beyond the original seed are required to accomplish the building of the space habitat framework.

We show in Table 4-6a Metzger's list of the assets to produce a self-replicating lunar industry. We show in Table 4-6b a list of the assets required for the restructuring probe. The mass of both probes is under the launch weight of a SpaceX Falcon Heavy (16.8 metric tons) [SpaceX 2018]. Our seed probe includes different jigs such as shovels, Fresnel lens, cutting, and test sensors. The number of jigs becomes a bottleneck in many of the restructuring processes. Some jigs can be produced using the mongrel alloys and the asteroid regolith. Many jigs are too sophisticated to build using the spiders, asteroid material, and available seed package tools. Some jigs are replaced with less sophisticated 18$^{th}$ century technology. One of those sophisticated jigs is the Fresnel lens. Its function will be performed with solar concentrators built from rods, panels, and a supply of silvered mylar.

We expect there could be additions to the asset list that will increase the weight. Refinement of the assets should reduce the individual weights of the tools. It is realistic to aim for a probe weight of 8 metric tons. Using an ion engine and a mission delta-v of 5 kilometers per second, we find we would need almost 1 metric ton of fuel for the one-way mission to a near-Earth asteroid. An eight-ton goal for the seed probe is desirable. It becomes possible to send two probes – one for

**Table 4-6 – Initial Probe Assets**

*a) Assets for Lunar Industry Probe*

| Metzger Assets | Qty. per set | Total Mass (kg) |
|---|---|---|
| Power Distrib & Backup | 1 | 2000 |
| Excavators (swarming) | 5 | 445 |
| Chem Plant 1 – Gases | 1 | 763 |
| Chem Plant 2 – Solids | 1 | 763 |
| Metals Refinery 1 | 1 | 1038 |
| Solar Cell Manufacturer | 1 | 188 |
| 3D Printer 1 – Small parts | 4 | 752 |
| 3D Printer 2 – Large parts | 4 | 1276 |
| Robonaut assemblers | 3 | 450 |
| Total per Set | | 7675 |

*Credit: Self Produced using [Metzger et al. 2012] [Facts]*

*b) Assets for Asteroid Restructuring Probe*

| Atira Asset | Qty. per set | Total Mass (kg) |
|---|---|---|
| Spider | 4 | 340 |
| Spider Control Units | 3000 | 2100 |
| Solar Cells | 6000 | 900 |
| Jigs | 720 | 3064 |
| Connectors | 200 | 400 |
| Base Unit | 1 | 1039 |
| Silvered Mylar | 2000 | 200 |
| Misc | 1107 | 563 |
| Total per Set | | 8606 |

*Credit: Self Produced*

the north pole and one for the south pole of the asteroid – with a single launch. The two probes would require about double the fuel and hopefully still meet the maximum payload weight for a single launch.

### *4.3.5 Technology Advancements*

The restructuring effort seed probe is smaller and less complicated than those in detailed historic studies [Freitas and Gilbreath 1982] [Metzger et al. 2012]. Our goal was to reduce (or eliminate) future transport of materials from the Earth. The earlier 1980 study developed a 100-ton seed probe that would require the launch equivalent of four Apollo missions to the Moon. The later 2012 study reduces that seed to 33.3 metric tons and could be accomplished in 5 Falcon Heavy launches. These early studies included material and support for astronaut crews. Self-replication systems that include crew support incur additional cost in materials, guidance, and technology.

The advancement of technology provides our restructuring effort with advantages over the 1980 study. The 1980 NASA study also included multiple robots with the seed unit. Mining robots, paving robots, repair robots, and transport vehicles were included and weighed over 18 metric tons. The 1980 estimate for the mining robot has been reduced from 4.4 tons to less than 100 kilograms [Metzger et al. 2012]. In 1980 each robot computer weighed 50 kilograms [Freitas and Gilbreath 1982]. Today such computers weigh less than a few kilograms. Also, a processor today has over 250,000 times the performance of a 1980 processor. The required power, size, price, cooling, and weight have all been reduced.

The restructuring process is different than the earlier approaches. We replace custom part manufacturing equipment with 3D printing. We replace teleoperation with modern day image processing and artificial intelligence. We replace large complex pieces of manufacturing equipment with many smaller and less efficient helper equipment designs.

We include in our probe enough supplies to build a modest number of productive replicators. Beyond that limited number, we do not attempt any additional self-replication. Our restructuring process relies on the production of tools and equipment to extend the capabilities of the replicators (spiders). With the limit on the number of spiders, our restructuring seed would only be about 8 metric tons and launched on one Space X Falcon launch. Ideally, our restructuring effort does not require any additional launches to complete its mission. A Falcon Heavy rocket can lift 16.5 tons to trans-Mars injection and could potentially deliver two of our seed probes to an asteroid. Two probes would reduce mission risks with redundancy and reduce the mission schedule.

### *4.3.6 Analysis Summary*

We have used mathematical and simulation analysis of the restructuring asteroid process. We extended a historic mathematical analysis to model replicators, helpers, and products. We provided the restructuring results using production rates of the replicators and helpers. We included our approach to analyze and simulate the entire re-structuring of the asteroid. We describe our simulators, the restructuring equipment, and the initial seed package. We also included a section describing technology advancements that could improve our restructuring process.

The analyses of those subsections show that the production and consumption rates have an impact on the system performance and on design decisions. We also saw different parts of the system becoming the bottleneck as the number of spiders and tools were produced. As more tools and activities are added, the analytic analysis and simulations become more complex. Fortunately, with large numbers of units and activities performing over longer times, our experience has shown that using average production and consumption rates will provide the insights we need to complete our analysis.

These early simulations identified different bottlenecks. Initially, there were not enough spiders to keep the base unit operating at 100% efficiency. With enough spiders, once the base system was operating at 100% efficiency, a surplus of output indicated that the spider-building spiders had become the bottleneck. More of these spiders addressed that bottleneck. We found that with many more spiders there would not be enough material to build the station in a timely fashion. The base-unit rod-production had become the bottleneck.

These simulations showed that there is a balance between rapidly finishing one task and starting other tasks. Additional units would reduce the time required to complete one task. Often though, there is other equipment that is needed to complete the station. For simple cases, mathematical analysis can provide the optimal balance. Our self-replication system will need to monitor progress and adjust assignments during the building of the station. It may be beneficial to have the restructuring asteroid system simulate future versions of itself to identify bottlenecks and provide guidance on tasks and assignments.

Our simulations have used the estimated resources and times. We have seen the overall system performance rates improving from the parallelism and the specialization. From the initial base unit production rates of only 7 tiles per hour and construction taking 37.4 billion years, it is exciting to see the scalability and productivity gains reducing this to about 100,000 hours (12 years).

## 4.4 Robots – Results

Our analysis in the previous subsections focused on the construction of spiders, tiles, and trusses. We present in this subsection our analysis and simulation of the entire restructuring of the asteroid.

### *4.4.1 Introduction*

Our initial probe lands on the asteroid Atira and only has 4 robotic systems (spiders). The probe serves as a base station. The first 4 spiders and the base station will build frames and assemble new spiders. The spiders will also build other tools and equipment from supplies and the asteroid material. Initially, the spiders will need to multitask on many different tasks. Some of the early tasks include:

- Mining regolith
- Loading the base unit with regolith
- Removing manufactured regolith tiles from base unit
- Assembling solar panels from frames and solar cells
- Assembling spiders from frames and modules
- Assembling trusses from tiles

We show in Figure 4-8 a chart that illustrates this early multitasking. This chart was produced early in our investigation and shows the activity of spiders over the first 4000 hours on the asteroid. The y-axis shows the count of the spiders. The graphed lines show the different spider activity. The x-axis shows the time in hours. During the first 500 hours in this simulation, the spiders alternate between collecting regolith, assembling solar panels, and assembling new spiders. This simulation produced many idle spiders. By hour 1000, there are only eight active spiders and 50 idle spiders. About this time the supply of connectors included with the probe are exhausted. Spiders begin to make new connectors (3D printing with metal grains) and the number of idle spiders drop significantly to support this effort. Groups of 20 spiders begin to build rod production units at time 2600. We then see more spiders begin to dig regolith to supply those units. This simulation includes spiders producing trusses beginning at about hour 3200. Near the end there are still almost 1000 spiders not being used in this simulation. The chart in Figure 4-8 begins to highlight the different spider activities and the changing bottlenecks. It also highlights the potential rapid growth from the self-replication of spiders with sufficient supplies.

### *4.4.2 Tool Construction*

The first year and a half shown in our full restructuring simulations are based on results from the discrete event simulations. The last 11 years of the simulations are based on the production and consumption rates of the station materials. This includes the trusses, panels, and fill. With larger numbers of entities in the simulation, we have reached a point where the Theory of Large Numbers applies. It becomes viable to use rates and averages instead of the discrete event simulation.

We review one of those simulations to understand the tool construction. This simulation found it took almost 12 years to construct the Atira station. This version of the space habitat had the elliptic outer torus with a major radius of 2332 meters and minor radii of 1105 meters and 368 meters. This version had two sets of inner circular tori and spokes. The inner tori have a radius of 1166 meters and minor radius of 75 meters.

The simulation limited the growth in spiders to 3000 units. This is limited by the number of electronic modules and solar cells sent with the probe. The spiders build 23,500 other pieces of equipment. The equipment built by the spiders will not have sophisticated electronic microprocessors; instead, they will have mechanical state machines to control their operation. Equipment and tools are supplied, emptied, and managed by the spiders. We found in our simulations that there is a limit when all spiders are occupied with managing the existing equipment. Additional equipment would not be used if built. We note that some earlier-built equipment is no longer needed near the end of the station construction. The retirement of that equipment will free some spiders. The retired equipment is typically replaced with other equipment necessary to finish the station.

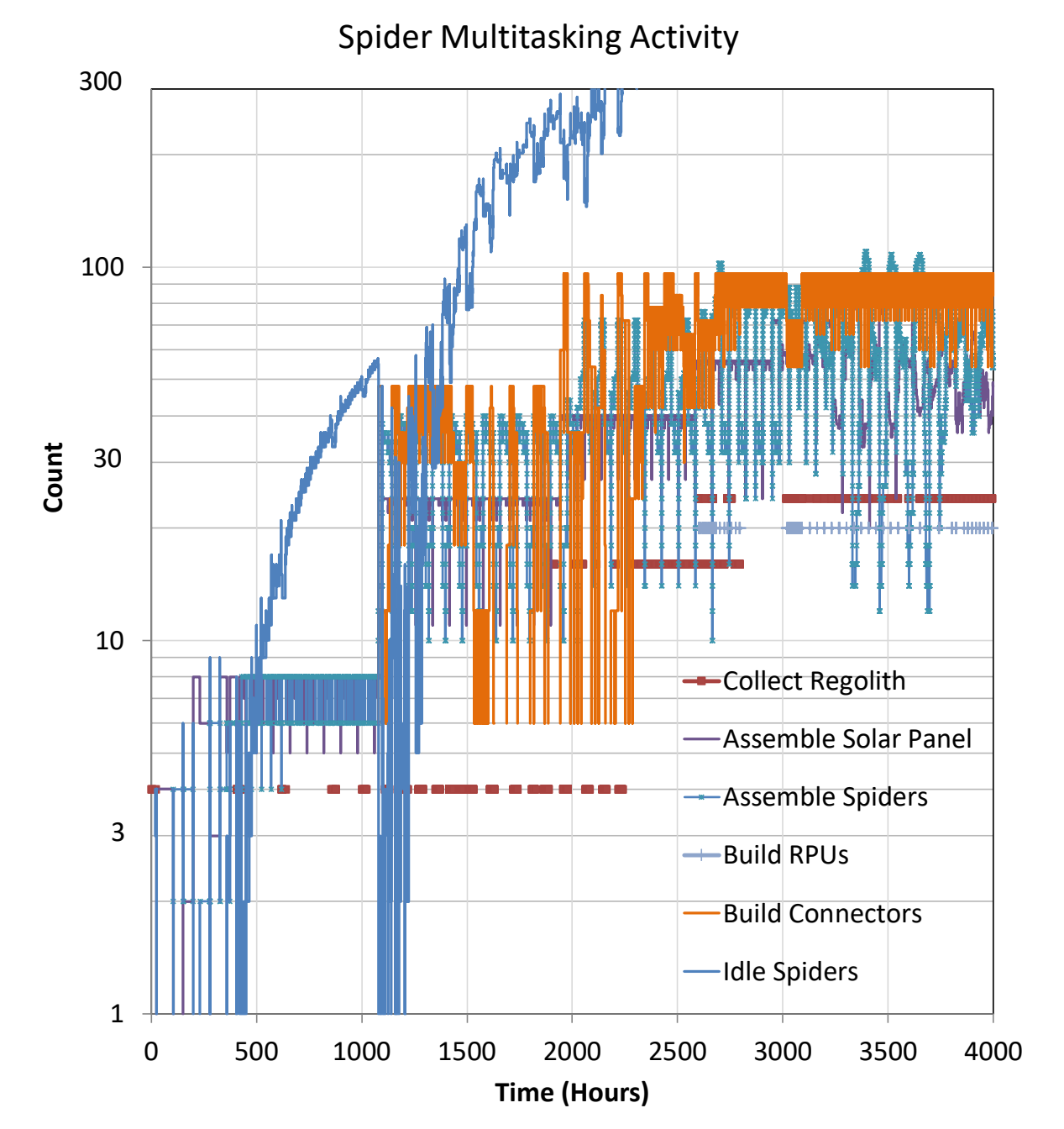

**Figure 4-8 – Early Spider Activity on Asteroid**

### *4.4.3 High Level Schedule*

The restructuring mission and process goes through a series of activities. Previous sections included rendered images of an example space station; see Figure 1-1 and Figure 3-22b. Those images help to illustrate these major activities and their associated components. These activities include:

1. Reach asteroid
2. Land seed probe
3. Start initial evaluation
4. Build web
5. Build spokes
6. Build shuttle bay
7. Build outer rim
8. Build inner torus
9. Build outer elliptic torus

Each of these restructuring tasks has been simulated in some fashion. The tasks begin with landing the seed unit and finish with attaching a skin of regolith panels to the exterior of the large outer elliptic torus. We have performed analysis of all the building tasks. Some analyses are detailed discrete event simulations, some are detailed rate driven simulations, and some are high level average rate estimates.

We include in Figure 4-9 a simple Gantt chart to illustrate the restructuring schedule. The chart shows the web of trusses being built over the asteroid in the first year and a half. The process starts building the spokes, and that requires almost 5 years to complete. No equipment or tools exist at the start of the restructuring process. It takes time to build the initial equipment and tools and then the progress on the station components improves. The shuttle bays at the north and south poles of the asteroid tie the spokes together. This schedule only assumes a single base station at one of the poles. Two probes and base stations (two poles) would double the number of spiders and improve the schedule. We see in the schedule that the outer rim is started early. Once the outer rim ties the station together, it is possible to begin rotating the structure and produce some centripetal gravity. Some tools require gravity (the Continuous Panel Production Unit in particular) and mandate this early rotation. The inner torus is started in the fifth year. Once sections of the inner torus are complete, a construction crew can arrive to make those sections habitable. The outer torus is built inward from the outer rim and requires over six years to complete. As segments of the outer torus are completed, the manufacturing crew can begin to make those segments habitable. More population arrives with the completion of the habitable segments.

For our example Atira asteroid model, this is an appropriate order of tasks. For different shape asteroids or station geometries, the order of the tasks may change. As an example, the outer rim and outer torus may not be built with smaller asteroid. For some shaped asteroids, the outer torus may be built before the inner torus. A dumbbell might only have a single hub and no tori structures. It may be appropriate to surround an asteroid with an ellipsoid station. The station could be built to one side of the rotating asteroid. We also note that there are issues with solar power, shadows, and rotation that need to be addressed.

Figure 4-9 also shows the immigration of the station population. The restructuring system is designed to work autonomously. The initial crew arrives in the seventh year. This is before the outer elliptic torus is complete and before the station has been spun up to full rotation speed. This crew would start making portions of the station airtight, producing an atmosphere, and creating an air processing system. Their work would initially be in a low gravity environment. The crew would start work in the inner torus or the shuttle bay since the outer torus would not be complete. Even without the crew arrival, the restructuring system continues to autonomously construct the station. We imagine that this initial crew would bring a refreshed set of technologies that would enable a new generation of robots, tools, and equipment to be built. The catalog of stored materials and volatiles would guide the research and selection of these new technologies.

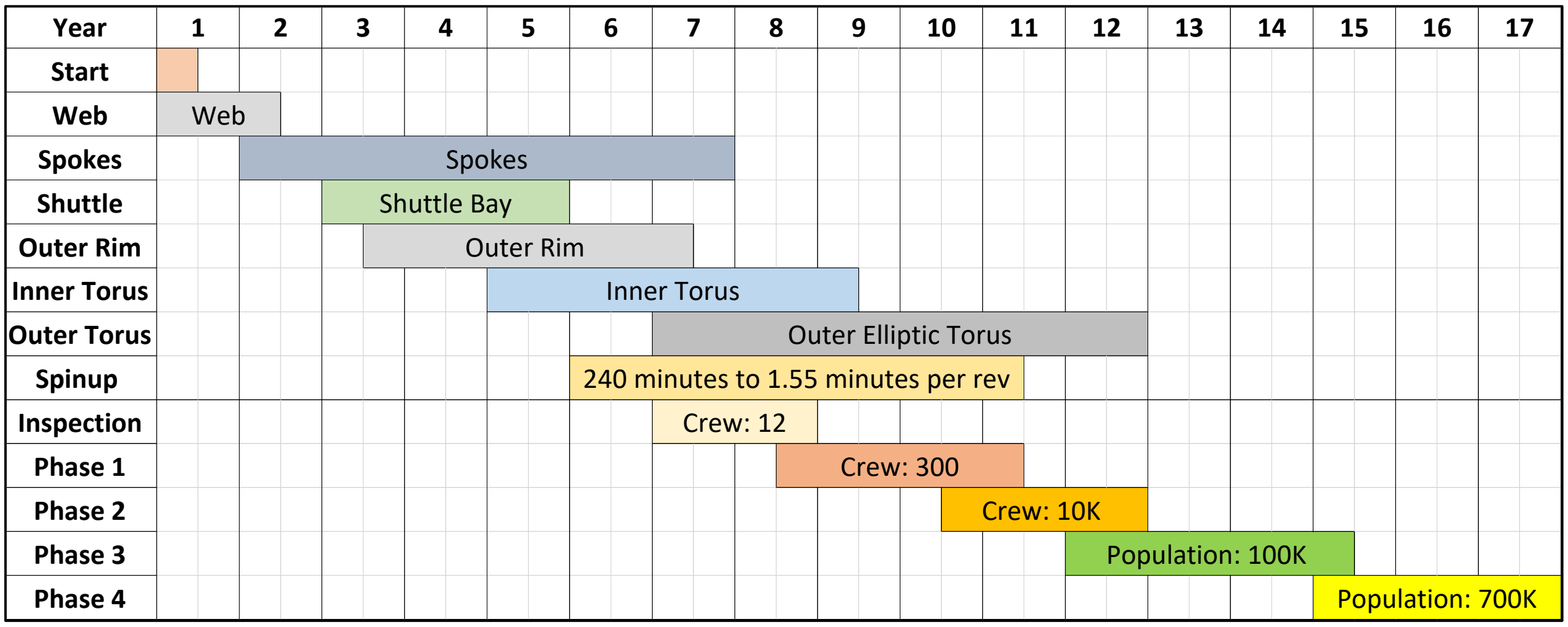

**Figure 4-9 – Schedule for Restructuring Asteroid Major Station Components**

### *4.4.4 Total Equipment Built*

Over the course of the simulation, the 3000 spiders were built. We show in Figure 4-10 a bar chart with the maximum number of key pieces of equipment created during this simulation. **Spiders** are the general-purpose robotic equipment. Four of these are sent with the original base station and electronics are sent to construct a total of 3000 spiders. Regolith is moved in **Trucks** that are covered to help reduce floating debris in the low gravity. The equipment called **Movers** is like a flatbed truck and used to move items other than regolith such as trusses, panels, and tiles. **Truss Building Units** (TBUs) are built to more efficiently build the standard trusses used throughout the restructuring effort. **Harvest Units** (HUs) have arms that gather loose regolith and convey to the trucks. The base station probe initially produces the rods and tiles from regolith; however, **Rod Processing Units** (RPUs) are created to build rods and tiles. We are assessing whether to replace the RPUs with **Laminate Processing Units** (LPUs), which are smaller equipment helpers that convert regolith to anhydrous laminates for construction. The restructuring process uses **Fill and Melt Units** (F&MUs) to fill walls and floors with ground regolith and melt layers for material containment and strength. **Skin Attachment Units** (SAUs) are mechanical automatons that attach panels to the shell, walls, and floors. As the name implies, the **Regolith Crushing Units** (RCUs) take large regolith pieces (cobbles) and crush them to pebbles and grains. **Diggers** use an attachment like jack hammer chisels to reduce boulders and flat rock regions for processing. These diggers hang onto the web and slowly and autonomously move across the asteroid surface. They leave behind a trail of smaller rocks. Early in the restructuring process, **Tile Placement Units** (TPUs) attach regolith tiles to structures to provide working surfaces, for material retention, and for strength. TPUs also create panels for equipment and floors. Once the outer rim of the outer torus begins to rotate and provide centripetal gravity, we begin to build **Continuous Panel Production Units** (CPPs). The CPPs are large complex units that produce larger panels of regolith to be attached to the shell, walls, and floors by the SAU. LPUs may supplant the need for CPPs. **Rotating Pellet Launchers** (RPLs) are spinning arms that launch regolith pellets at high speed [Johnson and Holbrow 1977]. The RPLs are used to launch and quickly move regolith from the asteroid to the outer rim. This is a faster technique to move regolith long distances instead of using trucks. We use this transfer of mass (and momentum) to spin up the outer rim (and ultimately the station) and reach the desired rotation rate.

About 26,500 pieces of equipment are constructed in the simulation. The chart does not include the subunits such as sun trackers, parabolic mirrors, connectors, and Stirling engines. Estimates for these subunits are included in the build time and volume estimates for the other equipment.

### *4.4.5 Total Equipment over Time*

Our simulations allow us to evaluate the quantity of equipment needed to complete the station. The initial probe with four spiders produce almost 26,500 pieces of equipment over twelve years. This count is another way to evaluate progress of the restructuring effort. We show in Figure 4-11 a running count of the 13 types of equipment over the twelve years. The y-axis shows the equipment count ranging from 0 to 30,000 pieces. The x-axis shows the passage of time in years ranging from 0 to 12 years. Initially only spiders and the base unit are available to do the work. Spiders use vitamins and in-situ material to construct other spiders. Spiders build the framework, connectors, subunits, tools, and equipment. Ultimately, the spiders no longer build, and instead, are occupied with managing the equipment. These simulations determine how many pieces of equipment can be managed without exceeding the available 3000 spiders. The stacked graph chart shows the quantity of units for each of the thirteen major pieces of equipment. The legend shows those units, and the stacked graphs are in the same order as the legend. We show the maximum quantity in parentheses in the legend for each piece of equipment produced in this simulation.

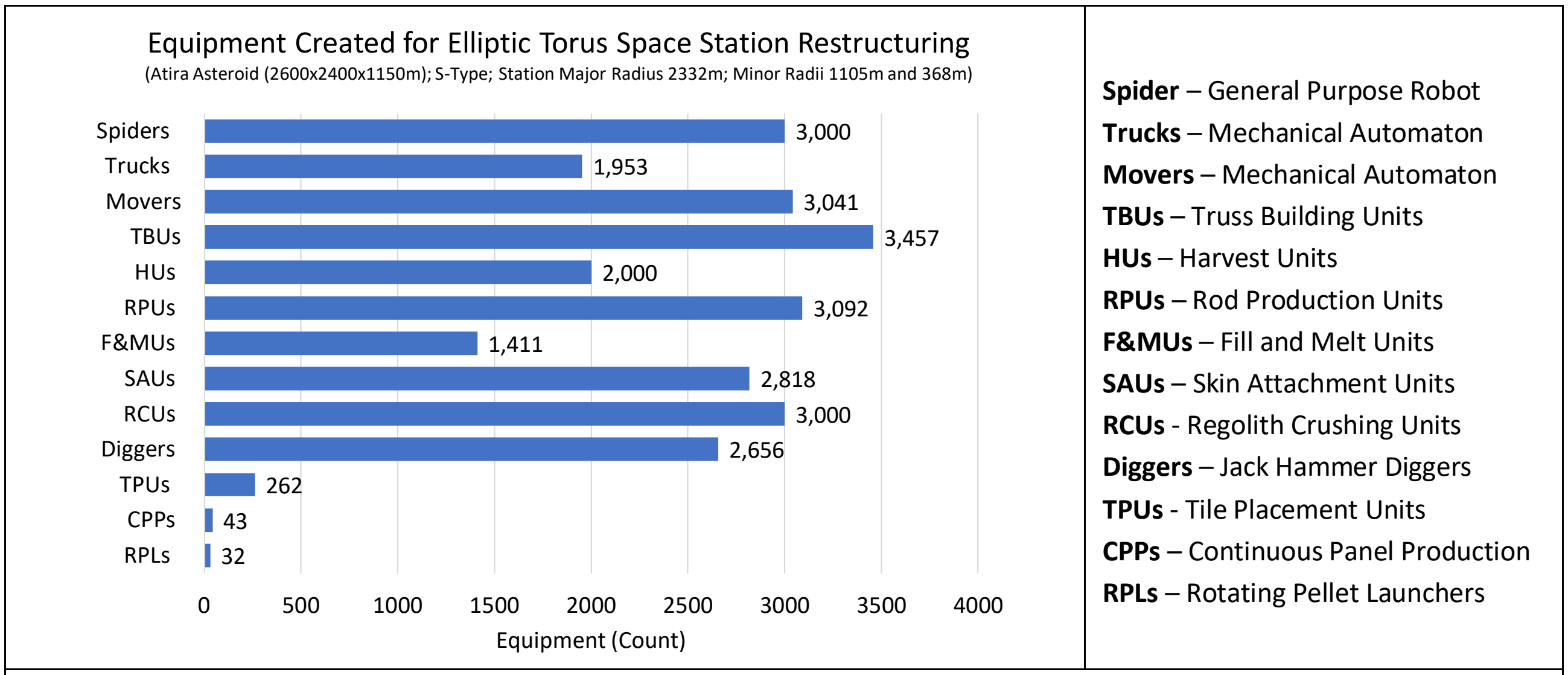

**Figure 4-10 – Total Count of Equipment Used During Simulation**

Exponential growth is characterized with slow initial growth followed by rapid (almost unbounded) growth. The graph in Figure 4-11 barely shows any activity during the first half year. It takes time to build the initial web of trusses and the first pieces of equipment with only four spiders and the base unit. It shows a modest growth spurt from the sixth to the nineth year. This is time spent preparing to build and starting construction on the enormous outer elliptic torus. In the nineth year, all 3000 spiders have been built and the self-replication ends. The last three years the spiders are fully occupied with managing the tools and equipment. A fairly constant set of equipment works on completing the outer torus. We expect that additional refinement would shorten the overall schedule. We describe the activity at several yearly milestones in the following paragraphs.

**Year 1: Web Production:** At the end of the first year, there are 405 spiders active on the asteroid and station. There are 43 Rod Production Units (RPUs) active. We see 20 Harvest Units (HUs) are digging regolith and many of the 362 Trucks are bringing that regolith to the RPUs. There is only 1 Regolith Crushing Unit (RCUs) active on the surface. Spiders and Harvest Units process the innate loose regolith pebbles and grains on the surface. We also find 48 Truss Building Units (TBUs) are creating trusses that are being moved by the 1524 Movers. Spiders are using those trusses to extend the web and build more equipment. Tile Placement Units (TPUs) start being built to provide panels for the RPU, Jaw Crushers, and other equipment.

**Year 3: Spoke and Shuttle Bay Production:** At the end of the third year, there are 716 spiders active on the asteroid and station. There are 74 Rod Production Units (RPUs) active. We see 115 Harvest Units (HUs) are gathering regolith and many of the 1892 Trucks are bringing that regolith to the RPUs. We also find 82 Truss Building Units (TBUs) are creating trusses that are being moved by the 2305 Movers. Spiders are using those trusses to build the station structure and build more equipment. There are 153 Tile Placement Units (TPUs) attaching tiles to the walls of the spokes and producing panels for other pieces of equipment. There are 259 Fill and Melt Units (F&MUs) working on the outer walls of the spokes.

**Year 5: Spoke, Outer Rim, and Inner Torus:** At the end of the fifth year, there are 840 spiders active on the asteroid and station. There are 277 Rod Production Units (RPUs) active. We see 462 Harvest Units (HUs) are gathering regolith and many of the 1953 Trucks continue to bring that regolith to the RPUs. We also find 262 Truss Building Units (TBUs) are creating trusses that are being moved by the 2311 Movers. Spiders are using those trusses to build the station structure and build more equipment. There are 257 Tile Placement

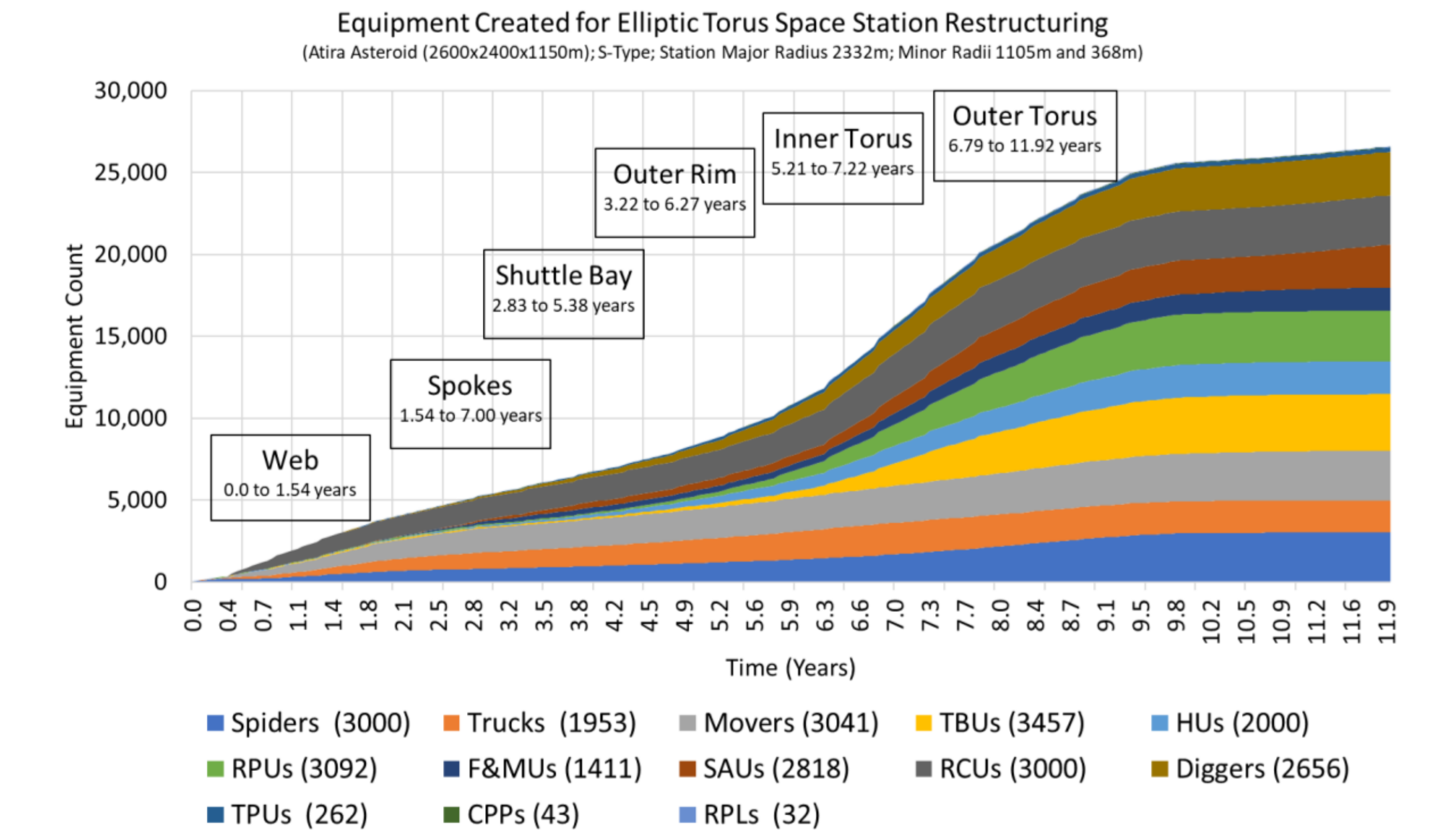

| | | |
|---|---|---|
| **Spider** – General Purpose Robot<br>**Truck** – Mechanical Automaton<br>**Mover** – Mechanical Automaton<br>**TBU** – Truss Building Unit<br>**HU** – Harvest Unit | **RPU** – Rod Production Unit<br>**F&MU** – Fill and Melt Unit<br>**SAU** – Skin Attachment Unit<br>**RCU** - Regolith Crushing Unit | **Diggers** – Jack Hammer Diggers<br>**TPU** - Tile Placement Unit<br>**CPP** – Continuous Panel Production Unit<br>**RPL** – Rotating Pellet Launcher |

**Figure 4-11 – Total cumulative equipment created during the restructuring process**

Units (TPUs) attaching tiles to the walls of the spokes and producing panels for other pieces of equipment. There are 372 Fill and Melt Units (F&MUs) filling the outer walls of those structures as the panels enclose them. As the outer rim becomes complete in the sixth year, some Regolith Pellet Launchers (RPLs) would be manufactured and begin launching pellets from the asteroid to the outer rim. The outer rim would begin to rotate faster than the asteroid, produce centripetal gravity, and some Continuous Panel Production Units (CPPs) and Skin Attachment Units (SAUs) would be built.

**Year 8: Inner Torus and Outer Torus:** At the end of the eighth year, there are 1904 spiders active on the asteroid and station. There is a significant increase in the number of equipment units to construct the large outer torus. There are 2320 Rod Production Units (RPUs) active. We see 1579 Harvest Units (HUs) are gathering regolith and many of the 1953 Trucks are bringing that regolith to the RPUs. There are also 28 Regolith Pellet Launchers (RPLs) moving regolith from the asteroid to the outer rim and torus. We also find 2617 Truss Building Units (TBUs) are creating trusses that are being moved by the 2864 Movers. There are 33 Continuous Panel Production (CPPs) systems operating on the rotating outer rim. The CPPs are producing panels that 476 Skin Attachment Units (SAUs) are welding to the shell, walls, and floors. Panels from the CPPs are used instead of those built by the Tile Placement Units (TPUs). As such, no additional TPUs are built. Much of the original loose regolith is gone from the asteroid and 2656 Diggers are creating the cobbles for 2825 Regolith Crushing Units.

**Year 10: Outer Torus:** At the end of the tenth year the equipment is only working on the outer torus. There are 2822 Spiders active on the asteroid and station. We find 1943 Trucks and 2949 Movers are active moving trusses, panels, and regolith. We see 2000 Harvest Units (HUs) are gathering regolith. There are also 32 Regolith Pellet Launchers (RPLs) moving regolith from the asteroid to the outer torus. There are still 2656 diggers creating the cobbles for 2909 Regolith Crushing Units. There are 42 Continuous Panel Production (CPPs) systems operating on the rotating outer rim. The CPPs are producing panels that 2023 Skin Attachment Units (SAUs) are welding to the shell, walls, and floors. There are 1016 Fill and Melt Units (F&MUs) filling the inner and outer walls of the outer torus as the panels enclose them.

**Year 12: Outer Torus:** The station is completed at the end of the twelfth year. All 3000 Spiders have been built and have built and are managing 23,500 pieces of equipment.

### *4.4.6 Equipment Working on Station Structures*

The chart in Figure 4-12 shows the total number of units building various parts of the Atira space station. The vertical axis shows the number of units, which includes spiders, trucks, movers, and the other support equipment. The horizontal axis shows the time in years when those units are working on the building tasks.

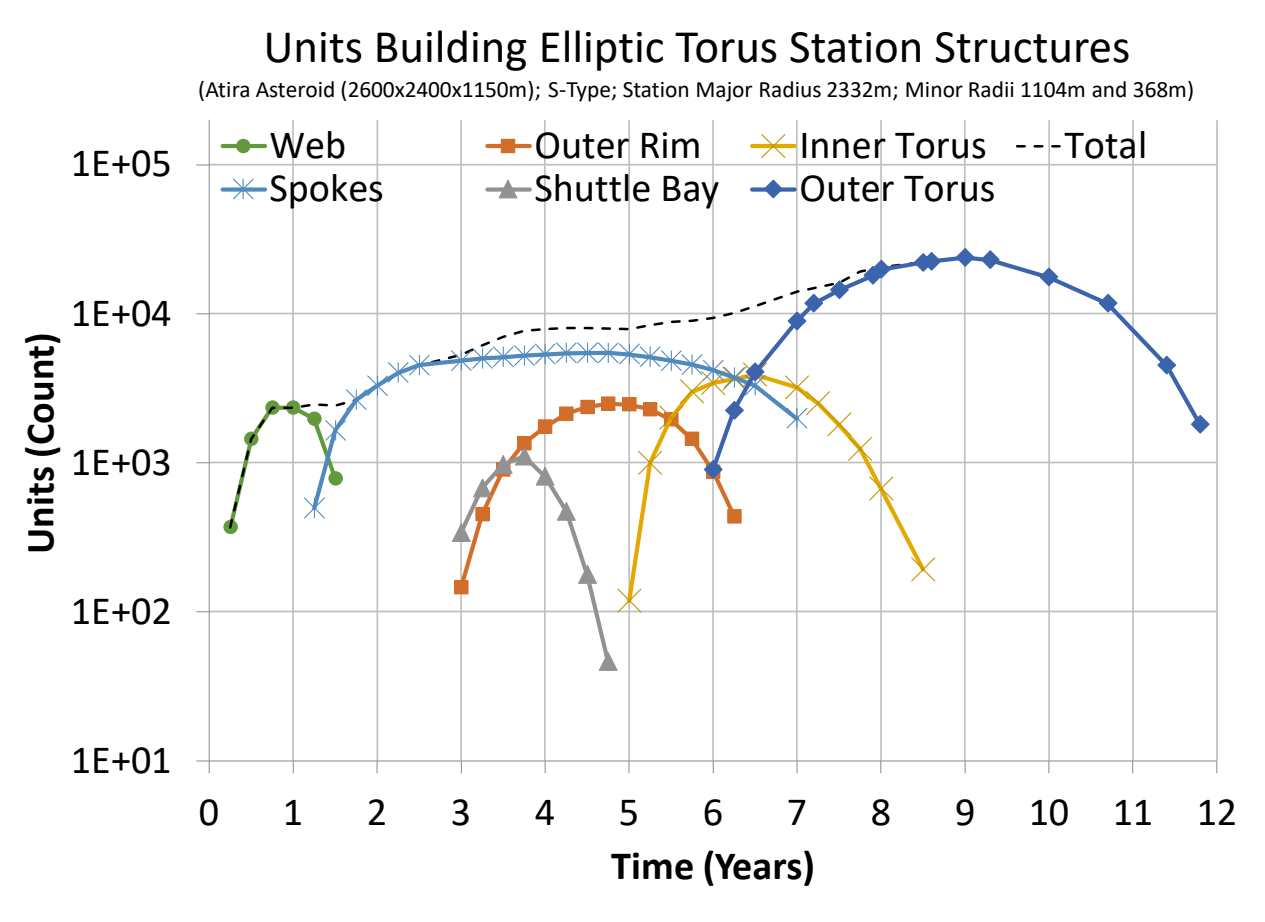

**Figure 4-12 –Units Building Major Station Components**

The units are building the web, spokes, shuttle bay, outer rim, inner torus, and outer torus. Each structure sees a similar pattern of increasing number of units, a constant number working, and then a decreasing number of units. This simulation uses the same rates and foundation as the previous simulation. The y-axis of this graph is a logarithmic scale while the y-axis of Figure 4-11 was linear. This chart includes recent refinements. In the future, we plan to harmonize the two simulations and include refinements and additional details.

This chart shows the order of the major building tasks. The web is built in the first two years. The spokes are started and then the inner torus. The shuttle bay displaces the initial seed unit and provides an axis point for the station rotation. The outer torus ties together shuttle bays, spokes, and inner tori on the north and south poles. At this point in the construction, the process includes a spin up task (not shown). This spin-up task increases the station rotation speed from one revolution every 240 minutes to one revolution every two minutes. This simulation shows the entire station could be completed in twelve years. We expect that improvements in our process (and simulations) will refine these schedule estimates.

### *4.4.7 Productivity Measure of Self Replication*

Our initial thought was to only use replicators to restructure the asteroid. Estimates quickly showed that it would take thousands of spiders and thousands of years to complete the effort. Many of the restructuring tasks are simple, repetitive, and constrained to well defined areas. This led to the concept of building tools to replace that type of repetitive work and reduce the number of required spiders. Launch costs limit the number of spiders while the number of tools and equipment is essentially unlimited using in-situ asteroid material.

Our restructuring process takes advantage of productivity gains from both parallelism and specialization. Parallelism, the number of replicators, is the traditional improvement of self-replication. We increase that parallelism with many tools and pieces of equipment. The specialized tools and equipment perform specific tasks much faster than the general-purpose replicators. The chart in Figure 4-11 illustrates the pieces of equipment being built over the construction of this space station. At the end of the restructuring effort, our baseline simulation shows many spiders (3000) working with even more support tools and equipment (over 23,500). This represents a high degree of parallelism. Instead of a single unit working on the space station, we have over 26,500 units working.

The specialization represents another productivity gain. We use the example from our system analysis example of the Replicator, Helper, and Product model; see §4.3.2. We estimate that two spiders could build a truss in 127 hours. The specialization of a Truss Building Unit (TBU) enables it to build a truss in an average of 3.05 hours. Spiders need to load rods or tiles into the TBU system, wind the clockwork springs, tune the Stirling engines, and adjust the mechanical programming as necessary. One spider on average can manage 5 TBUs. One spider by itself produces a truss in 254 hours on average. One spider with 5 TBUs can produce a truss in 0.61 hours on average. It would take 416 spiders by themselves to match the production rate of the single spider with 5 TBUs. We measure this specialization productivity gain as an "equivalent spider" metric of 416.

To compute the equivalent number of spiders, we must account for the tool productivity and the number of spiders supporting those tools. We compute the tool productivity by comparing the task time using only one spider to the time for that same task using the specialized equipment. We show in Table 4-7 this equivalent spider metric for the 13 pieces of specialized equipment.

We also include in Table 4-7 the number of spiders monitoring the equipment. This was estimated using the time to support the piece of equipment (position, program, load, and empty) compared to the time the equipment is autonomously performing its task. The reciprocal of this value would be the number of pieces of each type of equipment that a spider can monitor. Using the final equipment tally, we see that each spider is monitoring about 7 or 8 pieces of equipment. Many tools are mechanically programmed to perform a repetitive operation. Spiders only need to be involved to load, monitor, and adjust setting on those pieces of equipment. Some pieces of equipment need constant supervision and monitoring (e.g., CPP and HU). Other pieces need little support from spiders. They perform long duration tasks once they are positioned, programmed, loaded, and later emptied (e.g., Trucks and Diggers).

We also include in Table 4-7 a short description to provide some insight on this equipment estimate. The improvements shown in the table represent the decrease in spider-hours to perform the tasks. This also represents the number of spiders saved for each of the tasks. This could also be considered a multiplier for equivalent spiders working on the station. Just as an example, one of the most time-consuming tasks is the Fill & Melt of the regolith. The previous example simulation (Figure 4-10) used 1411 F&MUs to build the Atira station. It would have taken about 2 million spiders to perform the fill and melt tasks without those F&MUs. Instead, by using this support equipment, we only need about 290 spiders to monitor the 1411 F&MUs.

**Table 4-7 – Tools and Spider Equivalence**

| Equipment | Equivalent Spider | Monitor Spiders | Description |
|---|---|---|---|
| **Spider** – General Purpose Robot | 1 | 0 | Original units for comparison – moving, building, digging, throwing, etc. |
| **Trucks** – Mechanical Automaton | 10 | 1/10 | A spider can travel at 10 meters per minute and the truck travels at 1 meter per minute. |
| **Movers** – Mechanical Automaton | 50 | 1/10 | Able to move larger loads (5x) and it still moves at 10% the speed as a spider. |
| **TBU** – Truss Building Unit | 416 | 1/5 | 1 spider with 5 TBUs = 36.6 minutes / truss; 1 Spider = 254 hours/truss. |
| **HU** – Harvest Unit | 500 | 1 | Harvest Unit = 30 m3 / hour; Spider 0.0005 m3 shovel jig at 30 seconds * 60 sec/min * 60 min/hr or 0.06 m3/hour. |
| **RPU** – Rod Production Unit | 24 | 1/4 | 1 spider to spin, 1 spider to hold mirror, 1 to manage rod.<br>RPU builds 8x rods in same time. |
| **LPU** – Laminate Production Unit | 96 | 1/2 | 1 spider to spin, 1 spider to hold mirror, 1 to manage tile.<br>LPU builds 32x tiles in same time. |
| **F&MU** – Fill and Melt Unit | 1410 | 1/5 | Spider Only - 5.31 hours per m3 of fill.<br>Spider & FM&U - 0.0038 hours per m3 of fill. |
| **SAU** – Skin Attachment Unit | 240 | 1/24 | Spider Only - 0.703 hours per m2 of panel.<br>Spider and SAU - 0.00293 hours per m2 of panel. |
| **RCU** - Regolith Crushing Unit | 643 | 1/4 | Rock Crusher Unit - Jaw crusher – 0.9 m3 every hour; 3 Jaw Crushers per RCU = 2.7 m3 / hour. Spider 1 rock per hour = 8.4 kg = 4195.6 cc = 0.0042 m3 / hour; Result 2.7 m3/hour divided by 0.0042 m3/hour = 643x. |
| **Diggers** – Jack Hammer Diggers | 16 | 1/20 | One chisel produces 0.72 m3 per 3.4 hours. Digger has 4 chisels running at 4x the speed possible with a spider. Equivalent to 16x spiders.<br>1 shovel at 0.5 liter = 0.0005 m3 =10kg at 10 m/s = 100 kg m/s. |
| **TPU** - Tile Placement Unit | 18 | 1/3 | Spider Only - 0.877 hours per m2 of panel.<br>Spider and TPU - 0.0488 hours per m2 of panel. |
| **CPP** – Continuous Panel Production Unit | 8439 | 3 | Unit produces panels at 10 meters per minute (200 m2/minute). Spider Tiles at 1 m2 per 0.703 hours = 0.0237 m2/minute. Equivalent=200/0.0237 = 8439x. |
| **RPL** – Rotating Pellet Launcher | 2400 | 1/2 | 2 Spiders average to monitor RPL – load, program, adjust.<br>120 kg / minute over 2000 meters = 4000 kg-m/s (momentum).<br>Spider only handles 0.0005 m3 shovel jig. Using trucks and travel time to attain 2400x improvement. |

We use the equivalent spider metrics to determine how many total equivalent spiders would be required to build the Atira station. We use the maximum equipment counts in the chart in Figure 4-10. We combine that with the Equivalent Spider metric of Table 4-7. The Equivalent Spider total for each piece of equipment is in Figure 4-13. The chart shows the 3000 spiders that were built including the 4 spiders that were sent with the original launch. We find that the 1411 Fill & Melt Units perform the work of almost 2,000,000 spiders. The 3457 Truss Building Units are performing the work of 1.4 million spiders. These totals sum to 7,768,619 equivalent spiders. With today's launch costs, we could not justify sending almost 8 million robots into space. With 7,768,619 equivalent spiders given our 3,000 physical spiders, we find a productivity gain of 2590 using the tools and equipment. We harvest and process 6.57 billion kilograms of regolith with those 8 million equivalent robots over 12 years. This implies that each equivalent spider is processing 845.7 kilograms per year or 96.5 grams per hour – a very reasonable rate! We recall that the restructuring process began with a base unit, only 4 spiders, 3000 spider electromechanical modules, and a small assortment of jigs and supplies.

## 4.5 Robotics – Summary

Asteroid restructuring involves asteroids, space stations, and robotics. Previous sections covered asteroids and space stations. This section covered robotic technologies. We included a brief overview of the background and technologies of robots. This included example systems for space exploration and mining. To operate in the asteroid environment and perform the complicated restructuring process, the robots will need autonomous system software. The asteroid restructuring represents a huge construction task. We used self-replication to produce the labor to perform that construction. We introduced the concept of having the robots make specialized tools to improve their productivity. We described our simulations and their results show that asteroid restructuring is a viable approach to create a space station from an asteroid.

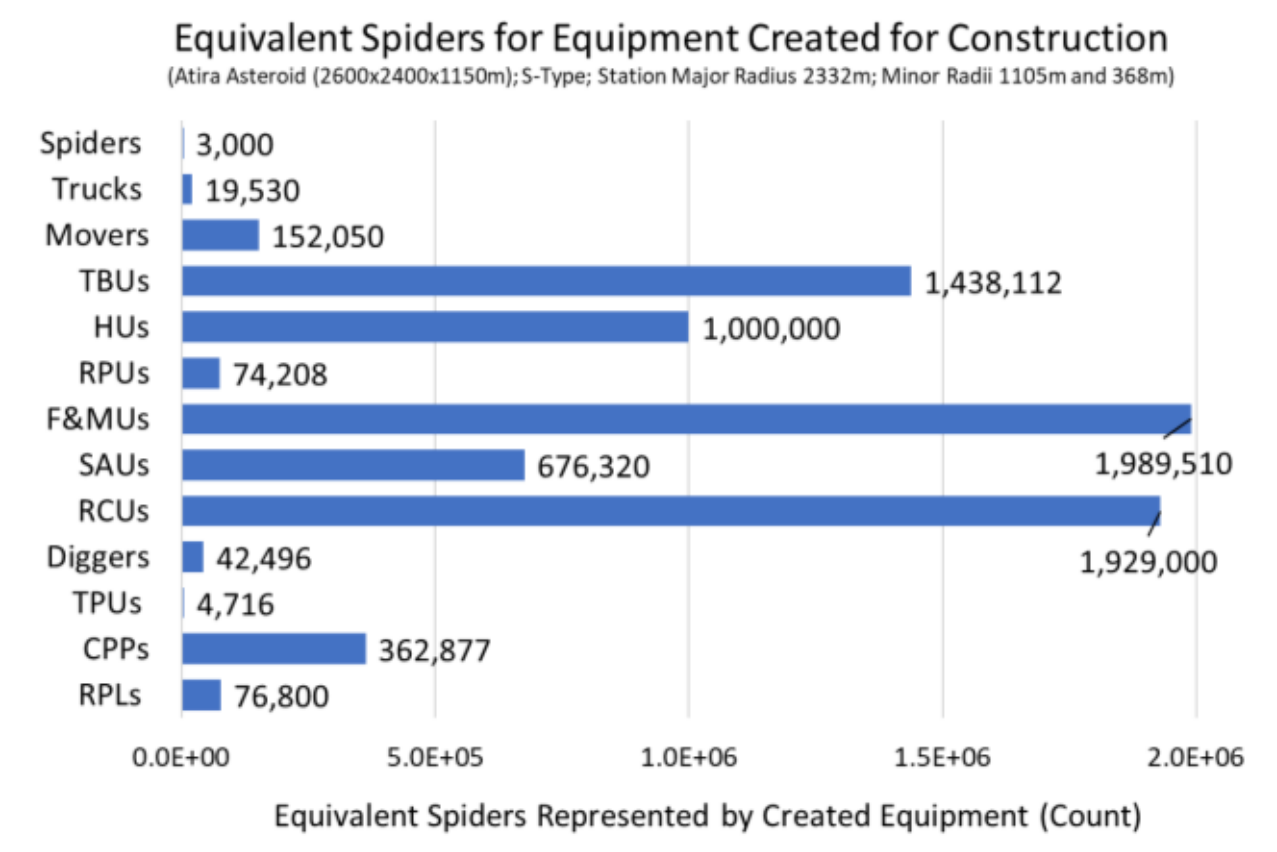

**Figure 4-13 – Total equivalent spider effort from equipment used during the restructuring process**

Our restructuring concept uses a form of limited self-replication called Productive Replicators [Freitas and Merkle 2004]. Our initial probe starts with a small set of 21st century robots, manufacturing capability, testing equipment, a modest number of robotic electronics and actuators, and some supplies. The initial set of robots and equipment complete the construction of robots using local material. These robots construct tools and equipment using technologies available during the first industrial revolution. Instead of large industrial complexes, overwhelming numbers of small mining and construction equipment restructure the asteroid into a space habitat. A hierarchy of cognitive architecture nodes cooperate to manage the restructuring process.

# 5 Asteroid Restructuring – System

Previous sections covered detail on asteroids, space stations, and robotic technologies. We use those details in this section to cover additional construction and system details. We review the torus space stations constructed from six asteroid sizes. We then overview the build time and population for those same asteroids and stations. We include in this section an estimate of the cost of a restructuring project. As a part of the system analysis, we include subsections on our quantitative and qualitative design.

## 5.1 Asteroids and Station Size

We varied the size of the space habitat and its components for different asteroids. In Figure 5-1, we show the material used for the major components of six space habitat versions. We assumed the same S-Type composition for all the asteroids. For each asteroid, we found the largest station radius using 30% of the available asteroid construction oxide material. We include the same 5 asteroids with mean diameters ranging from 490 meters to 4920 meters. For reference, we include the material required to construct a station the size of the Stanford Torus. Table 3-7 included the computed radii and populations for torus stations.

Except for the Stanford Torus, we design elliptic torus stations that are half filled with floors. The rotation major axis is 6.33 times the smaller minor axes in these stations. This ratio provides a comfortable gravity range over the floors. The gravity on the top floor (at the center) is 0.95g and on the bottom floor (outer rim) is 1.1g. We could build a station with a rotation radius of 205 meters and minor radius of 32 meters using the asteroid Bennu which has an equatorial radius of 241 meters. Ryugu, with a mean diameter of 896 meters, could produce a single torus with a rotation radius of 354 meters and minor radius of 56 meters. The Stanford Torus has a major radius of 830 meters and a minor radius of 65 meters [Johnson and Holbrow 1977]. For Moshup, with a diameter of 1317 meters, we can build a single torus with a radius of 937 meters. The Atira asteroid with a mean radius of 1928 meters would support a large outer torus. We evaluated the Atira station with a single, double, and triple set of inner tori and spokes. We found with the double set of shuttle bays, spokes, and inner torus; the elliptic outer torus would have a major radius of 2116 meters and minor axes of 334 and 1003 meters. In Figure 5-1 we only show the results for the double set. For a larger asteroid the size of Šteins with a mean diameter of 5160 meters, we describe a double station with a major radius of 4042 meters and minor radii of 1915 and 638 meters.

## 5.2 Build Time and Population

Most asteroid mining ventures focus on small asteroids – 10 meters to 100 meters. At this size, they consider towing it back to an Earth orbit and/or processing it in place and returning the mined product. Our restructuring goal is to convert the entire asteroid into the framework of a space habitat and leave that framework in the asteroid's orbit. The equipment and process are scalable and developed to restructure asteroids that are several kilometers in diameter and create a space station structure that can support a population of nearly one million people.

Most of our simulations consider the asteroid 163693 Atira. It contains almost 22 billion cubic meters of material, and its mass is over 40 trillion kilograms. The space habitat uses over 2.6 billion cubic meters of construction material. We find the mass of the station would be 5.3 trillion kilograms and would be about 13% of the mass of Atira. The restructuring process could also inventory and store 123 million cubic meters of volatiles and almost 341 million cubic meters of metal for the future inhabitants. This example station is a little larger than the Atira station in Figure 5-1. Its outer torus has a major radius of 2332 meters. The elliptical-shaped cross-section would have minor radii of 368 meters by 1104 meters. Again this is similar to the illustrations in Figure 1-1 and Figure 3-9. This station takes about 12 years to complete; see Figure 5-2. At the end of the 12 years, we have a habitat that is fully enclosed, is rotating to create Earth-like gravity, and has meters of shielding to protect the initial work crews and future colonists from radiation.

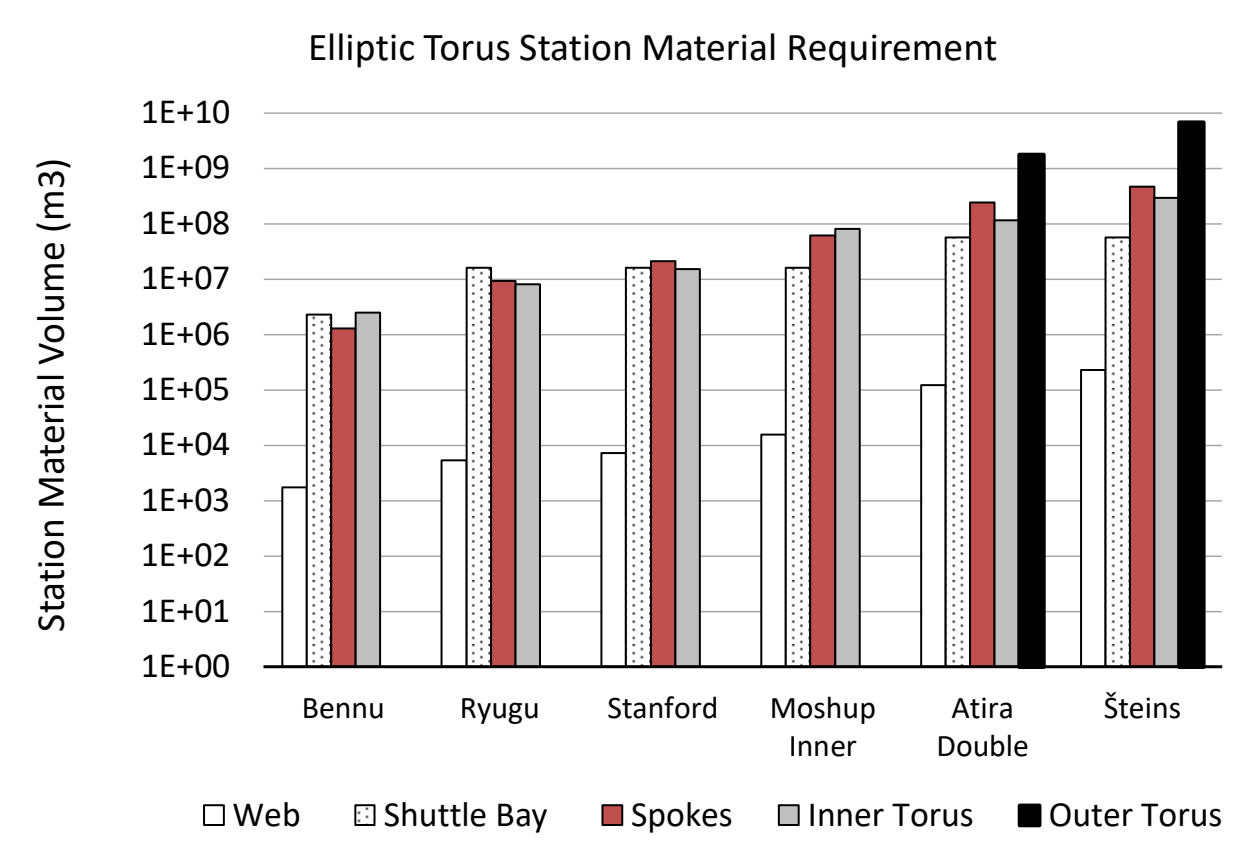

**Figure 5-1 – Material Allocation to Station Components**

We show in Figure 5-2 a chart with the population and build time of torus stations as a function of volume of material used in their construction. We include five asteroids and their corresponding space habitats. Two space habitat structures were considered in detail. The larger Atira station was analyzed and takes about 12 years to restructure. We also evaluated a smaller space habitat like the O'Neil Torus that can support a population of 13,000 people. The smaller station requires the material of an asteroid about the size of 162173 Ryugu (896 meters in diameter). The restructuring takes about 5 years and creates a single inner torus with a major radius of 354 meters. The other asteroid values are extrapolated from the analysis of those two asteroid designs.

The restructuring process scales to convert asteroids from modest sizes less than 1 kilometer in diameter to those of almost 10 kilometers in diameter. Large asteroids begin to take excessive time to restructure given our one seed and single rocket launch. The Šteins asteroid in Figure 5-2 takes 25 years to process. The 3000-spider limit in one seed package constrains the number of Replicators (spiders/productive replicators) and ultimately the number of managed Helpers (tools and equipment). Sending two base stations in the single launch or a second launch would increase the number of spiders and would decrease the processing time. Large structures will also be limited by the structural strength of the trusses and panels.

We plan that the process will be completely autonomous. As such it will be adaptable to different materials and construction issues. As an example, if the center of an asteroid were a solid fragment of iron, the process would leave that core for future inhabitants. There would be much less basalt materials and the overall size of the station would have to be reduced. In Figure 5-2, the analysis assumes about 30% of the asteroid oxide material will be usable for the station construction. A large percentage of that material will be used for shielding and will not need to be high quality.

## 5.3 Restructuring Cost

We compare this effort to O'Neill's estimates for his space cylinders [O'Neill 1974]. His estimates rely on a significant manual effort. He assumes 42 metric tons constructed per labor-year, which is comparable to large scale construction on Earth in the 1970s. The Atira station weighs 4.46e9 metric tons or 106 million labor years. This would be over $43 trillion in labor, which is not viable. We strive to reduce these costs with automation. The concept of applying automation to these efforts is not new. O'Neill stated:

> *"In the long run, space-colony construction is ideally suited to automation. A colony's structure consists mainly of cables, fittings and window panels of standard modular form in a pattern repeated thousands of times. The assembly takes place in a zero gravity environment free of the vagaries of weather. By the time that the colonies are evolving to low population density, therefore, I suspect that very few people will be involved in their construction."* [O'Neill 1974].

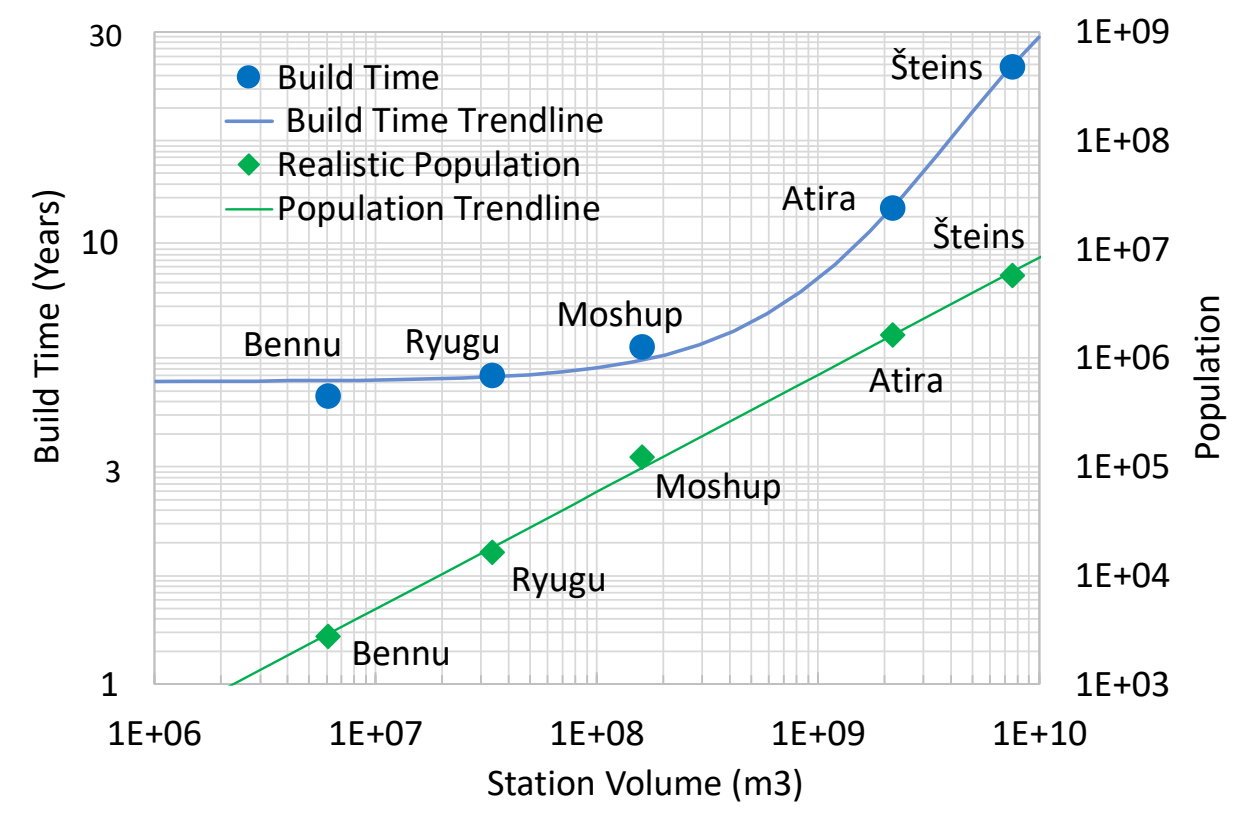

**Figure 5-2 – Station Material Volume and Resulting Population and Build Time**

The amount of automation used in the restructuring process is more than most published space habitat building approaches. We recall that we have an equivalent of 7.77 million spiders working at the end of 12 years on this project. Summing the spider-years over the 12 years results in 45 million spider years of continuous labor. It is comforting to see a rough similarity to the 106 million labor years using O'Neill's estimate. There are multiple scaling factors that should be applied to more accurately compare the two values. As an example, the restructuring process works more than 8 hours per day. As a first estimate this adds to our confidence that the restructuring approach using self-replication with parallelism and specialization is viable.

We created a table using data from O'Neill's Physics Today article [O'Neill 1974]; see Table 5-1. This table includes his details for the Model 2 Cylinder (population 150,000). In 1974, O'Neill was hoping his first station would be built in 1988. The second station would leverage the experience and capabilities of the first station and be built eight years later in 1996. With inflation, the 2019 dollars would be 6.1 times the 1972 dollars. The cost of this station would be about $200 billion in 2019 dollars. O'Neill noted that the Apollo project cost $33 billion in 1972 dollars. That would also be about $200 billion in 2019 dollars.

The cost of space missions have decreased since the 1970s. The estimated cost of the Artemis program [Artemis 2022] is $93 billion compared to the $200 billion for the Apollo program (in 2019 dollars). The cost of space probes have been decreasing too. The New Horizon mission to Pluto cost $700 million. The Osiris-REx mission to the asteroid Bennu has a program cost of $800 million. The Hayabusa 2 mission to the asteroid Ryugu cost $150 million. The recent Indian Chandrayaan lunar mission cost $145 million. Tiny CubeSats today can cost less than $1 million [Myers 2022].

We compare a self-replicating lunar industry [Metzger et al. 2012] and our restructuring effort to O'Neill's estimates; see Table 5-1. We use the Metzger's list of assets from Table 4-6a with his mission descriptions to produce a self-replicating lunar industry [Metzger et al. 2012]. We used the New Horizon costs for equipment in both the Atira and Lunar estimates. We multiplied by 6 for the Lunar Industry because of the 6 required launches. We used our list of probe assets from Table 4-6b to estimate our probe cost. The cost estimate for the Atira mission probe is about $105 million. The cost estimate for the six Lunar Industry payloads is $140 million. We include 10,000 labor-years for research, development, and support of the lunar industry. With less support, we include 8,000 labor-years for those activities for the restructuring effort. There are still significant labor and research efforts needed to see both missions to fruition. We also added costs to the Lunar Industry for teleoperation and for manned Moon support. This first initial estimate for the Atira restructuring mission is about $4.1 billion. Our estimates are "back-of-the-envelope" rough engineering estimates. We hope that experts could refine and lower these estimates in the future.

We offer in this paragraph several funding comparisons for the station. This station provides almost 1 billion square meters of floor space and 383 million square meters of residential space. There are many ways to fund the $4.1 billion cost. Each of the 1 billion square meters would cost $4.10 per square meter. Table 3-10 shows a population of 700,000

**Table 5-1 – Estimating Cost of Building Space Colonies (Compare at 2019 dollars)**

| | O'Neill Space Colony - Model 2 | | | Metzger Lunar Industry | | Atira Space Station | |
|---|---|---|---|---|---|---|---|
| | 1974 | Total 1972 | Total 2019 | 2012 | Total 2019 | 2022 | Total 2019 |
| Description Original Items | Unit cost | (in $\$10^9$) | (in $\$10^9$) | Description | (in $\$10^9$) | Description | (in $\$10^9$) |
| Launch vehicles | $\$0.5 \times 10^5$ | 1.5 | 9.2 | 6 Space X Launches | 0.3 | Space X Launch | 0.06 |
| Transport E→L5 | $250/lb | 11 | 67.2 | | n/a | | n/a |
| People E→L5 | $500/lb | 8.8 | 53.8 | | n/a | | n/a |
| Transport E→M | $500/lb | 2.2 | 13.4 | Lower launch vehicle costs | 1.3 | | n/a |
| Equipment for Moon | $400/lb | 1.8 | 11.0 | Similar to New Horizons (6x) | 4.2 | | n/a |
| Equipment for L5 | $400/lb | 2 | 12.2 | | n/a | Similar to New Horizons | 0.7 |
| Machines and tools (L5) | $625/lb | 2.8 | 17.1 | Assets: $140M Estimate | 0.14 | Probe: $105M Estimate | 0.11 |
| Salaries (L5) | 25% on Earth | 2 | 12.2 | Salaries on Moon | 12.2 | | n/a |
| Salaries (Earth) | 30,000 labor-year | 2 | 12.2 | Research, Support, & Tele-operation (10,000 labor-years) | 4.1 | Research & Support (8,000 labor years) | 3.3 |
| **Totals** | | 34.1 | 208.4 | | 22.3 | | 4.1 |

*Credit: Self produced using engineering estimates and concepts and data from [O'Neill 1974] [Facts] and [Metzger et al. 2012] [Facts]*

using 383 million square meters of residential space. The 700,000 people could invest in the $4.1 billion at $5857 apiece. Another alternative is for them to purchase their 547 square meters of residential space at a cost of $2243 and sell other regions of the station to research and industry. The population would need 18 million square meters of agriculture. This would cost $73.8 million in the station at $4.10 per square meter. An acre of agriculture space in the station would cost $16,592. This is a little more than the price of an acre of good Iowa farmland in 2022. These comparisons show that funding the station may not be excessive.

## 5.4 Station Quantitative Design Accuracy

In this project, we have researched and developed many engineering estimates. We used many different weights, speeds, rates, and costs. These metrics were used for equipment, processes, and missions. When possible we used comparable values or published engineering estimates. These are best effort values; however, we recognize that there is margin for error in those values.

We appreciated the comment in the foreword of a historic 1973 NASA document entitled "Feasibility Of Mining Lunar Resources for Earth Use: Circa 2000 A. D." [Nishioka et al. 1973]. This long memorandum details the technologies and systems required to establish the mining base, mine, refine, and return the lunar resources to earth for use. It also contains many estimates and metrics. We abbreviate the authors' foreword comment: "*Unfortunately, the quantitative results … determined in these types of studies dealing with … the future have a high probability of uncertainty and should thus be observed cautiously*." [Nishioka et al. 1973]. We recommend using the same caution with results of this paper.

## 5.5 Station Qualitative Design Evaluation

We again consider a torus station that has a major radius of 2400 meters and an elliptic cross-section with minor radii of 400 meters and 1200 meters. Figure 3-9 illustrated a cross-section of this habitat. Using square footage from all floors, this station could support over 10 million; see Figure 3-15.

We consider three views inside the large outer torus to address qualitative metrics. A picture is worth a thousand words and these inside views address qualitative criteria of the environmental design. We previously showed in Figure 3-22b an interior view of the space station. That rendering showed a wide field of view of one of the eight dividers in the station. That view is 3-kilometer-wide from almost 2 kilometers away. Each of the four entryways into the divider is as large as the Arc de Triomphe in Paris. Two of the spokes are within the divider. The divider includes 33 floors, each 15 meters in height. The dividers have over 1.5 million square meters of floor space – enough to support 10,000 nice size single home families. This scene has a fabulous view with a vertical height of 500 meters. We include two other views in Figure 5-3. The left picture is taken near one of the dividers and includes a two-story house (stick figure) for scale. The right picture is taken from the floor on the far-right side of the large elliptic torus. The rendering tool was programmed to paint a 100-meter square grid on the main floor – about the size of a city block. Along the side are some of the 376 two story homes rendered to help with scale. From this view we can see for 2.6 kilometers over the curved floor. In the scene is a 77-story tower built into one of the station spokes.

To further evaluate the potential of such a station, we consider other design criteria from the Stanford study [Johnson and Holbrow 1977]. The NASA SP-413 study explored physiological, environmental design, and organizational criteria for space stations. They felt these qualitative criteria must be met by a successful space habitat for the colonization of space. We show all three criteria in Table 5-2 updated with evaluations of the Atira station. This station has the potential to meet or exceed the design criteria of the Stanford station.

# 6 Asteroid Restructuring – Future

We offer in the following subsections several additional thoughts. We consider a geometry alternative, an alternative to shuttle bay landings, activities for the early colonists, thoughts about the Atira moon, and future projects for restructuring. This section ends with our conclusions for asteroid restructuring.

## 6.1 Geometry Alternative

As an afterthought in our research, we decided to better review the dumbbell structure. We noticed that the dumbbell nodes are quite large for the asteroids that we are

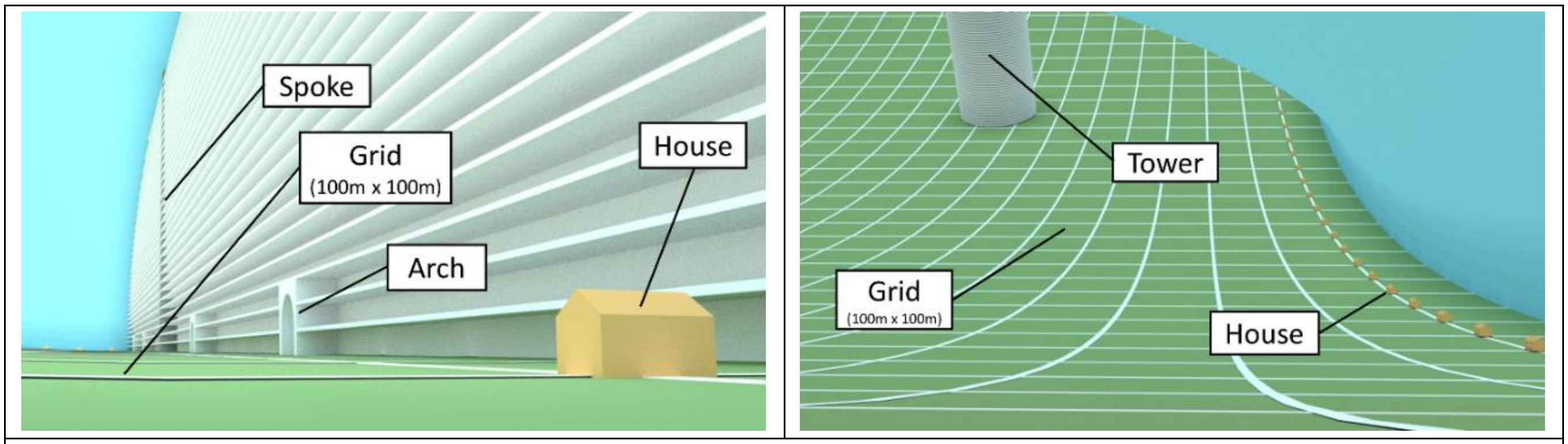

**Figure 5-3 – Qualitative Inside Views of Atira Station**

**Table 5-2 – Summary of Atira Station Design Criteria**

| Criteria | Metric | Atira Station Value / Comment |
|---|---|---|
| Summary of Physiological Criteria | Artificial Gravity (Centripetal Gravity) | Most living areas will be between 0.95g to 1.05g. Lowest levels will reach 1.05g and the inner torus will be 0.7g. |
| | Rotation rate | 0.62 rpm |
| | Radiation exposure | <5 rem/year |
| | Temperature | TBD: 23° +/- 8° C |
| | Atmospheric composition | TBD: Asteroid composition |
| Summary of Quantitative Environmental Design Criteria | Population | 700,000 comfortable to 6,500,000 maximum |
| | Projected area per person | 144.2 square meters with 5 meter floor spacing |
| | Projected volume per person | 627.6 cubic meters (without open volume) |
| | Open area per person | 18.7 square meters per person |
| | Agriculture area per person | 39 square meters |
| | Agriculture volume per person | 195 cubic meters |
| Summary of Qualitative Criteria of Environmental Design | Long lines of sight | About 3.5 kilometers |
| | Large overhead clearance | 500 meters on main floor |
| | Noncontrollable parts of environment | TBD |
| | External views of large natural objects | Large open areas designed for aesthetics |
| | Parts of interior out of sight of others | Multiple floors and dividers |
| | Natural light | Chevrons, fiber optic skylights, LED supplement |
| | Contact with the external environment | Large open areas designed for aesthetics |
| | Availability of privacy | Private homes |
| | Good internal communication | Modern terrestrial wireless capabilities |
| | Possible to physically isolate segments of habitat | Multiple floors and eight major segments in design |
| | Modular construction of habitat and of structures within habitat | Future design |
| | Flexible internal organization | Future design |
| | Details of interior design left to inhabitants | Left to future colonists |

*Credit: Self produced using NASA SP-413 [Johnson and Holbrow 1977] [NASA Report Public Domain]*

considering. We found that the dumbbell supports the largest population for a given asteroid mass; see Figure 3-20. The torus geometry is the next best to support a large population.

We offer a comparison in Figure 6-1 between the dumbbell and torus. The figure includes diagrams that are roughly to scale for a torus and a dumbbell station created from the Atira asteroid. We use our developed metrics and analysis to compare key features. Both stations were designed to use the same amount of building material. The Atira asteroid has 4.1e13 kilograms of mass and that quantity could support either the elliptic torus or the ellipsoid node dumbbell. We designed both stations to use 3.9e12 kilograms of material. A simple stress analysis finds that the tether or trusses of the dumbbell would need to have an equivalent truss cross section of at least 3500 square meters using a tensile strength of 5500 MPa (filled structure metric). This would likely require multiple truss tethers. We note additional analysis on the dumbbell station is required to address the rotational wobble and the tether stress strength.

For both stations, the gravity is 0.95g on the main floor and 1.1g on the outer rim. The torus would have about 60 floors and the dumbbell would have over 160 floors. For realistic populations, the dumbbell supports about 12% more people than the torus. Inhabitants on the main floor of the torus would have a vista looking down a long valley that is 2 kilometers wide and curves upward for 2.28 kilometers. The valley would have a ceiling of 334 meters. Inhabitants in the dumbbell would have a dome over them with a ceiling 820 meters above. The dumbbell would provide a smaller vista of 1.6 kilometers by 0.8 kilometers from the center of the main floor. Both designs are quite habitable and provide

**Elliptic Torus**

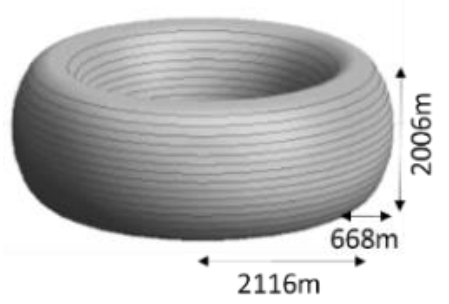

| | |
|---|---|
| Major Radius = 2116m | Realistic Population = 1.66 million |
| Minor Radii = 1003m and 334m | Rotation Rate = 0.63 RPM |
| Building Material = 3.86E+12 kg | Openness Ceiling = 334m |
| Habitable Volume = 7.49e9 m3 | Vista = 2282m by 1002m |

**Ellipsoid Node Dumbbell**

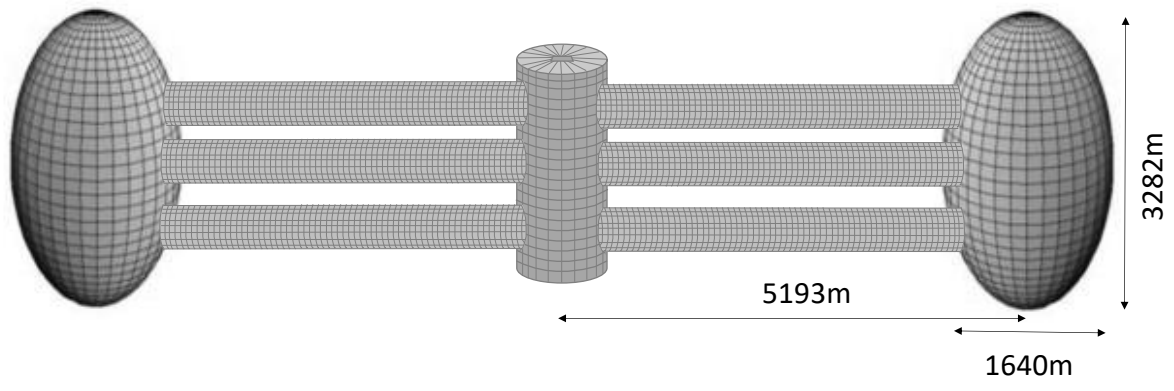

| | |
|---|---|
| Major Radius = 5193m | Realistic Population = 1.86 million |
| Minor Radii = 820m and 1641m | Rotation Rate = 0.40 RPM |
| Building Material = 3.86E+12 kg | Openness Ceiling = 820m |
| Habitable Volume = 9.24e9 m3 | Vista = 820m by 1640m |

**Figure 6-1 – Reconsidering Dumbbell Geometry**

open space for psychological well-being. It is possible that the dumbbell would be simpler to build. We need to identify difficulties and benefits between the two geometry designs.

## 6.2 Landing on Runways

In the future, space tourism could reach tens of thousands of launches per year - a rate comparable to the early decades of aviation [Globus 2006]. Tourism to a large space station may match international tourism to a city such as Washington DC. Over 2 million international tourists visit Washington DC each year and most are likely flying [DC 2017]. Assuming 40,000 flights to support this level of tourism, it is interesting to consider how to accommodate the 100 flights per day to a space station. A single docking station would have to accommodate over 4 flights per hour. This does not include flights supporting maintenance, imports, exports, and business. Most station designs use a central hub to support shuttle arrivals, servicing, and departures. This single central hub becomes a bottleneck for the passengers and trade

Runways on the surface of the rotating station have been historically avoided because of the perceived landing complexity. We have found that the shuttle does not need to perform a curved rotating approach to the station. The shuttle approach can be a straight vector. This minimizes fuel consumption and landing complexity. The touchdown on the rotating runway can be gentler than the touchdown of a large commercial aircraft landing on a terrestrial runway.

A small station could become imbalanced with the landing of a heavy shuttle. Some authors suggest landing two shuttles on opposite sides of the rotating station to reduce this imbalance impact. The Space Shuttle weighed about 2 million kilograms. The Space X Falcon 9 Heavy rocket weights 3 million kilograms. The smallest station in Table 3-8 was the Bennu torus and weighted 9,030 million kilograms. With our large station designs the shuttle mass is insignificant to the station mass and should not affect the rotation balance.

Runways can be positioned anywhere along a radial axis of the rotating space station. Dozens of runways could accommodate the higher traffic loads from tourism and from other space habitats. To introduce this concept, we offer a diagram and a chart in Figure 6-2. The diagram shows a cut through view of half of a rotating torus station. The rotation axis is along the right side of that diagram and show the rotation at $\omega$ radians per second. We include landing runways on the top (inner rim), on the middle side, and at the bottom (outer rim) of the torus tube.

We include a simulation result from a shuttle landing on a side runway on the station in Figure 6-2b. This station has a major axis (R=2116 meters) and a minor axis (h=334 meters). This is visually like the station shown in Figure 1-1. The side runway is illustrated in Figure 6-2a. The x-axis shows the horizontal distance, and the y-axis shows the vertical distance from the center of the rotating station. The chart shows the position of the runway at a radius of 2116 meters. This side runway would not completely encircle the torus. The chart in Figure 6-2b shows the shuttle approaching from the right side and the station rotating clockwise. It shows a portion of the runway at a radius of 2116 meters and the outer rim at 2450 meters. The shuttle approach in Figure 6-2b represents the last 7 seconds before landing. The shuttle is on a straight path towards its landing spot. It is decelerating at 1g for the flight towards the station. During the last moments of the approach, it decelerates at 2g in the y-direction. A force of 2g is less than most roller coasters or similar to a "hard" landing of a commercial air craft. The shuttle and runway edge both travel about 700 meters during the final approach. For most of the flight the shuttle is moving faster than the rotating runway and overtakes the runway edge. By design, the shuttle has a radial and tangential velocity (impact velocities) of 0 meters per second at touchdown.

We show detail of the extra landing deceleration in Figure 6-3. This view is near the landing point and details the final approach over the runway in Figure 6-2b. The graph axes show the horizontal and vertical distance from the station center. The markers on the shuttle and runway lines are spaced at 0.05 seconds apart. The runway floor rotates at 0.62 RPM with a tangential velocity of 137.2 meters per second. The shuttle approaches at a 14-degree angle; see Figure 6-2b. It is traveling at 211 meters per second (or a relative 73.8 meters per second) when it passes the contact point. It continues to decelerate at 9.8 meters per second squared. The shuttle decelerates at 2g for about the last two seconds. The shuttle speed reduces from 211 to 137.2 meters per second over six seconds. This angle and deceleration are designed to land the shuttle at the tangential velocity of the runway; as such, the shuttle lands at a relative 0 meters per second. This would be a very "soft landing."

The shuttle is initially moving faster than the runway floor. The runway edge continues to rotate as the shuttle moves towards the landing point. The chart in Figure 6-3 shows the position of the contact point after it has rotated during the shuttle landing as a red circle. When the shuttle lands, the shuttle is over 100 meters beyond the contact point (the edge of the runway). The shuttle lands with a forward velocity matching the tangential rotation speed; as such, there is minimal shuttle rolling or braking after landing.

**Approach Through Opening to Bottom Runway:** We also found that landings through openings along the outer perimeter of the rotating station are possible using this same straight approach. The shuttle passes through the opening, lands on the inner surface, and comes to a stop. Like the shuttle bay we enter through an opening and land inside the station. Unlike the shuttle bay, we have gravity and a runway to touchdown and decelerate like a commercial aircraft on Earth. The runway touchdown and deceleration are like the approach to other runway positions; see Figure 6-2b. Unlike the other runways, this one is inside the station and the shuttle enters through an opening. This landing intuitively seems like it would be complex and risky. This intuition appears to be wrong. The shuttle can simply approach on a straight path to the bottom of the station at a shallow angle. In the example of this section, the shuttle approaches the bottom runway at a radius 2450 meters at an angle of 16.5 degrees. We show the shuttle crossing the 20-meter-thick shell of the outer torus in Figure 6-4. The chart in Figure 6-4a shows the approach in the station coordinate system and shows two openings. The opening is shown at its position at 7.0 seconds before touchdown (yellow and to the upper-right). The opening (and station) is rotating clockwise and is also shown at its position 6.5 seconds before the shuttle touchdown (blue and to the lower-left). The shuttle enters and exits the opening at these times.

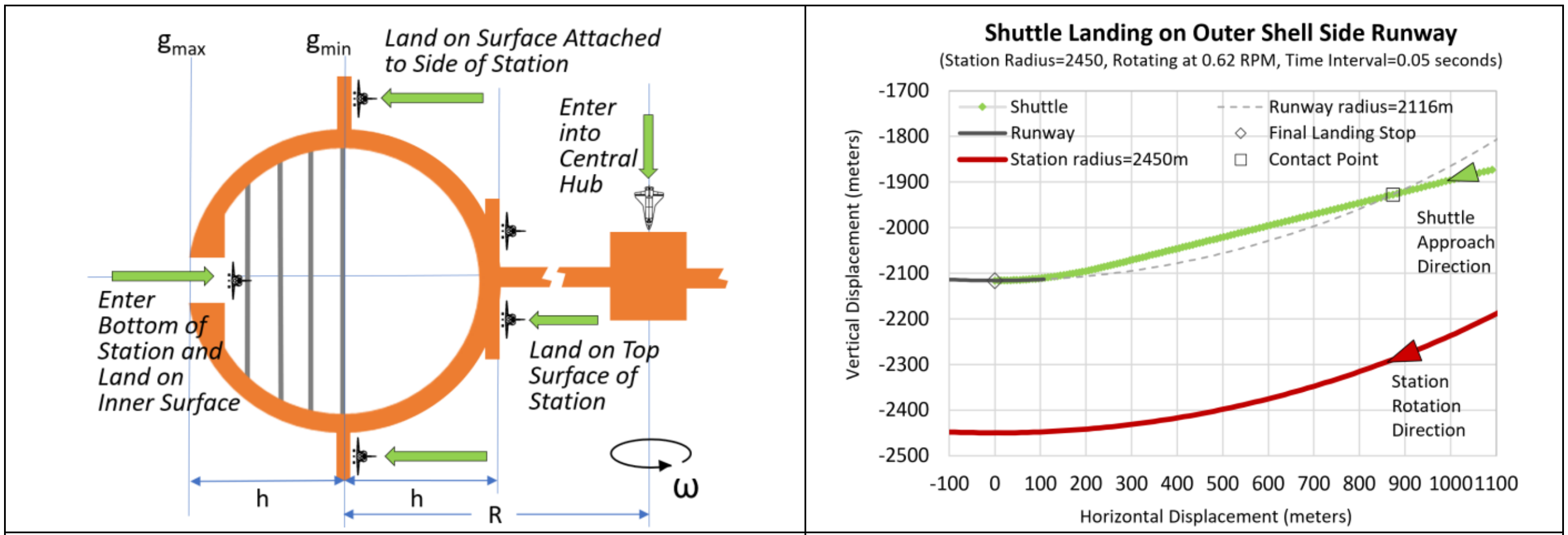

| a) Cross section of torus (or dumbbell) | b) Simulation result of shuttle landing |
|---|---|

**Figure 6-2 – Runways on rotating station**

We design the approach and opening locations so the shuttle enters at the center of the moving opening. The shuttle travels about 100 meters at 16.5 degrees through the opening. The opening rotates about 80 meters during that time. The opening is angled in the spinward direction. The entrance is designed so the shuttle exits at the center of the opening as it passes through the 20-meter-thick shell.

We include the approach in coordinates relative to the rotating landing point in Figure 6-4b. Relative to the landing spot, the opening is stationary. Figure 6-4b illustrates the same approach from the perspective of an observer near the landing point. They would see the shuttle enter, ascend upwards, and then descend to the landing spot. The distance to the runway edge and the maximum ascent height are clearly visible.

We designed the approach to land 100 meters from the opening edge. The shuttle approach appears to pass quite close to the opening edges in Figure 6-4a. This is misleading because the opening is rotating with the station. The same approach in the coordinate system of Figure 6-4b shows the shuttle passing about 40 meters from the edges of the opening.

This is one example of landing on the bottom runway. The outer rim could have multiple openings and support several simultaneous landings. The angle of the approach, the opening size, the opening distance from the landing point, and the height of the ascent can all be designed and be controlled. Even with smaller and larger radius stations, the same type of approach is possible. Simulation on multiple landing approaches for various station radii gives us confidence that this outer shell landing concept works. We recognize that additional study on the structural requirements of these runways and landings is necessary.

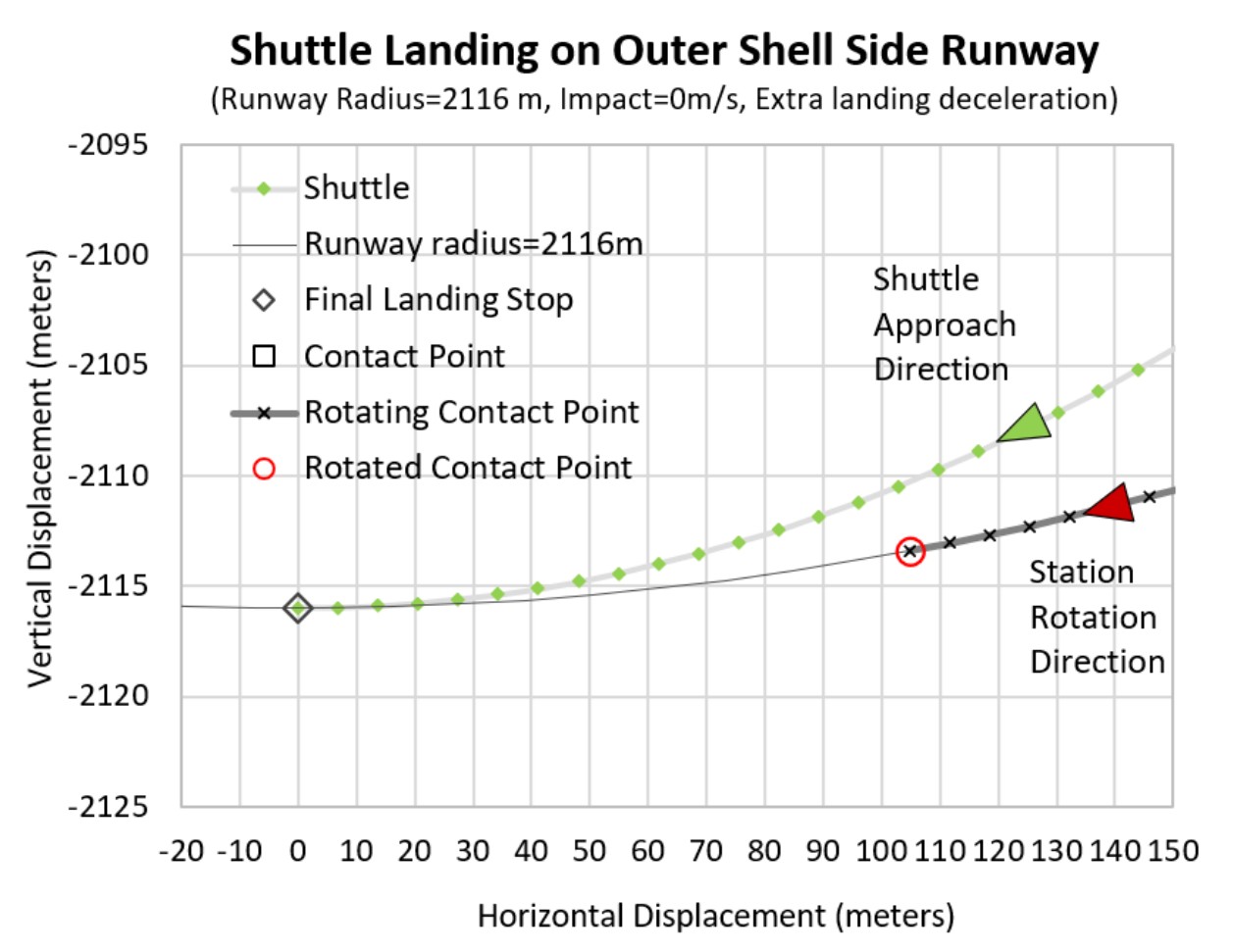

**Figure 6-3 – Final Landing Approach on Side Runway**

## 6.3 Early Colonists

The restructuring process only creates an enclosed rotating framework. The result is an environment that provides radiation protection and gravity for early crews and colonists. The process (currently) does not provide an atmosphere, light, heating, or cooling. The restructuring process produces regions of our example station ready for early colonists after 8 or 9 years. The Gantt chart in Figure 4-9 shows a crew of 12 arriving after 6 years. Portions of the station framework are complete by this time. Those areas are shielded from cosmic radiation and are beginning to rotate to provide some gravity. Those areas have no atmosphere, minimal light, and no heat or cooling. The job of this initial crew is to make those areas livable. The panels of the station shell are welded

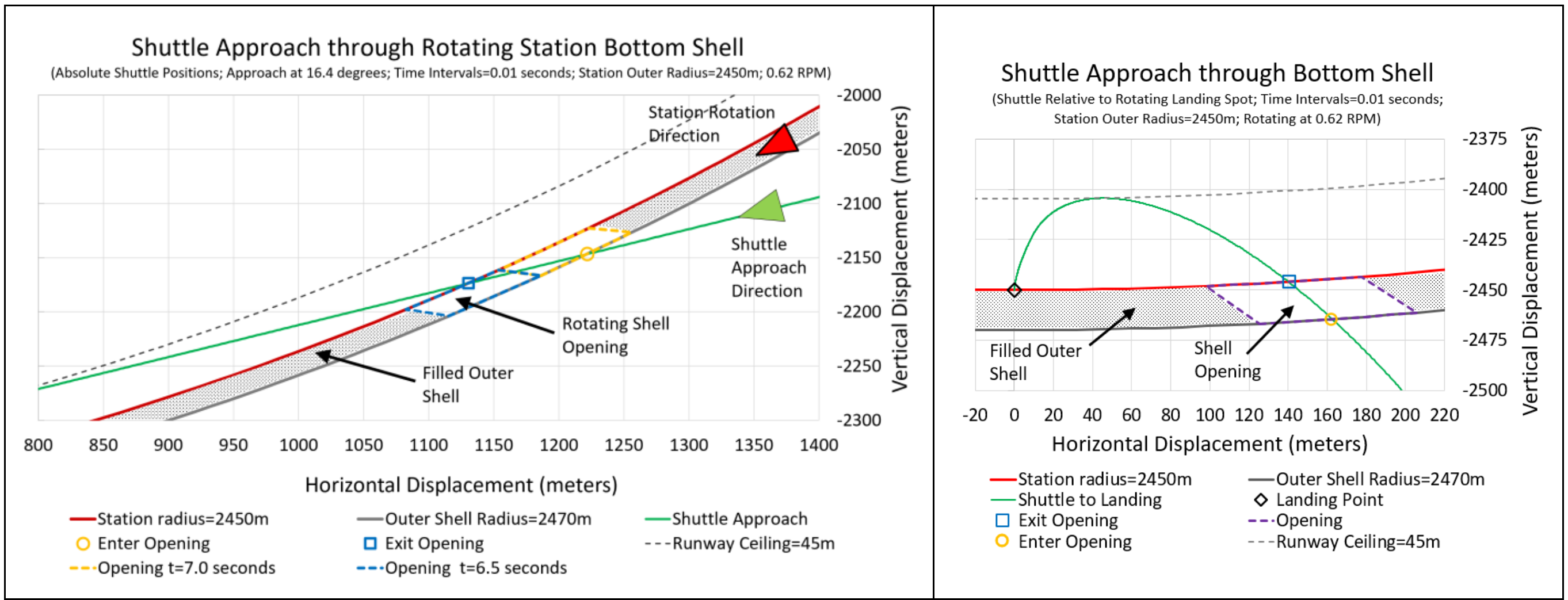

*a) Shuttle Passing Through Opening in Rotating Outer Shell* | *b) Relative Shuttle Position to Landing Spot*

**Figure 6-4 – Shuttle movement through bottom runway opening in rotating station**

together. The floors and walls are sealed with panels. Ideally this would provide an airtight seal but the crew may need to further seal the shell with an airtight coating. There will be conduit, lighting, and door openings to space in the framework. The crew must seal those openings with doors, chevrons, pipes, and light fibers. Producing atmosphere will be next on the agenda for this early crew. The restructuring process produces and inventories excess volatiles and metals. The stored frozen volatiles have been tested and inventoried and summaries sent to Earth. The initial crew will have a plan on which volatiles to retrieve first and where to place for sublimation. Spiders, movers, and trucks will be assigned to move the volatiles. Inside the station, the restructuring equipment will require a new source of power or access to sunlight. To melt the frozen inventory, heat will be needed.

Using the inventoried material available on the station, an early crew will build a Heating, Ventilation, and Air Conditioning (HVAC) system. Boiler systems could use parabolic mirrors mounted on the exterior of the station to heat liquids and transfer heat through pipes to the interior of the station. The spiders and movers could move and mount those mirrors to appropriate locations. With the right materials and new equipment brought by the early colonists, solar cells could be manufactured. These cells would produce electricity to drive lights, the HVAC system, and new equipment.

The early colonists will need to bring additional equipment and use the inventoried materials to spawn multiple industries. These industries are essential to make the station livable and may provide early exports from the station. These industries could include:

- Solar Cells
- Pipes and Wires
- Agriculture
- Heat/Boilers
- Fuel
- Metal

There will be advances in artificial intelligence and robotics while the first parts of the station are being restructured. The spiders will be reprogrammed occasionally during the restructuring process to take advantage of these advances. The Earth-bound support groups will develop new equipment and tools to take advantage of materials found during the early restructuring process. Early progress and discoveries during the restructuring process may motivate earlier support and retrieval missions. Perhaps we would see a migration like the 1849 California gold rush.

## 6.4 Atira Moon

Many asteroids have moons. We offer a rendering in Figure 6-5 of the Atira asteroid moon from the surface of Atira. A moon offers several future options for the colonists. This moon represents another 0.52 cubic kilometers or 1.42 trillion kilograms of material. Its proximity to the station (6 kilometers) and microgravity makes it a potential lab for station researchers. High risk industries could be stationed on the moon instead of in the space station habitat.

Spiders, trucks, and other equipment could be moved to the moon to begin a restructuring process on it. Building frameworks for labs on the surface of the Atira moon is an obvious extension to the Atira restructuring process. It might be valuable to spin up the lab created on the moon to provide gravity. In this paper, we showed that a one-kilometer diameter asteroid (the size of the Atira moon) has enough material to build a small station. Such a station could remain in orbit around Atira or moved further away as another colony or to provide an additional safety margin for high-risk industry.

A similar thought for the moon would be using it as a seed vessel. There will be a surplus of spiders and equipment after the Atira restructuring effort is complete. Some spiders and equipment will be needed to maintain the Atira station and to support the early crews and colonists. The surplus equipment and spiders could be moved to the moon and augmented with additional circuit boards and supplies. Rotary Pellet Launchers or fuel-based rockets could begin to move the moon out of orbit towards another asteroid for

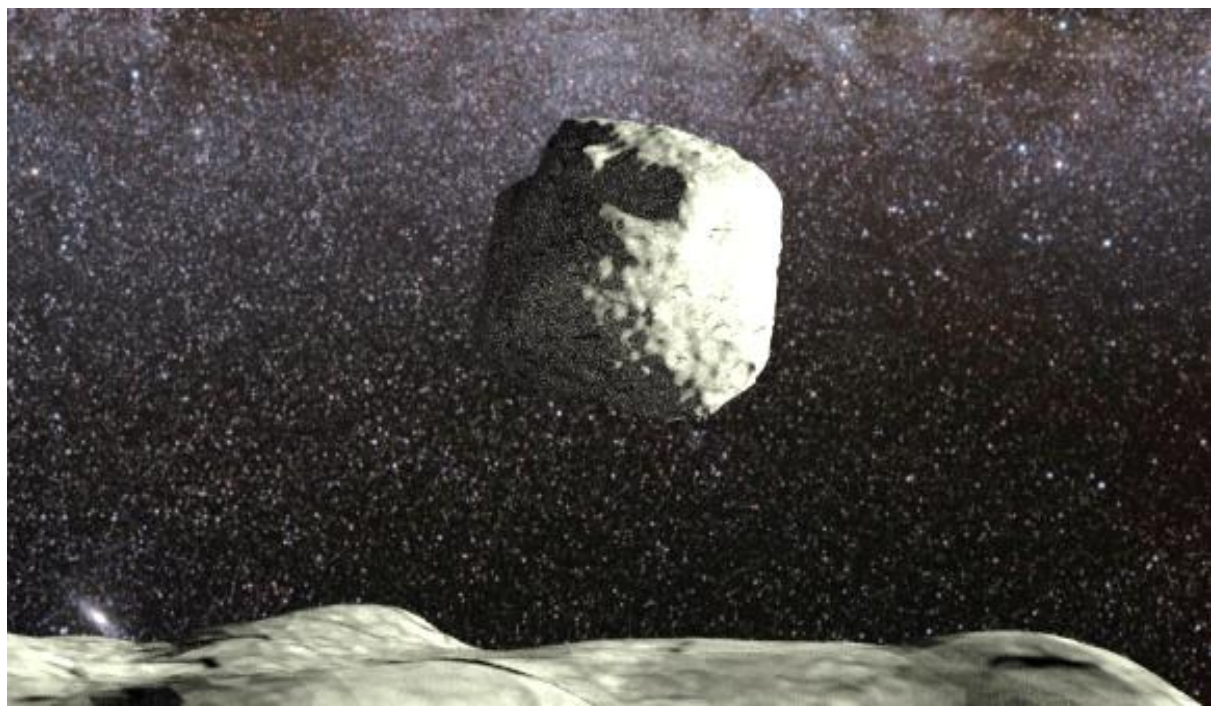

*Credits: Self produced with Blender using Milky Way Image: ESO/Brunier [Brunier 2009] [CC BY-4.0]; and Asteroid Model [Ellison 2018] [CC BY-4.0] modified/rescaled to match asteroid and moon dimensions*

**Figure 6-5 – Moonrise on Atira Asteroid**

restructuring. The RAMA project developed a similar idea [Dunn 2016]. The restructuring process would begin on the moon enroute to the new asteroid. Upon arrival at the new asteroid, a multitude of equipment would move to the new asteroid and begin its restructuring process.

Colonists could leave the moon alone. The view of the moonrise would be spectacular; see Figure 6-5. Observation decks on the exterior station wall could be fitted with electrochromic glass and electromagnetic shielding to reduce heat and radiation. With a lunar orbit of 15.5 hours and the station rotating once every two minutes, spectacular views of the moon, sun, and stars would be enjoyed frequently during a visit to an observation deck. One can envision these locations as centerpieces for restaurants and other tourist activities.

## 6.5 Restructuring Future

We have focused on using the restructuring process to convert an asteroid into a space station. If this process is viable, other asteroids could be restructured. The concept of using advanced technology robotic spiders (replicators) to build and manage many simple mechanical automata (helpers) can be used elsewhere. Reducing the risk to personnel and eliminating the expense of delivering equipment has ubiquitous value. We plan to apply and study these concepts for use in lunar, Titan, and Martian environments. There may also be opportunity in using this restructuring process on the Earth in less hospitable locations such as the Sahara, the Australian outback, and the Antarctic. In the 1970s Freeman Dyson described autonomous equipment to perform similar projects [Dyson 1979] [Freitas and Merkle 2004]. His projects included working on the Saturn moon Enceladus, Deserts, Terrestrial Industry, and Water projects. Our restructuring concepts can enable these projects too.

## 6.6 Conclusions

The asteroids of the solar system have sufficient material to build enough space stations to house more than the entire world’s population. There are many large near-Earth asteroids to restructure and each could house one million people. Researchers have been designing these stations for over 50 years. Two major obstacles have prevented the development of these stations. First, the high cost of launching probes and material into space. Second, the detrimental impact on people from low-gravity and radiation in space.

We introduce in this paper the restructuring process where a single modest-size probe lands on an asteroid and autonomously creates an enclosed space station framework. The probe has a small number of robots that eventually create thousands of robots, tools, and equipment. The restructuring process improves the productivity using self-replication parallelism and tool specialization. The probe would land on a large asteroid and potentially would take over a decade to convert the regolith into basalt rods, tiles, trusses, panels, and ultimately a complete space station. Metals and volatiles are found during the regolith processing. These valuable commodities are tested, inventoried, and stored for future use.

At the conclusion of the restructuring process, an enclosed space station framework is rotating in space and ready for crews and then colonists. The rotating shell provides near-Earth-like centripetal gravity. The station has many floors and provides space for a large population. A thick shell provides protection from radiation and space debris. The single launch and probe costs are small compared to the value of this real estate. The restructuring asteroid process directly addresses two major obstacles preventing the construction of space stations.

We are at the stage where it appears that the restructuring process is viable. Of course, we expect addition problems to be identified during reviews. We also expect that experts and future teams will be able to solve those problems and improve on the work already done. The restructuring process offers humanity the opportunity to truly become a space faring society.

# 7 Asteroid Restructuring – References

[Ackerman 2016] Project RAMA: Turning Asteroids Into Catapult-Powered Analog Spacecraft, Evan Ackerman, 20 Sep 2016, IEEE Spectrum, https://spectrum.ieee.org/tech-talk/aerospace/space-flight/project-rama-turning-asteroids-into-catapultpowered-analog-spacecraft

[ADA 2010] 2010 ADA Standards for Accessible Design, Department of Justice, 15 September 2010, Americans with Disabilities Act, https://www.ada.gov/regs2010/2010ADAStandards/2010ADAstandards.htm

[Albus 1997] 4-D/RCS Version 1.0: A Reference Model Architecture for Demo III, James S. Albus, 27 June 1997, NIST Interagency/Internal Report (NISTIR), National Institute of Standards and Technology, Gaithersburg, MD, https://tsapps.nist.gov/publication/get_pdf.cfm?pub_id=823355

[Angelo 2014] Encyclopedia of Space and Astronomy, Joseph A. Angelo, 14 May 2014, Infobase Publishing, Science, https://books.google.com/books/about/Encyclopedia_of_Space_and_Astronomy.html?id=VUWno1sOwnUC

[Artemis 2022] Artemis: Why we are Going to the Moon, National Aeronautics and Space Administration, NASA Official: Brian Dunbar, Accessed 11 November 2022, https://www.nasa.gov/specials/artemis/

[Barr 2018] The High Life? On the Psychological Impacts of Highrise Living, Jason M. Barr, 31 January 2018, https://buildingtheskyline.org/highrise-living/

[Bell and Hines 2012] Space Structures and Support Systems, Larry Bell and Gerald D. Hines, 2012, MS-Space Architecture, SICSA Space Architecture Seminar Lecture Series, Sasakawa International Center for Space Architecture (SICSA), Cullen College of Engineering, University of Houston, http://sicsa.egr.uh.edu/sites/sicsa/files/files/lectures/space-structures-and-support-systems.pdf

[Bender et al. 1979] Round-Trip Missions to Low- Delta-V Asteroids and Implications for Material Retrieval, David F. Bender, R. Scott Dunbar, and David J. Ross, 1979, NASA SP-428 Space Resources and Space Settlements, pp. 161-172, [O’Neill et al. 1979], https://ntrs.nasa.gov/api/citations/19790024054/downloads/19790024054.pdf

[Blacic 1985] Structural properties of lunar rock materials under anhydrous, hard vacuum conditions, J. C. Blacic, 1985, Papers Presented to the Symposium on Lunar Bases and Space Activities of the 21st Century, p. 487. NASA/Johnson Space Center, Houston., http://articles.adsabs.harvard.edu/cgi-bin/nph-iarticle_query?bibcode=1985lbsa.conf..487B

[Bock, Lambrou, and Simon 1979] Effect of Environmental Parameters on Habitat Structural Weight and Cost, Edward Bock, Fred Lambrou, Jr., and Michael Simon, 1979, NASA SP-428 Space Resources and Space Settlements, pp. 33-60, [O'Neill et al. 1979], https://ntrs.nasa.gov/api/citations/19790024054/downloads/19790024054.pdf

[Boyd 1987] Discourse on Winning and Losing, J.R. Boyd, August 1987, Air University Press, Curtis E. LeMay Center for Doctrine Development and Education, https://www.airuniversity.af.edu/Portals/10/AUPress/Books/B_0151_Boyd_Discourse_Winning_Losing.pdf

[Boyle 2020] Can a Moon Base Be Safe for Astronauts?, Rebecca Boyle, 22 October 2020, https://www.scientificamerican.com/article/can-a-moon-base-be-safe-for-astronauts/

[Bradley et al. 2012] Optimized Free Return Trajectories to Near-Earth Asteroids via Lunar Flyby, Nicholas Bradley, Sonia Hernandez, and Ryan P Russell. Published In Proceedings of the AAS/AIAA Astrodynamics Specialist Conference, pages 3353–3369, 2013, CiteSeerX, http://citeseerx.ist.psu.edu/viewdoc/download?doi=10.1.1.712.1087&rep=rep1&type=pdf

[Brody 2013] Not 'Elysium,' But Better 'Ringworld' Settlements Could Return Our Future to Its Past (Commentary), David Sky Brody, 10 August 2013, https://www.space.com/22326-elysium-movie-space-colonies-future.html

[Brooks 1985] A Robust Layered Control System for a Mobile Robot. Rodney A. Brooks, September 1985, MIT Artificial Intelligence Laboratory, A.I. Memo 864, https://people.csail.mit.edu/brooks/papers/AIM-864.pdf

[Brown 2002] Elements of Spacecraft Design, Charles D. Brown, AIAA Education Series, Reston, VA, 2002. https://www.worldcat.org/title/elements-of-spacecraft-design/oclc/850628450&referer=brief_results

[Brunier 2009] The Milky Way Panorama, Serge Brunier and European Southern Observatory, 14 September 2009, High-resolution images featured in the GigaGalaxy Zoom project, 360 Degree Photograph of the Milky way, Credit: ESO/Serge Brunier, [CC BY-4.0], https://www.eso.org/public/images/eso0932a/

[Carsley, Blacic, and Pletka 1992] Vacuum Melting and Mechanical Testing of Simulated Lunar Glasses, J. E. Carsley, J. D. Blacic, and B. J. Pletka, 31 June - 4 June 1992, Engineering, Construction, and Operations in Space III, Vol 2, pp.1219 1231, Eds: Willy Z. Sadeh, Stein Sture, and Russell J. Miller, American Society of Civil Engineers, https://archive.org/details/engineeringconst0000spac

[Copeland 2000] The Modern History of Computing (Stanford Encyclopedia of Philosophy), B. Jack Copeland, 18 December 2000, Stanford Encyclopedia of Philosophy, Retrieved 12 April 2018, https://plato.stanford.edu/entries/computing-history/

[Coriolis 1835] Sur les équations du mouvement relatif des systèmes de corps, (On the Equations of the Relative Movement of Systems of Bodies), Gaspard-Gustave de Coriolis, 1835, Journal de l'École Royale Polytechnique. 15: 144–154., http://www.bibnum.education.fr/physique/mecanique/sur-les-equations-du-mouvement-relatif-des-systemes-de-corps

[Cornall 2021] Sophie's Bionutrients develops world's first dairy-free micro-algae based milk alternative, Jim Cornall, 4 May 2021, NutraIngrediants.com, https://www.nutraingredients.com/Article/2021/05/04/Sophie-s-Bionutrients-develops-world-s-first-dairy-free-micro-algae-based-milk-alternative

[DC 2017] Destination DC, 2017 Visitor Statistics Washington, DC, 2017, 2017 Visitor Statistics, https://washington.org/sites/default/files/2021-02/2017%20Washington%2C%20DC%20Visitor%20Statistics.pdf

[Deller 2017] Hyper-Velocity Impacts on Rubble Pile Asteroid, Jakob Deller, 2017, Springer, 10.1007/978-3-319-47985-9, https://www.researchgate.net/publication/312008639_Hyper-Velocity_Impacts_on_Rubble_Pile_Asteroids, https://kar.kent.ac.uk/54352/1/78Deller_2015_PhDThesis_screen.pdf

[Doody 2011] Basics of Space Flight, David Doody, 7 May 2011, Prepared at NASA Jet Propulsion Lab (JPL), Public Domain, https://solarsystem.nasa.gov/basics/

[Drexler 1986] Engines of Creation: The Coming Era of Nanotechnology, K. Eric Drexler, 1986, Anchor Books, Random House, Inc., http://wfmh.org.pl/enginesofcreation/EOC_Chapter_1.html https://www.nanowerk.com/nanotechnology/reports/reportpdf/report47.pdf

[Dunn 2016] How We Want to Turn Asteroids Into Spacecraft, Jason Dunn, May 31, 2016, Made In Space, Newsroom, https://medium.com/made-in-space/how-we-want-to-turn-asteroids-into-spacecraft-e95d3214d787

[Dunn and Fagin 2017] Project RAMA: Reconstituting Asteroids into Mechanical Automata, Jason Dunn and Max Fagin, 2017, NASA Innovative Advanced Concepts, Made In Space, Inc., https://www.nasa.gov/sites/default/files/atoms/files/niac_2016_phasei_dunn_projectrama_tagged.pdf

[Dyson 1960] Search for Artificial Stellar Sources of Infra-Red Radiation, Freemann J. Dyson, 3 June 1960, Science. 131 (3414): pp. 1667-1668, Bibcode:1960Sci...131.1667D. doi:10.1126/science.131.3414, https://fermatslibrary.com/s/search-for-artificial-stellar-sources-of-infrared-radiation

[Dyson 1970] The twenty-first century, Vanuxem Lecture delivered at Princeton University, Freeman J. Dyson, 26 February 1970, Revised and reprinted as: Chapter 18 Thought Experiments [Dyson 1979], http://www.molecularassembler.com/KSRM/3.6.htm

[Dyson 1979] Disturbing the Universe, Freeman J. Dyson, 1 January 1979, Chapter 18, Thought Experiments, Harper and Row Publishers, New York, 1979, pp. 194-204., Extracts from [Freitas and Merkle 2004]. http://www.molecularassembler.com/KSRM/3.6.htm

[Eisele 2001] Concentration of Useful Minerals from Asteroids, Timothy C. Eisele Originally published 2001, Journal of the British Interplanetary Society, Vol. 54, Part 7/8, pp. 277-288, [CC BY-4.0], https://www.researchgate.net/publication/260976297_Concentration_of_Useful_Minerals_from_Asteroids

[Ellison 2018] Asteroid Ryugu (July 11th, 2018), Doug Ellison, 3D Model, Sketchfab, Data Credit - JAXA, University of Tokyo, Kochi University, Rikkyo University, Nagoya University, Chiba Institute of Technology, Meiji University, University of Aizu, AIST., https://sketchfab.com/3d-models/asteroid-ryugu-july-11th-2018-44876e2f0d314b05ba32b0472a1eddc6 [CC BY-4.0]

[Enos 2020] Enos - Bennu Pyramid, Heather Enos, 19 October 2020, OSIRIS-REx Science and Engineering Briefing, NASA/Goddard Space Flight Center, University of Arizona, https://svs.gsfc.nasa.gov/13738

[Fitzpatrick 2011] Newtonian Dynamics, Richard Fitzpatrick, 2011 March 31, Rigid Body Rotation, Rotational Stability, The University of Texas at Austin, https://farside.ph.utexas.edu/teaching/336k/Newton/node71.html

[Fitzpatrick 2023] Dumbbell Space Stability, Richard Fitzpatrick, 2023 June 28, Personal Correspondence with Professor University of Texas

[Freitas and Gilbreath 1982] Advanced Automation for Space Missions: Proceedings of the 1980 NASA/ASEE Summer Study held at the University of Santa Clara, Robert A. Freitas, Jr., and William P. Gilbreath, June 23-August 29, 1980; NASA Conference Publication 2255, 15 May 1981, NASA Ames Research Center, Moffett Field, Publication Date: 1 November 1982, https://ntrs.nasa.gov/archive/nasa/casi.ntrs.nasa.gov/19830007077.pdf

[Freitas and Merkle 2004] Kinematic Self-Replicating Machines, Robert A. Freitas Jr. and Ralph C. Merkle, 2004, Landes Bioscience, Georgetown, TX, https://epdf.pub/queue/kinematic-self-replicating-machines90c44fd3e88ba39ff48389263e3f55434462.html

[Fritz and Turkoglu 2017] Optimal Trajectory Determination and Mission Design for Asteroid/Deep-Space Exploration via Multibody Gravity Assist Maneuvers, Sean Fritz and Kamran Turkoglu, 2017, International Journal of Aerospace Engineering, Volume 2017, Article ID 6801023, 12 pages, http://downloads.hindawi.com/journals/ijae/2017/6801023.pdf

[Fu et al. 2016] How to establish a bioregenerative life support system for long-term crewed missions to the Moon and Mars, Fu, Y. L. Li, B. Xie, C. Dong, M. Wang, B. Jia, L. Shao, Y. Dong, S. Deng, H. Liu, G. Liu, B. Liu, D. Hu, and H. Liu. 2016., Astrobiology 16(12) DOI:10.1089/ast.2016.1477, https://www.researchgate.net/publication/311362350_How_to_Establish_a_Bioregenerative_Life_Support_System_for_Long-Term_Crewed_Missions_to_the_Moon_or_Mars

[García et al. 2016] Final Report: Starport, Enrique García and 37 individuals on the Starport 1 team, 2016, Space Studies Program, International Space University, https://isulibrary.isunet.edu/doc_num.php?explnum_id=1121

[Garrett 2019] A cinema, a pool, a bar: inside the post-apocalyptic underground future, Bradley L Garrett, 16 December 2019, The Guardian, https://www.theguardian.com/cities/2019/dec/16/a-cinema-a-pool-a-bar-inside-the-post-apocalyptic-underground-future

[Globus 1991] The Design and Visualization of a Space Biosphere, Al Globus, May 1991, Space Manufacturing 8, Energy and Materials from Space, Space Studies Institute, Princeton, NJ, pages 303-313, https://www.nas.nasa.gov/assets/nas/pdf/techreports/1991/rnr-91-018.pdf

[Globus 2006] Contest-Driven Development of Orbital Tourist Vehicles, Al Globus, American Institute of Aeronautics and Astronautics Space 2006, San Jose, California, 19-21 September 2006. https://pdfs.semanticscholar.org/4ca2/10f824b5c00c02b6a3f7ee8cef572f5bf649.pdf

[Globus and Hall 2017] Space Settlement Population Rotation Tolerance, Al Globus and Theodore Hall, June 2017, NSS Space Settlement Journal, https://space.nss.org/wp-content/uploads/NSS-JOURNAL-Space-Settlement-Population-Rotation-Tolerance.pdf

[Globus et al. 2007] The Kalpana One Orbital Space Settlement Revised, Al Globus, Nitin Arora, Ankur Bajoria, and Joe Strout,2007, American Institute of Aeronautics and Astronautics, http://alglobus.net/NASAwork/papers/2007KalpanaOne.pdf

[Good 2017] A Clockwork Rover for Venus, Andrew Good, 25 August 2017, NASA, Jet Propulsion Laboratory, California Institute of Technology, News, Technology, https://www.jpl.nasa.gov/news/a-clockwork-rover-for-venus

[Hall 1991] The Architecture of Artificial Gravity: Mathematical Musings on Designing for Life and Motion in a Centripetally Accelerated Environment, Theodore W. Hall, November 1991, Space Manufacturing 8 Energy and Materials from Space, Proceedings of the Tenth Princeton/AIAA/SSI Conference, 15-18 May 1991, https://citeseerx.ist.psu.edu/viewdoc/download?doi=10.1.1.551.3694&rep=rep1&type=pdf

[Hall 1993] The Architecture of Artificial Gravity: Archetypes and Transformations of Terrestrial Design, Theodore W. Hall, 12-15 May 1993, Space Manufacturing 9, The High Frontier Accession, Development and Utilization, Proceedings of the Eleventh SSI-Princeton Conference, p. 198-209, http://www.artificial-gravity.com/SSI-1993-Hall.pdf

[Hall 1997] Artificial Gravity and the Architecture of Orbital Habitats, Theodore W. Hall, 20 March 1997, Proceedings of 1st International Symposium on Space

Tourism, Daimler-Chrysler Aerospace GmbH., http://www.spacefuture.com/archive/artificial_gravity_and_the_architecture_of_orbital_habitats.shtml

[Hall 1999] Architectural considerations for self-replicating manufacturing systems. J. Storrs Hall, 1 September 1999, Nanotechnology 10:323, c, http://www.autogeny.org/selfrepsys/

[Hall 2006] Artificial Gravity Visualization, Empathy, and Design, Theodore Hall, 19-21 September 2006, AIAA 2006-7321, Session: SAS-3: Space Architecture Symposium: Gravity Regime Architecture and Construction, https://doi.org/10.2514/6.2006-7321, http://www.artificial-gravity.com/AIAA-2006-7321.pdf

[Hand and Finch 1998] Analytical Mechanics, Louis N. Hand and Janet D. Finch, 1998, Cambridge University Press. p. 267. ISBN 978-0-521-57572-0, https://www.iaa.csic.es/~dani/ebooks/Mechanics/Analytical%20mechanics%20-%20Hand,%20Finch.pdf

[Haskin 1992] Glass and Ceramics, Larry A. Haskin, 1992, NASA SP-509, Volume 3, Space Resources: Materials, pp.291-294, [McKay et al. 1992-v3], https://ntrs.nasa.gov/citations/19930007686

[Honeybee 2019] The World Is Not Enough Demonstrates the Future of Space Exploration, Honeybee Robotics, Inc., January 15, 2019, https://honeybeerobotics.com/wine-the-world-is-not-enough/, Wayback Machine Access: https://web.archive.org/web/20190215050619/https://honeybeerobotics.com/wine-the-world-is-not-enough/

[ICC 2009] International Building Code, International Code Council, February 2009, http://www.co.washington.ne.us/media/ICC-International_Building_Code_2009.pdf

[Janhunen 2018] Natural illumination solution for rotating space settlements, Pekka Janhunen, 14 September 2018, NSS Space Settlement Journal, Issue 4, https://arxiv.org/pdf/1806.09808.pdf

[JAXA MHU 2019]. The JAXA Mouse Habitat Unit (MHU), Japan Aerospace Exploration Agency (JAXA), Last Updated: July 26, 2019, https://iss.jaxa.jp/en/kiboexp/pm/mhu/

[Jensen 1993] Disk I/O in High Performance Computing System, David W. Jensen, 1993, University of Illinois at Champaign Urbana, Ph.D. Thesis, http://hdl.handle.net/2142/72078

[Jensen 1996] Observations Regarding Threat Command Agents, David W. Jensen, 31 July 1996, Ed: Major David Payne, Command Decision Modeling Technology Assessment, Compiled by The Us Army Artificial Intelligence Center for the National Simulation Center, https://apps.dtic.mil/docs/citations/ADA334926

[Jensen 2008] Developing System-on-Chips with Moore, Amdahl, Pareto, and Ohm, David W. Jensen, 2008, Proc. Int'l Conf. IEEE Electro/Information Technology, IEEE Press, 2008, pp. 13-18, https://ieeexplore.ieee.org/document/4554260

[Johnson and Holbrow 1977] NASA SP-413: Space Settlements: A Design Study, Eds: Richard D. Johnson and Charles Holbrow, 1977, Scientific and Technical Information Office, https://ntrs.nasa.gov/citations/19770014162

[JPL SB Mission Design Tool] JPL Small-Body Mission-Design Tool, NASA Jet Propulsion Laboratory, Solar System Dynamics, https://ssd.jpl.nasa.gov/?mdesign_server

[JPL SBD Search Engine] JPL Small-Body Database Search Engine, NASA Jet Propulsion Laboratory, Solar System Dynamics, https://ssd.jpl.nasa.gov/sbdb_query.cgi

[JPL SBD Small Body Lookup] JPL Small-Body Database Lookup, NASA Jet Propulsion Laboratory, Solar System Dynamics, https://ssd.jpl.nasa.gov/tools/sbdb_lookup.html#/

[Kayser 2011] Solar Sinter, Markus Kayser, 2011, Kayser Works, Royal College of Art, London, https://kayserworks.com/798817030644

[Keeter 2020] Long-Term Challenges to Human Space Exploration, Bill Keeter, 4 September 2020, National Aeronautics and Space Administration, https://www.nasa.gov/centers/hq/library/find/bibliographies/Long-Term_Challenges_to_Human_Space_Exploration

[Kersch 2015] Answer: How many people can you feed per square-kilometer of farmland?, Worldbuilding User: ckersch, 3 February 2015, https://worldbuilding.stackexchange.com/questions/9582/how-many-people-can-you-feed-per-square-kilometer-of-farmland

[Kesson 1975] Mare basalts - Melting experiments and petrogenetic interpretations, S. E. Kesson, Lunar Science Conference, 6th, Houston, Tex., March 17-21, 1975, Proceedings. Volume 1. (A78-46603 21-91) New York, Pergamon Press, Inc., 1975, p. 921-944. SAO/NASA Astrophysics Data System (ADS), https://adsabs.harvard.edu/full/1975LPSC....6..921K

[King 2002] Clockwork Prayer: A Sixteenth-Century Mechanical Monk, Elizabeth King, Spring 2002, Blackbird Archive, Online Journal of Literature and the Arts, https://blackbird.vcu.edu/v1n1/nonfiction/king_e/prayer_introduction.htm

[Koelle and Williams 1959] Project Horizon: Volume II, Technical Considerations & Plans, H. H. Koelle and F. L. Williams, 9 June 1959, A U. S. Army Study For The Establishment of a Lunar Outpost, United States Army, https://history.army.mil/faq/horizon/Horizon_V2.pdf

[Lackner and Wendt 1995] Exponential growth of large self-reproducing machine systems, Klaus S. Lackner and Christopher H. Wendt, May 1995, Mathematical and Computer Modelling, Volume 21, Issue 10, Pages 55-81, https://doi.org/10.1016/0895-7177(95)00071-9, https://www.sciencedirect.com/science/article/pii/0895717795000719

[Lesser and Corkill 2014] Comprehensive History of the Multi-Agent Systems Lab at the University of Massachusetts Amherst, 1978-2014, Victor Lesser and Daniel Corkill, https://mas.cs.umass.edu/Documents/LabHistory_Web-Article.pdf

[Lewis-Weber 2016] Lunar-Based Self-Replicating Solar Factory, Justin Lewis-Weber, 2016, New Space, vol. 4, issue 1, pp. 53-62, https://www.researchgate.net/publication/297746685_Lunar-Based_Self-Replicating_Solar_Factory

[Liu et al. 2020] Computing Systems for Autonomous Driving: State-of-the-Art and Challenges, Liangkai Liu, Sidi Lu, Ren Zhong, Baofu Wu, Yongtao Yao, Qingyang Zhang, and Weisong Shi, 7 December 2020, Accepted to IEEE Internet of Things Journal, https://arxiv.org/abs/2009.14349

[Lucas 2019] What Are Centrifugal & Centripetal Forces?, Jim Lucas, 10 May 2019, Live Science Contributor, https://www.livescience.com/52488-centrifugal-centripetal-forces.html

[Maynard and Sevier 1966] General Mission Summary and Configuration Description, Owen E. Maynard and John R. Sevier, 25 June 1966, NASA Technical Memorandum NASA TM X-58006, Apollo Lunar Landing Mission Symposium, p.49, Fig. 2, https://www.hq.nasa.gov/alsj/LunarLandingMIssionSymposium1966_1978075303.pdf

[Mazanek et al. 2014] Asteroid Redirect Mission Concept: A Bold Approach for Utilizing Space Resources, Daniel D. Mazanek, Raymond G. Merrill, John R. Brophy, and Robert P. Mueller, 29 September 2014, 65th International Astronautical Congress, Toronto, Canada, IAC-14-D3.1.8, https://www.nasa.gov/sites/default/files/files/IAC-14-D3-Mazanek.pdf

[McKay et al. 1992-v3] NASA SP-509 Volume 3: Space Resources, Materials, 1992, Eds: Mary Fae McKay, David S. McKay, and Michael B. Duke, Scientific and Technical Information Office, https://ntrs.nasa.gov/citations/19930007686

[McKendree 1995] Implications of Molecular Nanotechnology Technical Performance Parameters on Previously Defined Space System Architectures, Thomas Lawrence McKendree, 9-11 November 1995, The Fourth Foresight Conference on Molecular Nanotechnology, Palo Alto, California, http://www.zyvex.com/nanotech/nano4/mckendreePaper.html

[Metzger 2015] Resources for 3D Additive Construction on the Moon, Asteroids and Mars, Philip Metzger, 24 August 2015. University of Central Florida, Keck Institute for Space Studies, https://kiss.caltech.edu/workshops/3d/presentations/Metzger.pdf

[Metzger et al. 2012] Affordable, rapid bootstrapping of space industry and solar system civilization, Philip T. Metzger, Anthony Muscatello, Robert P. Mueller, and James Mantovani, January 2013, Journal of Aerospace Engineering, Vol 26, Issue 1, pp. 18-29, DOI: 10.1061/(ASCE)AS.1943-5525.0000236, https://arxiv.org/ftp/arxiv/papers/1612/1612.03238.pdf

[Misra 2010] The "Tesla" Orbital Space Settlement, Gaurav Misra, 2010, Barcelona, Spain, 40th International Conference on Environmental Systems, American Institute of Aeronautics and Astronautics, Inc, AIAA 2010-6133, https://doi.org/10.2514/6.2010-6133, http://spacearchitect.org/pubs/AIAA-2010-6133.pdf

[Moore 1962] Machine Models of Self- Reproduction, Edward F. Moore, 1962 Reprinted 1965, Mathematical Problems in the Biological Sciences, Proceedings of Symposia in Applied Mathematics, Volume 14, 1962, http://is.ifmo.ru/autograph/moore-article.pdf

[Mordanicus 2014] Space Colonization: The Dumbbell Design, Mordanicus, 6 December 2014, Lagrangian Republican Association, https://republicoflagrangia.org/2014/12/06/the-dumbbell-design/

[Moses et al. 2014] An architecture for universal construction via modular robotic components, Matthew S. Moses, Hans Ma, Kevin C. Wolfe, Gregory S. Chirikjian, 2014, Robots and Autonomous Systems, 62, pp. 945–965, https://rpk.lcsr.jhu.edu/wp-content/uploads/2014/08/Moses13_An-Architecture.pdf

[Mueller 2017] Construction with Regolith, The Technology and Future of In-Situ Resource Utilization (ISRU), Robert P. Mueller, 6 March 2017, A Capstone Graduate Seminar, Orlando, FL March 6, 2017, http://ntrs.nasa.gov/archive/nasa/casi.ntrs.nasa.gov/20170002067.pdf

[Mueller et al. 2009]. Lightweight Bulldozer Attachment for Construction and Excavation on the Lunar Surface, Robert P. Mueller, Allen R. Wilkinson, Christopher A. Gallo, Andrew J. Nick, Jason M. Schuler, and Robert H. King, 2009, Proc., AIAA Space 2009 Conference, Reston, VA. https://ntrs.nasa.gov/archive/nasa/casi.ntrs.nasa.gov/20130012987.pdf

[Mueller et al. 2013] Regolith Advanced Surface Systems Operations Robot (RASSOR), Robert P. Mueller, Rachel E. Cox, Tom Ebert, Jonathan D. Smith, Jason M. Schuler, and Andrew J. Nick, March 2013, 2013 IEEE Aerospace Conference, 2-9 March 2013, https://ntrs.nasa.gov/archive/nasa/casi.ntrs.nasa.gov/20130008972.pdf

[Myers 2022] CubeSat: The little satellite that could, Andrews Myers, 8 June 2022, Stanford Engineering, Technology and Society, https://engineering.stanford.edu/magazine/cubesat-little-satellite-could

[NASA 2004] Mars Mice, NASA, 20 January 2004, NASA Science – Share the Science, science.nasa.gov, https://science.nasa.gov/science-news/science-at-nasa/2004/20jan_marsmice/

[Nau et al. 2003] SHOP2: An HTN Planning Environment., Dana Nau, T. Au, O. Ilghami, Ugur Kuter, J. William Murdock, D. Wu, and Fusun Yaman, 2003, Journal of Artificial Intelligence Research (JAIR), https://arxiv.org/pdf/1106.4869.pdf

[Nishioka et al. 1973] Feasibility of mining lunar resources for earth use: Circa 2000 AD. Volume 2: Technical discussion, K. Nishioka, R D. Arno, A. D. Alexander, and R. E. Slye, 1 August 1973, NASA Ames Research Center Moffett Field, CA, United States, NASA TM Document ID: 19730022067, https://ntrs.nasa.gov/api/citations/19730022067/downloads/19730022067.pdf?attachment=true

[Noordung 1929] The Problem of Space Travel - The Rocket Motor, Herman Potočnik Noordung, Originally: Richard Carl Schmidt & Co., Berlin W62, 1929; Digital edition of English translation, Tanja N. Zhelnina, December 2010, https://history.nasa.gov/SP-4026.pdf

[O'Leary et al. 1979] Retrieval of Asteroidal Materials, Brian O'Leary, Michael Gaffey, David Ross, and Robert Salkeld, 1979, NASA SP-428 Space Resources and Space Settlements, pp.173-189, [O'Neill et al. 1979], https://ntrs.nasa.gov/api/citations/19790024054/downloads/19790024054.pdf

[O'Neill 1974] The Colonization of Space, Gerard K. O'Neill, September 1974, Physics Today. 27 (9): 32-40. Bibcode:1974PhT....27i..32O. doi:10.1063/1.3128863, https://physicstoday.scitation.org/doi/pdf/10.1063/1.3128863

[O'Neill 1976] The High Frontier: Human Colonies in Space, Gerard K. O'Neill, Morrow New York 1976, ISBN-13:978-1896522678; https://archive.org/details/highfrontierhuma0000onei_y9i8/mode/1up

[O'Neill and Driggers 1976] Observable Effects In and Human Adaptation To Rotating Environments, Gerard K. O'Neill and Gerald W. Driggers, Space-Based Manufacturing From Nonterrestrial Materials, p. 173–176. Edited by Gerard K. O'Neill and Brian O'Leary, AIAA, August 1977. Volume 57, Progress in Astronautics and Aeronautics: technical papers derived from the 1976 Summer Study at NASA Ames Research Center, ISBN (print): 978-0-915928-21-7, https://arc.aiaa.org/doi/10.2514/5.9781600865312.0173.0176

[O'Neill et al. 1979] NASA SP-428, Space Resources and Space Settlements, Gerard O'Neill, Eds: John Billingham, William Gilbreath, and Brian O'Leary, 1 January 1979, NASA, Technical papers derived from the 1977 Summer Study at NASA Ames Research Center, Moffett Field, California, https://ntrs.nasa.gov/api/citations/19790024054/downloads/19790024054.pdf

[Oberth 1923] Die Rakete zu den Planetenräumen, Hermann Oberth, 1923, The Rocket into Interplanetary Space, Verlag R. Oldenbourg, München und Berlin, https://vdoc.pub/documents/the-rocket-into-planetary-space-442l1h2fh480

[Poljak, Klindzic and Kruljac 2018] Effects of exoplanetary gravity on human locomotor ability, Nikola Poljak, Dora Klindzic and Mateo Kruljac, Department of Physics, Faculty of Science, University of Zagreb, Croatia, Dated: August 23, 2018, https://arxiv.org/pdf/1808.07417.pdf

[Prado and Fraser 2018] Construction Materials from Minimally Processed Bulk Lunar and Asteroidal Soils, Mark Prado and Sam Fraser, 2018, The Permanent Space Development Foundation, Inc, https://www.permanent.com/space-industry-bulk-construction.html

[Queijo et al. 1988] Some Operational Aspects of a Rotating Advanced-Technology Space Station for the Year 2025, M.J. Queijo, A.J. Butterfield, W.F. Cuddihy, C.B. King, R.W. Stone, J.R. Wrobel, and P.A. Garn, June 1988, The Bionetics Corporation, NASA Contractor Report 181617, Report 19880017013, National Aeronautics and Space Administration, Langley Research Center, https://ntrs.nasa.gov/api/citations/19880017013/downloads/19880017013.pdf

[Rivera-Valentin et al. 2017] Discovery Announcement of Binary System (163693) Atira, E. G. Rivera-Valentin, P. A. Taylor, A. Virkki, and B. Aponte-Hernandez, 20 January 2017, Planetary, Arecibo Observatory, http://outreach.naic.edu/ao/blog/discovery-announcement-binary-system-163693-atira

[Ross 2001] Near-Earth Asteroid Mining, Shane D. Ross, Control and Dynamical Systems Caltech 107-81, Pasadena, CA, December 14, 2001 Space Industry Report, https://space.nss.org/media/Near-Earth-Asteroid-Mining-Ross-2001.pdf

[Ruess, Schaenzlin, and Benaroya 2006] Structural Design of a Lunar Habitat, F. Ruess, J. Schaenzlin, and H. Benaroya, July 2006, Journal of Aerospace Engineering. American Society of Civil Engineers. 19 (3): 138. doi:10.1061/(ASCE)0893-1321(2006)19:3(133), http://coewww.rutgers.edu/~benaroya/publications/Ruess%20et%20al%20ASCE%20JAE.pdf

[Sanders and Larson 2012] Progress made in lunar in situ resource utilization under NASA's exploration technology and development program, G.B. Sanders and W.E. Larson, 2012, Journal of Aerospace Engineering. 26.1, 5-17. https://ntrs.nasa.gov/archive/nasa/casi.ntrs.nasa.gov/20120006109.pdf

[Sauder 2017] Automaton Rover for Extreme Environments, Jonathan Sauder, 2017, NASA Innovative Advanced Concepts (NIAC) Phase I Final Report, California Institute of Technology, https://www.nasa.gov/sites/default/files/atoms/files/niac_2016_phasei_saunder_aree_tagged.pdf

[Savard 2012] A Space Habitat Design, John J. G. Savard, quadibloc web page, http://www.quadibloc.com/science/spaint.htm

[Schrunk et al. 2008] The Moon: Resources, Future Development and Settlement, Second Edition, David Schrunk, Burton Sharpe, Bonnie L. Cooper, and Madhu Thangavelu, Published: 4 October 2007, Springer Science & Business Media, https://epdf.pub/queue/moon-resources-future-development-and-settlement.html

[Sifakis 2018] Autonomous Systems – An Architectural Characterization, Joseph Sifakis, November 2018, arXiv:1811.10277, https://arxiv.org/ftp/arxiv/papers/1811/1811.10277.pdf

[Soilleux 2019] Orbital Civil Engineering: Waste Silicates Reformed into Radiation-shielded Pressure Hulls, Richard Soilleux, July 2019, Journal of the British Interplanetary Society (JBIS), Volume 72, No 7, https://www.bis-space.com/membership/jbis/2019/JBIS-v72-no07-July-2019%20-%20Subscription%20Copy.pdf

[Soilleux and Gunn 2018] Environmental Control and Life Support (ECLSS) for Large Orbital Habitats: Ventilation for Heat and Water Transport and Management, Richard J. Soilleux and Stephen D. Gunn, January 2018, NSS Space Settlement Journal, Issue 3, November 2017-September 2018, https://space.nss.org/wp-content/uploads/NSS-JOURNAL-ECLSS-for-Large-Orbital-Habitats-Ventilation-and-Heat-Transport.pdf

[SpaceX 2018] SpaceX, Space Exploration Technologies Corp, 2018, https://www.spacex.com/

[Stanley 2018] Biosphere 2 and Closed Ecological Systems, Systems Biology for Sustainable Life Outside Earth, Greg Stanley, 12 May 2018, Presentation to National Space Society North Houston, Nhttps://gregstanleyandassociates.com/successstories/Biosphere2/Bio2-r1.pdf

[Sterritt and Hinchey 2005] SPAACE:: Self-Properties for an Autonomous & Autonomic Computing Environment, Roy Sterritt and Michael G. Hinchey, June 2005, Proceedings of the International Conference on Software Engineering Research and Practice, SERP 2005, Las Vegas, Nevada, USA, https://www.researchgate.net/publication/221610408_SPAACE_Self-Properties_for_an_Autonomous_Autonomic_Computing_Environment

[Stryk 2008] Šteins Asteroid Image Credit, Ted Stryk, September 2008, Processing European Space Agency data, ESA 2008 MPS for OSIRIS Team, MPS/UPD/LAM/IAA/RSSD/INTA/UPM/DASP/IDA; https://commons.wikimedia.org/wiki/File:2867_%C5%A0teins_by_Rosetta_(reprocessed).png

[Tosaka 2008] Continuous Casting (Tundish and Mold), Wikipedia User: Tosaka, 25 July 2008, Kontinuerlig gjutning, Creative Commons Attribution 3.0 Unported (CC BY 3.0), Artwork Changes: Labels Added, https://en.wikipedia.org/wiki/Continuous_casting#/media/File:Continuous_casting_(Tundish_and_Mold)-2_NT.PNG

[Tsiolkovsky 1883] First space ship draft, Konstantin Tsiolkovsky, 1883, Manuscript "Free space" (Свободное пространство), https://en.wikipedia.org/wiki/File:Chertrg_Tsiolkovsky.jpg

[Tsiolkovsky 1920] Beyond the Planet Earth, Konstantin Tsiolkovsky, 1920, originally published under the title Вне Земли by the Kaluga Society for Natural History and Local Studies, Translated by Kenneth Syers, Foreword by B. N. Vorobyev, 1960, New York, Pergamon Press, https://archive.org/details/beyondplaneteart00tsio

[Vandenbos 2006] Asteroid mining, Jan Vandenbos, 2006, Slideshare, Published on Apr 26, 2012, https://www.slideshare.net/jvandenbos/asteroid-mining-2006

[Volpe 2018] 2018 Robotics Activities at JPL, Richard Volpe, Jet Propulsion Laboratory, 2018, California Institute of Technology, https://www-robotics.jpl.nasa.gov/media/documents/volpe%20paper,%20iSAIRAS%202018.pdf

[Volpe 2018a] The LEMUR Robots, Jet Propulsion Laboratory, Retrieved 2018-04-11, Curator: Richard Volpe, Webmaster: David Martin, https://www-robotics.jpl.nasa.gov/how-we-do-it/systems/the-lemur-robots/

[Volpe 2018b] The ATHLETE Rover. Jet Propulsion Laboratory, Retrieved 2018-04-11, Curator: Richard Volpe, Webmaster: Kristy Kawasaki, https://www-robotics.jpl.nasa.gov/systems/system.cfm?System=11

[von Neumann 1966] The Theory of Self-reproducing Automata, von Neumann, John, A. Burks (ed.). Urbana, IL: Univ. of Illinois Press. ISBN 978-0-598-37798-2, 1966, https://archive.org/details/theoryofselfrepr00vonn_0

[Wall 2017] Building Huge Structures in Space: 'Archinaut' Takes Another Step, Mike Wall, Posted August 14, 2017, Retrieved April 18, 2018, Space.com, https://www.space.com/37767-made-in-space-archinaut-vacuum-chamber-test.html

[Webster 2020] Asterank, Ian Webster, 2020, Asterank Database and Website, http://www.asterank.com/

[Wheeler 2017] Agriculture for Space: People and Places Paving the Way, Raymond M. Wheeler, 15 January 2017, Open Agriculture, 2017, Volume 2, pp. 14-32, 10.1515/opag-2017-0002, https://www.researchgate.net/publication/313732842_Agriculture_for_Space_People_and_Places_Paving_the_Way/fulltext/58a68b7a4585150402ee09ce/Agriculture-for-Space-People-and-Places-Paving-the-Way.pdf

[Williams 2016] Tools used by the pioneers to tame the wilderness, Doug Williams, 12 Sept 2016, Outdoor Revival, https://m.outdoorrevival.com/instant-articles/tools-used-by-the-pioneers-to-tame-the-wilderness.html

[Williams et al. 1979] Mining and Beneficiation of Lunar Ores, Richard J. Williams, David S. Mckay, David Giles, and Theodore E. Bunch, 1979, 1979, NASA SP-428 Space Resources and Space Settlements, pp.275-288, [O'Neill et al. 1979], https://ntrs.nasa.gov/archive/nasa/casi.ntrs.nasa.gov/19790024054.pdf

[Yale 2013] Non-brittle glass possible: In probing mysteries of glass, researchers find a key to toughness, Yale University, 26 February 2013, Science Daily, https://www.sciencedaily.com/releases/2013/02/130226114023.htm

# 8 License Types

[CC BY-4.0] Attribution 4.0 International (CC BY 4.0), Creative Commons, You are free to share and adapt for any purpose, even commercially. You can copy and redistribute the material in any medium or format. You can remix, transform, and build upon the material. You must give appropriate credit, provide a link to the license, and indicate if changes were made, https://creativecommons.org/licenses/by/4.0

[CC BY-NC-ND 4.0] Attribution-NonCommercial-NoDerivatives 4.0 International (CC BY-NC-ND 4.0), Creative Commons, You are free to copy and redistribute the material. You can copy and redistribute the material in any medium or format. You must give appropriate credit, provide a link to the license, and indicate if changes were made. You may not use for commercial purposes. You may not remix, transform, or build upon the material. https://creativecommons.org/licenses/by-nc-nd/4.0/

[CC BY-SA 2.0] Attribution-ShareAlike 2.0 Generic (CC BY-SA 2.0), Creative Commons, Free to Share and Adapt; You must give appropriate credit, provide a link to the license, and indicate if changes were made; You may adapt the material for any purpose, even commercially; ShareAlike: You must distribute your contributions under the same license as the original, https://creativecommons.org/licenses/by-sa/2.0/

[CC BY-SA 4.0] Attribution-ShareAlike 4.0 International (CC BY-SA 4.0), Creative Commons, Free to Share and Adapt; You must give appropriate credit, provide a link to the license, and indicate if changes were made; You may adapt the material for any purpose, even commercially; ShareAlike: You must distribute your contributions under the same license as the original, https://creativecommons.org/licenses/by-sa/4.0/

[ESA Standard License] Most images have been released publicly from ESA. You may use ESA images or videos for educational or informational purposes. The publicly released ESA images may be reproduced without fee, on the following conditions: Credit ESA as the source of the images. https://www.esa.int/ESA_Multimedia/Copyright_Notice_Images

[Facts] Facts and data are generally not eligible for copyright. "Charts, graphs, and tables are not subject to copyright protection because they do not meet the first requirement for copyright protection, that is, they are not "original works of authorship," under the definitions of 17 U.S.C. § 102(a). Essentially, that means that a graph, chart, or table that expresses data is treated the same as the underlying data. Facts, data, and the representations of those facts and data are excellent examples of things that require much "sweat of the brow" to create, but yet still do not receive copyright protection." https://deepblue.lib.umich.edu/bitstream/handle/2027.42/83329/copyrightability_of_tables_charts_and_graphs.pdf

[JPL Image Public Domain] JPL Image Use Policy, Unless otherwise noted, images and video on JPL public web sites (public sites ending with a jpl.nasa.gov address) may be used for any purpose without prior permission, subject to the special cases noted below. Special cases include JPL logo, identifiable person, or third party images and video materials, https://www.jpl.nasa.gov/jpl-image-use-policy

[NASA Image Public Domain] National Aeronautics and Space Administration, Accessed 21 November 2021, NASA content is generally not subject to copyright in the United States. You may use this material for educational or informational purposes, including photo collections, textbooks, public exhibits, computer graphical simulations and Internet Web pages, https://images.nasa.gov/, https://www.nasa.gov/multimedia/guidelines/index.html

[NASA Report Public Domain] NASA Scientific and Technical Information Program, Accessed 21 November 2021, Generally, United States government works (works prepared by officers and employees of the U.S. Government as part of their official duties) are not protected by copyright in the U.S. (17 U.S.C. §105) and may be used without obtaining permission from NASA, https://ntrs.nasa.gov/, https://sti.nasa.gov/disclaimers/

[Public Domain] The public domain consists of all the creative work to which no exclusive intellectual property rights apply. Those rights may have expired, been forfeited, expressly waived, or may be inapplicable.